\documentclass[a4paper,12pt,oneside]{book}

\usepackage[T1]{fontenc}
\usepackage[ansinew]{inputenc}
\usepackage{amsmath}
\usepackage{amssymb}
\usepackage{graphicx}

\pdfoutput=1

\usepackage{color}
\usepackage{euscript}
\usepackage[noautoscale]{youngtab}
\newlength{\defbaselineskip}
\newcommand{\setlinespacing}[1]%
           {\setlength{\baselineskip}{#1 \defbaselineskip}}

\newcommand{\schro}{Schr$\ddot{\textrm{o}}$dinger }
\setlinespacing{1.66}
\begin{document}
\thispagestyle{empty}
 \setlinespacing{1.66}
 \begin{center}
\vskip 2cm \Huge Finite Element Analysis of the \schro Equation\\

\large \vskip 1cm Avtar Singh Sehra
\\

\vskip 5cm
\begin{center}
\begin{minipage}{4cm}
\begin{center}
        \includegraphics[width=3cm]{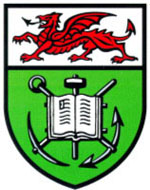}
\end{center}
\end{minipage}
\end{center}
\vskip 4cm
\small SUBMITTED TO THE UNIVERSITY OF WALES\\
IN FULFILMENT OF THE REQUIREMENTS OF\\
MASTERS IN COMPUTATIONAL RESEARCH\\
AT\\
SCHOOL OF ENGINEERING\\
UNIVERSITY OF WALES SWANSEA\\
SINGLETON PARK SWANSEA\\
SA2 8PP\\
\end{center}

\frontmatter \large
\chapter{Abstract}
 \normalsize \setlinespacing{1.66}

The purpose of this work is to test the application of the finite element method to quantum mechanical problems, in particular for solving the \schro equation.  

We begin with an overview of quantum mechanics, the \schro equation and numerical techniques used to solve quantum mechanical problems.  We note that one of the most important aspects of using the Crank-Nicolson method in solving the \schro equation is that the numerical time stepping equations are unitary, thus they inherently conserve probability, which is an important factor of quantum physics.   

We give an introduction to finite element analysis using the diffusion equation as an example.  We consider three numerical time evolution methods: the (tried and tested) Crank-Nicolson method, the continuous space-time method, and the discontinuous space-time method.  Once a numerical background is established we apply these techniques to quantum mechanical problems: a wave packet trapped in an infinite quantum well, and a wave packet trapped in an infinite well with a finite barrier.  

The first point of interest is that the finite element equations associated with the continuous and discontinuous space-time methods are not unitary. We show that the explicit part of the continuous method is unstable, and the implicit part is heavily damped, so both are as bad as using the forward or backward Euler methods.  However, it is also shown that when the implicit and explicit parts are combined by taking their average we obtain the Crank-Nicolson method, which is stable and unitary.  It is also shown that the discontinuous space-time method suffers from a small amount of damping, which can be controlled by the timestep size, however this comes at the cost of greater computation time.  From this we conclude that the standard Galerkin space-time methods are not as good as the Crank-Nicolson method.  

On the other hand the Crank-Nicolson method also has its limitations.  When a particle interacts with a barrier i.e. when a change of potential occurs, a fluctuation in probability conservation also occurs.  This is shown to happen due to the wave packet splitting into a reflected and transmitted part, which increases the complexity of the wave function.  It is shown that these fluctuations can be controlled by resolving the wave function more accurately, which is achieved by increasing the number of spatial elements used.  This is further explored when we show that slight "damping" takes place when we model a particle in a sinusoidal potential, which is a result of the packet undergoing a chain of  reflections and transmissions.  In this case a much finer resolution ($\geq 3000$ spatial elements) is required in order to accurately model the wave function at later times.
\thispagestyle{empty}

\vskip 3cm \hskip 7cm \large \textit{for my family and teachers.}

\normalsize

\thispagestyle{empty}
\begin{center}
\Large UNIVERSITY OF WALES SWANSEA
\end{center}
\vskip 1.2cm Author: \hskip 1.95cm \textbf{Avtar Singh Sehra}
\\
Title: \hskip 2.4cm \textbf{Finite Element Analysis of the} 

\hskip 2.95cm \textbf{\schro Equation}
\\
Department: \hskip 1.1cm \textbf{School of Engineering}
\\
Degree: \hskip 2.05cm \textbf{MRes}
\\
Year: \hskip 2.45cm \textbf{25 August 2006}
\\
\\
\\
\\
\normalsize \setlinespacing{1} This work has not previously been accepted in substance for any degree and is not being
concurrently submitted in candidature for any degree.

\begin{flushright}\includegraphics[width=6cm]{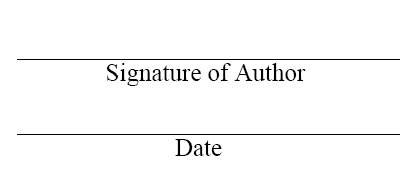}\end{flushright}
\vskip 0.8cm
This thesis is the result of my own investigations, except where otherwise stated.  Other sources are
acknowledged by explicit references.  A bibliography is appended.

\begin{flushright}\includegraphics[width=6cm]{sign}\end{flushright}
\vskip 0.8cm
I hereby give consent for my thesis, if accepted, to be available for photocopying and for inter-library loan,
and for the title and summary to be made available to outside organisations.

\begin{flushright}\includegraphics[width=6cm]{sign}\end{flushright}
\vskip 0.8cm \normalsize \setlinespacing{1.66}

\chapter{Acknowledgments}

There are many people to thank for their support and encouragement,
without whom this work would not have been possible. Firstly I want to thank my
supervisors, Prof. Djordje Peric and Dr. Wulf Dettmer, for their support, guidance, and stimulating discussions; but most of all for their broad technical insight and sense of scientific curiosity and adventure.  Also, thanks to my colleagues and friends in the Civil and Computational Engineering Center, with whom I spent many evenings in the local all-you-can-eat restaurants!  And a great big thanks to Aurora Trivini for her kindness and support -- particularly for all the hours she spent listening to me ramble on about everything from atoms to art.   

I would like to thank both Swansea School of Engineering and EPSRC for
providing resources and funding for my research throughout this work.
\\
\\
Finally thanks to my family -- I cannot show
enough gratitude for their support and encouragement, but more importantly for their infinite patience.

\setlinespacing{1.66}

 \tableofcontents

\mainmatter
{\typeout{Introduction}
\chapter{Introduction}
\label{introduction}
The development of quantum mechanics in the early part of the 20th century led to a greater understanding of the atomic and subatomic world.  On the nano-scale quantum theory is a more fundamental theory than Newtonian mechanics and classical electromagnetism as it provides a more accurate description of many phenomena which are never observed in the macroscopic world.  Everything from understanding the discrete nature of observable properties such as energy and momentum to the conceptual leap of modelling particles as waves has all had a profound effect on the way we see and use the microscopic world.

One of the biggest applications of quantum mechanics was in the mid 20th century, as it formed the framework for understanding and development of semiconductor materials and devices, such as transistors, diodes, solar-cells, lasers, and microprocessors.  However, the ongoing research in quantum mechanics and its application to solid-state physics has led to far more complex technologies, such as quantum dots and quantum wires.  These new technologies offer important opportunities as the building blocks for the next generation of electronic and opto-electronic devices ranging from ultra-fast optical switching to ultra-dense memories.  Some recent devices which inherently employ such quantum technologies are quantum-well lasers, which have led to the development of the compact blue laser used in high storage optical disc players like the Playstation 3, and high performance solar-cells.  With the huge potential of quantum wires and dots they are under immense active experimental and theoretical investigation.

In order to test new theories, devices and applications it has become more popular from the business point of view to construct computational models before any physical experimentation takes place.  Sometimes it is also useful to do practical and numerical tests side by side in order to compare results.  So as the development of more complex quantum devices continues we will require more efficient and accurate numerical techniques.    

In the past models such as drift-diffusion formed the basis for simulating semiconductor devices, but such techniques are not adequate to model the new breed of quantum devices where the quantum effects of a single electron can play a significant part in a device's operation. However, a concise quantum mechanical simulation of an entire semiconductor device is not feasible from the numerical point of view. So it has been stated that to model more complicated quantum devices we could use the fact that in many semiconductor devices quantum effects take place in a localized region (micro-structure), for example within the active zone of a quantum well laser, whereas the rest of the device (macro-structure) can be described by classical models \cite{wias}. Therefore, it would be possible to follow a strategy where we can couple quantum mechanical and macroscopic models (similar to multi-scale modelling in other areas of engineering).

An important aspect of quantum phenomena is the conservation of probability.  For this reason the promising time evolution numerical technique seems to be the Crank-Nicolson method, which is not only unconditionally stable but the time-stepping equations associated with it are unitary and thus inherently conserve probability\footnote{The unitary property and probability conservation will be discussed in more detail in the next chapter.}.  Then for the numerical solution of the \schro equation one can simply use a finite-difference method for space discretization and then apply the Crank-Nicolson method for time evolution.  In previous work, \cite{watanabe}, more sophisticated techniques such as the finite element method have been used for spatial discretization.  By using finite elements combined with high-order spatial discretisation we can then model the wave function of a particle more accurately.  Combining this system with the Crank-Nicolson time evolution method we can then model the time evolution of a wave function efficiently and accurately, rather than simply using a simple finite difference method \cite{ramdas}.  One step beyond this is the use of finite elements in space and time -- known as the space-time finite element method.  

The most important aspect of our work will be the comparison of the Crank-Nicolson and finite element method to the space-time finite element method.  We will work with simple micro-structure models such as particles (represented by wave-packets) in quantum wires (a one dimensional quantum well).  The Crank-Nicolson and space-time models will then be compared for their efficiency, conservation of probability and accuracy.
\\
In Chapter \ref{quantum mechanics} we will begin with an overview of quantum mechanics, the derivation of the \schro equation, and some simple numerical solution methods.  Chapter \ref{Finite Element} will form the basis of our work on the \schro equation.  We will begin with an introduction to the finite element method, then we will go on to deriving the element equations for the diffusion equation\footnote{Due to its similar structure to the \schro equation we are able to use the results in later chapters.} by first using the Crank-Nicolson method, then the continuous space-time method and finally the discontinuous space-time method.  In Chapter \ref{Dependent Schrodinger} we go on to model the \schro equation using the previously discussed methods.  We model simple systems such as infinite quantum wells, and a quantum well with a barrier.  We show that the Crank-Nicolson method is by far more efficient and accurate to the space-time method, however we also show the limitations of the Crank-Nicolson method.  In Chapter \ref{space varying potential} we go on to analyse the limitations of the Crank-Nicolson method by applying it to a packet trapped in an infinite well with a sinusoidal potential.  Finally in Chapter \ref{summary} we give a summary of the work.

\subsubsection{Note on Notation}
In this work we will be dealing with a large number of matrices and operators.  To keep things consistent we will use the following notation:
\\
\begin{center}
\begin{tabular}{cc}
$\textbf{N}$& Finite element basis operator.\\
$\textbf{B}$& Finite element derivative operator.\\
$\textbf{I}$& Diagonal identity matrix.\\
$\tilde{\textbf{X}}$ or $\tilde{\textbf{X}}'$&  Intermediate step in matrix $\textbf{X}$.\\
$\textbf{X}'$& Final real matrix form of $\textbf{X}$.\\
\end{tabular}
\end{center}}

{\typeout{Quantum Mechanics}
\chapter{Quantum Mechanics}
\label{quantum mechanics}

In general, quantum physics is concerned with processes which involve discrete energies and quanta (i.e. single particles such as the photon).  The motion and behaviour of quantum processes can be described by the \schro  equation.  The use of the \schro  equation to study quantum phenomena is known as \textit{Quantum Mechanics}, akin to classical mechanics being the tool to study classical physics.  In this chapter we will give a brief overview of quantum mechanics.  Beginning with the postulates of quantum mechanics, we will go on to discuss the derivation of the \schro equation and give some simple applications.  We will end this chapter with a description of the current numerical techniques used to solve the \schro equation.

\section{Postulates of Quantum Mechanics.}

\begin{itemize}
	\item \textbf{Postulate 1}
The state of a quantum mechanical system is completely specified by a function $\psi(\textbf{x},t)$, which depends on the space and time coordinates of the particle. This function, called the wave function or state function, has the important property that its norm $\psi^{*}(\textbf{x},t)\psi(\textbf{x},t)dv$ is the probability that the particle lies in the volume element $dv$ located at $\textbf{x}$ at time $t$.
The wavefunction must satisfy certain mathematical conditions because of this probabilistic interpretation. For the case of a single particle, the probability of finding it somewhere is 1.  So we have the normalization condition:
\begin{equation}
\int^{\infty}_{-\infty}	\psi^{*}(\textbf{x},t)\psi(\textbf{x},t)dv=1
\label{norm-equation}
\end{equation}
The wavefunction must also be single-valued, continuous, and finite.

	\item \textbf{Postulate 2}
To every observable, $A$, in classical mechanics (e.g. energy and momentum) there corresponds a linear Hermitian operator, $\hat{A}$, in quantum mechanics.

	\item \textbf{Postulate 3}
In any measurement of the observable associated with operator $\hat{A}$, the only values that will ever be observed are the eigenvalues $a$, which satisfy the eigenvalue equation:
\begin{equation}
\hat{A}\psi_a=a\psi_a
\end{equation}
where $\psi_a$ is the eigenfunction associated with the eigenvalue $a$ of the operator $\hat{A}$.
This postulate captures the central point of quantum mechanics that values of dynamical variables can be quantized. If the system is in an eigenstate of $\hat{A}$ with eigenvalue $a$, then any measurement of the quantity $A$ will yield $a$.
Although measurements must always yield an eigenvalue, the state does not have to be an eigenstate of $\hat{A}$.  An arbitrary state can be expanded in the complete set of eigenvectors of $\hat{A}$ ($ \hat{A}\psi_i=a_i\psi_i$) as:
\begin{equation}
\psi=\sum^n_i c_i \psi_i
\end{equation} 
In this case we only know that the measurement of $A$ will yield one of the values $a_i$ with a probability $|c_i|^2$ 

	\item \textbf{Postulate 4}
If a system is in a state described by a normalized wave function $\psi$, then the average value of the observable corresponding to $\hat{A}$ is given by:
\begin{equation}
<A>=\int^{\infty}_{-\infty}	\psi^{*}(\textbf{x},t)\hat{A}\psi(\textbf{x},t)dv
\end{equation}

	\item \textbf{Postulate 5}
The wavefunction of a system evolves in time according to the time-dependent \schro equation:
\begin{equation}
i \hbar \frac{\partial}{\partial t}\psi(\textbf{x},t)=-\frac{\hbar^2}{2 m}\nabla^2 \psi(\textbf{x},t) +V(\textbf{x})\psi(\textbf{x},t)
\label{schrodinger}
\end{equation}
where $m$ is the mass of the particle, $\hbar=\frac{h}{2\pi}$ is the Planck constant and $V(\textbf{x})$ is a real function representing the potential energy of the system.  Although the time-independent \schro equation can be derived through elementary methods (discussed in the next section), the time-dependent version can not be derived so must be accepted as a fundamental postulate of quantum mechanics. 
\end{itemize}

\section{The \schro Equation}
In 1925 Erwin \schro developed a method of quantum mechanics involving partial differential equations.  This method differed to the one developed earlier by Werner Heisenberg which employed matrices.  These differential and matrix based methods were later shown to be mathematically equivalent\cite{wiki-quantum}.

\subsection{Time-Independent \schro Equation}
One of the fundamental concepts of quantum physics is that of wave-particle duality: that is waves can behave like particles and particles like waves.  For example, Einstein showed that a photon, which is considered to be a wave packet, has momentum just like a particle moving with the same energy, Appendix~\ref{background}.
The dynamical behaviour of these quantum waves/particles can be described in a non-relativistic\footnote{For the relativistic description of particles and waves we require the Dirac equation for spin $\frac{1}{2}$ particles, the Klein Gordon equation for spin $0$ particles.  This is all encompassed more generally in the study of Quantum Field Theory.} manner through the use of wave mechanics.  The single-particle three-dimensional time-dependent \schro equation is given in Eqn.~(\ref{schrodinger}).  Before we consider the full time-dependent equation, which must be accepted as a postulate of Quantum Mechanics, we will give a brief derivation of the time-independent version, which has a conceptual derivation linked to the wave equation.
\subsubsection{Derivation of the Time-Independent \schro Equation}
Starting with the one-dimensional classical wave equation,
\begin{equation}
\frac{\partial^2 u}{\partial x^2}	= \frac{1}{v^2} \frac{\partial^2 u}{\partial t^2},	
\end{equation}
and using separation of variables, 
\begin{equation}
u(x,t)=\psi(x)f(t),
\end{equation}
we obtain
\begin{equation}
f(t) \frac{d^2 }{d x^2} \psi(x)	= \frac{1}{v^2} \psi(x) \frac{d^2 }{d t^2}f(t)	
\end{equation}
Then, using a standard solution of the wave equation, $f(t)=e^{i \omega t}$, we obtain
\begin{equation}
\frac{d^2 }{d x^2} \psi(x)	= -\frac{\omega^2}{v^2}\psi(x)
\label{time-independent}
\end{equation}
This gives an ordinary differential equation describing the spatial amplitude of the matter wave as a function of position.  This can be put in the standard form for the \schro equation by using the fact that the energy of a particle is the sum of kinetic and potential parts,
\begin{equation}
E=\frac{p^2}{2 m} +V(x),
\end{equation}
Finally, using $\omega=2\pi\nu$, $v=\nu\lambda$, and $h=p\lambda$ we have
\begin{equation}
\frac{\omega^2}{v^2}=\frac{4\pi^2\nu^2}{v^2}=\frac{4\pi^2}{\lambda^2}=\frac{2m[E-V(x)]}{\hbar^2}
\end{equation}
which when combined with Eqn.~(\ref{time-independent}) gives
\begin{equation}
\frac{d^2 }{d x^2} \psi(x)+ \frac{2m}{\hbar^2}[E-V(x)]\psi(x)=0
\label{time-independent-schrodinger}
\end{equation}
This single-particle one-dimensional equation can be extended to the case of three dimensions, where after rearranging it becomes
\begin{equation}
-\frac{\hbar^2}{2m}\nabla^2 \psi(\textbf{x})+ V(\textbf{x})\psi(\textbf{x})=E\psi(\textbf{x})
\label{time-independent-schrodinger-final}
\end{equation}
The solutions to this equation then represent the state function of a particle of mass $m$ in a potential $V(\textbf{x})$.

\subsection{Time-Dependent \schro Equation}
As stated in the previous section, although the time-independent \schro equation can be derived analytically, the time-dependent \schro equation cannot be derived using such methods and is therefore generally considered as a postulate of quantum mechanics \cite{liboff}.  However, we are able to show that the time-dependent equation is a reasonable model of the dynamic evolution of a particle's states function even though it is not derivable.  As before, using separation of variables, 
	\[
	\psi(\textbf{x},t)=\psi(\textbf{x}) f(t),
\]
and substituting this into Eqn.~(\ref{schrodinger}) we have
\begin{equation}
\frac{i\hbar}{f(t)}\frac{df}{dt}=\frac{1}{\psi(\textbf{x})}\left[-\frac{\hbar^2}{2m} \nabla^2+V(\textbf{x})\right]\psi(\textbf{x})	
\end{equation}
Now, as the left-hand side is a function of $t$ only and the right hand side is a function of $\textbf{x}$ only, the two sides must be equal to a constant. Assigning this constant as $E$, as the right-hand side clearly has dimensions of energy, we can then extract two ordinary differential equations:
\begin{equation}
\frac{1}{f(t)}\frac{d f(t)}{dt}=-\frac{i E}{\hbar}
\label{shro-time}
\end{equation}
and where the other is the time-independent \schro equation, Eqn.~(\ref{time-independent-schrodinger-final}).  Simply solving Eqn.~(\ref{shro-time}) we have
\begin{equation}
f(t)=e^{-iEt/\hbar}
\end{equation}
The energy operator, given by Eqn.~(\ref{time-independent-schrodinger-final}), known as the Hamiltonian is a Hermitian operator, therefore its eigenvalues are real, so $E$ is real. This means that the solutions of Eqn.~(\ref{shro-time}) are purely oscillatory.  Therefore, if
\begin{equation}
\psi(\textbf{x},t)=\psi(\textbf{x})e^{-iEt/\hbar},
\label{stationary}
\end{equation}
then the total wave function $\psi(\textbf{x},t)$ differs from $\psi(\textbf{x})$ only by a phase factor of constant magnitude.  This then implies that the probability, or the norm, of the particle state is time independent,
\begin{equation}
|\psi(\textbf{x},t)|^2=\psi^*(\textbf{x},t)\psi(\textbf{x},t)= e^{iEt/\hbar}\psi^*(\textbf{x})e^{-iEt/\hbar}\psi(\textbf{x})=\psi^*(\textbf{x})\psi(\textbf{x})=|\psi(\textbf{x})|^2
\end{equation}
It also implies that the expectation value for any time-independent operator is also time-independent,
 \begin{equation}
<A>=\int^{\infty}_{-\infty}	\psi^{*}(\textbf{x},t)\hat{A}\psi(\textbf{x},t)dv=\int^{\infty}_{-\infty}	\psi^{*}(\textbf{x})\hat{A}\psi(\textbf{x})dv
\end{equation}
For this reason the states described by the wavefunction in Eqn.~(\ref{stationary}) are called stationary states.  However, even though the probablity distribution described by $\psi(\textbf{x},t)$ is stationary, the particle it describes is not.  This could be conceptually understood by having a particle in a box.  In such a case the particle will be moving around in the box: bouncing off the walls etc.  However, the probability distribution of the particle within the box will be constant in time.  Thus, if the probability is $0.5$ in the middle of the box, and $0$ at the box edges, this implies that if we check for the particle in 100 identical boxes we will find it in the middle in 50 of them.

\section{Analytical Solutions}
\subsection{Particle in a Box}
As a simple example we consider a particle constrained to move in a single dimension under the influence of a potential $V(x)$ which is zero for $0\leq x \leq L$ and infinite elsewhere, Fig.~\ref{Infinitewell}.
\begin{figure}
    \begin{center}
        \includegraphics[width=0.6\textwidth]{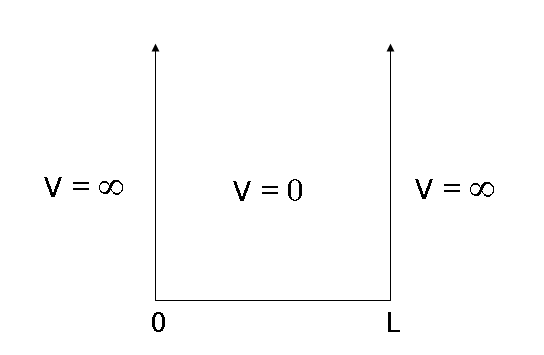}\\
        \caption{Infinite Square well.}
        \label{Infinitewell}
    \end{center}
\end{figure}
Since the wavefunction is not allowed to become infinite, it must have a value of zero where $V(x)$ is infinite i.e. $\psi(0)=\psi(L)=0$, so $\psi(x)$ is nonzero only within $[0,L]$. The Schrödinger equation for this simple case is
\begin{equation}
-\frac{\hbar^2}{2m}\frac{d^2 \psi(x)}{d x^2}=E\psi(x) \qquad \qquad 0\leq x \leq L
\end{equation}
Solving this and applying the normalization condition, $|\psi(x)|^2=1$, we obtain the eigenfunctions
\begin{equation}
\psi_n(x)=\sqrt{\frac{2}{L}}Sin \left(\frac{n\pi x}{L}\right) \qquad \qquad n=1,2,3,...
\end{equation}
and the corresponding eigenvalues
\begin{equation}
E_n=\frac{\hbar^2 \pi^2 n^2}{2 L^2  m} \qquad \qquad n=1,2,3,...
\end{equation}
\begin{figure}
    \begin{center}
        \includegraphics[width=0.6\textwidth]{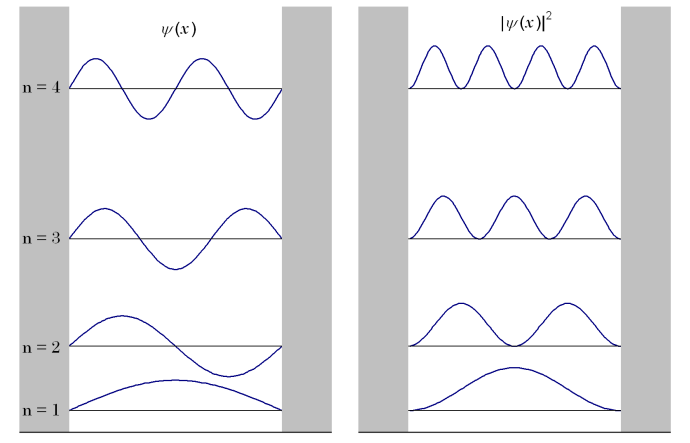}\\
        \caption{Particle in a box wavefunctions.}
        \label{box}
    \end{center}
\end{figure}
\subsection{Harmonic Oscillator }
We can now consider a particle in a classic spring like potential,
\begin{equation}
V(x)=\frac{1}{2}k x^2.
\end{equation}
The time-independent \schro equation with this potential is 
 \begin{equation}
-\frac{\hbar^2}{2m}\frac{d^2 \psi(x)}{d x^2}+\frac{1}{2}k x^2\psi(x)=E\psi(x),
\label{harmonic}
\end{equation}
we can note that if a reduced mass $\mu=\frac{m_1 m_2}{m_1 + m_2}$ is used we can model the behaviour of a chemical bond between two atoms of mass $m_1$ and $m_2$.  
A simple solution to this \schro equation is given by the fact that as  the derivative of the wavefunction must give back the square of $x$ plus a constant times the original function, the solution takes the form
\begin{equation}
\psi(x)=Ce^{-\alpha x^2/2}
\end{equation}
However, the most general normalized form of the solution is
\begin{equation}
\psi_n(x)=\frac{\alpha}{\pi}^{1/4}\frac{1}{\sqrt{2^n n!}} H_n(y)e^{- y^2/2} \qquad \qquad n=1,2,3,...
\end{equation}
with the energy eigenvalues
\begin{equation}
E_n=\hbar\omega(n+1/2),
\end{equation}
where $y=\sqrt{\alpha}x$, $\alpha=\frac{m\omega}{\hbar}$, and $H_n(y)$ are the Hermite polynomials given in Fig.~\ref{hermite}.
\begin{figure}
    \begin{center}
        \includegraphics[width=0.5\textwidth]{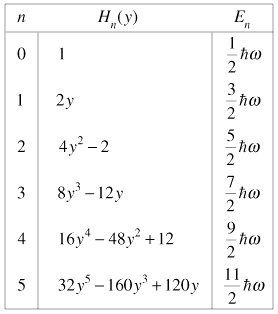}\\
        \caption{Hermite polynomials and the corresponding energy eigenstates.}
        \label{hermite}
    \end{center}
\end{figure}
These quantum harmonic oscillator states are shown in Fig.~\ref{sho_psi}.
\begin{figure}
    \begin{center}
        \includegraphics[width=0.6\textwidth]{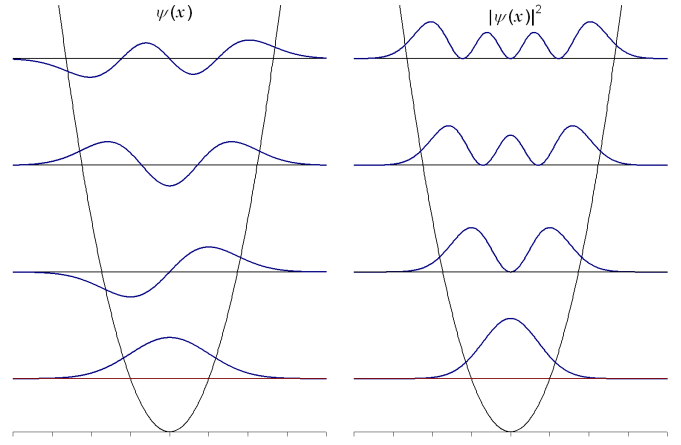}\\
        \caption{Quantum harmonic oscillator eigenstates.}
        \label{sho_psi}
    \end{center}
\end{figure}

\section{Finite-Difference Discretization}
\subsection{Time Independent Problems}
In the case for complicated potential fields, $V(x)$ and particle scattering models the numerical finite difference method has been used for many years to solve the \schro equation \cite{chen}.  For the time-independent case we can simply discretise the \schro equation and put it into matrix form, which can then be numerically solved.  For the one dimensional case, and ignoring the potential, the \schro equation at each point along $x$ can be written as
\begin{equation}
E\psi_{x_n}=-\frac{\hbar^2}{2m}\left(\frac{d^2\psi}{d x^2} \right)_{x_n}
\label{fin-schro}
\end{equation}
Now using the basic finite-difference approximation,
\begin{equation}
\frac{d^2\psi(x)}{dx^2}_{x=x_n}=\frac{\psi_{x_{n+1}}-2\psi_{x_n}+\psi_{x_{n-1}}}{a^2},
\label{space-difference}
\end{equation}
where $a$ is the spatial interval spacing, we can write Eqn.~(\ref{fin-schro}) as
\begin{equation}
E\psi_{x_n}=k\left(2\psi_{x_n}-\psi_{x_{n+1}}-\psi_{x_{n-1}}\right),
\end{equation}
where $k=\frac{\hbar^2}{2ma^2}$.  This can now be written in matrix form as
\begin{equation}
E\left(\begin{array}{c}
\psi_1\\
\psi_2\\
\psi_3\\
\vdots\\
\psi_{N-1}\\
\psi_N\\
\end{array}
\right) =\left(
\begin{array}{cccccc}
	2k & -k & 0 & 0 & 0 & \cdots \\
	-k & 2k & -k & 0 & 0 & \cdots \\
	0 & -k & 2k & -k  & 0 & \cdots \\
  \vdots &  &   & \ddots &  &  \vdots \\
  \cdots & 0 & 0 & -k & 2k & -k \\
  \cdots & 0& 0& 0 & -k & 2k\\
\end{array}
\right)
\left( \begin{array}{c}
\psi_1\\
\psi_2\\
\psi_3\\
\vdots\\
\psi_{N-1}\\
\psi_N\\
\end{array}
\right)
\end{equation}
This can also be written in operator form as
\begin{equation}
E\textbf{I}\bar{\psi}=\textbf{H}\bar{\psi},
\end{equation}
where $\textbf{I}$ is the identity matrix.  This eigenvalue problem can be solved numerically, and the corresponding eigenvectors, which represent the eigenstate of the particle, can be found.
In order to implement a potential, $V(x)$, we can simply add a diagonal matrix $\textbf{V}$ to $\textbf{H}$, where the diagonal components of $\textbf{V}$ are equal to the potential at the nodes:  $V_{nn}=V(x_n)$\footnote{In finite element analysis, rather than taking the nodal values of the potential, the average over the element is taken.}.
 
\subsection{Time-Dependent Problems}
\subsubsection{Explicit Method}
The finite-difference discretization of the time-dependent \schro equation can be simply done using the explicit method.  As before we can discretise the spatial part of Eqn.~(\ref{schrodinger}) using the approximation in Eqn.~(\ref{space-difference}).  Then applying the explicit time-difference approximation,
\begin{equation}
\left.\frac{d\psi(x,t)}{dt} \right|_{t=t_j,x=x_n}=\frac{\psi_{x_n}^{t_{j+1}}-\psi_{x_n}^{t_{j}}}{b},
\label{time-difference}
\end{equation}
where $b$ is the temporal interval spacing, we are able to construct the explicit finite-difference approximation to the \schro equation:
\begin{equation}
\psi_{x_n}^{t_{j+1}}=\psi_{x_n}^{t_j}-\frac{ib}{\hbar}\left[ -\frac{\hbar^2}{2m a^2}\left(\psi_{x_{n+1}}^{t_j}-2\psi_{x_n}^{t_j}+\psi_{x_{n-1}}^{t_j} \right) + V_{x_n}\psi_{x_n}^{t_j} \right]
\end{equation}
In operator form this can be written
\begin{equation}
\bar{\psi}^{t_{j+1}}=\left(\textbf{I} -\frac{i}{\hbar} b \textbf{H}\right)\bar{\psi}^{t_j},
\end{equation}
where as before $\textbf{H}$ is the discretized Hamiltonian (with the potential matrix $\textbf{V}$ absorbed) and $\textbf{I}$ is the unit matrix.  The problem with this approach is that it is numerically unstable and also, more importantly, the operator $\textbf{I} -\frac{i}{\hbar} b \textbf{H}$ is not unitary, which is a required property in order to conserve probability, i.e. $\int\psi^*\psi dx=1$.

\subsubsection{Implicit Methods}
Now conducting an implicit discretization we have
\begin{equation}
\psi_{x_n}^{t_j}=\psi_{x_n}^{t_{j+1}}+\frac{ib}{\hbar}\left[ -\frac{\hbar^2}{2m a^2}\left(\psi_{x_{n+1}}^{t_{j+1}}-2\psi_{x_n}^{t_{j+1}}+\psi_{x_{n-1}}^{t_{j+1}} \right) + V_{x_n}\psi_{x_n}^{t_{j+1}} \right].
\end{equation}
Which can also be put into operator form as
\begin{equation}
\bar{\psi}^{t_{j+1}}=\left(\textbf{I} +\frac{i}{\hbar} b \textbf{H}\right)^{-1}\bar{\psi}^{t_{j}},
\end{equation}
Even though this numerical solution is stable it still does not correspond to a unitary transformation, and thus leads to unphysical quantum results.

\subsubsection{Cayley's Form}
A numerical finite-difference technique that produces a stable and unitary discretized operator is called the Cayley's Form.  For this we use a centered-time-difference or Crank-Nicolson Scheme to construct the temporal discretization:
\begin{eqnarray}
i\hbar\frac{\psi_{x_n}^{t_{j+1}}-\psi_{x_n}^{t_j}}{b}=&\frac{1}{2}&\left\{\left[-\frac{\hbar^2}{2m}\frac{d^2\psi(x,t)}{dx^2}+V(x)\psi(x,t)\right]^{t_{j+1}}_n\right.\nonumber\\
&+&\left.\left[-\frac{\hbar^2}{2m}\frac{d^2\psi(x,t)}{dx^2}+V(x)\psi(x,t)\right]^{t_j}_n\right\}
\end{eqnarray}
After spatial discretization we have
\begin{eqnarray}
i\hbar\frac{\psi_{x_n}^{t_{j+1}}-\psi_{x_n}^{t_j}}{b}=&\frac{1}{2}&\left\{\left[-\frac{\hbar^2}{2m a^2}\left(\psi_{x_{n+1}}^{t_{j+1}}-2\psi_{x_n}^{t_{j+1}}+\psi_{x_{n-1}}^{t_{j+1}} \right) + V_{x_n}\psi_{x_n}^{t_{j+1}} \right]\right.\nonumber\\
&+&\left.\left[-\frac{\hbar^2}{2m a^2}\left(\psi_{x_{n+1}}^{t_j}-2\psi_{x_n}^{t_j}+\psi_{x_{n-1}}^{t_j} \right) + V_{x_n}\psi_{x_n}^{t_j} \right]\right\}\nonumber\\
\end{eqnarray}
After rearranging we obtain
\begin{eqnarray}
& &\psi^{t_{j+1}}_{x_n}+if_{x_n}\psi^{t_{j+1}}_{x_n}+2ig\psi^{t_{j+1}}_{x_n}-ig\psi^{t_{j+1}}_{x_{n+1}}-ig\psi^{t_{j+1}}_{x_{n-1}}\nonumber\\
&=&\psi^{t_{j}}_{x_n}-if_{x_n}\psi^{t_{j}}_{x_n}-2ig\psi^{t_{j}}_{x_n}+ig\psi^{t_{j}}_{x_{n+1}}+ig\psi^{t_{j}}_{x_{n-1}},
\end{eqnarray}
where $f_{x_n}=\frac{b}{2\hbar}V_{x_n}$ and $g=\frac{b \hbar}{4 m a^2}$.  Simplifying further we have
\begin{eqnarray}
& &(1+if_{x_n}+2ig)\psi^{t_{j+1}}_{x_n}-ig\psi^{t_{j+1}}_{x_{n+1}}-ig\psi^{t_{j+1}}_{x_{n-1}}\nonumber\\
&=&(1-if_{x_n}-2ig)\psi^{t_{j}}_{x_n}+ig\psi^{t_{j}}_{x_{n+1}}+ig\psi^{t_{j}}_{x_{n-1}},
\end{eqnarray}
This can then be put in matrix form as
\begin{equation}
\left(\textbf{I}+i\textbf{H}\right)
\left( \begin{array}{c}
\psi_1\\
\psi_2\\
\psi_3\\
\vdots\\
\psi_{N-1}\\
\psi_N\\
\end{array}
\right)^{t_{j+1}}
=\left(\textbf{I}-i
\textbf{H}\right)\left( \begin{array}{c}
\psi_1\\
\psi_2\\
\psi_3\\
\vdots\\
\psi_{N-1}\\
\psi_N\\
\end{array}
\right)^{t_{j}}
\label{fd-schro}.
\end{equation}
As before $\textbf{I}$ is the unit matrix, but now \textbf{H} is given by
\begin{equation}
\textbf{H}=\left(
\begin{array}{cccccc}
	f_1+2g & -g & 0 & 0 & 0 & \cdots \\
	-g & f_2+2g & -g & 0 & 0 & \cdots \\
	0 & -g & f_3+2gk & -g  & 0 & \cdots \\
  \vdots & &  & \ddots & &  \vdots \\
  \cdots & 0 & 0 & -g & f_{N-1}+2g & -g \\
  \cdots & 0& 0& 0 & -g & f_N+2g\\
\end{array}
\right)
\end{equation}
So we have the numerical difference equation in the Cayley's form:
\begin{equation}
\bar{\psi}^{t_{j+1}}=\frac{\textbf{I}-i\textbf{H}}{\textbf{I}+i\textbf{H}}\bar{\psi}^{t_{j}}
\label{discret-shro}
\end{equation}
The temporal operator that relates $\psi^j$ to $\psi^{j+1}$ is now not only numerically stable but also unitary; this can simply be shown as
\begin{equation}
\left(\frac{\textbf{I}-i\textbf{H}}{\textbf{I}+i\textbf{H}}\right)^{*} \left(\frac{\textbf{I}-i\textbf{H}}{\textbf{I}+i\textbf{H}}\right) =\left(\frac{\textbf{I}+i\textbf{H}}{\textbf{I}-i\textbf{H}}\right)
\left(\frac{\textbf{I}-i\textbf{H}}{\textbf{I}+i\textbf{H}}\right)=\textbf{I}.
\end{equation}
Through this unitary property Eqn.~(\ref{discret-shro}) then satisfies conservation of probability as required,
\begin{equation}
\sum_{x_n} \bar{\psi}^{t_{j+1}*} \bar{\psi}^{t_{j+1}} a=\sum_{x_n} \bar{\psi}^{t_j*} \bar{\psi}^{t_j} a=1.
\end{equation}
where $a$ is, as before, the spatial interval spacing $\Delta x$.\footnote{As the spatial axis is discretized we are using summation instead of integration, however this is equivalent to the continuum equation given in Eqn.~(\ref{norm-equation}).}}

{\typeout{Space-Time Finite Element Analysis}
\chapter{Finite Element Analysis}
\label{Finite Element}
Development of the Finite Element Method (FEM) can be traced back to the 1940's.  However, it wasn't until the late 1950's and 1960's that it emerged as a useful tool in engineering.  Then, when a rigorous mathematical foundation was developed in the early 1970's it became a dominant method in applied mathematics for numerical modelling of physical systems in many engineering and scientific disciplines, e.g. electromagnetic and fluid dynamics as well as civil and aeronautical engineering \cite{wiki-finite}.  Olek Zienkiewicz, from University of Wales Swansea, originally an expert in finite difference methods (FDM) was one of the pioneers in bringing FEM to the wider scientific and engineering community through the first book on the subject \cite{zink}.

Even though FEM is a little more complicated to implement compared to FDM, one of its biggest advantages is its ability to handle complicated geometries (and boundaries) with relative ease.  However, even though handling complex geometries in FEM is theoretically straight forward, the problem of computational time is strongly influenced by the ability to precondition the problem i.e. by choosing the most appropriate element type for the most efficient computational performance.  

As a sideline we can also note that FDM is a subset of the FEM approach. This can be seen through choosing basis (shape) functions as either piecewise constant or Dirac delta function; then the stiffness matrix $K$ can be interpreted as a difference operator \cite{brenner}.  Then, by using a uniform mesh the FE equations reduce to FD equations.

There are two specific techniques for the application of FEM to a problem, the variational and Galerkin.  The variational approach requires a FE discretization of the functional associated with the problem (or, if it can be defined, the Lagrangian of a system).  The discretization is done in the standard way using basis functions for each element of the domain considered.  Then, by minimising the discretized functional and assembling the system for all the individual elements we are able to obtain the required FE equation of the system.  This is a powerful method as it takes into consideration the physics of a system\footnote{Lagrangian of a system for conservative systems or virtual work for the general case} in order to simplify and solve the problem.

As opposed to the variational method the Galerkin approach is directly applied to the differential equations of the problem, and then the equation can be discretized and assembled in order to obtain the FE equations of the system.  In this work we will use the Galerkin method as this eliminates the work of finding the functional associated with the problem.

In the rest of this chapter we will lay the foundation for the application of FEM to the types of problems we will encounter in the quantum context.  Beginning with a simple example of a one-dimensional eigenvalue problem, we will then go on to discuss time-dependent problems, i.e. the use of a combination of FEM, to solve for the spatial part, and FDM, to solve for the temporal part.  Finally, we will give a brief introduction to the use of space-time FEM to solve spatial and temporal parts of a problem together.  

\section{Eigenvalue Problems}
\label{eigen-value}
Even though in this work we will be dealing solely with time-dependent problems we will never the less include a brief summary of the general eigenvalue problem and the use of FEM for their solution.  In quantum mechanics the eigenvalue problem is one of the most important aspects: in order to determine energy levels and the associated eigen-functions.  For the case of complicated potentials, as in irregular lattices, and in quantum dots, numerical computation of eigenfunctions and eigenvalues is of great importance.

\subsection{FEM for Eigenvalue Problems}
\label{eigen-value-classic}
To lay down the method of the Galerkin approach to eigenvalue FE problems we will consider the case of torsional vibrations of a uniform circular-cross-section \cite{bick}.  The differential equations and boundary conditions required to determine the mode shapes and natural frequencies are
\begin{eqnarray}
JG\frac{d^2\psi}{dx^2}+\omega^2\rho J \psi&=&0\nonumber\\
\psi(0)&=&0\nonumber\\
\psi'(L)&=&0.
\end{eqnarray}
Rearranging this we can put it into an eigen-value form
\begin{equation}
\frac{d^2\psi}{dx^2}+\gamma\psi=0,
\label{torsion-eigen}
\end{equation}
where $\gamma=\frac{\omega^2 \rho}{G}$.  To apply the Galerkin method we multiply Eqn.~(\ref{torsion-eigen}) by a test function $\phi$ and integrate it by parts,
\begin{eqnarray}
\int^L_0\{\frac{d^2\psi}{dx^2}+\gamma\psi \}\phi dx&=&\nonumber\\
\int^L_0 \frac{d}{dx}\left\{\frac{d\psi}{dx}\phi\right\} dx -\int^L_0 \frac{d\phi}{dx}\frac{d\psi}{dx} dx + \int^L_0\gamma\psi\phi dx &=&\nonumber\\
\underbrace{\left. \frac{d\psi}{dx}\phi\right|^L_0}_{=0}  -\int^L_0 \frac{d\phi}{dx}\frac{d\psi}{dx} dx + \int^L_0\gamma\psi\phi dx &=&0
\end{eqnarray}
Therefore, once we eliminate the first term on the LHS (due to boundary conditions) we obtain 
\begin{equation}
-\int^L_0 \frac{d\psi}{dx}\frac{d\phi}{dx}dx+\int^L_0\gamma\psi\phi dx =0
\label{torque-equation}
\end{equation}
The next step is to implement a FE approximation using a set of basis functions, $N_i$:
\begin{equation}
\psi=\sum_i\psi_i N_i=\textbf{N}\bar{\psi} \qquad \qquad \phi=\sum_i\phi_i N_i=\textbf{N}\bar{\phi}
\label{fe-function}
\end{equation}
where the basis operator, $\textbf{N}$, in one dimension is given as
\begin{equation}
\textbf{N}=[N_1, N_2] \qquad N_1=\frac{1-\xi}{2} \qquad N_2=\frac{1+\xi}{2} 
\end{equation}
The coordinate transformation is then simply given as
\begin{equation}
x=\sum_i x_i N_i.
\end{equation}
Using this information we can write the derivative transformation as\footnote{In this 1D case the derivative transformation is very basic, however when we go on to work in 2D (for space-time FEM) we will require the more complicated 2D coordinate Jacobian.}
\begin{eqnarray}
\frac{dx}{d\xi}&=&\frac{x_2-x_1}{2}=\frac{l_e}{2} \\
\frac{d\psi}{d\xi}&=&\frac{d\psi}{dx}\frac{dx}{d\xi} \qquad \Rightarrow \qquad \frac{d\psi}{d\xi}=\frac{2}{l_e}\frac{d\psi}{dx}\nonumber\\
\therefore \qquad \frac{d}{dx}&=&\frac{2}{l_e}\frac{d}{d\xi}.
\label{dbydxeq2overle}
\end{eqnarray}
Or in matrix notation we can write
\begin{eqnarray}
\frac{d\psi}{d\xi}&=&-\frac{\psi_1}{2}+\frac{\psi_2}{2}=\frac{1}{2}\left[-1 \quad 1\right]\left[ \begin{array}{c}
\psi_1\\
\psi_2\\
\end{array}
\right]\nonumber\\
\therefore \qquad \frac{d}{d\xi}&=&\frac{1}{2}\left[-1 \quad 1\right]
\end{eqnarray}
Combining this with Eqn.~(\ref{dbydxeq2overle}) we have
\begin{equation}
\frac{d}{dx}=\frac{1}{l_e}\left[ -1 \quad 1 \right]=\textbf{B}.
\end{equation}
This $\textbf{B}$ operator then gives the FE approximation of the derivative of a function $\psi$:
\begin{equation}
\frac{d\psi}{dx}=\frac{1}{l_e}\left[-1 \quad 1\right]\bar{\psi}=\textbf{B}\bar{\psi},
\label{B-operator}
\end{equation}
where $\bar{\psi}$ is the FE approximation vector of the continuous function $\psi$.
Now, using Eqns.~(\ref{B-operator}) and (\ref{fe-function}) in (\ref{torque-equation}) we obtain
\begin{eqnarray}
-\sum_e\int^1_{-1} \bar{\phi}^{\dagger} \textbf{B}^{\dagger} \textbf{B}\bar{\psi}\frac{l_e}{2}d\xi+\sum_e\frac{l_e \gamma}{2}\int^1_{-1}\bar{\phi}^{\dagger}\textbf{N}^{\dagger}\textbf{N}\bar{\psi}d\xi&=&0\\
\nonumber\\
-\frac{1}{2 l_e}\sum_e\int^1_{-1} \bar{\phi}^{\dagger} \left[ \begin{array}{c}
-1\\
1\\
\end{array}
\right] \left[ -1 \quad 1 \right]\bar{\psi}d\xi&+&\nonumber\\
\sum_e\frac{l_e \gamma}{2}\int^1_{-1}\bar{\phi}^{\dagger} \left[ \begin{array}{c}
N_1\\
N_2\\
\end{array}
\right] \left[ N_1 \quad N_2 \right]\bar{\psi}d\xi&=&0\nonumber\\
\end{eqnarray} 
where the sum is taken over all the elements.  The first term can be simplified as
\begin{equation}
-\sum_e \frac{1}{l_e}\bar{\phi^{\dagger}}
\left[
\begin{array}{cc}
	1 & -1 \\
	-1 & 1 \\
\end{array}
\right]
\bar{\psi}
\end{equation}
and the second term, after a little more computation, can be written as
\begin{eqnarray}
& &\sum_e\frac{l_e \gamma}{2}\int^1_{-1}\bar{\phi}^{\dagger} \left[ \begin{array}{c}
N_1\\
N_2\\
\end{array}
\right] \left[ N_1 \quad N_2 \right]\bar{\psi}d\xi\nonumber\\
&=& \sum_e\frac{l_e \gamma}{2}\int^1_{-1}\bar{\phi}^{\dagger} 
\left[
\begin{array}{cc}
	N_1N_1 & N_1N_2 \\
	N_2N_1 & N_2N_2 \\
\end{array}
\right]
\bar{\psi}d\xi\nonumber\\
&=& \sum_e\frac{l_e \gamma}{6}\bar{\phi}^{\dagger} 
\left[
\begin{array}{cc}
	2 & 1 \\
	1 & 2 \\
\end{array}
\right]
\bar{\psi}
\end{eqnarray}
The total FE model is then
\begin{equation}
-\sum_e \bar{\phi^{\dagger}}\left\{
\left[
\begin{array}{cc}
	1 & -1 \\
	-1 & 1 \\
\end{array}
\right]-
\frac{l_e^2 \gamma}{6}
\left[
\begin{array}{cc}
	2 & 1 \\
	1 & 2 \\
\end{array}
\right]
\right\}
\bar{\psi}=0
\end{equation}
Now, as this will be true for all test functions $\bar{\phi}$, we can write the two element approximation as
\begin{equation}
\left\{
\left[
\begin{array}{ccc}
	1 & -1 & 0 \\
	-1 & 2 & -1 \\
   0 & -1 & 1 \\
\end{array}
\right]-
\lambda
\left[
\begin{array}{ccc}
	2 & 1 & 0 \\
	1 & 4 & 1\\
   0 & 1 & 2 \\
\end{array}
\right]
\right\}
\bar{\psi}=0
\label{torque-fe-matrix}
\end{equation}
where $\lambda=\frac{l_e^2 \gamma}{6}$ is the eigenvalue, and using boundary conditions $\bar{\psi}=[0 \quad \psi_1 \quad \psi_2]$ is the eigenvector.  This matrix torsional vibration equation, with two elements, is now in the form of a generalized eigenvalue problem.  The complexities in solving this equation will be dealt with when dealing with such problems in the context of quantum mechanics.  Just to note, if we wanted to solve Eqn.~(\ref{torque-fe-matrix}) analytically we would write it as
\begin{equation}
\textbf{H}\bar{\psi}=
\left[
\begin{array}{cc}
	2-4\lambda & -1-\lambda \\
	-1-\lambda & 1-2\lambda \\
\end{array}
\right]
\left[ \begin{array}{c}
\psi_1\\
\psi_2\\
\end{array}
\right] 
\end{equation}
where we have eliminated the first row and column as $\psi_0=0$. Therefore, a solution exists when the determinant of $\textbf{H}$ vanishes.  In this way we obtain the eigenvalues and eigenvectors of the problem.\footnote{For a full solution and explanation of this method see \cite{bick}}

\subsection{Application to the \schro Equation}
In quantum mechanics the very basic eigen-value problem consists of solving
\begin{equation}
-\frac{\hbar^2}{2m}\frac{d^2\psi}{dx^2}+V(x)\psi=E\psi.
\end{equation}
The aim is to determine the energy level configurations for particles in various potentials and spaces.  We will first consider a particle in an infinite well where $V=0$, this will then be extended to a general well $V(x)$.    

\subsubsection{Model of Infinite Potential Well}
To model an infinite potential well using FEM we begin with the \schro equation with zero potential,
\begin{equation}
\frac{d^2\psi}{dx^2}+\gamma\psi=0
\end{equation}
where $\gamma=\frac{2mE}{\hbar^2}$.  Now using the FEM construction described in Sec.~\ref{eigen-value-classic} we obtain the \schro equation FE approximation:
\begin{equation}
\sum_e \left\{
\left[
\begin{array}{cc}
	1 & -1 \\
	-1 & 1 \\
\end{array}
\right]-
\frac{l_e^2 \gamma}{6}
\left[
\begin{array}{cc}
	2 & 1 \\
	1 & 2 \\
\end{array}
\right]
\right\}
\bar{\psi}=0
\label{schro-time-ind}
\end{equation}
From the boundary conditions of an infinite potential well we know that the nodal approximations at the edges of the well are zero: $\psi_1=\psi_{N+1}=0$ (where $N$ is the number of elements).  Therefore, assembling for $N$ elements we have the generalized eigenvalue problem:
\begin{equation}
\left\{
\left[
\begin{array}{ccccc}
	2 & -1 & 0 &  \cdots & 0  \\
	-1 & 2 & -1 & & \vdots  \\
	0 & -1 & 2 &   & 0  \\
  \vdots &  &  & \ddots & -1 \\
  0 & \cdots & 0 & -1 & 2 \\
\end{array}
\right]-
\frac{l_e^2 \gamma}{6}
\left[
\begin{array}{ccccc}
	4 & 1 & 0 &  \cdots & 0  \\
	1 & 4 & 1 & & \vdots  \\
	0 & 1 & 4 &   & 0  \\
  \vdots &  &  & \ddots & 1 \\
  0 & \cdots & 0 & 1 & 4 \\
\end{array}
\right]
\right\}
\left[
\begin{array}{c}
	\psi_2  \\
  \psi_3  \\
  \psi_4	\\
	 \vdots \\
   \psi_{N} \\
\end{array}
\right]
=\textbf{0}
\label{time-independent-no-potential-equation}
\end{equation}

\subsubsection{Model of a General Potential $V(x)$}
To take into account a general potential we need to model
\begin{equation}
\frac{d^2\psi}{dx^2}-V(x)\psi+\gamma\psi=0.
\end{equation}
The extra potential term is incorporated through the following FE approximation
\begin{equation}
\sum_e\frac{2m\bar{V}_e}{\hbar^2}\left(\int^1_{-1}\textbf{N}^{\dagger}\textbf{N}d\xi\right)\bar{\psi}=\sum_e\frac{m l_e\bar{V}_e}{\hbar^2}
\left[
\begin{array}{cc}
	2 & 1 \\
	1 & 2 \\
\end{array}
\right]
\bar{\psi}
\end{equation}
where $\bar{V}_e$ is the average potential within the element $e$.  Assembling for $N$ elements we obtain the full \schro FE approximation\footnote{An in-depth study of the formulation and solution of quantum eigen-value problems can be found in \cite{ramdas}.}:
\begin{equation}
\textbf{0}=\left\{\textbf{A}'+\frac{ml_e^2}{\hbar^2}\textbf{B}'-\frac{l_e^2 \gamma}{6}\textbf{C}'\right\}\bar{\psi}
\end{equation}
where the matrices are given as:
\begin{eqnarray}
\textbf{A}'&=&\left[
\begin{array}{ccccc}
	1 & -1 & 0 &  \cdots & 0  \\
	-1 & 2 & -1 & & \vdots  \\
	0 & -1 & 2 &   & 0  \\
  \vdots &  &  & \ddots & -1 \\
  0 & \cdots & 0 & -1 & 1 \\
\end{array}
\right],\nonumber\\
\textbf{B}'&=&
\left[
\begin{array}{ccccc}
	\bar{V}_1 & -\bar{V}_1 & 0 &  \cdots & 0  \\
	-\bar{V}_1 & \bar{V}_1+\bar{V}_2 & -\bar{V}_2 & & \vdots  \\
	0 & -\bar{V}_2 & \bar{V}_2+\bar{V}_3 &   & 0  \\
  \vdots &  &  & \ddots & -\bar{V}_N \\
  0 & \cdots & 0 & -\bar{V}_N & \bar{V}_N \\
\end{array}
\right],\nonumber\\
\textbf{C}'&=&
\left[
\begin{array}{ccccc}
	2 & 1 & 0 &  \cdots & 0  \\
	1 & 4 & 1 & & \vdots  \\
	0 & 1 & 4 &   & 0  \\
  \vdots &  &  & \ddots & 1 \\
  0 & \cdots & 0 & 1 & 2 \\
\end{array}
\right],
\end{eqnarray}
and the nodal vector is
\begin{equation}
\bar{\psi}=
\left[
\begin{array}{c}
	\psi_1  \\
  \psi_2  \\
  \psi_3	\\
	 \vdots \\
   \psi_{N+1} \\
\end{array}
\right].
\end{equation}  

\section{Time-Dependent Problems}
\label{time-descrete}
Before discussing space-time FEM we will first give a basic example of the solution of time-dependent problems using FEM/FDM.  In this example we will use the one-dimensional diffusion equation with initial boundary conditions, which take the form
\begin{eqnarray}
kA\frac{\partial^2 \psi}{\partial x^2}&=&\rho c_p A\frac{\partial\psi}{\partial t}\qquad 0\leq x \leq L, 0\leq t\nonumber\\
\psi(0,t)&=&\psi_0 \qquad 0\leq t\nonumber\\
\psi(L,t)&=&0 \qquad  0\leq t\nonumber\\
\psi(x,0)&=&0 \qquad  0\leq x\leq L\nonumber\\
\label{diffusion-bound}
\end{eqnarray}
So the problem we will solve can be written as
\begin{equation}
\gamma \frac{\partial^2 \psi}{\partial x^2}=\frac{\partial \psi}{\partial t}
\label{diffusion-eq}
\end{equation}
where $\gamma=\frac{K}{\rho c_p}$.  We will solve the spatial part of this problem using the previous FEM approach, but then the temporal part will be dealt with using FDM approach.

\subsection{FEM Spatial Discretization}
As in the case of torsional vibrations we begin by multiplying Eqn.~(\ref{diffusion-eq}) with a test function $\phi$ and then integrating by parts
\begin{eqnarray}
\underbrace{\left. \gamma\phi\frac{\partial \psi}{\partial x}\right|^L_0}_{=0}-\gamma\int^L_0\frac{\partial\phi}{\partial x}\frac{\partial\psi}{\partial x}dx&=&\int^L_0\phi\frac{\partial\psi}{\partial t}dx\nonumber\\
-\gamma\int^L_0\frac{\partial\phi}{\partial x}\frac{\partial\psi}{\partial x}dx&=&\int^L_0\phi\frac{\partial\psi}{\partial t}dx
\label{diffusion-galerkin}
\end{eqnarray}
Now, using Eqns.~(\ref{B-operator}) and (\ref{fe-function}) in (\ref{diffusion-galerkin}) we obtain
\begin{equation}
-\gamma \sum_e\int^1_{-1}\bar{\phi}^{\dagger}\textbf{B}^{\dagger}\textbf{B}\bar{\psi}\frac{l_e}{2}d\xi
=\sum_e\int^1_{-1}\bar{\phi}^{\dagger}\textbf{N}^{\dagger}\textbf{N}\dot{\bar{\psi}}\frac{l_e}{2}d\xi
\end{equation}
where the sum is again over all the elements, and the vector $\bar{\psi}$ is now time-dependent.  Simplifying and integrating we obtain
 \begin{eqnarray}
-\frac{l_e\gamma}{2 }\sum_e \bar{\phi}^{\dagger} \left[ \int^1_{-1}\textbf{B}^{\dagger}\textbf{B} d\xi \right] \bar{\psi}&=&\frac{l_e}{2}\sum_e \bar{\phi}^{\dagger} \left[ \int^1_{-1}\textbf{N}^{\dagger}\textbf{N}d\xi\right] \dot{\bar{\psi}}\nonumber\\
-\frac{2\gamma}{l_e^2}\sum_e\bar{\phi}^{\dagger}\left[
\begin{array}{cc}
	1 & -1 \\
	-1 & 1 \\
\end{array}
\right]\bar{\psi}&=&\sum_e\bar{\phi}^{\dagger}
\left[
\begin{array}{cc}
	2/3 & 1/3 \\
	1/3 & 2/3 \\
\end{array}
\right]
\dot{\bar{\psi}}
\end{eqnarray}
Therefore, we have
\begin{equation}
\sum_e\bar{\phi}^{\dagger}\left\{
\lambda\left[\begin{array}{cc}
	1 & -1 \\
	-1 & 1 \\
\end{array}
\right]\bar{\psi}+
\left[
\begin{array}{cc}
	2 & 1 \\
	1 & 2\\
\end{array}
\right]
\dot{\bar{\psi}}
\right\}=0,
\end{equation}
where $\lambda=\frac{6\gamma}{l_e^2}$.  Then, as this is true for all test functions, $\phi$, we have the element equation
\begin{equation}
\lambda\left[\begin{array}{cc}
	1 & -1 \\
	-1 & 1 \\
\end{array}
\right]\bar{\psi}+
\left[
\begin{array}{cc}
	2 & 1 \\
	1 & 2\\
\end{array}
\right]
\dot{\bar{\psi}}=0
\end{equation}
When assembled for four elements, we obtain
\begin{equation}
\lambda
\left[
\begin{array}{ccccc}
	1 & -1 & 0 & 0 & 0 \\
	-1 & 2 & -1 & 0 & 0 \\
	0 & -1 & 2 & -1 & 0 \\
	0 & 0 & -1 & 2 & -1 \\
	0 & 0 & 0 & -1 & 2 \\
\end{array}
\right]\bar{\psi}
+
\left[
\begin{array}{ccccc}
	2 & 1 & 0 & 0 & 0 \\
	1 & 4 & 1 & 0 & 0 \\
	0 & 1 & 4 & 1 & 0 \\
	0 & 0 & 1 & 4 & 1 \\
	0 & 0 & 0 & 1 & 4 \\
\end{array}
\right]\dot{\bar{\psi}}=0
\label{full-four}
\end{equation}
From the initial conditions, in Eqns.~(\ref{diffusion-bound}), we know that the first component of $\bar{\psi}$ is a constant ($\psi_0=constant$) and the last component is always zero ($\psi_4=0$).  Using this information we can reduce Eqn.~(\ref{full-four}) to
\begin{equation}
\lambda\left[
\begin{array}{ccc}
	 2 & -1 & 0  \\
	-1 & 2 & -1  \\
	 0 & -1 & 2  \\
\end{array}
\right]\bar{\psi}\nonumber\\
+\left[\begin{array}{ccc}
	 4 & 1 & 0  \\
	 1 & 4 & 1  \\
	 0 & 1 & 4 \\
\end{array}
\right]\dot{\bar{\psi}}=\left[
\begin{array}{c}
\lambda \psi_0\\
0\\
0\\
\end{array}
\right]
\label{reduced-four}
\end{equation}
where $\bar{\psi}=\left[\psi_1 \quad \psi_2 \quad \psi_3 \right]$.  Writing this in a more convenient notation we have
\begin{equation}
\lambda\tilde{ \textbf{A}} \bar{\psi}+\tilde{\textbf{B}}\dot{\bar{\psi}}=\tilde{\textbf{C}}
\label{reduced-four-simple}
\end{equation}

\subsection{FD Time-Integration}
The next step is to use FD techniques to carry out time-integration.  This can be done in many ways, but here we will only consider two techniques: the explicit Euler and the implicit Crank-Nicolson methods.
\subsubsection{Explicit Euler Time-Integration} 
This method is the simplest to implement, however it can be unstable.   We begin with the following approximation
\begin{equation}
\dot{\bar{\psi}}=\frac{\bar{\psi}^{n+1} - \bar{\psi}^{n}}{\Delta t}
\end{equation}
Implementing this into Eqn.~(\ref{reduced-four-simple}) and simplifying we have
 \begin{eqnarray}
\Delta t\lambda \tilde{\textbf{A} }\bar{\psi}^{n}+\tilde{\textbf{B}}\left[ \bar{\psi}^{n+1} - \bar{\psi}^{n}\right]&=& \Delta t\tilde{\textbf{C}}\nonumber\\
\left[\lambda'\textbf{A}'-\textbf{B}'\right]\bar{\psi}^n
+\textbf{B}'\bar{\psi}^{n+1}&=&\textbf{C}'
\end{eqnarray}
where $\textbf{A}'=\tilde{\textbf{A}}$, $\textbf{B}'=\tilde{\textbf{B}}$, $\textbf{C}'=\Delta t \tilde{\textbf{C}}$ and $\lambda ' = \Delta t \lambda$.  Using the initial condition vector $\bar{\psi}^{0}$ we can determine $\bar{\phi}^{1}=\textbf{B}'\bar{\psi}^{1}$, which can be solved to obtain $\bar{\psi}^{1}$.  This process can be continued in order to obtain $\bar{\psi}^{2}$ from $\bar{\psi}^{1}$ and so on.
\subsubsection{Implicit Crank-Nicolson Time-Integration}
The explicit Euler method is mathematically and computationally very simple, however it can be unstable.  On the other hand, the implicit Crank-Nicolson method in unconditionally stable, even though it is slightly more complicated and computationally intensive.  In order to use this method we begin with the following approximation
\begin{equation}
\tilde{\textbf{B}}\dot{\bar{\psi}}=\frac{1}{2}\left\{\left[\tilde{\textbf{C}}-\lambda \tilde{\textbf{A}}\bar{\psi} \right]^n+\left[\tilde{\textbf{C}}-\lambda \tilde{\textbf{A}} \bar{\psi} \right]^{n+1}\right\}
\end{equation}
Therefore we have
\begin{eqnarray}
\tilde{\textbf{B}}\bar{\psi}^{n+1}-\tilde{\textbf{B}}\bar{\psi}^{n}&=& \frac{\Delta t}{2}\left[ \tilde{\textbf{C}}^n+\tilde{\textbf{C}}^{n+1}\right]-\frac{\Delta t\lambda}{2}\tilde{\textbf{A}}\bar{\psi}^n-\frac{\Delta t\lambda}{2}\tilde{\textbf{A}}\bar{\psi}^{n+1}\nonumber\\
\left[\textbf{B}'+\lambda'\textbf{A}'\right]\bar{\psi}^{n+1}&=&
\left[\textbf{C} ^{'n} +\textbf{C}^{'n+1}\right]+\left[\textbf{B}'
-\lambda'\textbf{A}'\right]\bar{\psi}^n
\end{eqnarray}
where $\textbf{A}'=\tilde{\textbf{A}}$, $\textbf{B}'=\tilde{\textbf{B}}$, $\textbf{C}'=\frac{\Delta t}{2}\tilde{\textbf{C}}$ and $\lambda'=\frac{\Delta t \lambda}{2}$ .  As stated previously this method is unconditionally stable; so even though oscillations occur and the accuracy may suffer for large step sizes $\Delta t$, the oscillations never become unbounded.\footnote{For further details and example of time-integration techniques see \cite{bick}}.

\section{Space-Time Finite Element Method}
In order to solve Eqn.~(\ref{diffusion-eq}) using space-time FEM the $x,t$ domain will be discretised into rectangular elements, labeled as in Fig.~\ref{square-element}.  
\begin{figure}
    \begin{center}
        \includegraphics[width=0.6\textwidth]{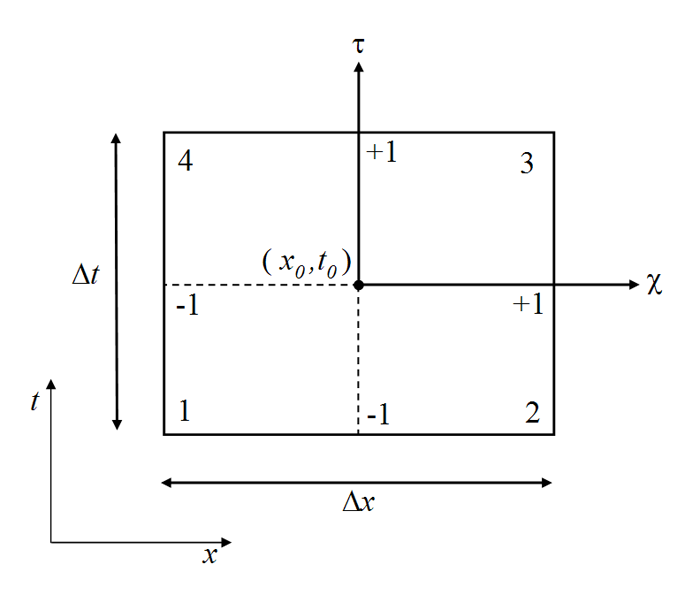}\\
        \caption{Node numbering and coordinates of rectangular element.}
        \label{square-element}
    \end{center}
\end{figure}
The approximate solution can then be modelled using the linear, dimensionless, local basis functions:
\begin{eqnarray}
N_1(\chi,\tau)&=&\frac{(1-\chi)(1-\tau)}{4}\qquad N_2(\chi,\tau)=\frac{(1+\chi)(1-\tau)}{4}\nonumber\\
\nonumber\\
N_3(\chi,\tau)&=&\frac{(1+\chi)(1+\tau)}{4}\qquad N_4(\chi,\tau)=\frac{(1-\chi)(1+\tau)}{4}
\label{space-time-shape-functions}
\end{eqnarray}
The other important factor to consider is that of continuous or discontinuous boundaries between the temporal elements $t_n$ and $t_{n+1}$.  In the discontinuous method the approximate solution of $\psi$ is continuous within the elements but discontinuous at the boundaries, so each time step forms a slab in space. In the continuous case the solution flows uninterrupted from one time step to another, like a continuous surface in the $x,t$ domain.  It has been shown (in \cite{det}) that the linear discontinuous method has a higher accuracy compared to the continuous method.  This is due to the extra nodal degree of freedom available in each time step.  However, in order to implement the discontinuous method an extra jump term has to be included in the discretization process.  To demonstrate the space-time FEM process we will show the linear continuous discretization, and then demonstrate the discontinuous case.

\subsection{Linear Continuous Discretization}
\label{linear-cont-discrete}
We begin by writing the solution as a function of the local coordinates
\begin{equation}
\psi=\sum_1^4 N_i \psi_i =\bar{N}\bar{\psi}
\end{equation}
where $\bar{N}=[N_1, N_2, N_3, N_4]$ and $\bar{\psi}=[\psi_1, \psi_2, \psi_3, \psi_4]$.  The corresponding differentials are
\begin{eqnarray}
\frac{\partial \psi}{\partial \chi}&=&\frac{\partial \psi}{\partial x} \frac{\partial x}{\partial \chi}+ \frac{\partial \psi}{\partial t}\frac{\partial t}{\partial \chi}\nonumber\\
\frac{\partial \psi}{\partial \tau}&=&\frac{\partial \psi}{\partial x} \frac{\partial x}{\partial \tau}+ \frac{\partial \psi}{\partial t}\frac{\partial t}{\partial \tau}
\end{eqnarray}
Writing as a matrix this becomes
\begin{equation}
\left[
\begin{array}{c}
\frac{\partial \psi}{\partial \chi}\\
\frac{\partial \psi}{\partial \tau}\\
\end{array}
\right]=
\left[\begin{array}{cc}
	 \frac{\partial x}{\partial \chi} & \frac{\partial t}{\partial \chi}  \\
	 \frac{\partial x}{\partial \tau}&  \frac{\partial t}{\partial \tau}  \\
\end{array}
\right]\left[
\begin{array}{c}
\frac{\partial \psi}{\partial x}\\
\frac{\partial \psi}{\partial t}\\
\end{array}
\right]=
\textbf{J}\left[
\begin{array}{c}
\frac{\partial \psi}{\partial x}\\
\frac{\partial \psi}{\partial t}\\
\end{array}
\right],
\end{equation}
where $\textbf{J}$ is the Jacobian of the transformation.  Using the coordinate transformations
\begin{equation}
x=\sum_1^4 x_i N_i=\bar{N}\bar{x} \qquad t=\sum_1^4 t_i N_i=\bar{N}\bar{t} 
\end{equation}
and the nodal numbering described in Fig.~\ref{square-element}, the Jacobian reduces to the simple form
 \begin{equation}
\textbf{J}=
\left[\begin{array}{cc}
	 \frac{\Delta x}{2} & 0  \\
	 0 &  \frac{\Delta t}{2}  \\
\end{array}
\right]
\end{equation}
The inverse of this, which will be required later, is simply given as
 \begin{equation}
\textbf{J}^{-1}=
\left[\begin{array}{cc}
	 \frac{2}{\Delta x} & 0  \\
	 0 &  \frac{2}{ \Delta t}  \\
\end{array}
\right]
\end{equation}
Therefore we now have
\begin{equation}
\left[
\begin{array}{c}
\frac{\partial \psi}{\partial x}\\
\frac{\partial \psi}{\partial t}\\
\end{array}
\right]=
\textbf{J}^{-1}
\left[
\begin{array}{c}
\frac{\partial \psi}{\partial \chi}\\
\frac{\partial \psi}{\partial \tau}\\
\end{array}
\right],
\label{inverse-derivative}
\end{equation}
The local derivatives with respect to $\chi$ and $\tau$ can be discretised as
\begin{eqnarray}
\frac{\partial \psi}{\partial \chi}&=&-\frac{1}{4}\left[\psi_1(1-\tau)-\psi_2(1-\tau)-\psi_3(1+\tau)+\psi_4(1+\tau)\right]\nonumber\\
\frac{\partial \psi}{\partial \tau}&=&-\frac{1}{4}\left[\psi_1(1-\chi)+\psi_2(1+\chi)-\psi_3(1+\chi)-\psi_4(1-\chi)\right]
\end{eqnarray}
In matrix form this can be written as
\begin{equation}
\left[
\begin{array}{c}
\frac{\partial \psi}{\partial \chi}\\
\frac{\partial \psi}{\partial \tau}\\
\end{array}
\right]=
-\frac{1}{4}
\left[\begin{array}{cccc}
(1-\tau) & -(1-\tau) & -(1+\tau) &(1+\tau)\\
(1-\chi) & (1+\chi) & -(1+\chi) & -(1-\chi)\\
\end{array}
\right]
\left[
\begin{array}{c}
\psi_1\\
\psi_2\\
\psi_3\\
\psi_4\\
\end{array}
\right]=\textbf{A}\bar{\psi}.
\label{a-matrix}
\end{equation}
Now, combining Eqns.~(\ref{inverse-derivative}) and (\ref{a-matrix}) we have
\begin{equation}
\left[
\begin{array}{c}
\frac{\partial \psi}{\partial x}\\
\frac{\partial \psi}{\partial t}\\
\end{array}
\right]=
\textbf{J}^{-1}\textbf{A}\bar{\psi}=\textbf{B}\bar{\psi},
\end{equation}
where
\begin{equation}
\textbf{B}=-\frac{1}{2}
\left[\begin{array}{cccc}
\frac{(1-\tau)}{\Delta x} & \frac{-(1-\tau)}{\Delta x} & \frac{-(1+\tau)}{\Delta x} &\frac{(1+\tau)}{\Delta x}\\
\frac{(1-\chi)}{\Delta t} & \frac{(1+\chi)}{\Delta t} & \frac{-(1+\chi)}{\Delta t} & \frac{-(1-\chi)}{\Delta t}\\
\end{array}
\right]
\label{b-matrix}
\end{equation}
In order to use this in the discretisation of the diffusion equation we write 
\begin{equation}
\frac{\partial \psi}{\partial x}=\textbf{B}_1 \bar{\psi} \qquad \frac{\partial \psi}{\partial t}=\textbf{B}_2 \bar{\psi},
\end{equation}
where $\textbf{B}_1$ and  $\textbf{B}_2$ are the upper and lower rows of $\textbf{B}$.  The continuous space-time discretisation begins with Eqn.~(\ref{diffusion-galerkin}), and then we apply the discrete derivative and function approximations to obtain\footnote{Note that as we are transforming from a global $(x,y)$ coordinate basis to a local $(\chi, \tau)$ basis we have implemented the volume transformation $dxdt=|\textbf{J}|d\chi d\tau$.}
\begin{eqnarray}
\bar{\phi}^{\dagger}\left[\int^{+1}_{-1}\left( \textbf{N}^{\dagger}\textbf{B}_2 + \gamma \textbf{B}_1^{\dagger}\textbf{B}_1 \right) \frac{\Delta x \Delta t}{4}d\chi d\tau \right]\bar{\psi} 
\end{eqnarray}
Carrying out the simple integrals, and noting that this is true for all test functions, we obtain the following set of equations\footnote{Where the extra factor of $2$ in the second term comes from integrating over $\tau$ i.e. $\int^{+1}_{-1}d\tau=2$, even though there are no explicit $\tau$ variables.}
\begin{equation}
\textbf{0}=
\left\{
\left[
\begin{array}{cccc}
2&1&-1&-2\\
1&2&-2&-1\\
1&2&-2&-1\\
2&1&-1&-2\\
\end{array}
\right]- \frac{2\gamma\Delta t}{\Delta x^2}
\left[
\begin{array}{cccc}
 2&-2&-1& 1\\
-2& 2& 1&-1\\
-1& 1& 2&-2\\
 1&-1&-2& 2\\
\end{array}
\right]
\right\}
\left[
\begin{array}{c}
\psi_1\\
\psi_2\\
\psi_3\\
\psi_4\\
\end{array}
\right]
\label{full-equation-spacetime-continuous}
\end{equation}
Taking advantage of boundary conditions we can note that for each time step $t_n$ to $t_{n+1}$ the nodal values for $t_n$ are known.  In this way we can reduce the $4\times4$ element equation to a $2\times 2$ one.  However, in order to pick the correct set of equations in Eqn.~(\ref{full-equation-spacetime-continuous}) we need to look at the structure of the test function:
\begin{equation}
\phi=\bar{N}\bar{\phi}=\left[\begin{array}{cccc}N_1 & N_2 & N_3 & N_4\end{array}\right]\left[
\begin{array}{c}
\phi_1\\
\phi_2\\
\phi_3\\
\phi_4\\
\end{array}
\right].
\end{equation}
Here we can note that the shape functions for $N_1$ and $N_2$ are $1$ at time $t_n$, and $N_3$ and $N_4$ are $1$ at time $t_{n+1}$.  Thus if we use the first two rows in Eqn.~(\ref{full-equation-spacetime-continuous}) (associated with $N_1$ and $N_2$) we obtain an explicit numerical method which is weighted on information from $t_n$, however if we use the second two rows (associated with $N_3$ and $N_4$) we obtain an implicit numerical method which is weighted on information from $t_{n+1}$.  Thus, going for the second two rows, and rearranging the results we have  
\begin{eqnarray}
\textbf{0}&=&\sum_{n_e}\left\{\left[
\begin{array}{cc}
1 & 2\\
2 & 1\\
\end{array}
\right]+
\frac{4\gamma \Delta t}{\Delta x^2}
\left[
\begin{array}{cc}
-1 & 1\\
1 & -1\\
\end{array}
\right]
\right\}
\left[
\begin{array}{c}
\psi_3\\
\psi_4\\
\end{array}
\right]\nonumber\\
\nonumber\\
&-&\left\{\left[
\begin{array}{cc}
2 & 1\\
1 & 2\\
\end{array}
\right]-
\frac{2\gamma\Delta t}{\Delta x^2}
\left[
\begin{array}{cc}
1 & -1\\
-1 & 1\\
\end{array}
\right]
\right\}
\left[
\begin{array}{c}
\psi_1\\
\psi_2\\
\end{array}
\right].
\end{eqnarray}
A space-time difference stencil can now be obtained by assembling the above equations for two neighbouring elements, which can then be easily extended for larger number of elements.\footnote{For a detailed discussion and analysis of space-time FEM see \cite{det}}

\subsection{Linear Discontinuous Discretization}
\label{linear-discontinuous-discrete}
For this method the solution is linear within each time step and discontinuous at the temporal boundaries, Fig.~\ref{space-time-elements}.
\begin{figure}
    \begin{center}
        \includegraphics[width=0.5\textwidth]{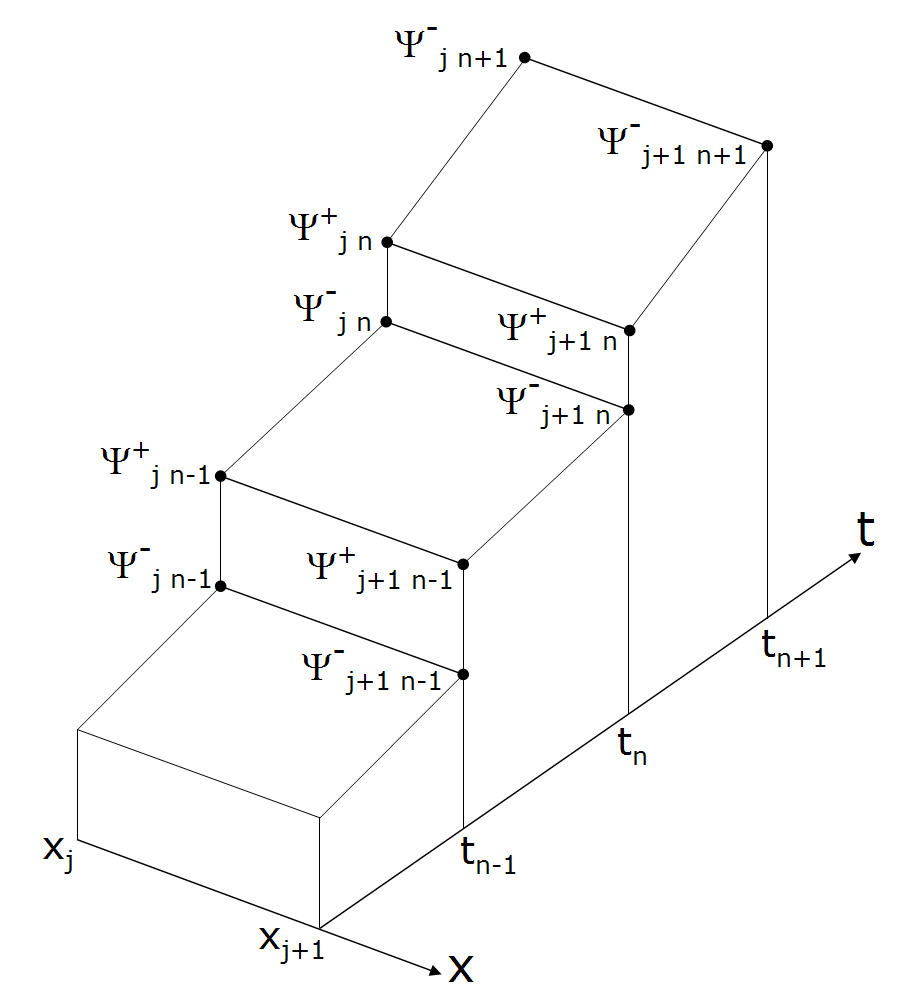}\\
        \caption{Linear discontinuous finite elements in space-time.}
        \label{space-time-elements}
    \end{center}
\end{figure}
In order to implement this method we need to include a jump term,
\begin{equation}
\sum_e\left[\int^{x_{j+1}}_{x_j} \phi^+_{t_n}\left(\psi^+_{t_n}-\psi^-_{t_n}\right)dx\right],
\label{full-jump-term}
\end{equation}
into the space-time discretization process.  
To begin the discontinuous space-time discretization we first write the approximate element solution for the time interval $t_n$ to $t_{n+1}$ as
\begin{eqnarray}
\psi=\bar{N}\bar{\psi}&=&\left[\begin{array}{cccc}N_1 & N_4 & N_2 & N_3\end{array}\right]\left[
\begin{array}{c}
\psi_1\\
\psi_4\\
\psi_2\\
\psi_3\\
\end{array}
\right]\nonumber\\
&=&\left[\begin{array}{cccc}N_{j,n} & N_{j,n+1} & N_{j+1,n} & N_{j+1,n+1}\end{array}\right]\left[
\begin{array}{c}
\psi^+_{j,n}\\
\psi^-_{j,n+1}\\
\psi^+_{j+1,n}\\
\psi^-_{j+1,n+1}\\
\end{array}
\right],
\label{space-time-nodal-numbering}
\end{eqnarray}
where the shape functions $N$ are as in Eqn.~(\ref{space-time-shape-functions}), and the relabelling is done to make the final assembly process simpler (to take account of this relabelling we will also need to rearrange the rows and columns of Eqn.~(\ref{full-equation-spacetime-continuous}) for the discontinuous space-time element equation).  Using this relabelled approximate solution we can discretise the jump term as
\begin{eqnarray}
& &\int^{x_{j+1}}_{x_j}\bar{\phi}^{+{\dagger}}\bar{N}^{+{\dagger}}\left(\bar{N}^+\bar{\psi}^+-\bar{N}^-\bar{\psi}^-\right)dx\nonumber\\
&=&\bar{\phi}\int^{1}_{-1}
\left\{\left[
\begin{array}{c}
N_{j,n}\\
0\\
N_{j+1,n}\\
0\\
\end{array}
\right]\left(\left[\begin{array}{cccc}N_{j,n} & 0 & N_{j+1,n} & 0\end{array}\right]\left[
\begin{array}{c}
\psi^+_{j,n}\\
\psi^-_{j,n+1}\\
\psi^+_{j+1,n}\\
\psi^-_{j+1,n+1}\\
\end{array}
\right]\right.\right.\nonumber\\
&-&\left.\left.\left[\begin{array}{cccc}0 & N_{j,n} & 0 & N_{j+1,n}\end{array}\right]\left[
\begin{array}{c}
\psi^+_{j,n-1}\\
\psi^-_{j,n}\\
\psi^+_{j+1,n-1}\\
\psi^-_{j+1,n}\\
\end{array}
\right]\right) \frac{\Delta x}{2}d\chi\right\}.
\end{eqnarray}
Calculating the tensor products, and for $t_n$ we set $\tau=-1$ in $\bar{N}$, we have
\begin{eqnarray}
& &\bar{\phi}^{\dagger}\frac{1}{4}\int^{1}_{-1}
\left\{
\left[
\begin{array}{cccc}
(1-\xi)(1-\xi)&0&(1-\xi)(1+\xi)&0\\
0&0&0&0\\
(1+\xi)(1-\xi)&0&(1+\xi)(1+\xi)&0\\
0&0&0&0\\
\end{array}
\right]\left[
\begin{array}{c}
\psi^+_{j,n}\\
\psi^-_{j,n+1}\\
\psi^+_{j+1,n}\\
\psi^-_{j+1,n+1}\\
\end{array}
\right]\right.\nonumber\\
&-&\left.\left[
\begin{array}{cccc}
0&(1-\xi)(1-\xi)&0&(1-\xi)(1+\xi)\\
0&0&0&0\\
0&(1+\xi)(1-\xi)&0&(1+\xi)(1+\xi)\\
0&0&0&0\\
\end{array}
\right]\left[
\begin{array}{c}
\psi^+_{j,n-1}\\
\psi^-_{j,n}\\
\psi^+_{j+1,n-1}\\
\psi^-_{j+1,n}\\
\end{array}
\right] \right\}\frac{\Delta x}{2}d\chi.\nonumber\\
\end{eqnarray}
Carrying out the integrals we obtain
\begin{equation}
\bar{\phi}^{\dagger}
\frac{\Delta x}{6}
\left\{\left[
\begin{array}{cccc}
2&0&1&0\\
0&0&0&0\\
1&0&2&0\\
0&0&0&0\\
\end{array}
\right]\left[
\begin{array}{c}
\psi^+_{j,n}\\
\psi^-_{j,n+1}\\
\psi^+_{j+1,n}\\
\psi^-_{j+1,n+1}\\
\end{array}
\right]-\left[
\begin{array}{cccc}
0&2&0&1\\
0&0&0&0\\
0&1&0&2\\
0&0&0&0\\
\end{array}
\right]\left[
\begin{array}{c}
\psi^+_{j,n-1}\\
\psi^-_{j,n}\\
\psi^+_{j+1,n-1}\\
\psi^-_{j+1,n}\\
\end{array}
\right] \right\}.
\end{equation}
We now multiply this by $-\frac{12}{\Delta x}$ so it can be added to Eqn.~(\ref{full-equation-spacetime-continuous}).   However, we must also rearrange the matrices in Eqn.~(\ref{full-equation-spacetime-continuous}) so that they correspond to the nodal ordering in Eqn.~(\ref{space-time-nodal-numbering}).  After doing this we have the full discontinuous space-time element equation
\begin{eqnarray}
\textbf{0}&=&
\left\{
\left[
\begin{array}{cccc}
2&-2&1&-1\\
2&-2&1&-1\\
1&-1&2&-2\\
1&-1&2&-2\\
\end{array}
\right]- \frac{2\gamma\Delta t}{\Delta x^2}
\left[
\begin{array}{cccc}
 2&1&-2& -1\\
1& 2& -1&-2\\
-2& -1& 2&1\\
 -1&-2&1& 2\\
\end{array}
\right]\right.\nonumber\\
&-&\left.2\left[
\begin{array}{cccc}
 2&0&1&0\\
	0& 0& 0&0\\
1& 0& 2&0\\
 0&0&0& 0\\
\end{array}
\right]
\right\}
\left[
\begin{array}{c}
\psi^+_{j,n}\\
\psi^-_{j,n+1}\\
\psi^+_{j+1,n}\\
\psi^-_{j+1,n+1}\\
\end{array}
\right]
+
2\left[
\begin{array}{cccc}
0&2&0&1\\
0&0&0&0\\
0&1&0&2\\
0&0&0&0\\
\end{array}
\right]\left[
\begin{array}{c}
\psi^+_{j,n-1}\\
\psi^-_{j,n}\\
\psi^+_{j+1,n-1}\\
\psi^-_{j+1,n}\\
\end{array}
\right].\nonumber\\
\label{full-equation-spacetime-discontinuous}
\end{eqnarray}
Writing this in operator form we have
\begin{equation}
\left(\textbf{A}'+\textbf{B}'+\textbf{C}'\right)\left[
\begin{array}{c}
\psi^+_{j,n}\\
\psi^-_{j,n+1}\\
\psi^+_{j+1,n}\\
\psi^-_{j+1,n+1}\\
\end{array}
\right]=\textbf{D}'
\left[
\begin{array}{c}
\psi^+_{j,n-1}\\
\psi^-_{j,n}\\
\psi^+_{j+1,n-1}\\
\psi^-_{j+1,n}\\
\end{array}
\right]
\end{equation}
where $\textbf{A}'$, $\textbf{B}'$, $\textbf{C}'$, and $\textbf{D}'$ are the operators of the respective matrices in Eqn.~(\ref{full-equation-spacetime-discontinuous}).}

{\typeout{Time-Dependent Schrodinger Equation}
\chapter{Time-Dependent Analysis}
\label{Dependent Schrodinger} 
In this chapter we will begin by discretising the time-dependent \schro equation by the use of FE for the spatial part and Crank-Nicolson for the temporal part.  As there are extensive results and  literature on this method (\cite{watanabe} and references therein) we will have a comparable benchmark.  We will first do this analysis for a Gaussian wave-packet in an infinite potential well, and then we will conduct a similar analysis but with a potential barrier located within the well.  The next step will be to discretise the time-dependent \schro equation using the space-time FE approximation.  The results of the space-time method can then be compared to those of the first method.

\section{Crank-Nicolson and Finite Element Analysis}
\label{Crank-Nicolson and Finite Element Analysis}
\subsection{Infinite Potential Well}
\label{infinite-potential-well}
We will begin with the simple case of a particle in an infinite well, Fig.~\ref{Infinitewell}.  This will then form the basis for modelling a particle in an infinite well with a finite potential barrier.  

\subsubsection{Crank-Nicolson Temporal Approximation}
We begin by discretising the equation
\begin{equation}
\frac{\partial \psi}{\partial t}-i b\frac{\partial^2\psi}{\partial x^2}=0
\end{equation}
where $b=\frac{\hbar}{2m}$.  Now, following the steps of Sec.~\ref{time-descrete}, we first apply spatial FE discretisation:
\begin{equation}
\int^{+1}_{-1}\textbf{N}^{\dagger}\textbf{N}d\xi\dot{\bar{\psi}}+i\frac{\hbar}{2m}\int^{+1}_{-1}\textbf{B}^{\dagger}\textbf{B}d\xi\bar{\psi}=\textbf{0}
\end{equation} 
After computing the integrals we obtain
\begin{equation}
\left[
\begin{array}{cc}
	2 & 1 \\
	1 & 2 \\
\end{array}
\right]\dot{\bar{\psi}}+i\frac{6\hbar}{2ml_e^2}\left[
\begin{array}{cc}
	1 & -1 \\
	-1 & 1 \\
\end{array}
\right]\bar{\psi}=\textbf{0}.
\end{equation}
In operator form we have
\begin{equation}
\tilde{\textbf{A}}\dot{\bar{\psi}}+i\tilde{\textbf{B}}\bar{\psi}=\textbf{0},
\end{equation}
where
\begin{equation}
\tilde{\textbf{A}}=\left[
\begin{array}{cc}
	2 & 1 \\
	1 & 2 \\
\end{array}
\right] \qquad \textrm{and} \qquad \tilde{\textbf{B}}=\frac{6\hbar}{2ml_e^2}\left[
\begin{array}{cc}
	1 & -1 \\
	-1 & 1 \\
\end{array}
\right]
\end{equation}
Now applying the Crank-Nicolson approximation we obtain
\begin{equation}
\tilde{\textbf{A}}\bar{\psi}^{n+1}-\tilde{\textbf{A}}\bar{\psi}^{n}=-\left(\frac{\Delta t}{2}i\tilde{\textbf{B}}\bar{\psi}^{n}+\frac{\Delta t}{2}i\tilde{\textbf{B}}\bar{\psi}^{n+1}\right)
\end{equation}
and rearranging we have
\begin{equation}
\left(\tilde{\textbf{A}}+i\tilde{\textbf{B}}\right)\bar{\psi}^{n+1}=\left(\tilde{\textbf{A}}-i\tilde{\textbf{B}}\right)\bar{\psi}^{n}.
\label{cn-fd-schro}
\end{equation}
The constants $\frac{\Delta t}{2}$ have been absorbed into $\tilde{\textbf{B}}$, which gives
\begin{equation}
\tilde{\textbf{B}}=\frac{6\hbar\Delta t}{4 m l_e^2}
\left[
\begin{array}{cc}
	1 & -1 \\
	-1 & 1 \\
\end{array}
\right].
\end{equation}
Here $\Delta t$ is temporal difference and $l_e$ is the spatial element size, $m$ is the mass of the particle and $\hbar$ is the Planck constant.  It can also be seen that the transformation in Eqn.~(\ref{cn-fd-schro}) is unitary, and as an extra confirmation of its validity it takes the same form as the full finite difference approximation in Eqn.~(\ref{fd-schro}).

\subsubsection{Construction of the Numerical Method}
\label{infinite-numerical-method}
Our aim now is to solve Eqn.~(\ref{cn-fd-schro}) for $\bar{\psi}^{n+1}$ given $\bar{\psi}^{n}$ i.e. find $\psi(t)$ knowing the initial condition $\psi(0)$.  In order to conduct this iterative computational calculation we must first simplify Eqn.~(\ref{cn-fd-schro}) so that the complex values can be easily handled.  The element state vector $\bar{\psi}$ currently takes the form
\begin{equation}
\bar{\psi}=\left[
\begin{array}{c}
	\psi_1 \\
	\psi_2 \\
\end{array}
\right]=
\left[
\begin{array}{c}
	a^1+ib^1 \\
	a^2+ib^2 \\
\end{array}
\right],
\end{equation}
where $\psi_1$ is the complex left nodal value and $\psi_2$ is the complex right nodal value.  However, we can write this complex two-component vector as a real four-component vector:
\begin{equation}
\bar{\psi}=\left[
\begin{array}{c}
	Re[\psi_1] \\
	Im[\psi_1] \\
	Re[\psi_2] \\
	Im[\psi_2] \\
\end{array}
\right]=
\left[
\begin{array}{c}
	a^1	\\
	b^1 \\
	a^2 \\
	b^2 \\
\end{array}
\right].
\label{4-component-element-vector}
\end{equation}
Using this real four-component element vector we can write the complex element equations in a totally real form as
\begin{equation}
\left(\textbf{A}'+\alpha\textbf{B}'\right) \bar{\psi}^{n+1}=
\left(\textbf{A}'-\alpha\textbf{B}'\right) \bar{\psi}^n,
\label{real-schro-crank-equation}
\end{equation}
where $\alpha=\frac{6\hbar\Delta t}{4 m l_e^2}$, and the real matrices are given as
\begin{equation}
\textbf{A}'=
 \left[
\begin{array}{cccc}
	2 & 0 & 1& 0 \\
	0 & 2 & 0& 1 \\
	1 & 0 & 2& 0 \\
	0 & 1 & 0& 2 \\
\end{array}
\right]
\qquad 
\textbf{B}'=
 \left[
\begin{array}{cccc}
	0 & -1 & 0& 1 \\
	1 & 0 & -1& 0 \\
	0 & 1 & 0& -1 \\
	-1 & 0 & 1& 0 \\
\end{array}
\right]
\label{real-matrices-A-B}
\end{equation}
Eqn.~(\ref{real-schro-crank-equation}) can now be assembled using the normal FE method.  For example, considering two elements the nodal vector becomes 
\begin{equation}
\bar{\psi} ^T=\left[
\begin{array}{cccccc}
a^1 & b^1& a^2& b^2 &a^3& b^3\\
\end{array}\right],
\end{equation}
and the matrices take the form:
\begin{equation}
\textbf{A}'=
 \left[
\begin{array}{cccccc}
	2 & 0 & 1& 0 &0 &0 \\
	0 & 2 & 0& 1 &0 &0 \\
	1 & 0 & 4& 0 &1 &0 \\
	0 & 1 & 0& 4 &0 &1 \\
	0 & 0 & 1& 0 &2 &0 \\
	0 & 0 & 0& 1 &0 &2 \\
\end{array}
\right]
\qquad 
\textbf{B}'=
 \left[
\begin{array}{cccccc}
	0 & -1 & 0& 1 &0 &0 \\
	1 & 0 & -1& 0 &0 &0 \\
	0 & 1 & 0& -2 &0 &1 \\
	-1 & 0 & 2& 0 &-1 &0 \\
	0 & 0 & 0& 1 &0 &-1 \\
	0 & 0 & -1& 0 &1 &0 \\\end{array}
\right]
\end{equation}

\subsubsection{Initial State Function}
\label{initial-state-function-section}
At the initial time $t=0$ we can assume that a particle is placed into an infinite potential well at position $x_0$ with a momentum $k_0$.  The initial particle state can then be modelled as a Gaussian wave packet, as described in Appendix~\ref{gaussian-packet}:
\begin{equation}
\bar{\psi}(x)= 
\left[
\begin{array}{c}
	Re[\psi(x)] \\
	Im[\psi(x)] \\
\end{array}
\right].
\end{equation}
The value of this wave packet at each nodal position $j$ can then be written as
\begin{equation}
\bar{\psi}_j= 
\left[
\begin{array}{c}
	Re[\psi_j] \\
	Im[\psi_j] \\
\end{array}
\right]=
\left(\frac{1}{2\pi\sigma^2}\right)^{\frac{1}{4}}e^{-(x_j-x_0)^2/4\sigma^2}
\left[\begin{array}{c}
\cos(k_0 x_j)\\
\sin(k_0 x_j)\\
\end{array}
\right].
\label{Gaussian-wave-packet-model}
\end{equation}
Doing this for each node of the spatial domain we can construct the inital state vector:
\begin{equation}
\left(\bar{\psi}^0\right)^T=\left[
\begin{array}{ccccccc}
Re[\psi_1] & Im[\psi_1]& Re[\psi_2] & Im[\psi_2]& \ldots & Re[\psi_{n+1}] & Im[\psi_{n+1}]\\
\end{array}\right],
\label{full-state-vector}
\end{equation}
where $n$ is the number of elements.

We can note that if we use $\sigma=2$ the wave-packet in Eqn.~(\ref{Gaussian-wave-packet-model}) is automatically normalized:

\begin{equation}
\int^{-\infty}_{-\infty} \bar{\psi}^{*}\bar{\psi}dx= \int^{-\infty}_{-\infty}\left(\frac{1}{8\pi}\right)^{\frac{1}{2}}e^{-(x-x_0)/8}dx=1.
\end{equation}
This then removes the added task of normalizing the final results.

\subsubsection{Numerical Solution}
\label{infinite-well-numerical-solution}
We can now determine the time evolution of the state vector in Eqn.~(\ref{full-state-vector}).  We begin by first finding
\begin{equation}
\bar{\phi}^0=\left(\textbf{A}'-\alpha\textbf{B}'\right)\bar{\psi}^0.
\end{equation}
Once we evaluate this simple matrix multiplication we obtain the vector $\bar{\phi}^0$.  The next step is to evaluate the following system of equations for the unknown vector $\bar{\psi}^1$:
\begin{equation}
\left(\textbf{A}'+\alpha\textbf{B}'\right)\bar{\psi}^1=\bar{\phi}^0,
\end{equation}
which can be achieved by using the simple LU decomposition method.  Once the solution for $\bar{\psi}^1$ is obtained, we repeat the process to find $\bar{\phi}^1$ and in turn $\bar{\psi}^2$ and so on until we reach the solution for $\bar{\psi}$ at time $t$.

\subsubsection{Numerical Results}
After running the numerical simulations (code described in Appendix~\ref{Code}) with the parameters: $250$ elements, time-step $dt=0.5$, infinite well size of $-20\leq x \leq 20$, and the wave packet initially centered at $x_0=0$, we obtain data for the real and imaginary parts of the time evolution of the wave packet.  A selection of these results and their square-sums ($Re^2+Im^2$) are plotted in Figs~\ref{infinite-packet-timestep-1} and \ref{infinite-packet-timestep-2}.  In these plots it can be seen that the wave packet moves to the right until it collides with the infinite potential barrier on the right.  It is reflected, and then it continues to the left side of the well, where it again rebounds to head back to the right.  On collision with the infinite walls the Gaussian envelope undergoes a distortion.  This is due to the fact that the real and imaginary parts of the wave-packet, even though they are not physically observable, undergo phase changes on reflection.  If this simulation is run long enough the Gaussian envelope will spread out until it covers the entire well \cite{liboff}\footnote{This can be seen in Fig.~\ref{infinite-packet-timestep-k=low-barrier} where a low energy wave-packet is placed in an infinite well divided by a finite barrier (which is larger relative to the energy of the packet by a factor of $2.5$).  As the wave-packet is trapped on the left side, and very little is transmitted to the right side, it eventually spreads and covers the entire left half of the well.}.

\begin{figure}
\begin{center}
\begin{tabular}{cc}
\includegraphics[width=0.5\textwidth]{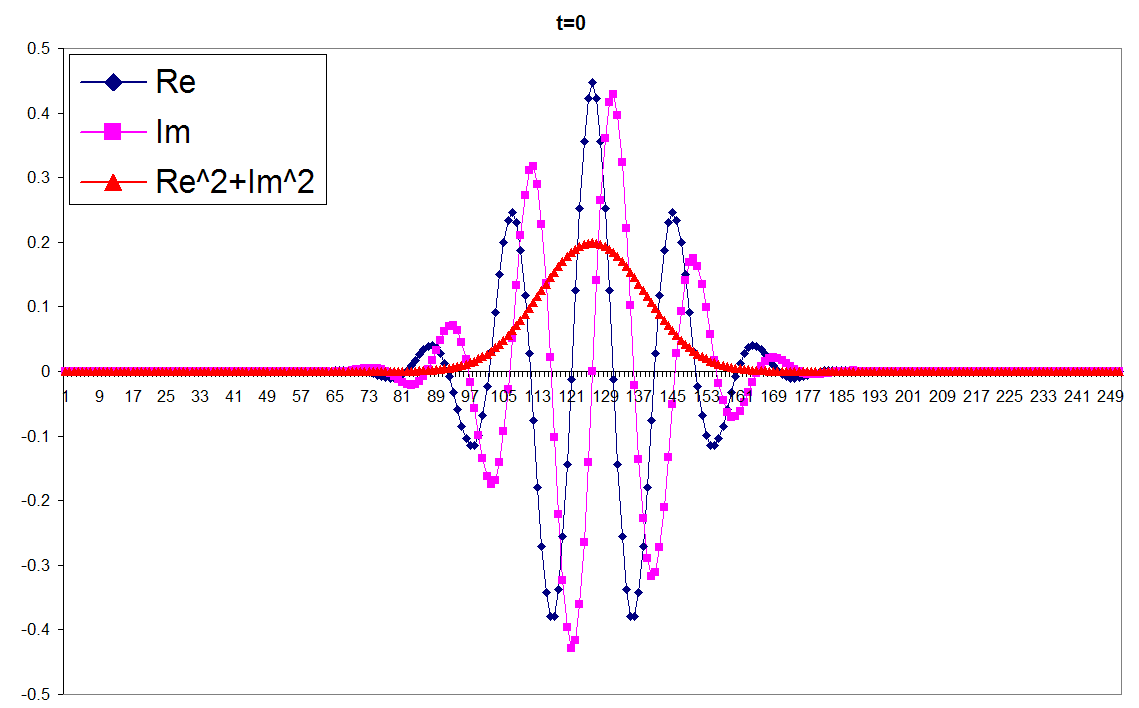}&
\includegraphics[width=0.5\textwidth]{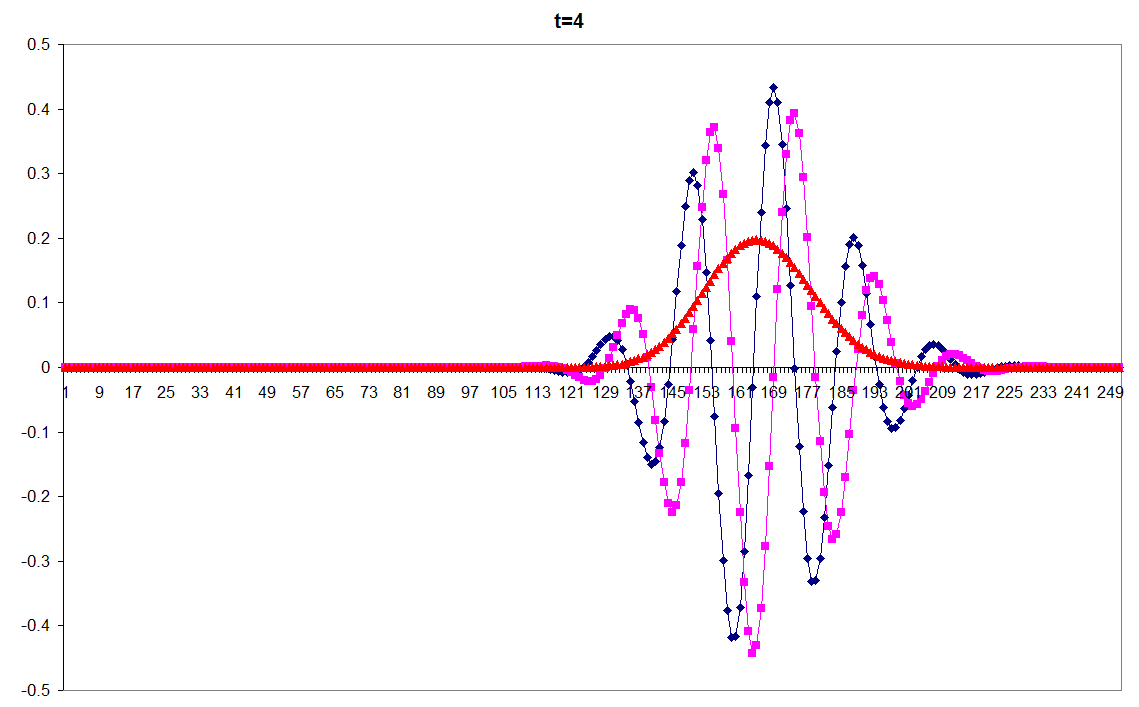}\\
\includegraphics[width=0.5\textwidth]{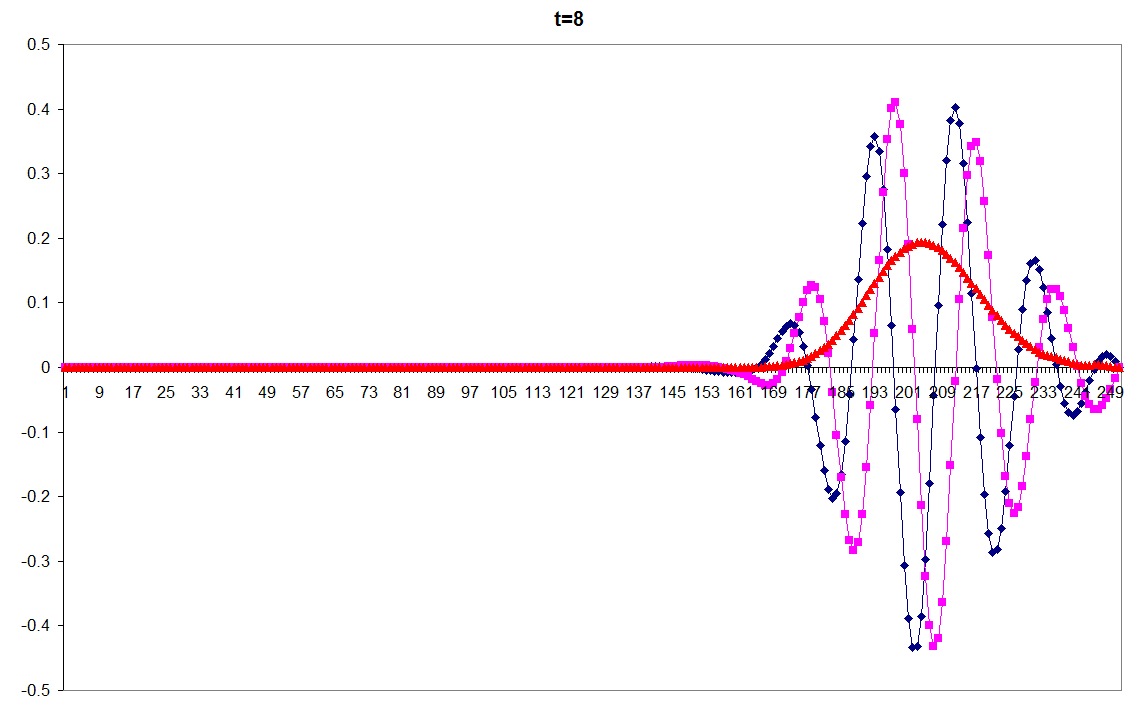}&
\includegraphics[width=0.5\textwidth]{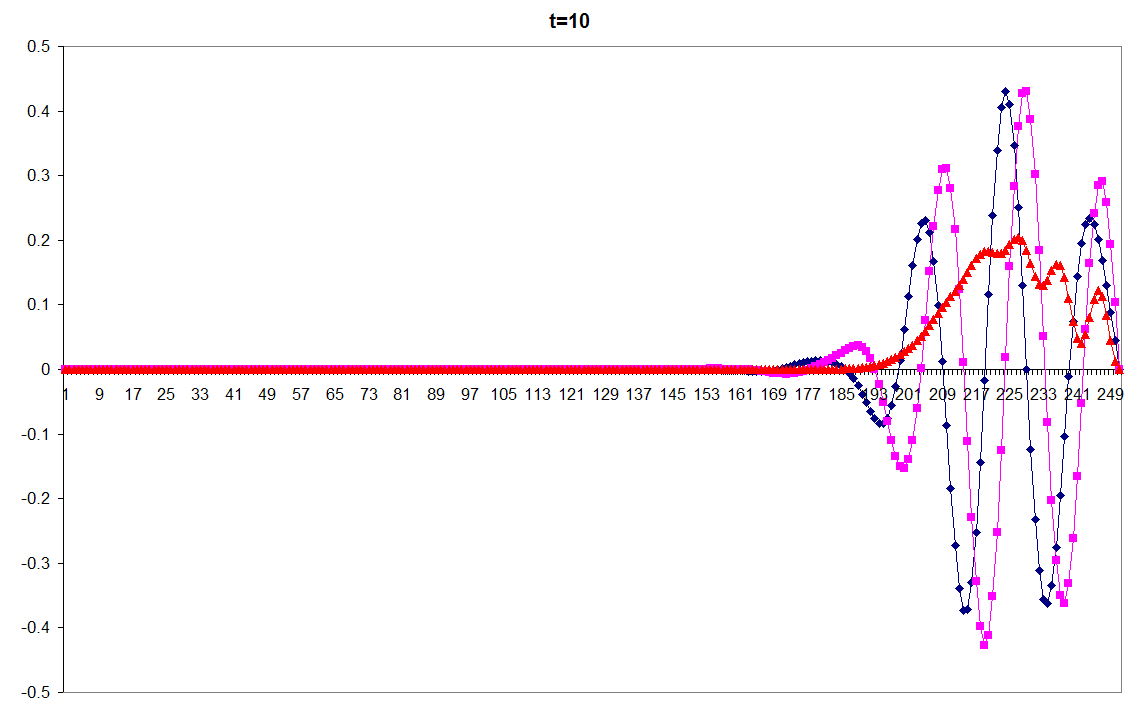}\\
\includegraphics[width=0.5\textwidth]{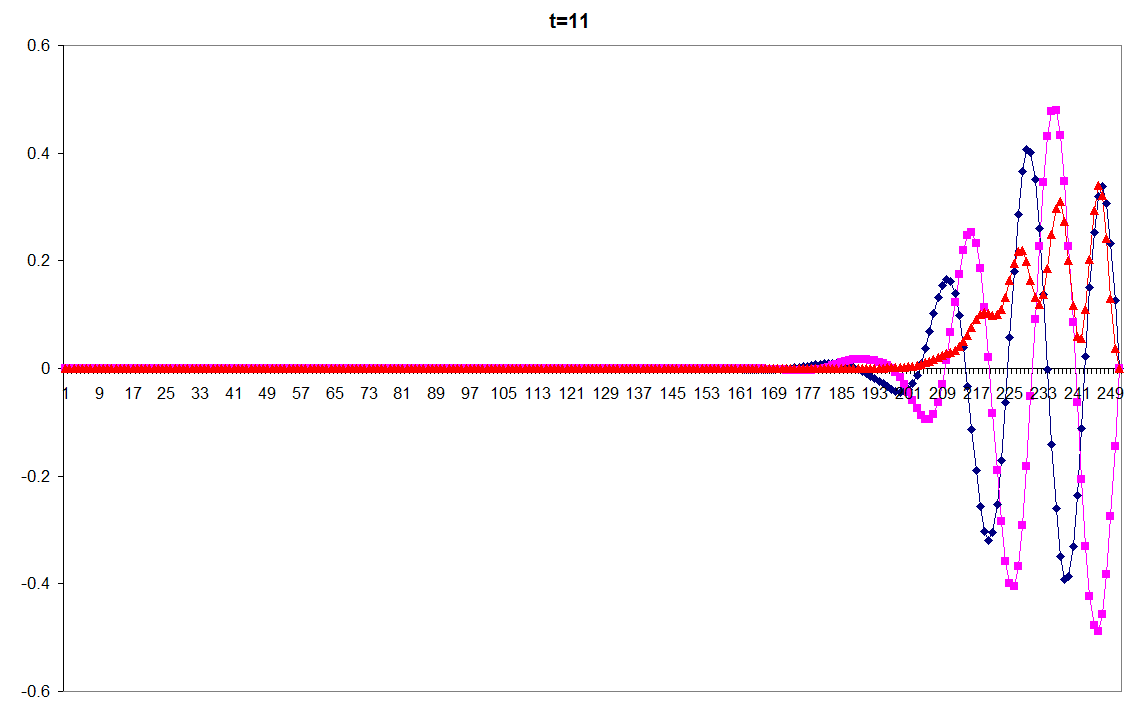}&
\includegraphics[width=0.5\textwidth]{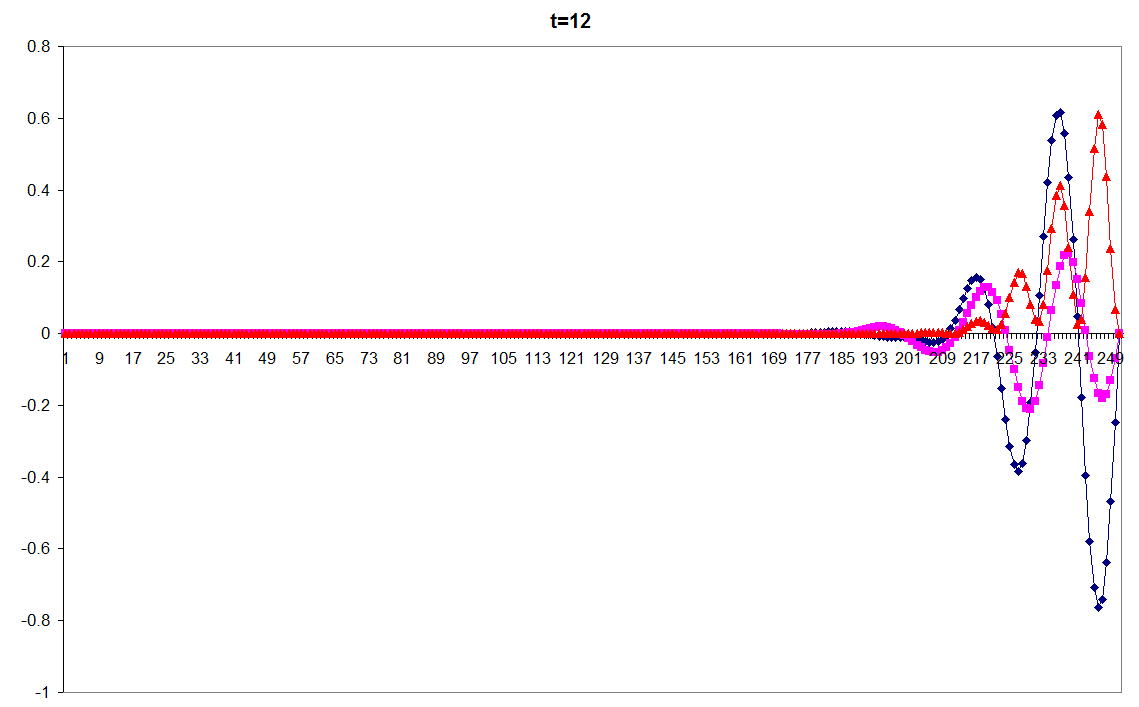}\\
\includegraphics[width=0.5\textwidth]{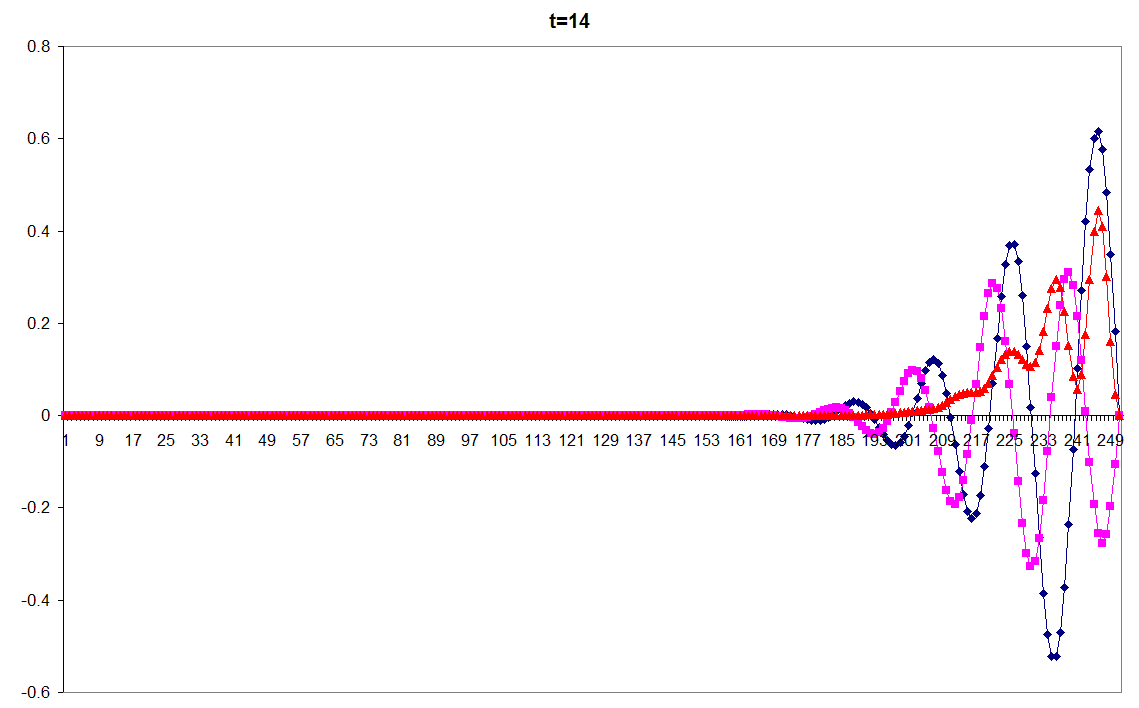}&
\includegraphics[width=0.5\textwidth]{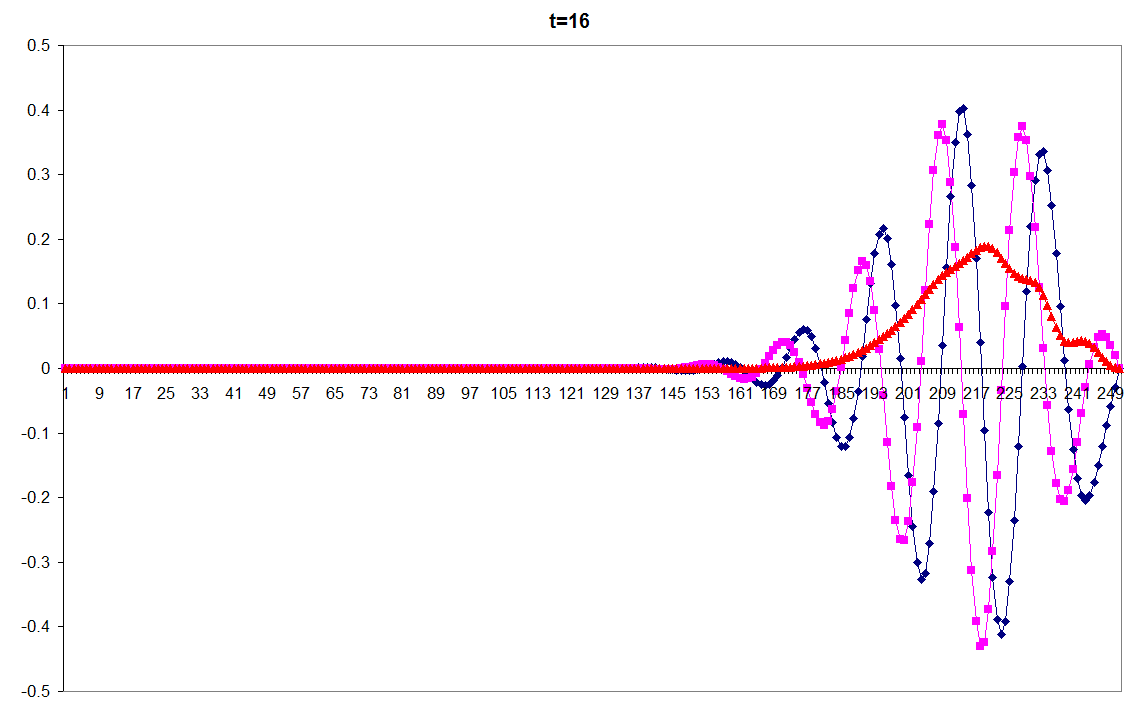}\\
\includegraphics[width=0.5\textwidth]{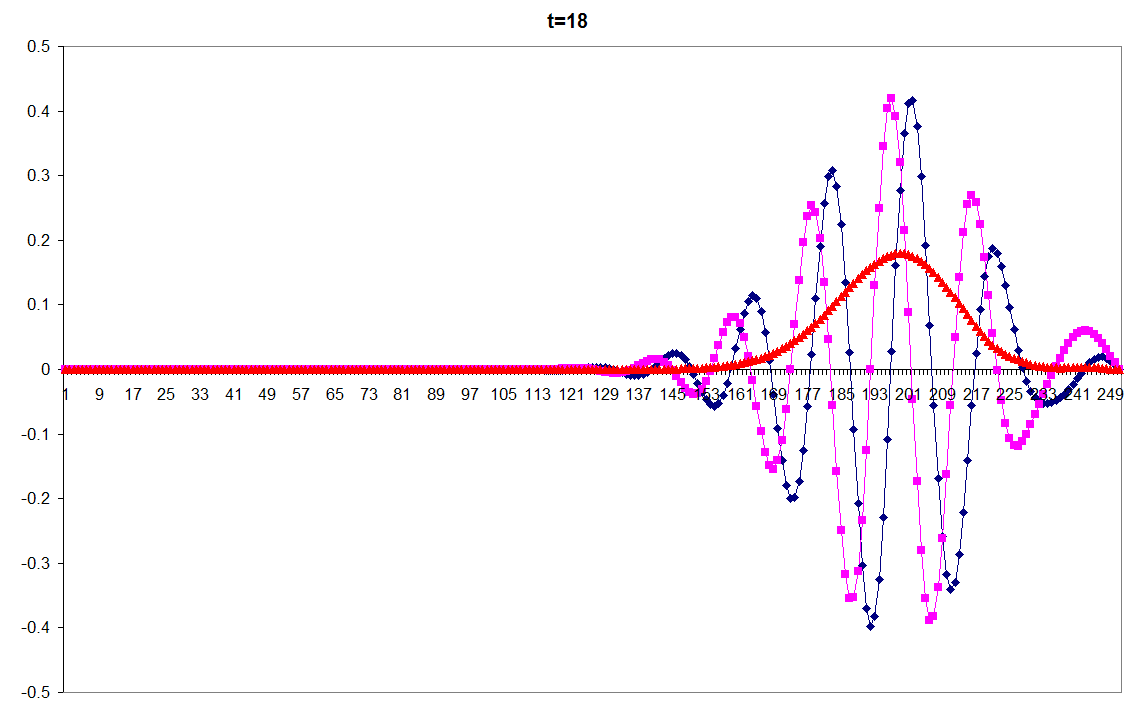}&
\includegraphics[width=0.5\textwidth]{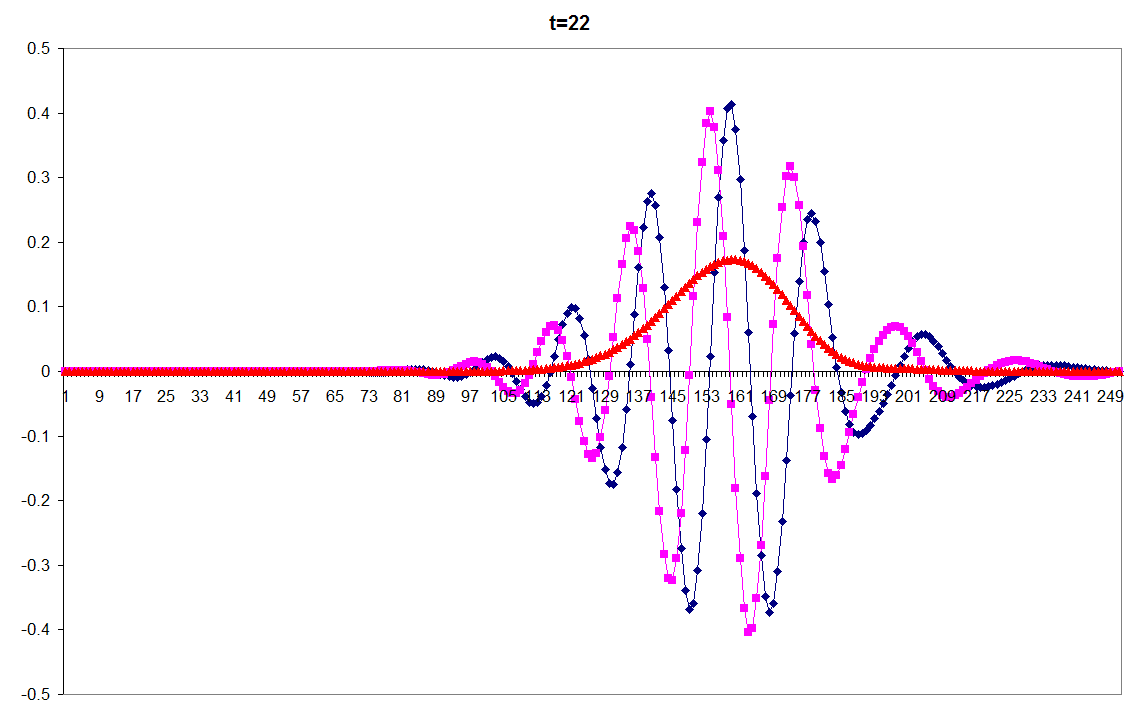}\\
\end{tabular}
\caption{Wave packet trapped in a infinite well (cont. on next page).}
\label{infinite-packet-timestep-1}
\end{center}
\end{figure}

\begin{figure}
\begin{center}
\begin{tabular}{cc}
\includegraphics[width=0.5\textwidth]{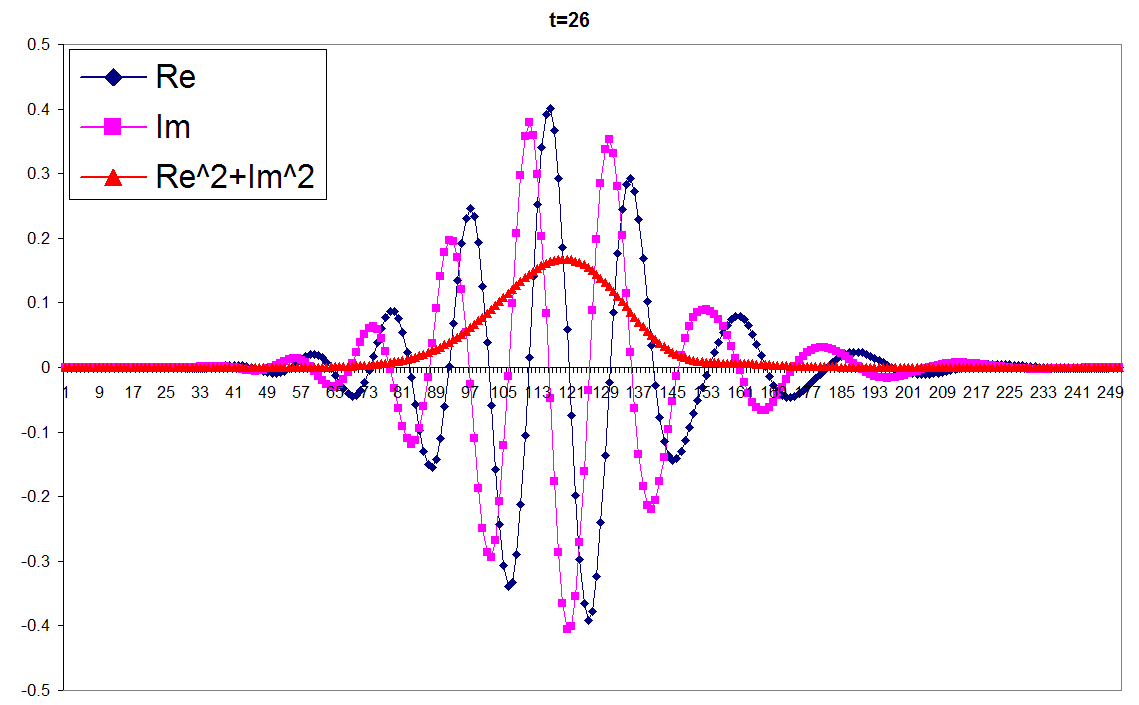}&
\includegraphics[width=0.5\textwidth]{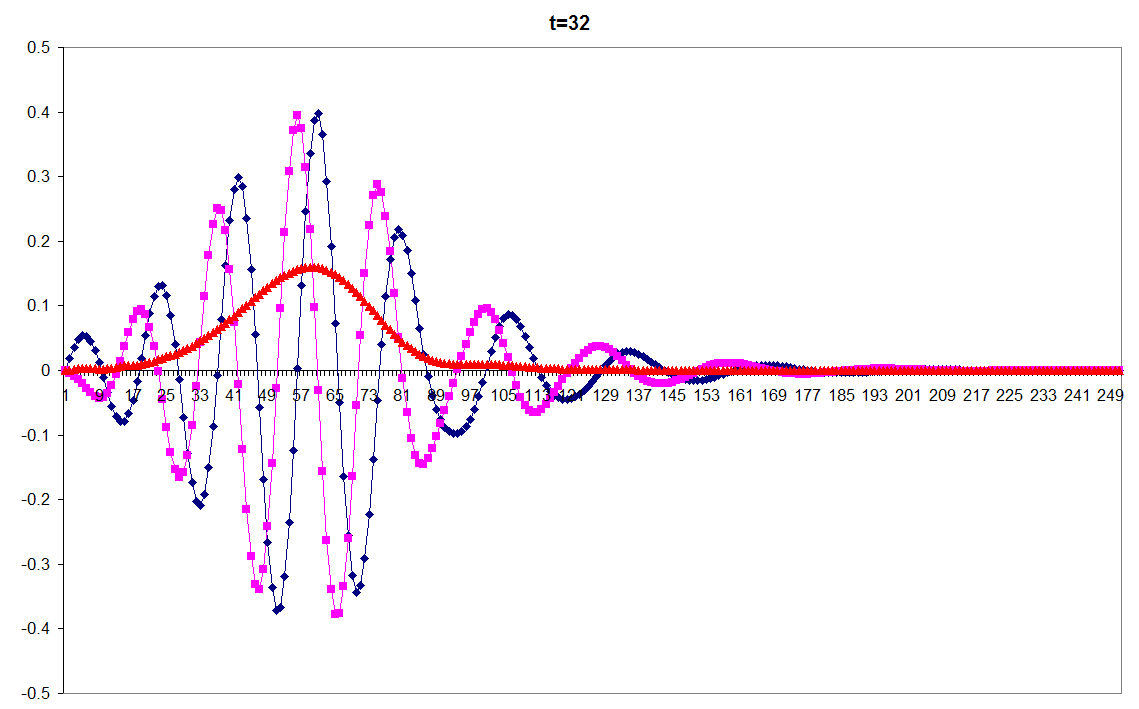}\\
\includegraphics[width=0.5\textwidth]{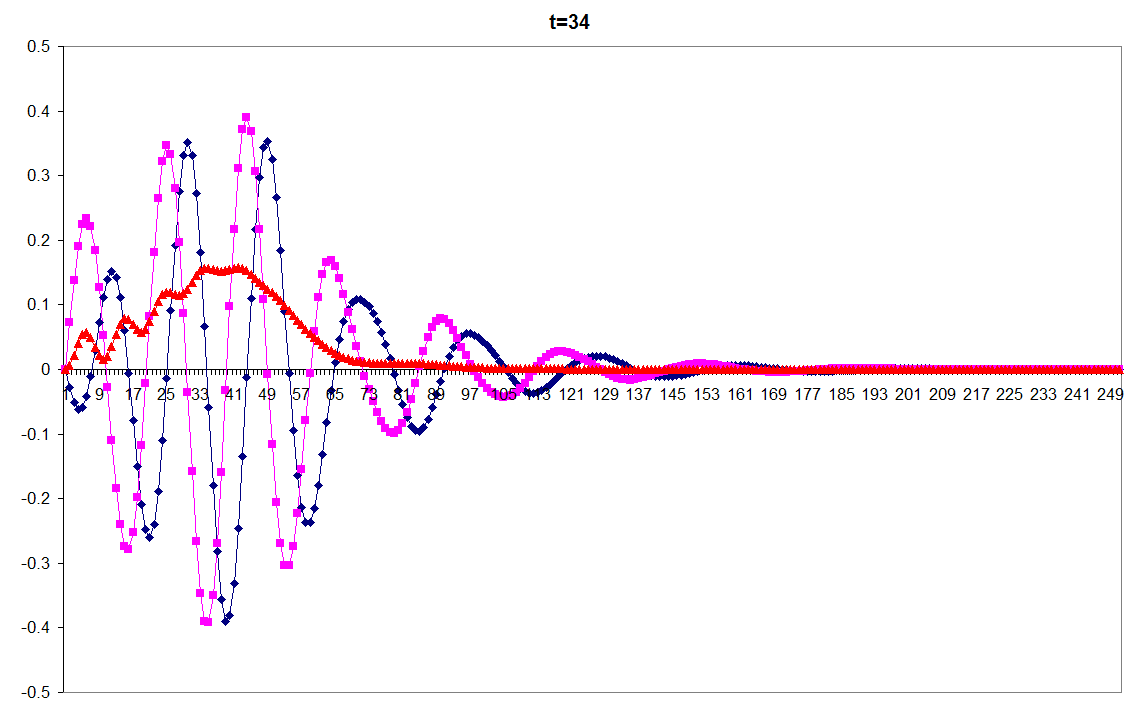}&
\includegraphics[width=0.5\textwidth]{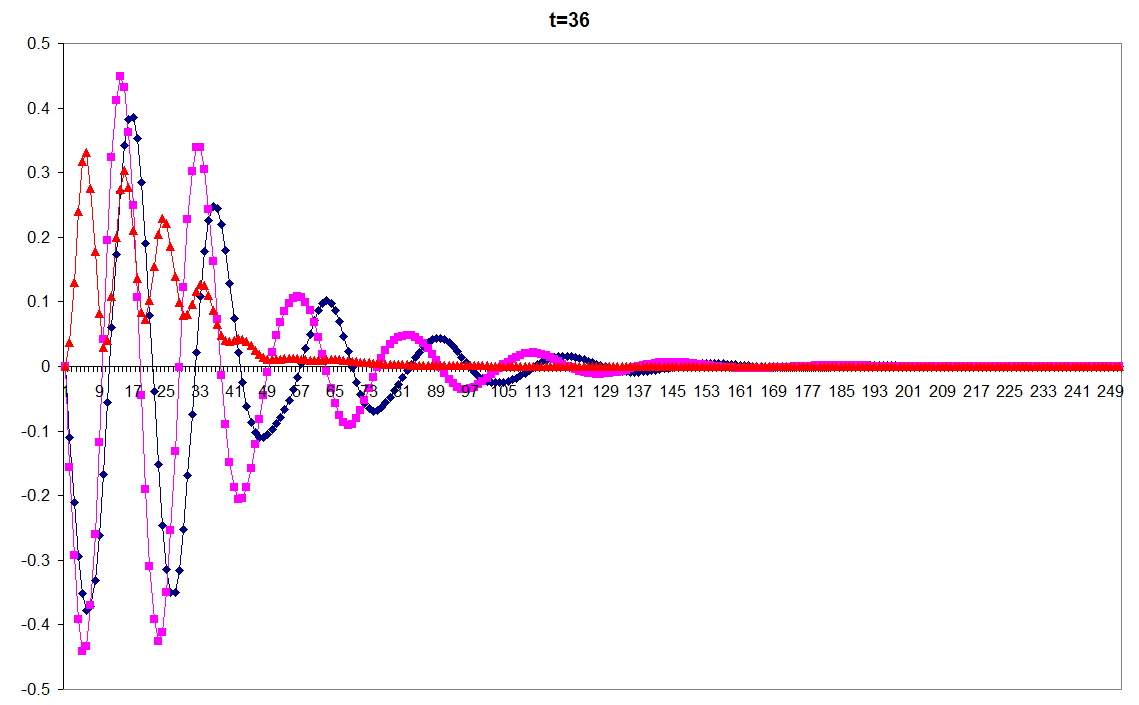}\\
\includegraphics[width=0.5\textwidth]{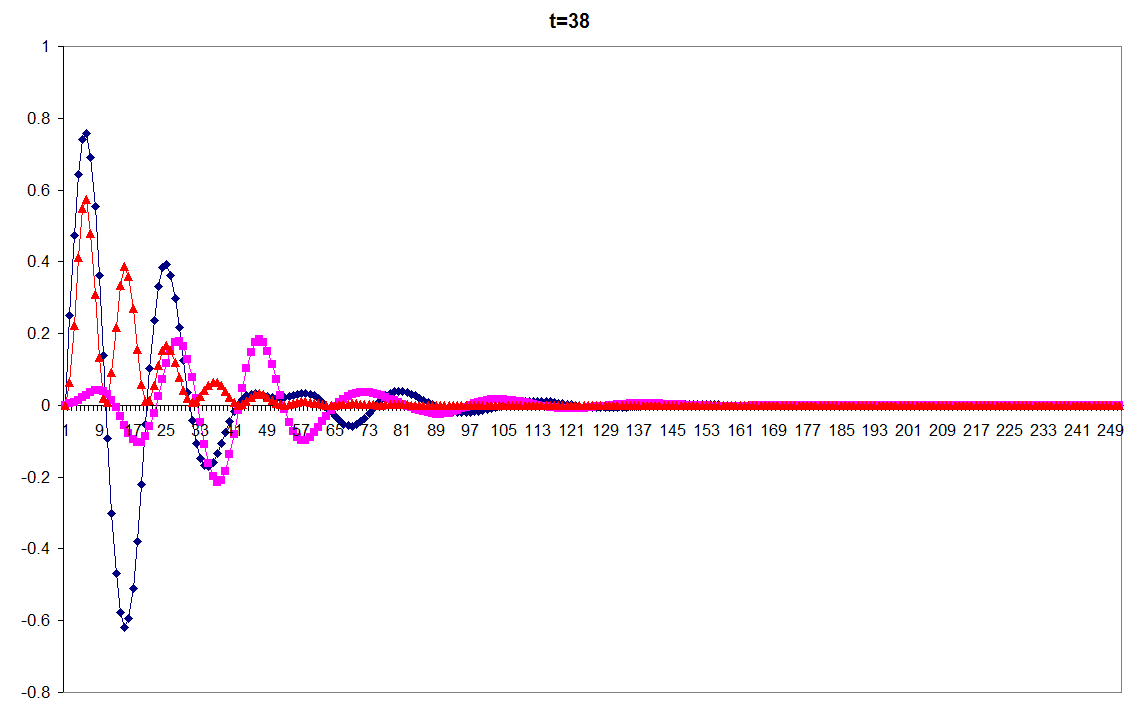}&
\includegraphics[width=0.5\textwidth]{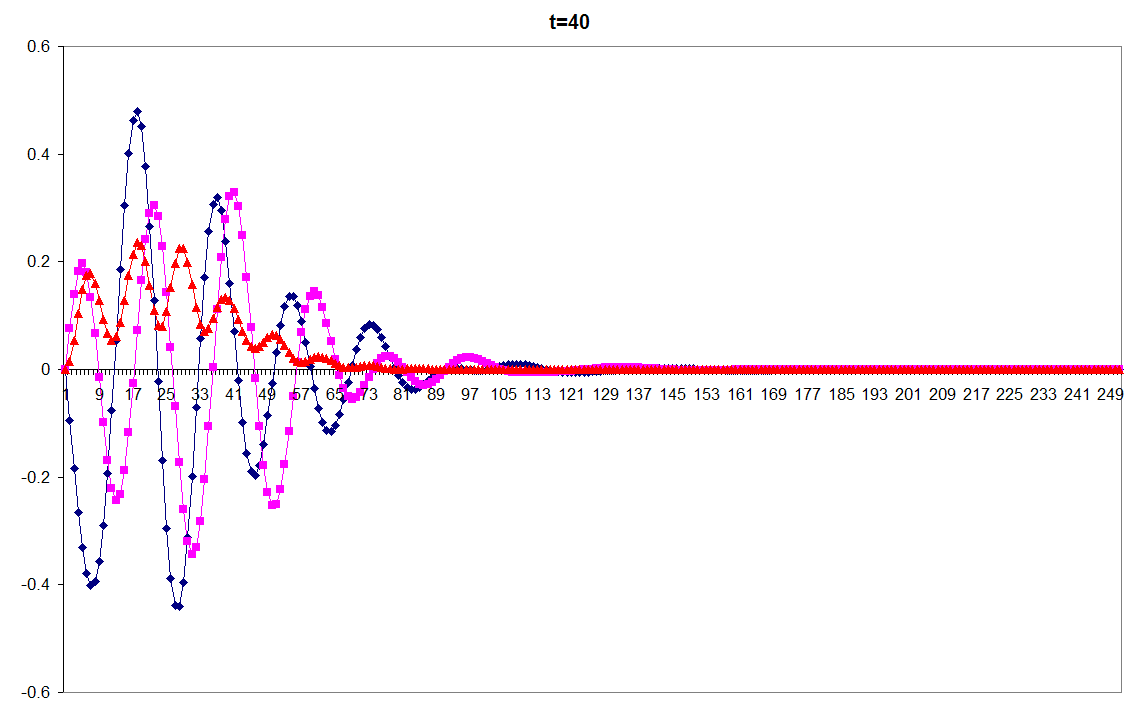}\\
\includegraphics[width=0.5\textwidth]{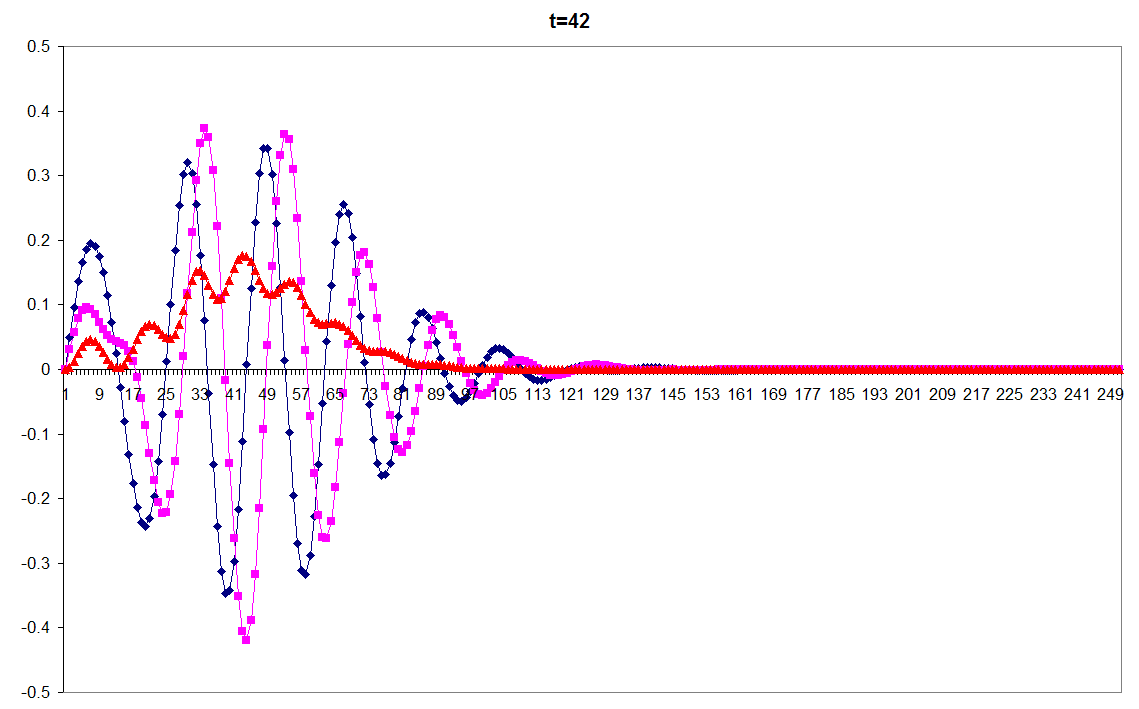}&
\includegraphics[width=0.5\textwidth]{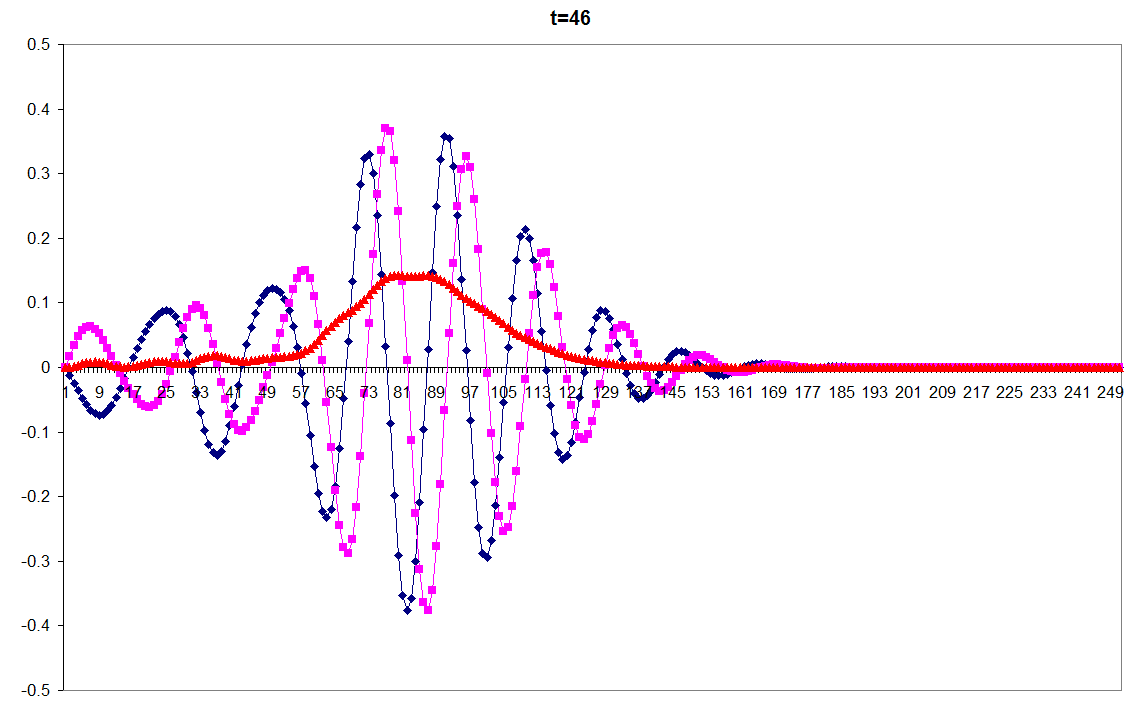}\\
\includegraphics[width=0.5\textwidth]{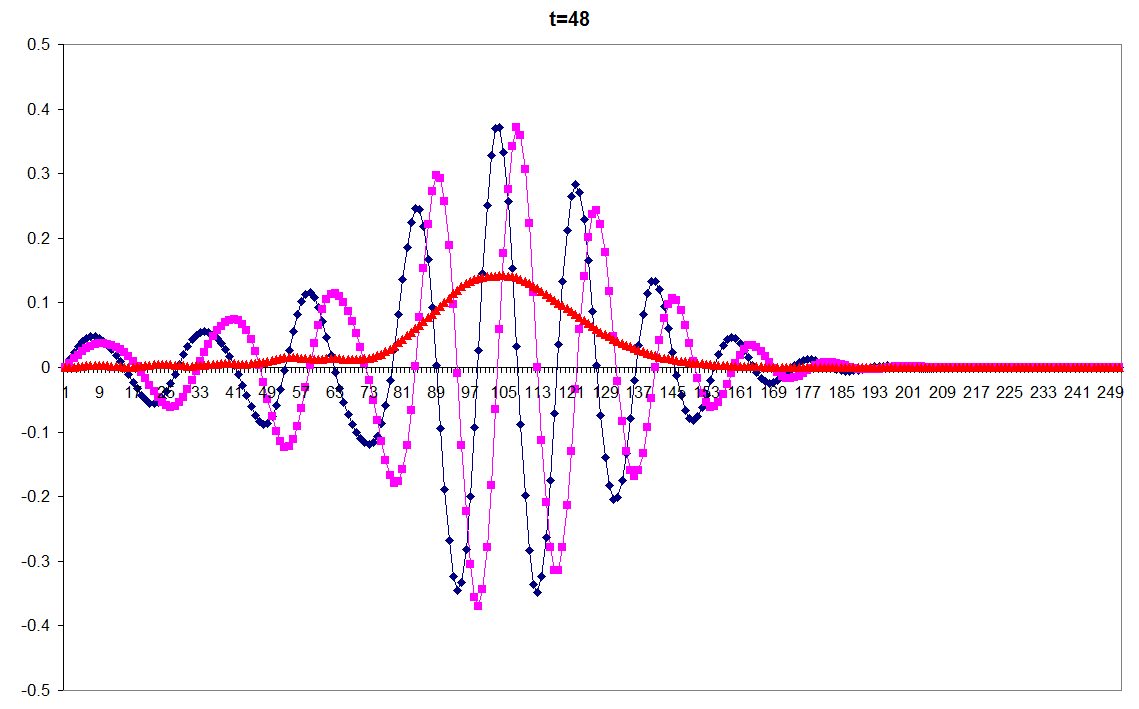}&
\includegraphics[width=0.5\textwidth]{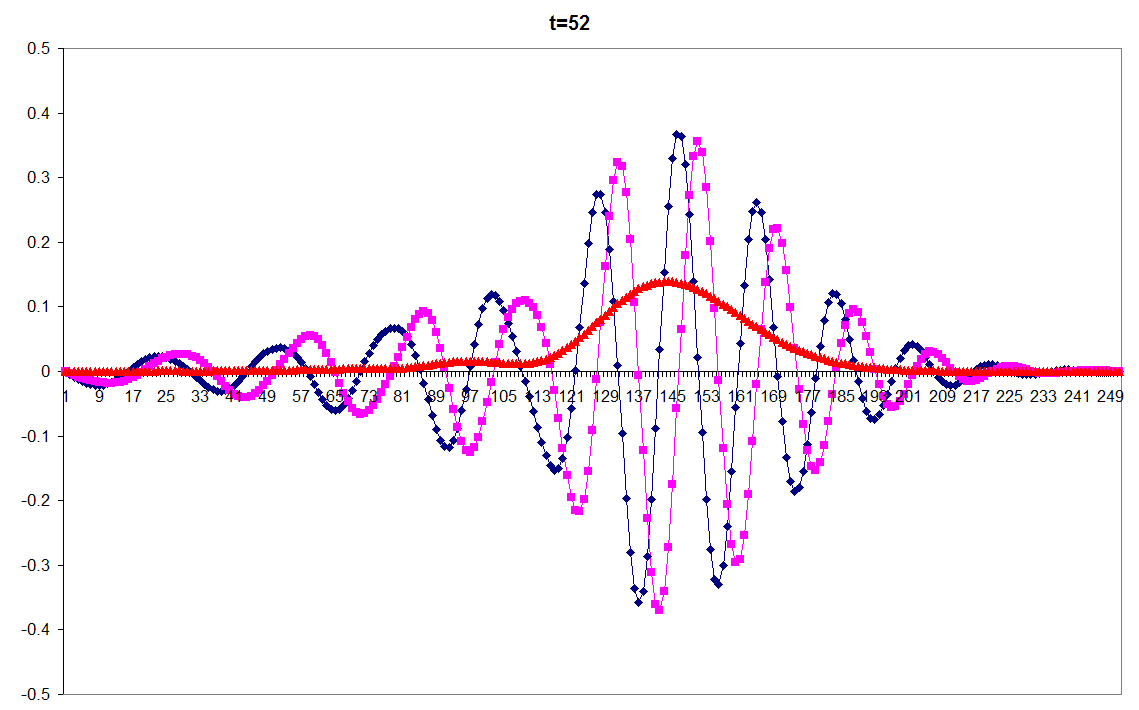}\\
\end{tabular}
\caption{(Cont. from previous page) Wave packet trapped in an infinite well.}
\label{infinite-packet-timestep-2}
\end{center}
\end{figure}

In order to study the conservation of probability we incorporated a simple trapezoidal rule to sum the areas under the elements at each time step - this was done for varying element sizes in order to study the accuracy.  In Fig.~\ref{area-conservation-Infinite} it can be seen that when 50 elements or more were used the total area after each time step remains constant at 1, as expected.  Using only 5 elements the area drops to $\approx0.43193$, however it remains constant at each time step.  The difference in areas for the varying number of elements can simply be accounted for by the fact that with a lower number of elements the shape of the solution can not be resolved accurately, hence the area under the curves is not representative of the actual area.  On the other hand, as the area is always constant irrespective of the number of elements, it implies the conservation of probability property is maintained.  
\begin{figure}
    \begin{center}
     \includegraphics[width=0.8\textwidth]{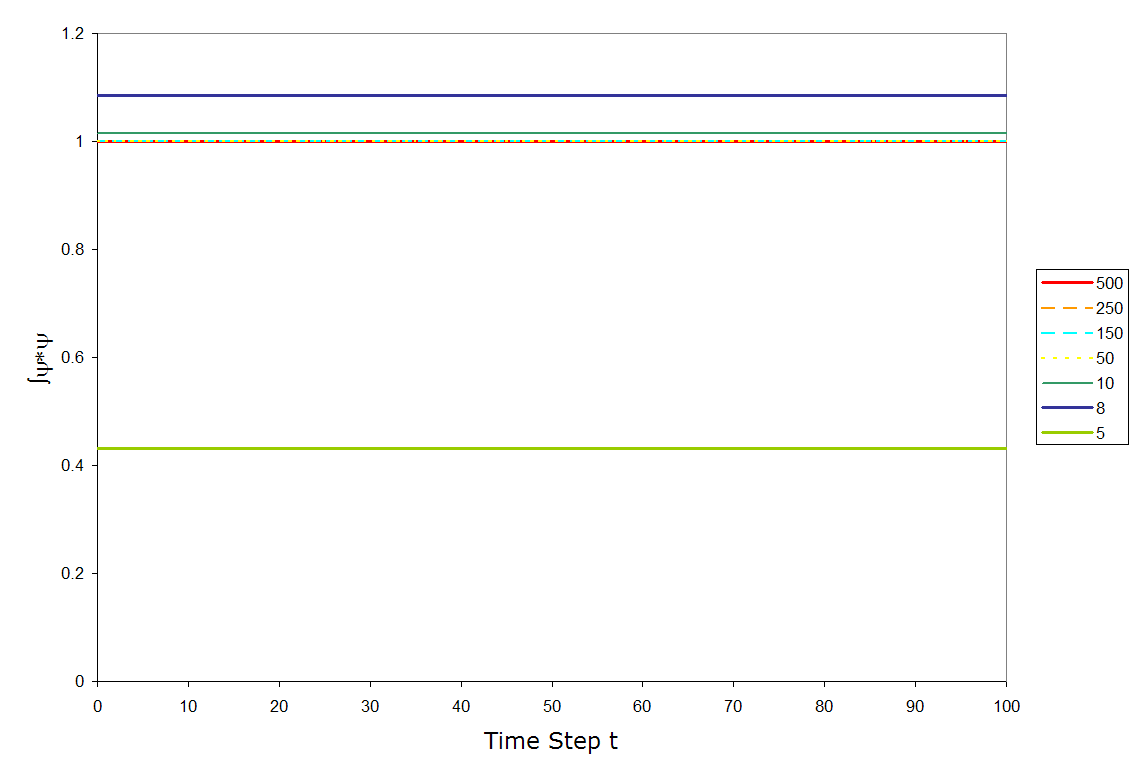}\\
        \caption{Conservation of area at each time step for varying number of elements.}
        \label{area-conservation-Infinite}
    \end{center}
\end{figure}

\subsubsection{Wave-Packet with $k=0$}
If we set the packet wave-number, $k_0$, in Eqn.~(\ref{Gaussian-wave-packet-model}) to zero, we have a stationary wave-packet, which represents a particle at rest.  The numerical results in this case behave as expected, as the packet disperses and spreads over the infinite well, Fig.~(\ref{infinite-packet-k=0}). 
\begin{figure}
\begin{center}
\begin{tabular}{cc}
\includegraphics[width=0.5\textwidth]{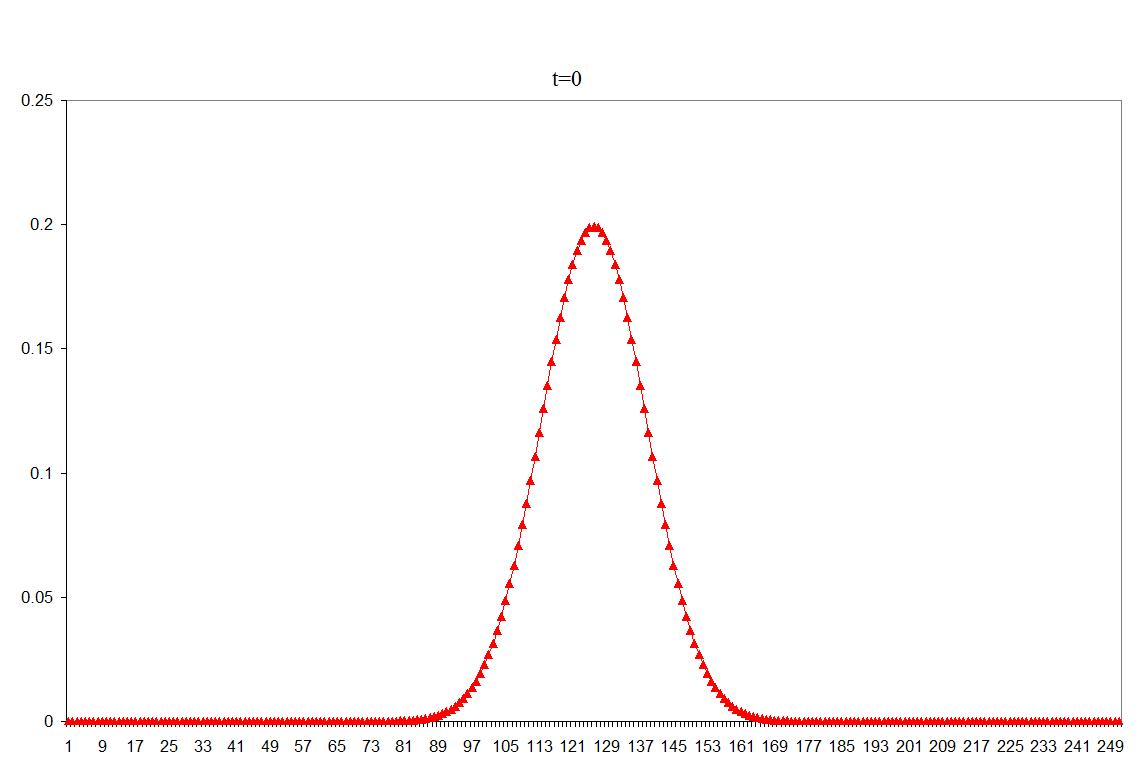}&
\includegraphics[width=0.5\textwidth]{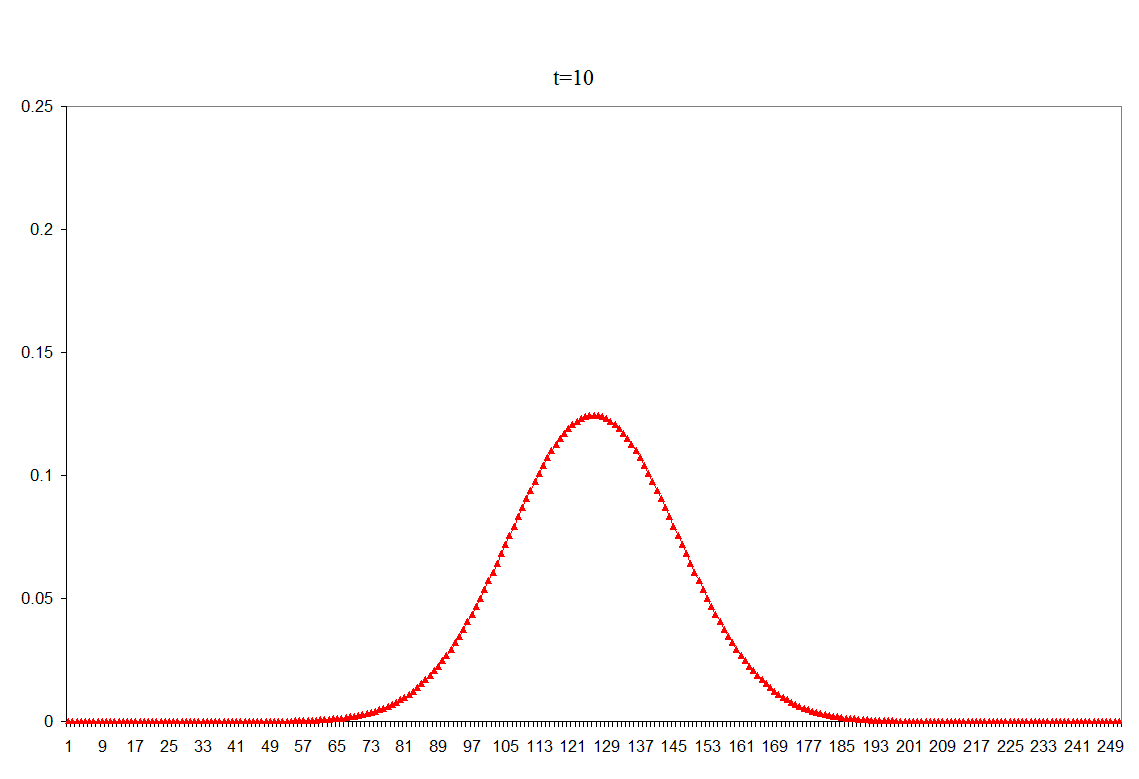}\\
\includegraphics[width=0.5\textwidth]{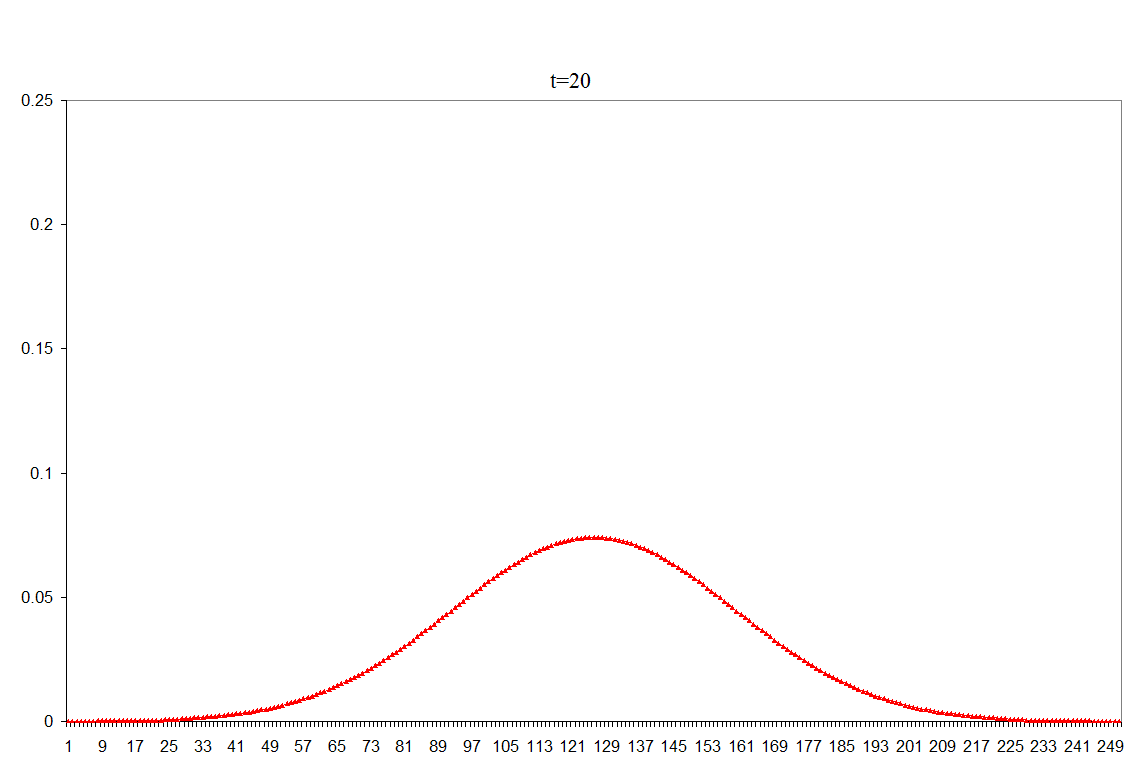}&
\includegraphics[width=0.5\textwidth]{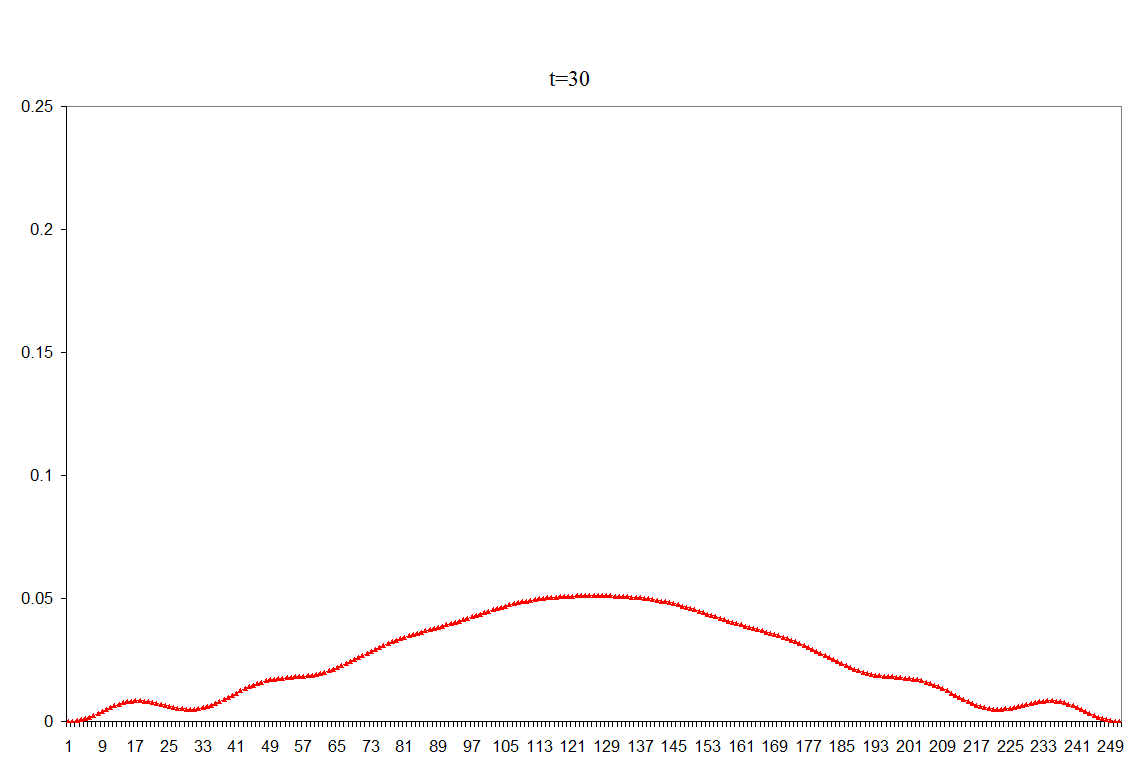}\\
\includegraphics[width=0.5\textwidth]{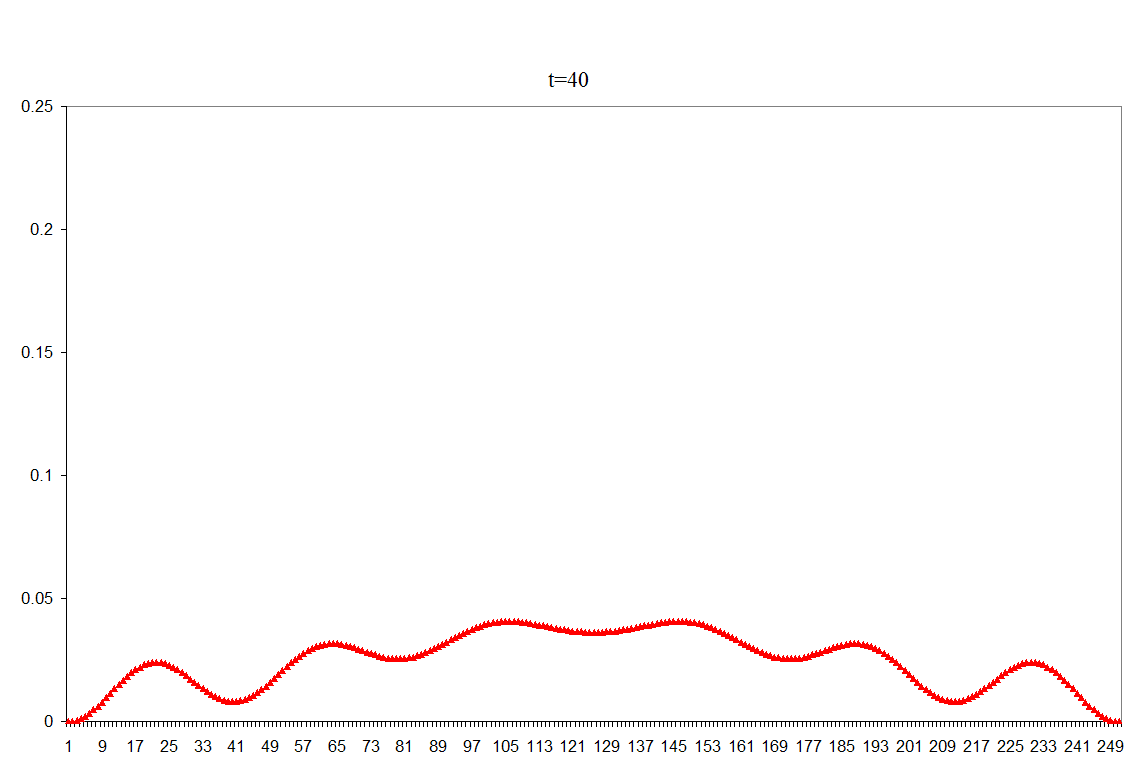}&
\includegraphics[width=0.5\textwidth]{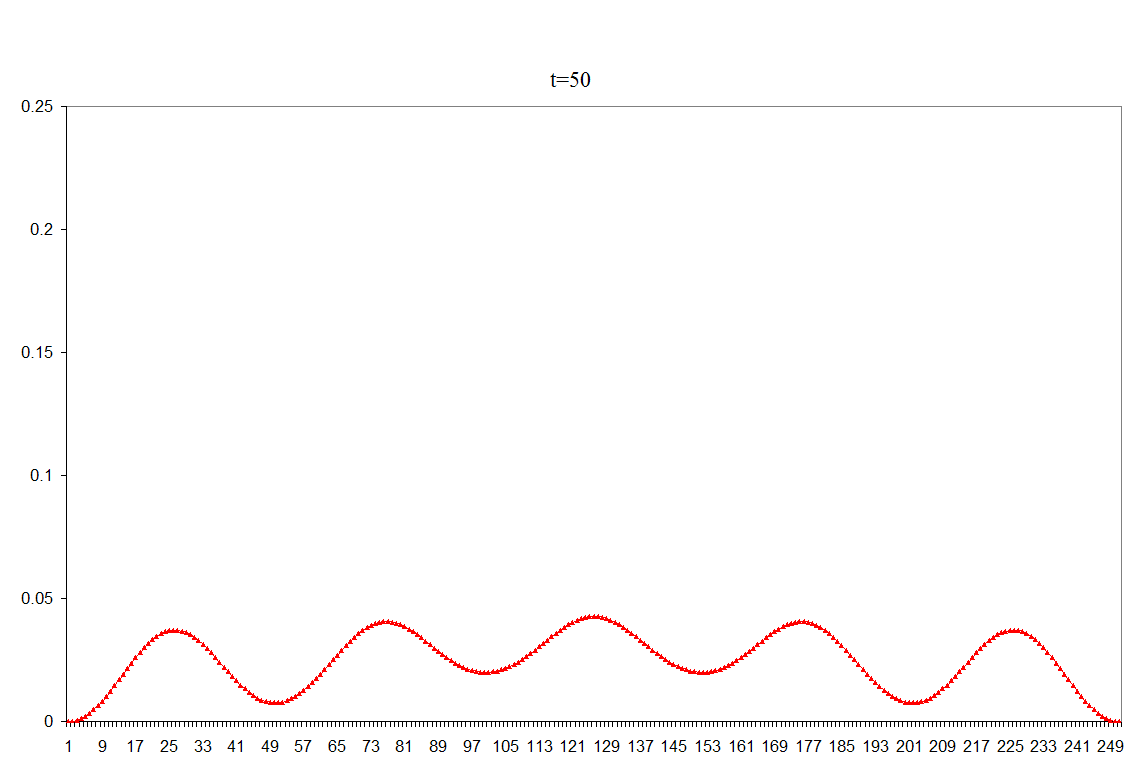}\\
\includegraphics[width=0.5\textwidth]{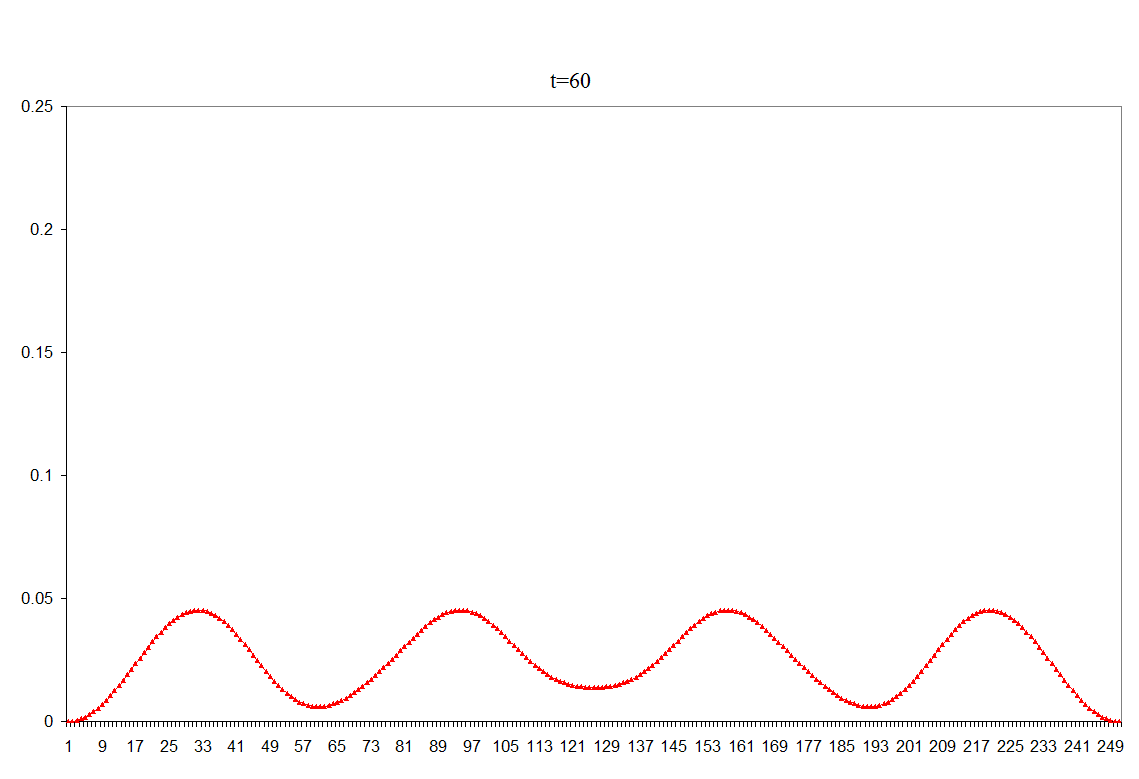}&
\includegraphics[width=0.5\textwidth]{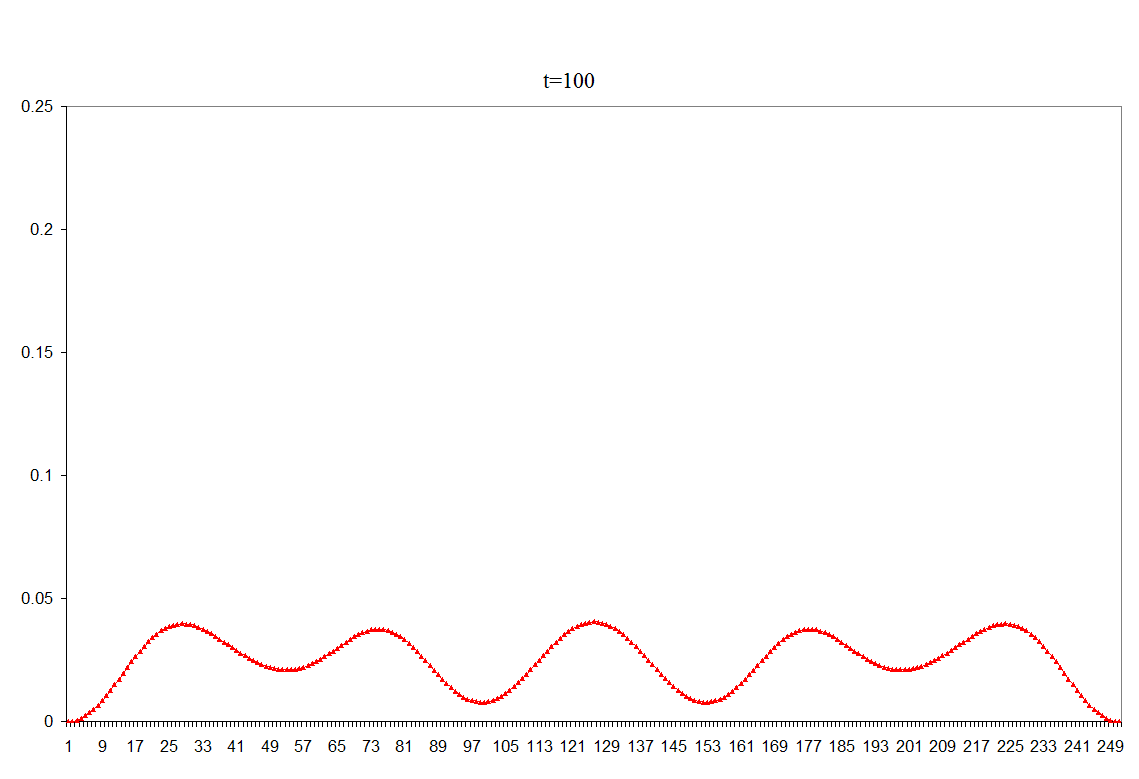}\\
\end{tabular}
\caption{Wave packet with $k_0=0$ spreads over the infinite well ($Re^2+Im^2$ are shown).}
\label{infinite-packet-k=0}
\end{center}
\end{figure}  
We can also note that as the packet spreads over the well with time, the area remains constant, Fig.~(\ref{area-conservation-infinite-k=0}).
\begin{figure}
    \begin{center}
     \includegraphics[width=0.7\textwidth]{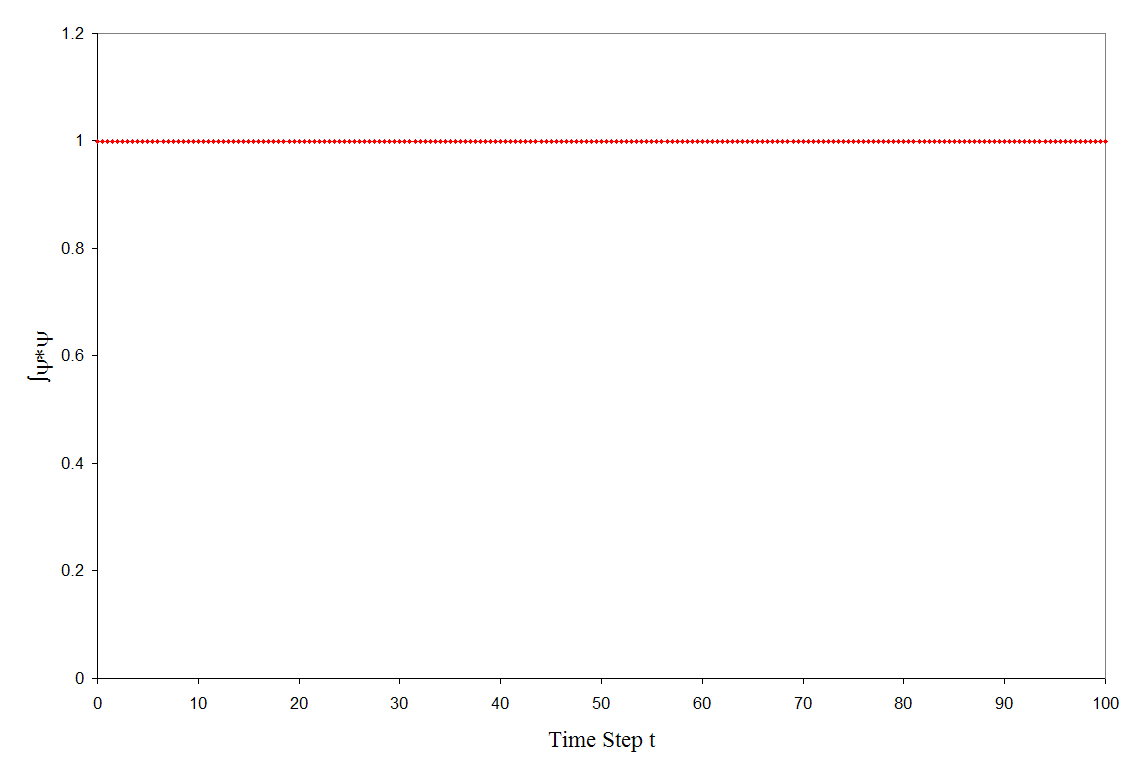}\\
        \caption{Conservation of area at each time step for packet with $k_0=0$.}
        \label{area-conservation-infinite-k=0}
    \end{center}
\end{figure}

\subsection{Infinite Potential Well with Barrier}
\label{infinite-well-wave-packet-equation}
To model the infinite potential well with a barrier, Fig.~\ref{Infinitebarrierwell},
\begin{figure}
    \begin{center}
     \includegraphics[width=0.6\textwidth]{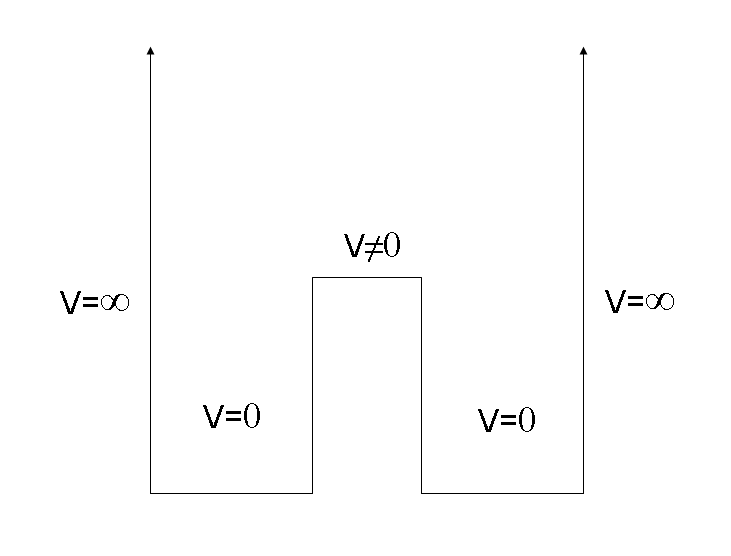}\\
        \caption{Infinite potential well with a barrier.}
        \label{Infinitebarrierwell}
    \end{center}
\end{figure}
we will discretise the \schro equation with the potential term:
\begin{equation}
\frac{\partial \psi}{\partial t}-i\frac{\hbar}{2m}\frac{\partial^2 \psi}{\partial t^2}=-\frac{i}{\hbar}V\psi,
\label{constant-potential-schro}
\end{equation}
where $V$ is a constant potential for the barrier and zero everywhere else in the well.  
The LHS of Eqn.~(\ref{constant-potential-schro}) will be discretised as before, so we will only consider the element equation for the RHS.  Following the procedure of Sec.~\ref{eigen-value}, due to the similarity of the potential term and the eigenvalue term, we derive the potential element equation through the Galerkin technique as
\begin{eqnarray}
-\frac{i}{\hbar}\int^L_0 V(x)\phi \psi dx&=&-\frac{i}{2\hbar}\sum_e l_e V_e\int^{1}_{-1}\bar{\phi}^{\dagger}\textbf{N}^{\dagger}\textbf{N}\bar{\psi}d\xi\nonumber\\
&=&-\frac{i}{2\hbar}\sum_e l_e V_e\int^1_{-1}\bar{\phi}^{\dagger} \left[ \begin{array}{c}
N_1\\
N_2\\
\end{array}
\right] \left[ N_1 \quad N_2 \right]\bar{\psi}d\xi\nonumber\\
&=&\bar{\phi}^{\dagger}\left\{-\frac{i}{6\hbar}\sum_e l_e V_e \left[ \begin{array}{cc}
2 & 1\\
1 & 2\\
\end{array}
\right] \bar{\psi}\right\}
\label{standard-potential}
\end{eqnarray}
We can then write the total \schro element equation, including the potential term, as
\begin{equation}
\left[
\begin{array}{cc}
	2 & 1 \\
	1 & 2 \\
\end{array}
\right]\dot{\bar{\psi}}+i\left\{\frac{6\hbar}{2ml_e^2}\left[
\begin{array}{cc}
	1 & -1 \\
	-1 & 1 \\
\end{array}
\right]+\frac {V_e}{\hbar} \left[ \begin{array}{cc}
2 & 1\\
1 & 2\\
\end{array}
\right] \right\}\bar{\psi}=\textbf{0}.
\label{full-schro-element-equation}
\end{equation}
Writing this in operator form we have
\begin{equation}
\tilde{\textbf{A}}\dot{\bar{\psi}}+i\left\{ \tilde{\textbf{B}}+\tilde{\textbf{C}}\right\}\bar{\psi}=\textbf{0},
\end{equation}
where
\begin{equation}
\tilde{\textbf{A}}=\left[
\begin{array}{cc}
	2 & 1 \\
	1 & 2 \\
\end{array}
\right]\textrm{,} \qquad \tilde{\textbf{B}}=\frac{6\hbar}{2ml_e^2}\left[
\begin{array}{cc}
	1 & -1 \\
	-1 & 1 \\
\end{array}
\right]
\qquad
\textrm{and}
\qquad \tilde{\textbf{C}}=\frac {V_e}{\hbar}\left[
\begin{array}{cc}
	2 & 1 \\
	1 & 2 \\
\end{array}
\right].
\end{equation}
Applying the Crank-Nicolson approximation, as before, we obtain
\begin{equation}
\left(\tilde{\textbf{A}}'+i\left\{\tilde{\textbf{B}}' + \tilde{\textbf{C}}'\right\}\right) \bar{\psi}^{n+1}=
\left(\tilde{\textbf{A}}'-i\left\{\tilde{\textbf{B}}'+\tilde{\textbf{C}}'\right\}\right)\bar{\psi}^{n},
\label{cn-fd-schro-potential}
\end{equation}
where $\tilde{\textbf{A}}'= \tilde{\textbf{A}}$, but $\tilde{\textbf{B}}'$ and $\tilde{\textbf{C}}'$ are given as
\begin{equation}
\tilde{\textbf{B}}'=\frac{6\hbar\Delta t}{4 m l_e^2}
\left[
\begin{array}{cc}
	1 & -1 \\
	-1 & 1 \\
\end{array}
\right]\qquad
\textrm{and}
\qquad \tilde{\textbf{C}}'=\frac {V_e\Delta t}{2\hbar}\left[
\begin{array}{cc}
	2 & 1 \\
	1 & 2 \\
\end{array}
\right].
\end{equation}

\subsubsection{Numerical Construction}
\label{barrier-numerical-method}
In order to compute Eqn.~(\ref{cn-fd-schro-potential}) numerically we separate the real and complex parts as in Sec.~\ref{infinite-numerical-method}.  In this way we obtain the four component element vector as in Eqn.~(\ref{4-component-element-vector}), and the element equation becomes
\begin{equation}
\left(
\textbf{A}'+\alpha\textbf{B}'+\beta\textbf{C}'
\right)\bar{\psi}^{n+1}=
\left(
\textbf{A}'-\alpha\textbf{B}'-\beta\textbf{C}'
\right) \bar{\psi}^n,
\label{real-schro-barrier-crank-equation}
\end{equation}
where $\alpha=\frac{6\hbar\Delta t}{4 m l_e^2}$ and $\beta=\frac{V_e\Delta t}{2\hbar}$, and the real matrices $\textbf{A}'$ and $\textbf{B}'$ are as in Eqn.~(\ref{real-matrices-A-B}), and $\textbf{C}'$ is 
\begin{equation}
\textbf{C}'=
 \left[
\begin{array}{cccc}
	0 & -2 & 0& -1 \\
	2 & 0 & 1& 0 \\
	0 & -1 & 0& -2 \\
	1 & 0 & 2& 0 \\
\end{array}
\right]
\label{real-matrix-c}
\end{equation}
Using the initial state function as in Sec.~\ref{initial-state-function-section} the numerical solution of Eqn.~(\ref{real-schro-barrier-crank-equation}) now follows the same procedure as in Sec.~\ref{infinite-well-numerical-solution}.  The most important difference is that for the elements corresponding to the potential barrier $\beta\neq0$, but for all other elements where $\beta=0$ we have the original element equation.

\subsubsection{Numerical Results}
To implement the barrier modifications to the initial infinite well we begin, as before, with an infinite well of size $-20\leq x \leq 20$, and then we incorporate a potential barrier of height $V=2.5$ at the center of the well: $-0.8\leq x \leq 0.8$.  Therefore, if we consider a total of $250$ elements the barrier is located at elements $120\leq x_e \leq 130$; we can then assemble Eqn.~(\ref{real-schro-barrier-crank-equation}) with $V_e=0$ from elements $1$ to $119$, then with $V_e=2.5$ from $120$ to $130$, then again with $V_e=0$ from $131$ to $250$.  Then using a time step $dt=0.5$ and beginning with the initial wave-packet at $x_0=-13$ (located to the left of the barrier) we can obtain the data for the real and imaginary parts of the time evolution of the initial wave packet as before.  A selection of these results and their square-sums are plotted in Figs~\ref{infinite-packet-timestep-1-barrier} and \ref{infinite-packet-timestep-2-barrier}.  In these plots it can be seen that the wave-packet moves to the right until it hits the barrier.  A small part of it is transmitted through the barrier and the rest is reflected back.  The reflected and transmitted parts continue moving until they collide with the infinite walls of the well and return back to the barrier.

\begin{figure}
\begin{center}
\begin{tabular}{cc}
\includegraphics[width=0.5\textwidth]{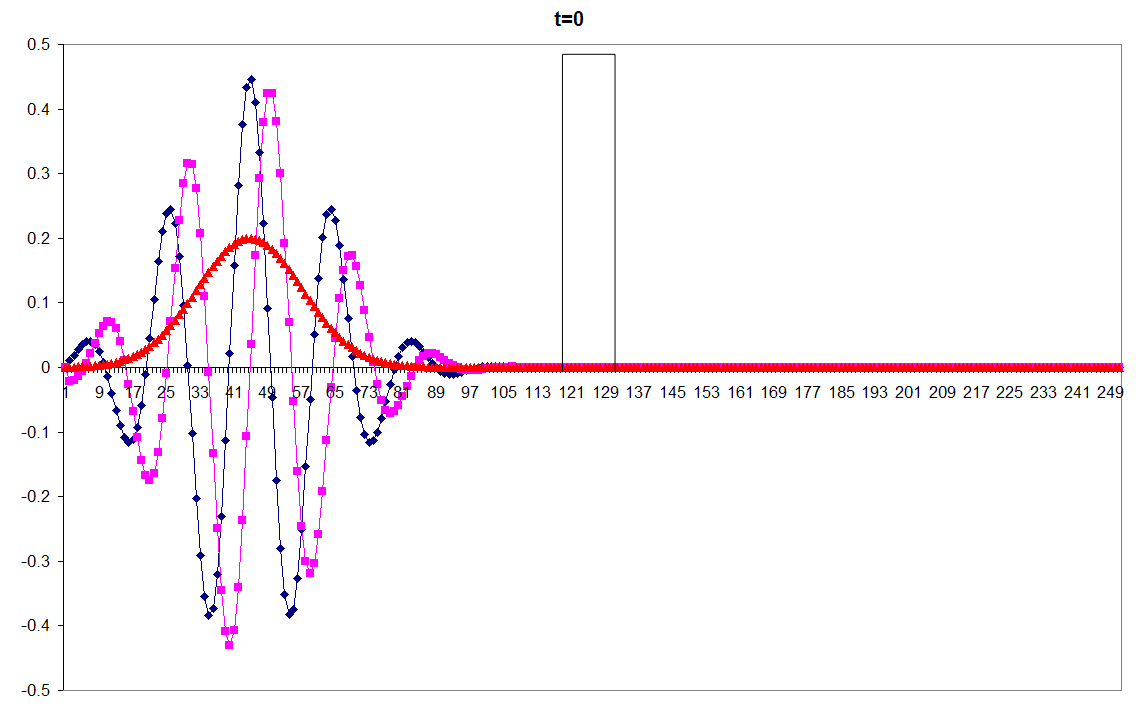}&
\includegraphics[width=0.5\textwidth]{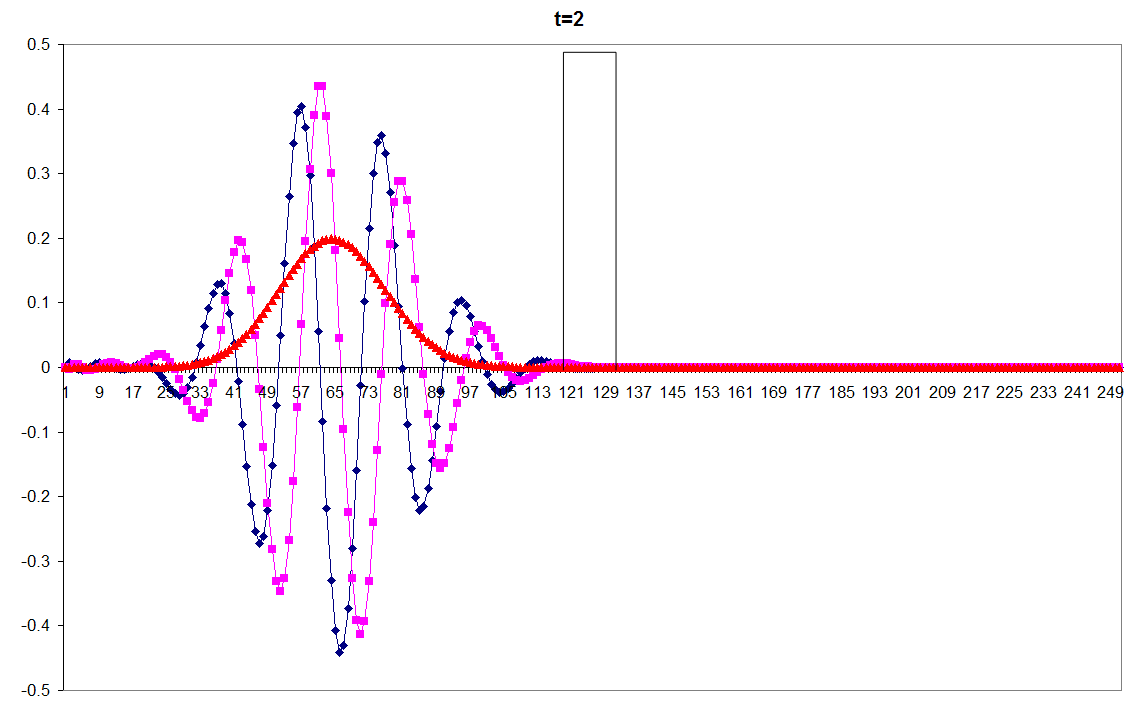}\\
\includegraphics[width=0.5\textwidth]{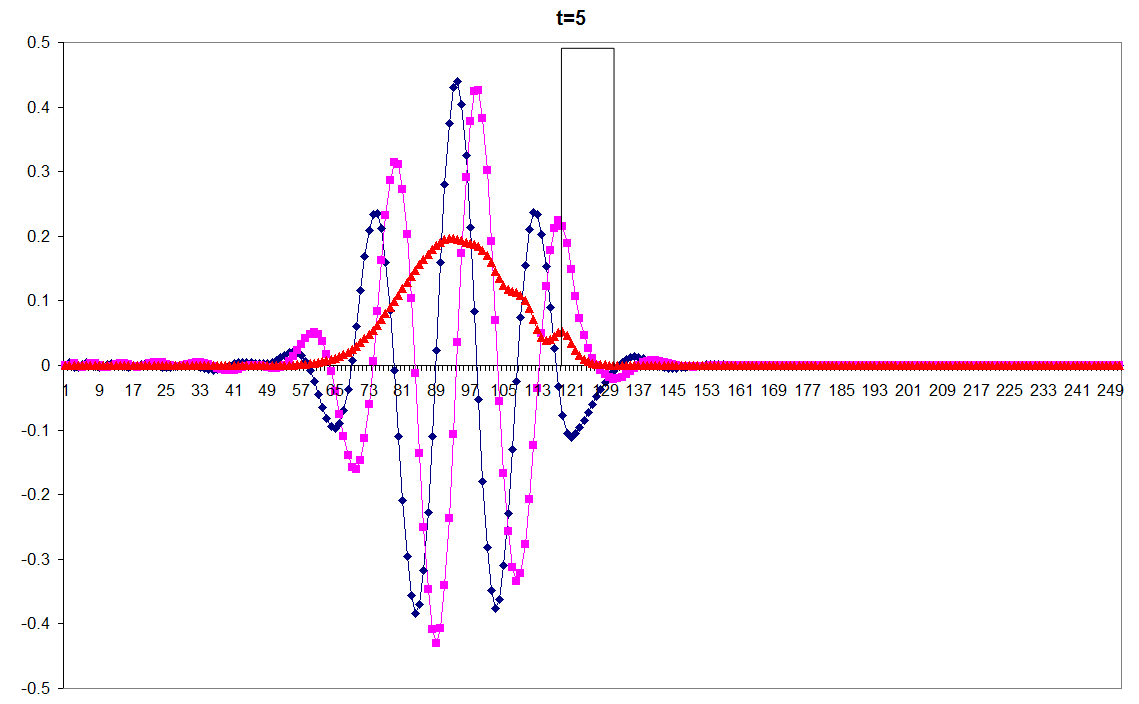}&
\includegraphics[width=0.5\textwidth]{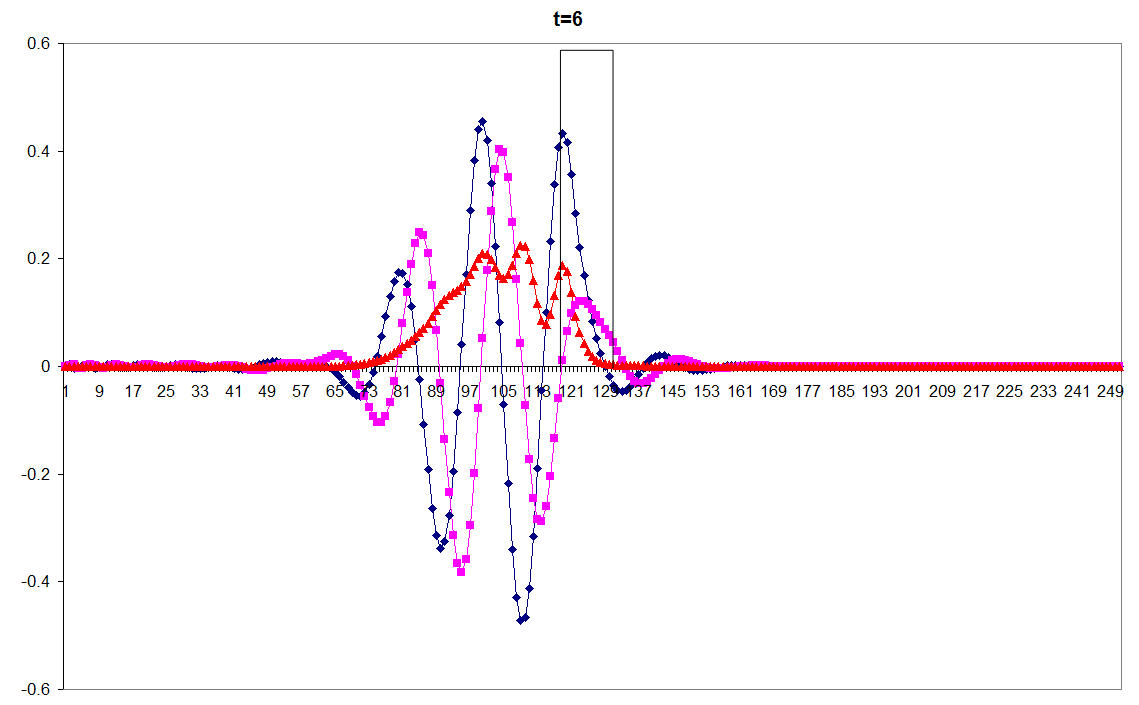}\\
\includegraphics[width=0.5\textwidth]{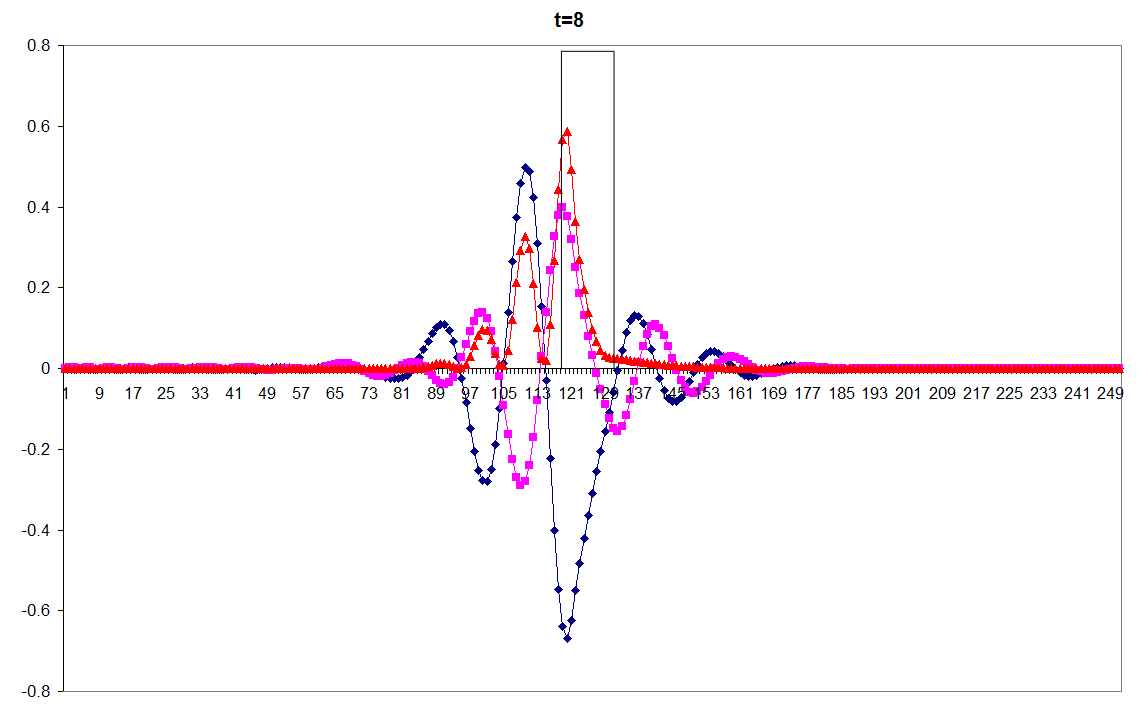}&
\includegraphics[width=0.5\textwidth]{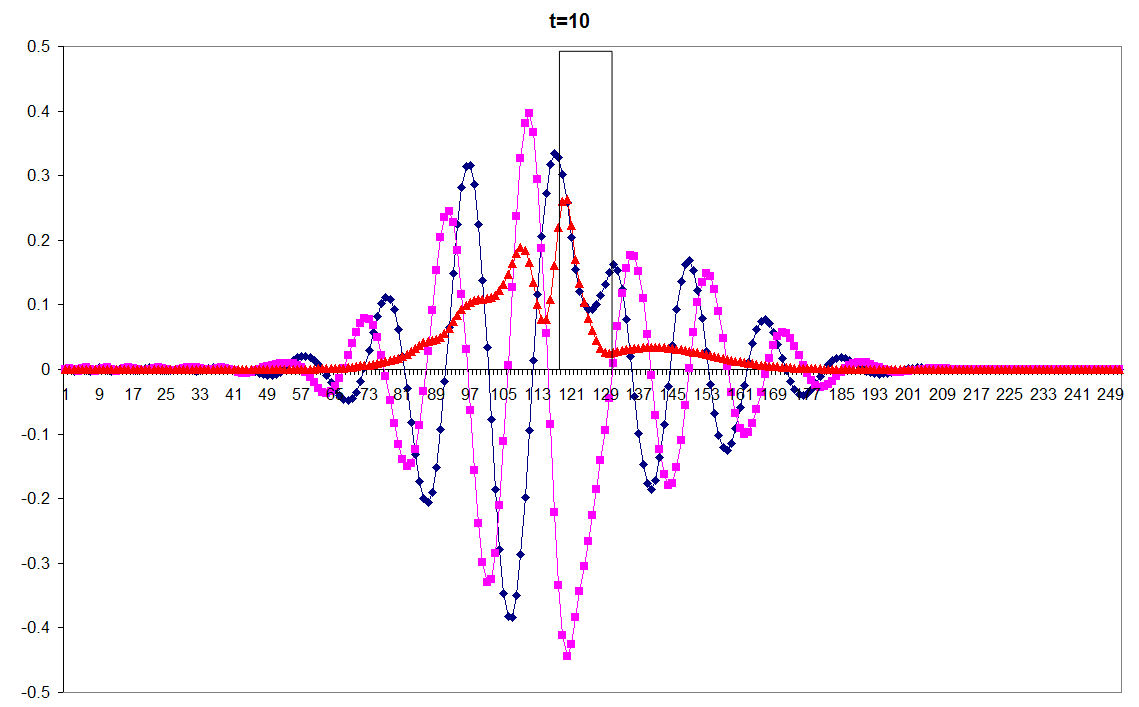}\\
\includegraphics[width=0.5\textwidth]{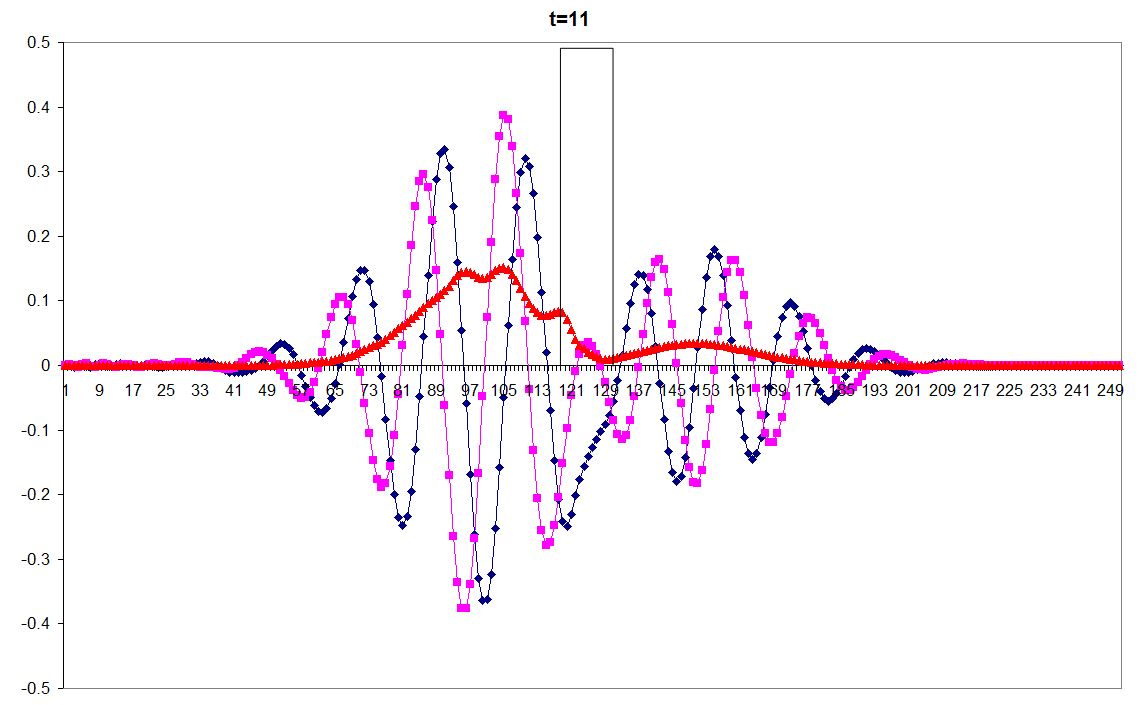}&
\includegraphics[width=0.5\textwidth]{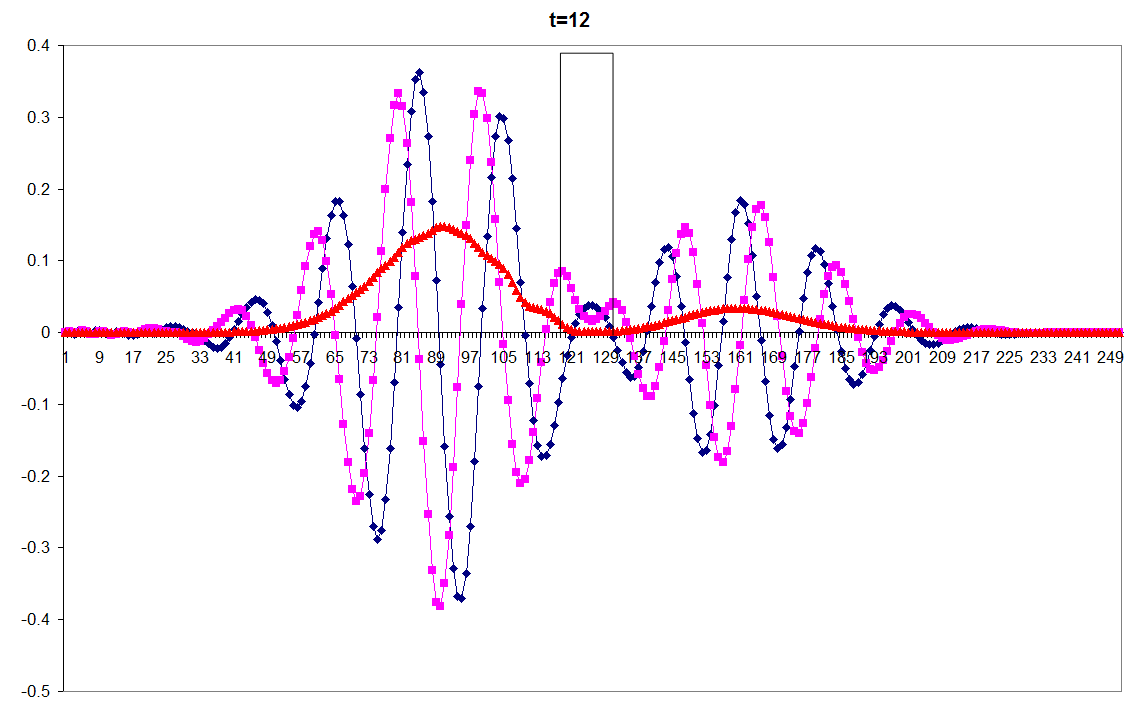}\\
\includegraphics[width=0.5\textwidth]{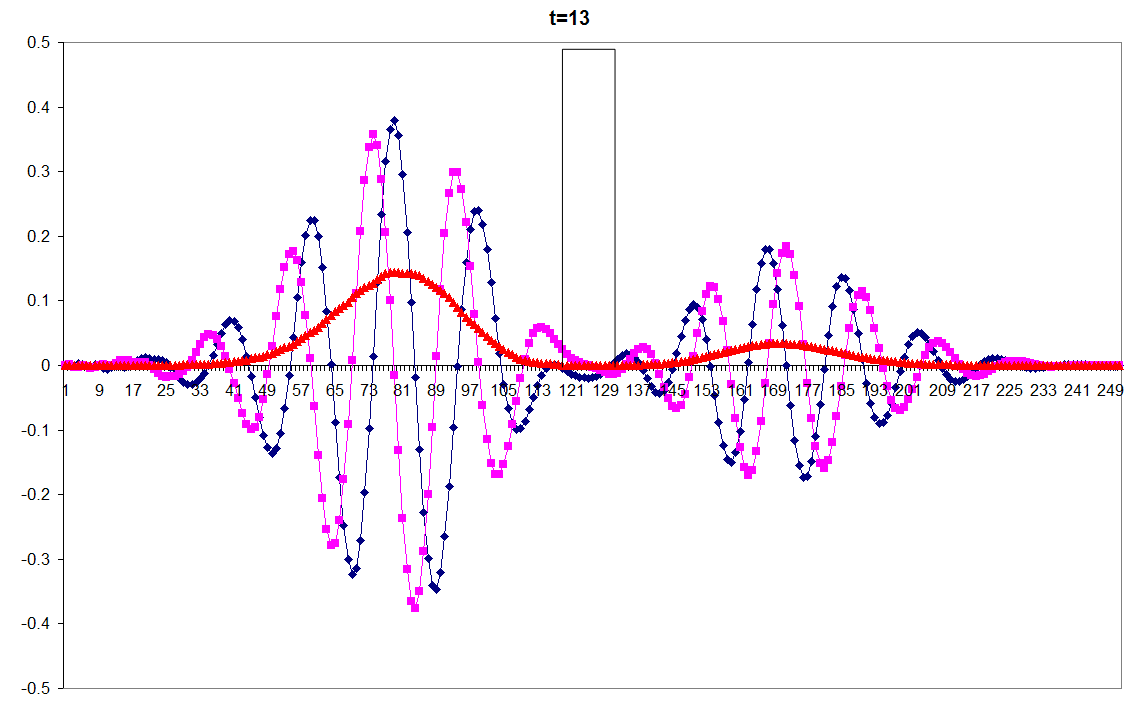}&
\includegraphics[width=0.5\textwidth]{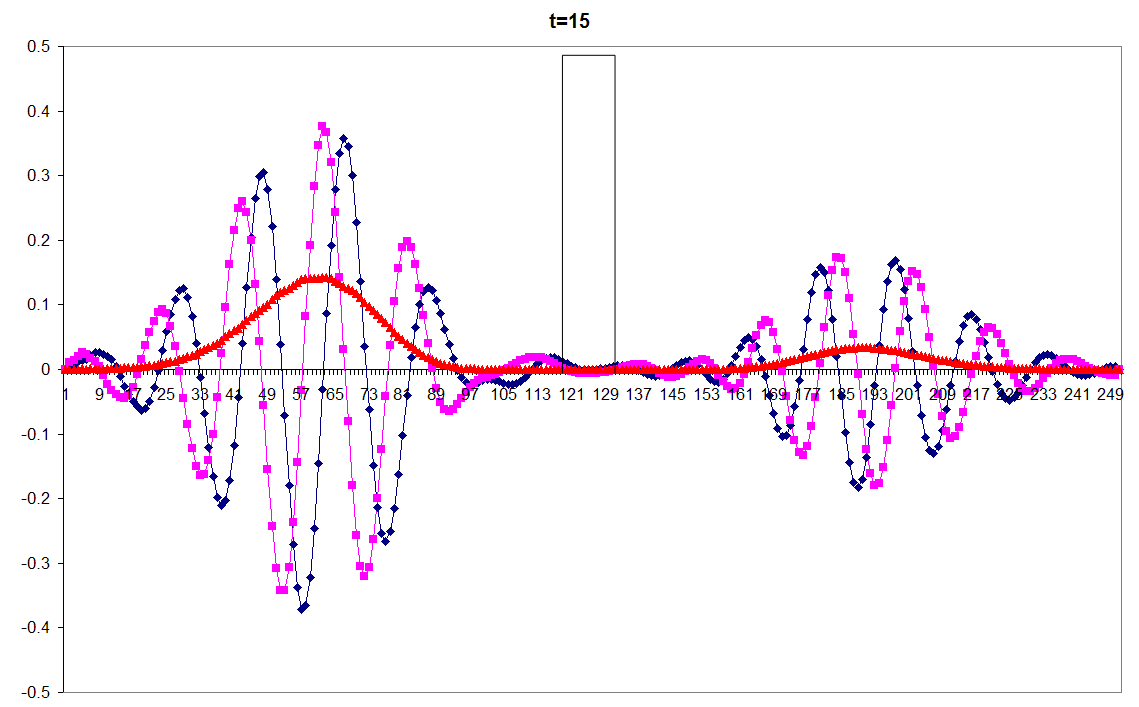}\\
\end{tabular}
\caption{Wave packet trapped in a infinite well with barrier at elements $120\leq x_e \leq 130$ (cont. on next page).}
\label{infinite-packet-timestep-1-barrier}
\end{center}
\end{figure}

\begin{figure}
\begin{center}
\begin{tabular}{cc}
\includegraphics[width=0.5\textwidth]{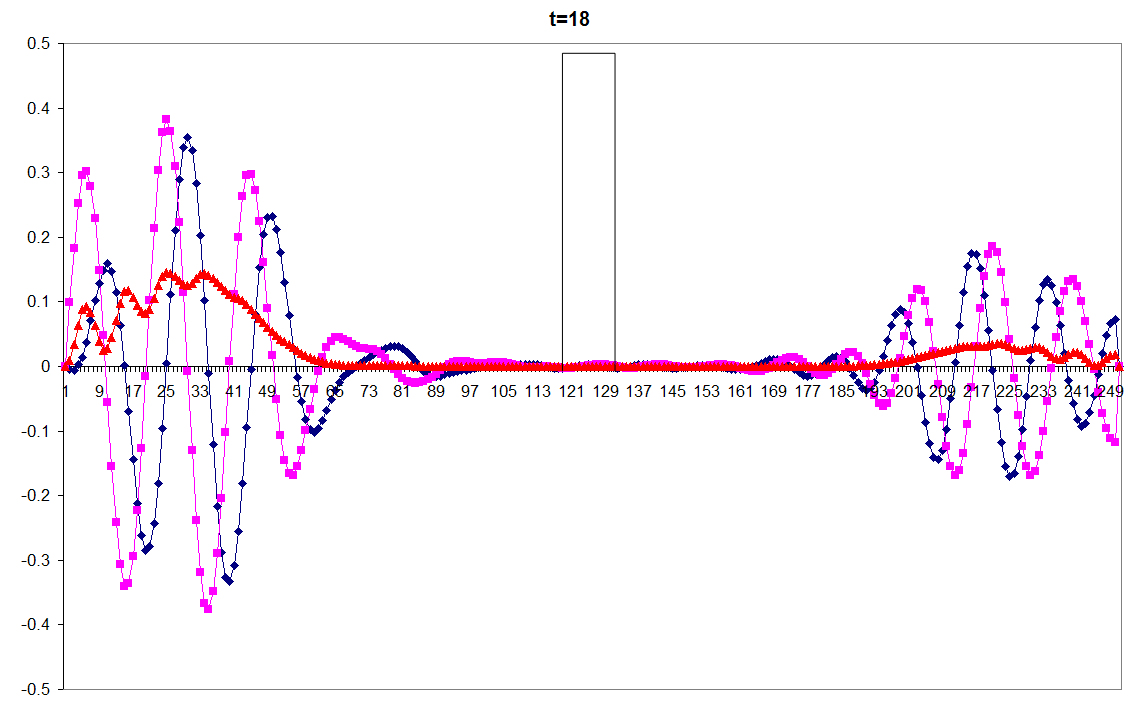}&
\includegraphics[width=0.5\textwidth]{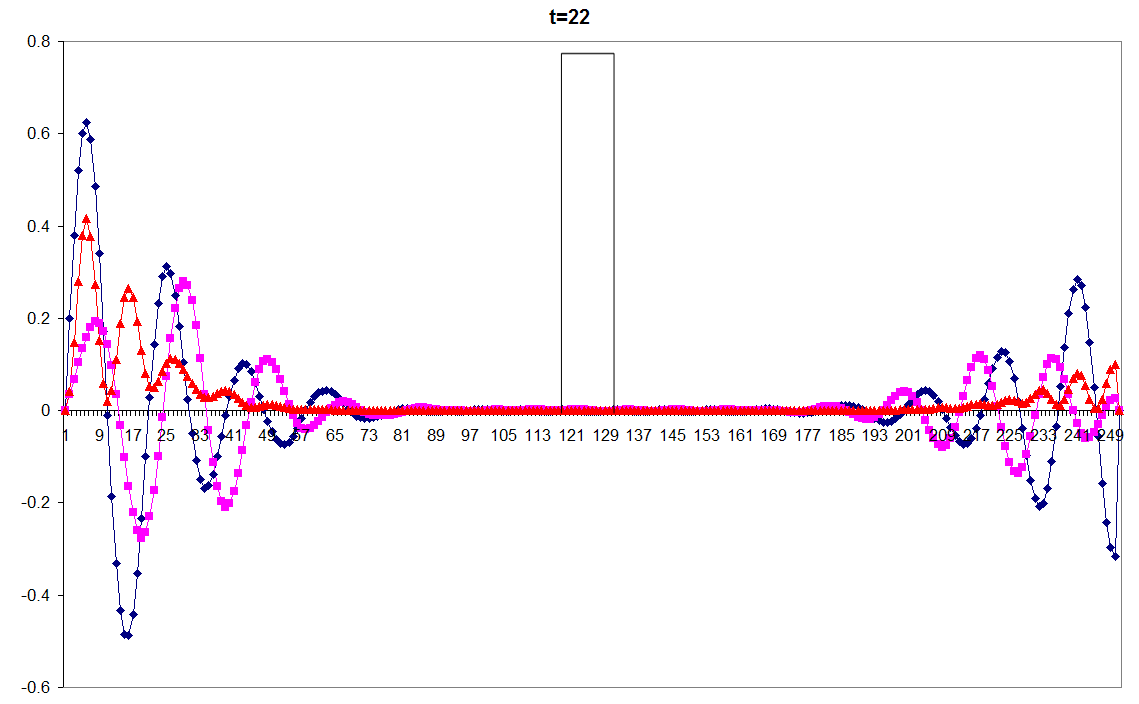}\\
\includegraphics[width=0.5\textwidth]{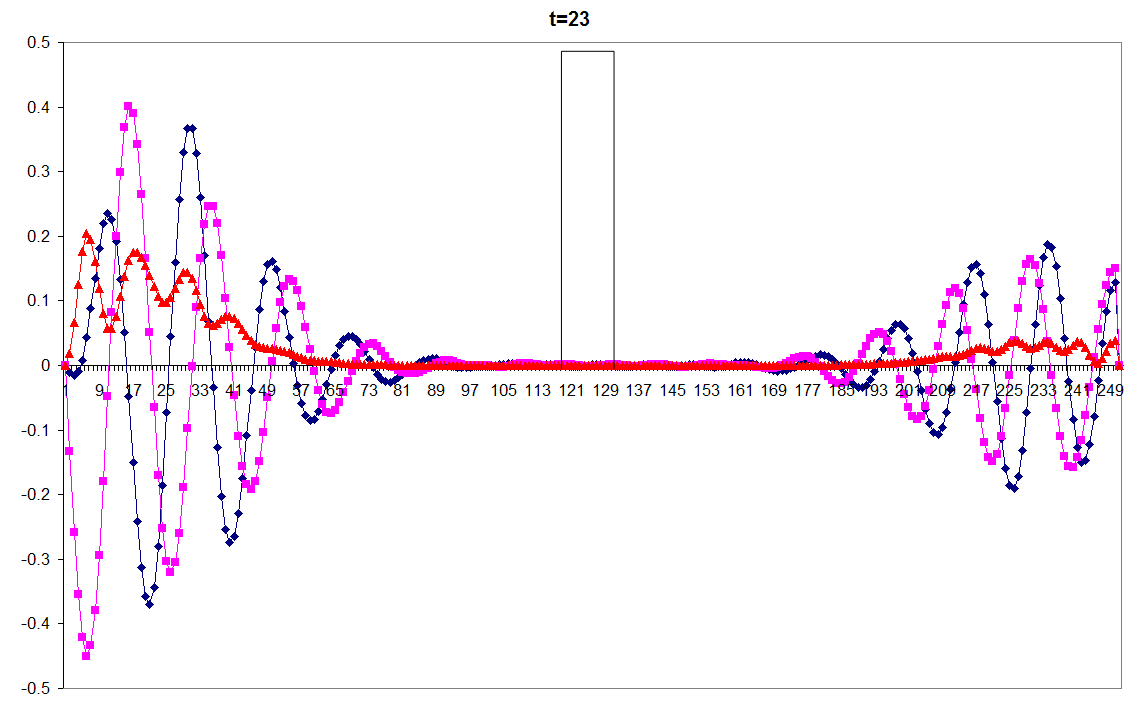}&
\includegraphics[width=0.5\textwidth]{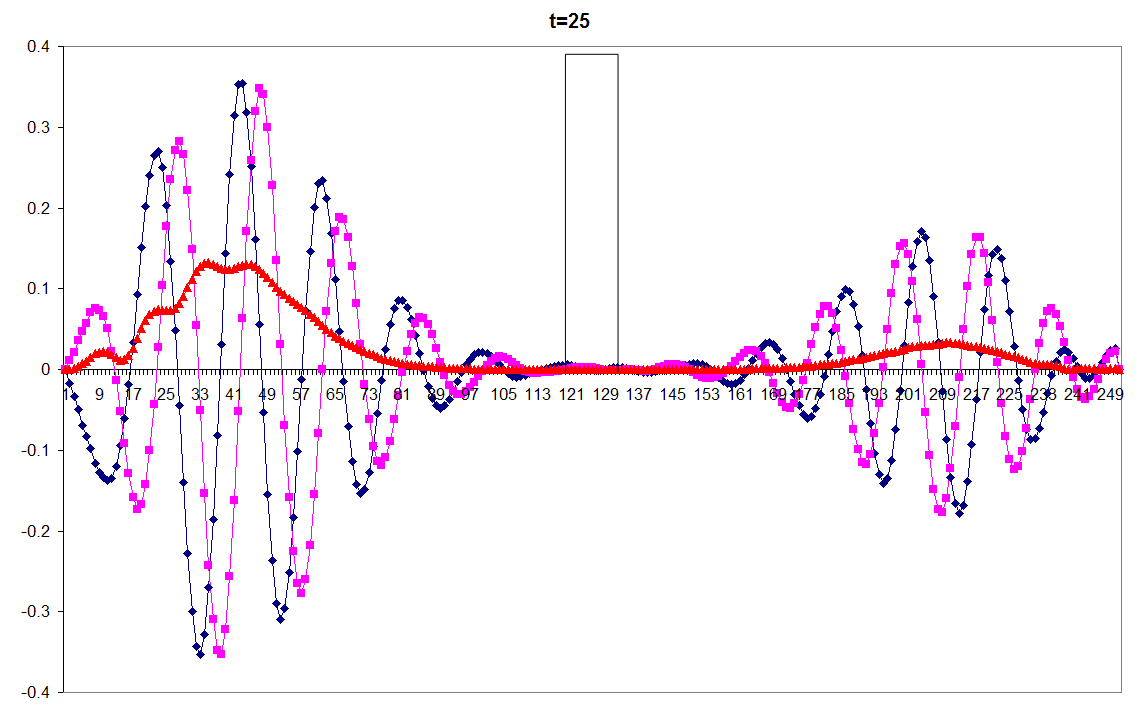}\\
\includegraphics[width=0.5\textwidth]{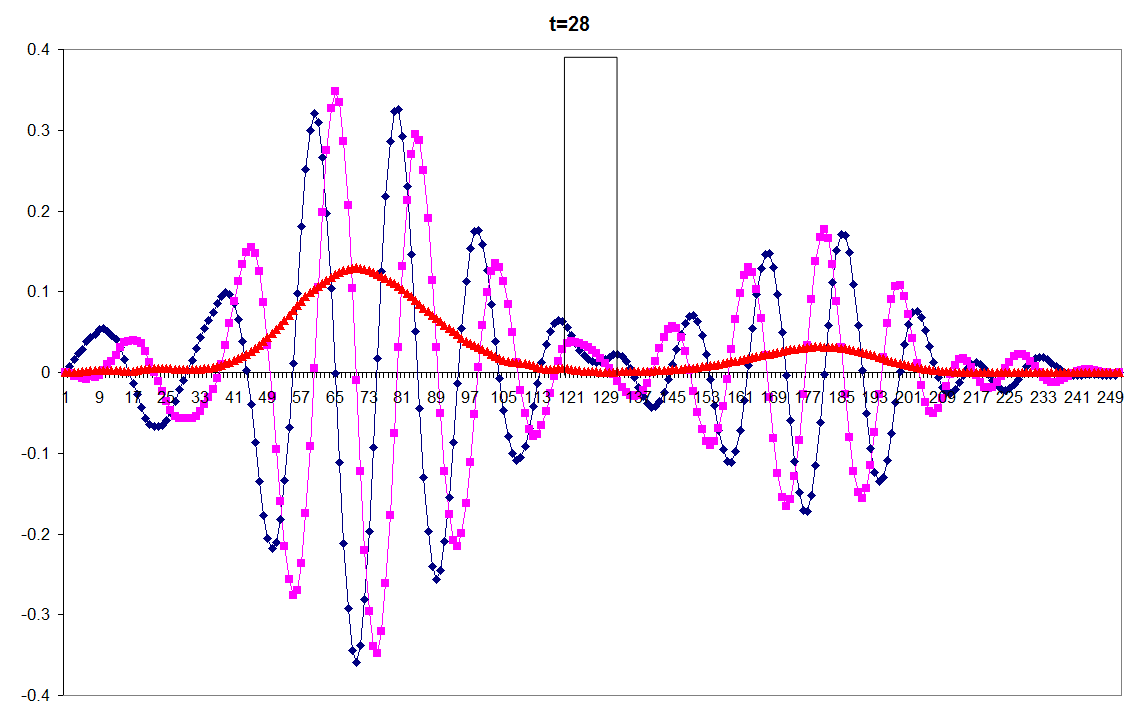}&
\includegraphics[width=0.5\textwidth]{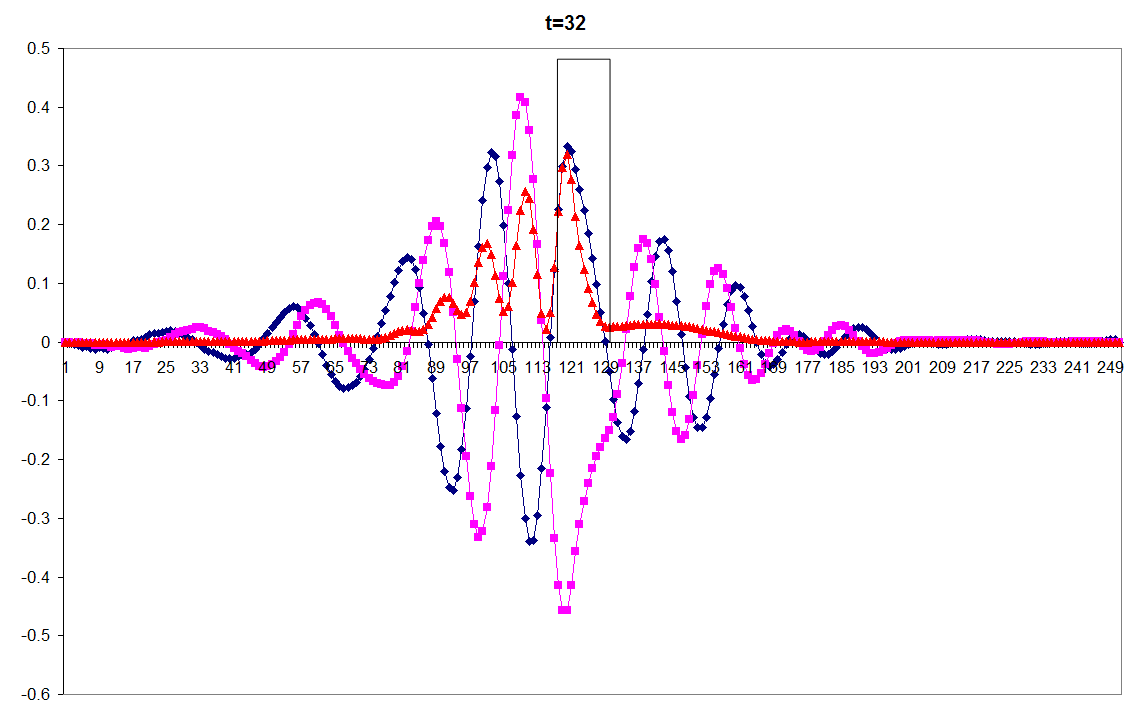}\\
\includegraphics[width=0.5\textwidth]{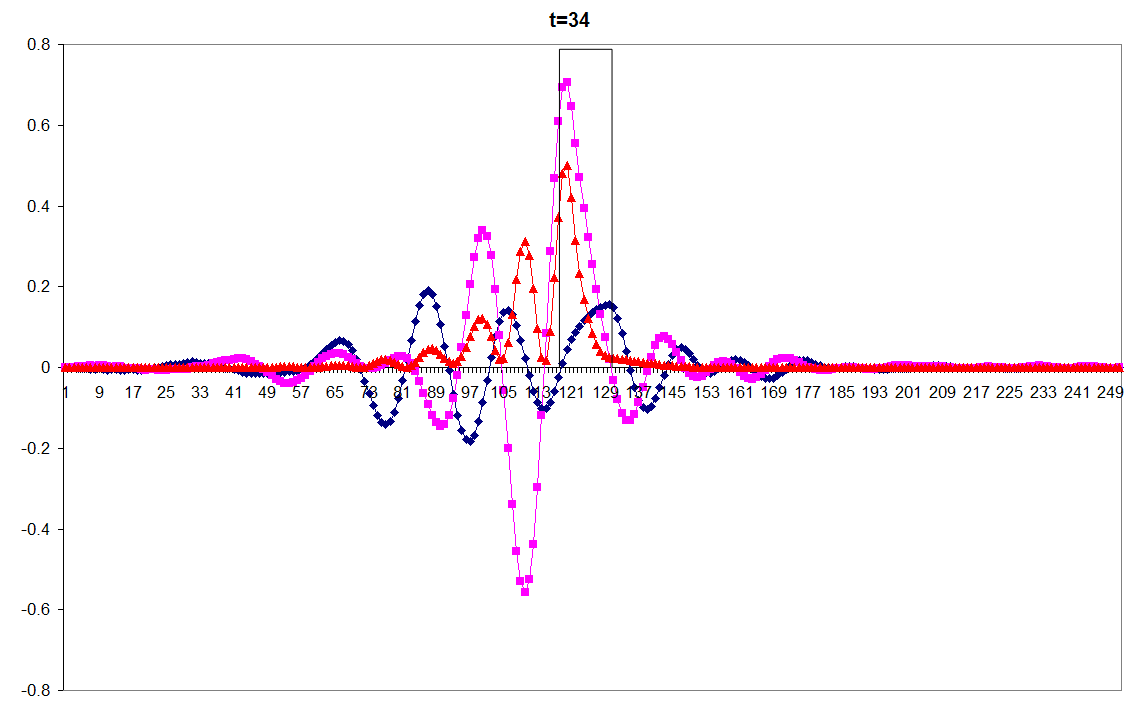}&
\includegraphics[width=0.5\textwidth]{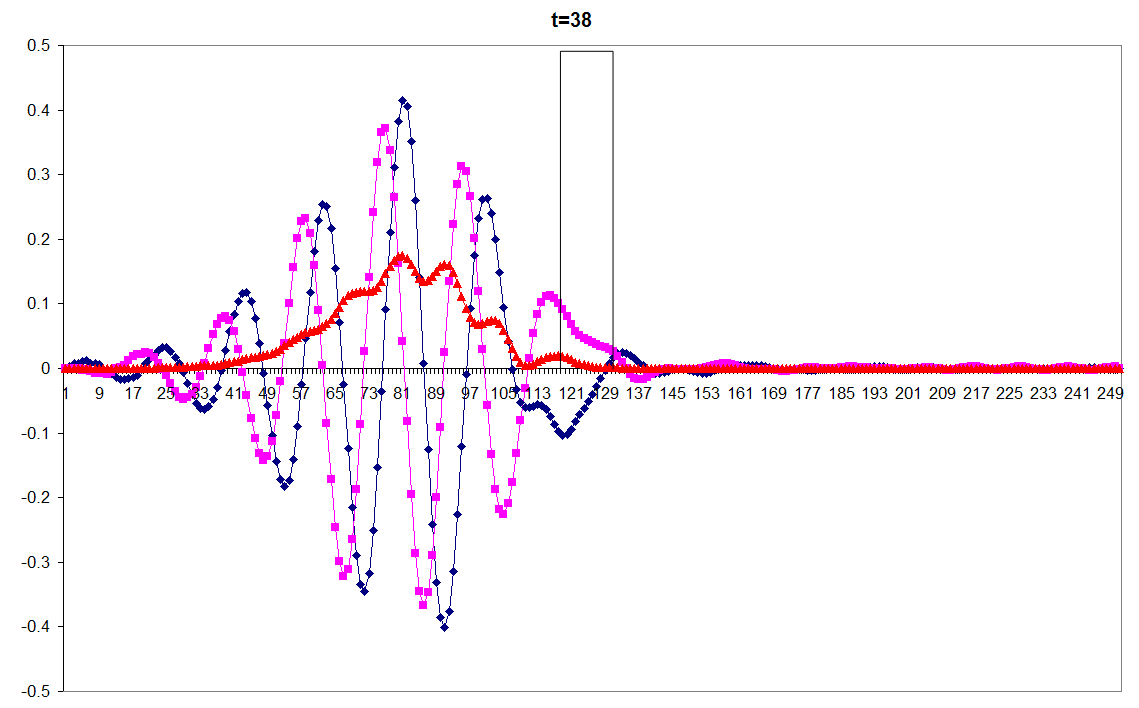}\\
\includegraphics[width=0.5\textwidth]{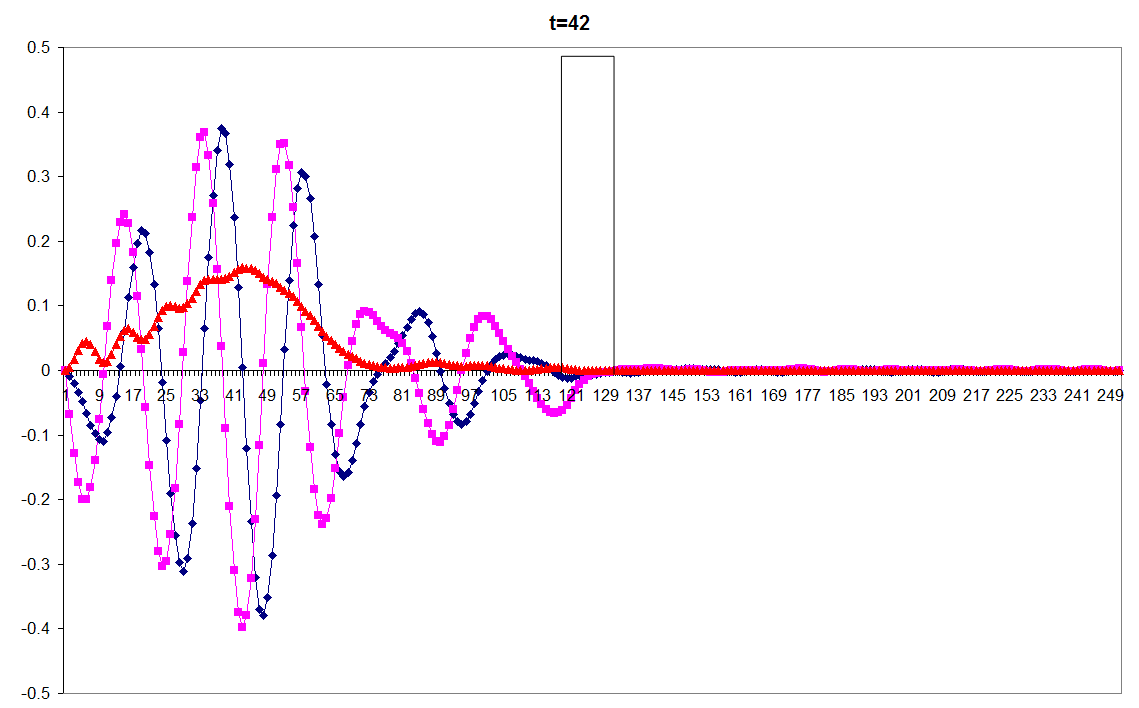}&
\includegraphics[width=0.5\textwidth]{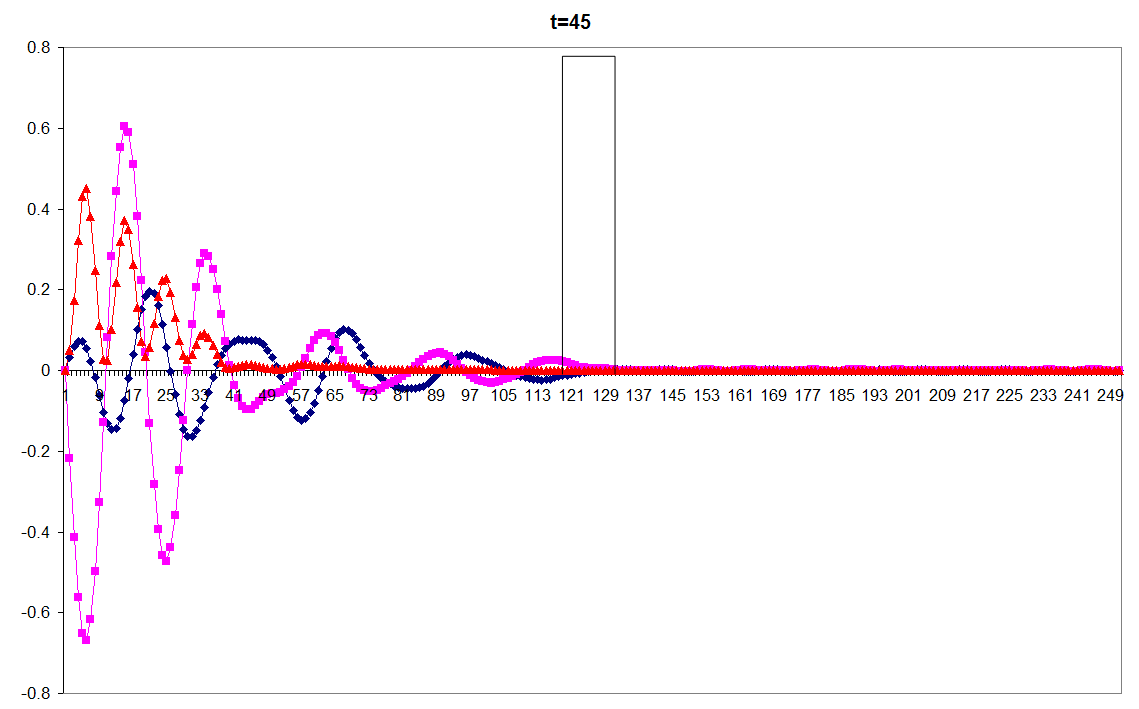}\\
\end{tabular}
\caption{(Cont. from previous page) Wave packet trapped in an infinite well with barrier at elements $120\leq x_e \leq 130$.}
\label{infinite-packet-timestep-2-barrier}
\end{center}
\end{figure}

When we look at the conservation of probability property we find that with the introduction of a finite barrier the conservation of total area now fluctuates around the previous value of $1$, Fig.~\ref{area-conservation-barrier-space}.  Comparing Figs.~\ref{infinite-packet-timestep-1-barrier} and \ref{area-conservation-barrier-space} (for 250 elements) we can conclude that the first fluctuation (dip at $5\leq t \leq 10$), in Fig.~\ref{area-conservation-barrier-space}, occurs when the wave-packet first interacts with the barrier, in Fig.\ref{infinite-packet-timestep-1-barrier}.  The second fluctuation  (peak at $15\leq t \leq 25$) occurs when the reflected and transmitted packets interact with the well walls.  However, after the second fluctuation (dip at  $28\leq t \leq 35$, the transmitted and reflected waves recombine, in this case when the packet interacts with the well walls no  peak fluctuation occurs.  The next fluctuation is again a dip when the packet interacts with the barrier again.   These fluctuations then continue, with the dips representing an interaction with the barrier and the peaks representing interactions with the well walls (when reflected and transmitted waves exist).  We can argue that when the problem involves reflected and transmitted waves we require greater resolving power in order to determine the shape of the total probability distribution $\psi$.  We can show this to be the case by increasing the number of elements in the simulation and keeping the timestep $dt$ constant.  In Fig.~\ref{area-conservation-barrier-space} it can be seen that if we increase the number of elements the peak and dip fluctuations begin to diminish -- for $150$ elements the fluctuations have a maximum value of $2.2\%$, but for $1000$ elements the fluctuations fall to a maximum value of $0.2\%$. 

Also, from Fig.~\ref{area-conservation-barrier-time} we can see that if we decrease the timestep ($dt$) but keep the number of elements constant we can slightly decrease the fluctuations.  However, for time-steps smaller than $dt=0.1$ there seems to be no change in the fluctuations.  Taking these observations into account we conducted the barrier simulation for $500$ elements and a time step of $dt=0.1$.  From Figs.~\ref{area-conservation-barrier-space} and \ref{area-conservation-barrier-time} we can see that this is slightly more effective than just using $500$ elements (and time step $dt=0.5$) or a time-step of $0.1$ (and $250$ elements) alone.   

\begin{figure}
    \begin{center}
     \includegraphics[width=0.8\textwidth]{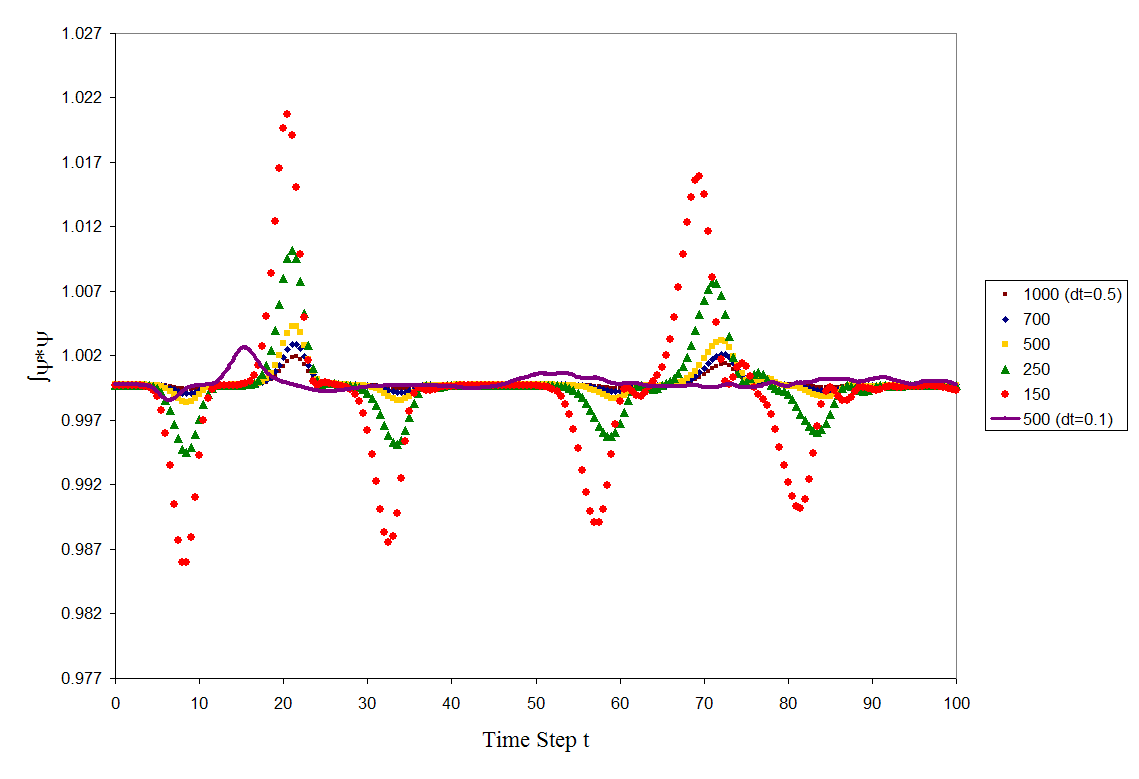}\\
        \caption{Conservation of area at each time step for varying number of elements. (Also on the same plot is area conservation for 500 elements and time step dt=0.1)}
        \label{area-conservation-barrier-space}
    \end{center}
\end{figure}

\begin{figure}
    \begin{center}
     \includegraphics[width=0.8\textwidth]{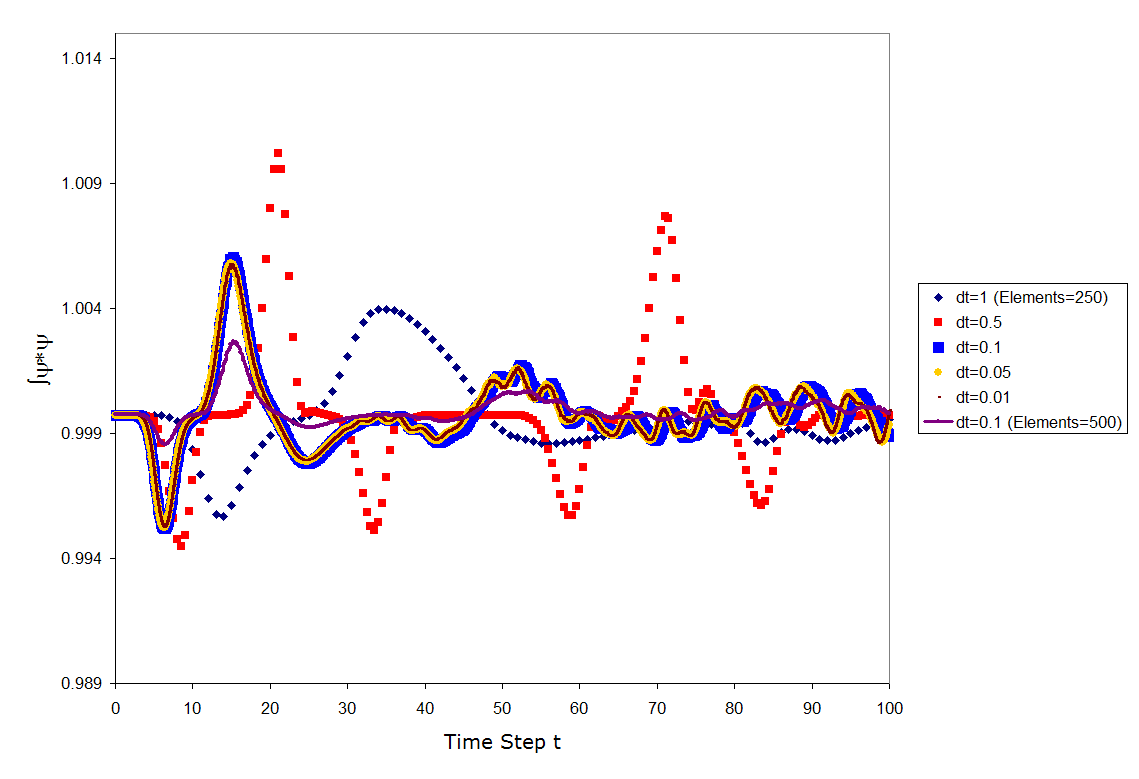}\\
        \caption{Conservation of area at each time step for varying size of dt. (Also on the same plot is area conservation for 500 elements and time step dt=0.1)}
        \label{area-conservation-barrier-time}
    \end{center}
\end{figure}

In order to study these fluctuations in detail we conducted further simulations, but this time by varying the initial wave-vector $k_0$, given in Eqn.~(\ref{Gaussian-wave-packet-model}).  All the previous simulations were performed with $k_0=2$, and the finite potential barrier was of height $V=2.5$.  So if we used $k_0<<V$ the low energy wave packet would be effectively trapped in the left side of the well -- with very little being transmitted.  On the other hand if we used $k_0>>V$, the high energy wave-packet would move around the infinite well unhindered by the barrier -- so very little reflection would occur.  As these cases will be very similar to our first set of results (of the wave-packet in an infinite well) the fluctuations due to the interaction with the barrier should disappear.  In Fig.~\ref{infinite-packet-timestep-k=high-barrier} we see that using a value of $k_0=3$ the majority of the wave-packet is transmitted.  And in Fig.~\ref{infinite-packet-timestep-k=low-barrier} where we used $k_0=1$ very little is transmitted, so the packet is effectively trapped in the left side.  When we plot the areas using these wave vectors, and also of simulations with $k_0=0.1$, $4$, and $12$ in Fig.\ref{area-conservation-barrier-k0} we find that with a very small and very high $k_0$ the fluctuations do indeed decrease (for the case of $k_0=0.1$ and $12$ they almost vanish).  From this we can conclude that when the energy of the wave-packet is comparable to the potential barrier, and so any interaction between the two becomes significant, we require greater number of elements to take account of the finer resolution changes in the wave-packets $Re$ and $Im$ components.  So the fluctuations in probability conservation are more to do with the spatial element discretisation rather than the size of the time step $dt$. 

\begin{figure}
\begin{center}
\begin{tabular}{cc}
\includegraphics[width=0.5\textwidth]{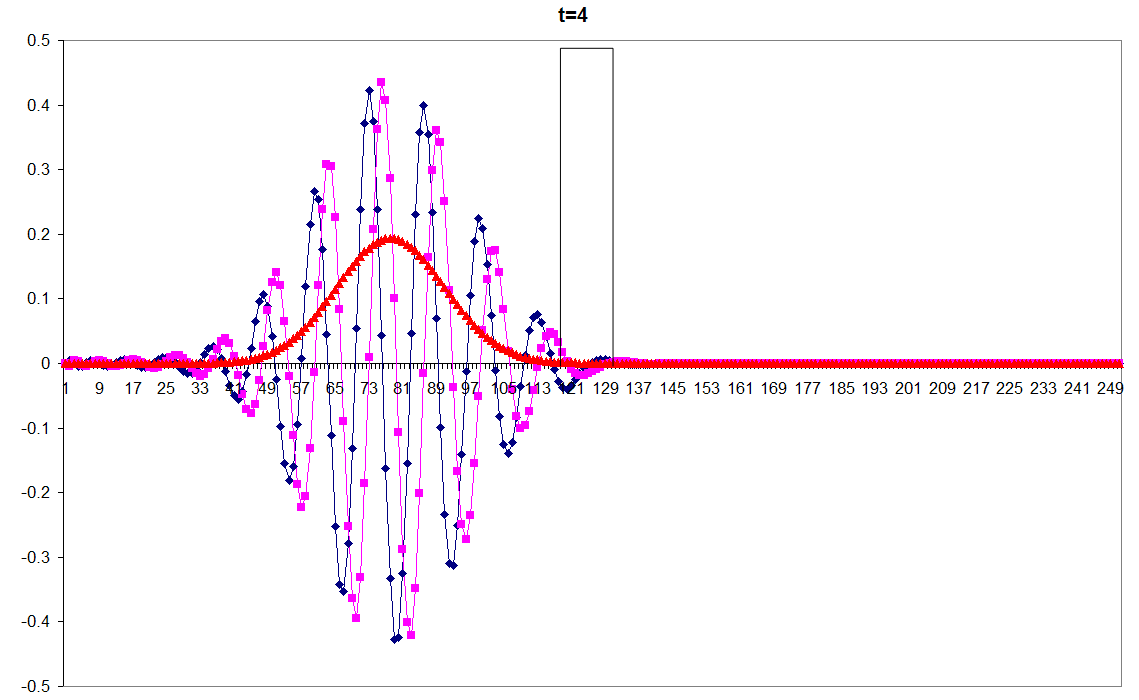}&
\includegraphics[width=0.5\textwidth]{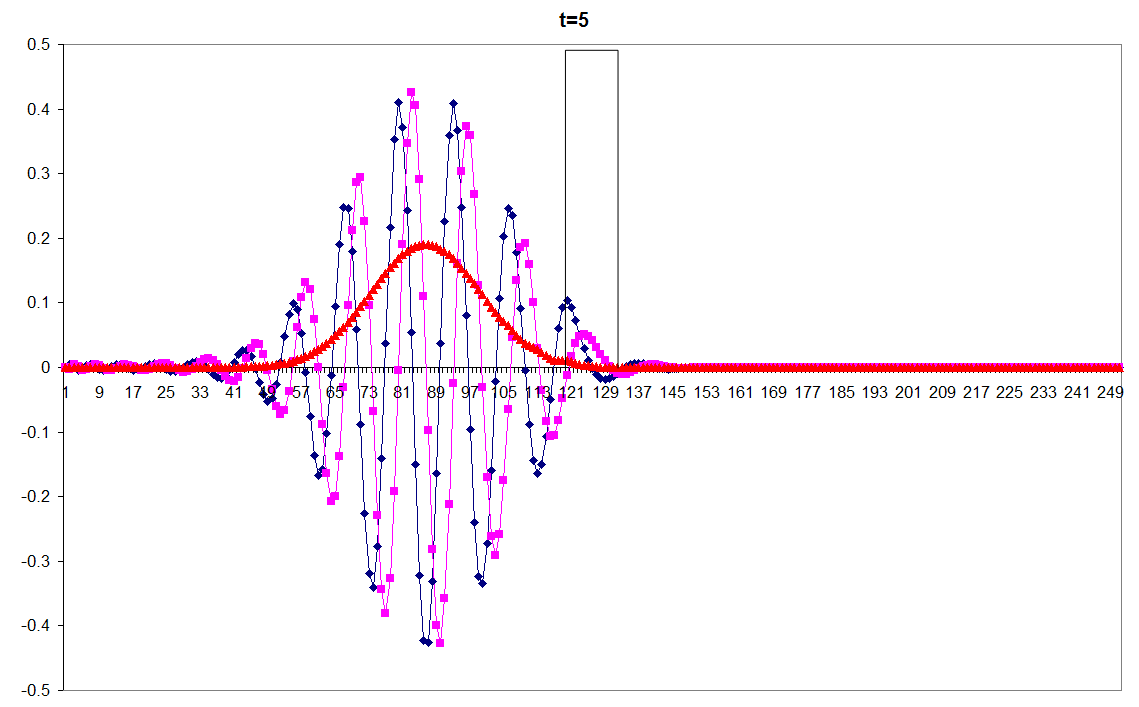}\\
\includegraphics[width=0.5\textwidth]{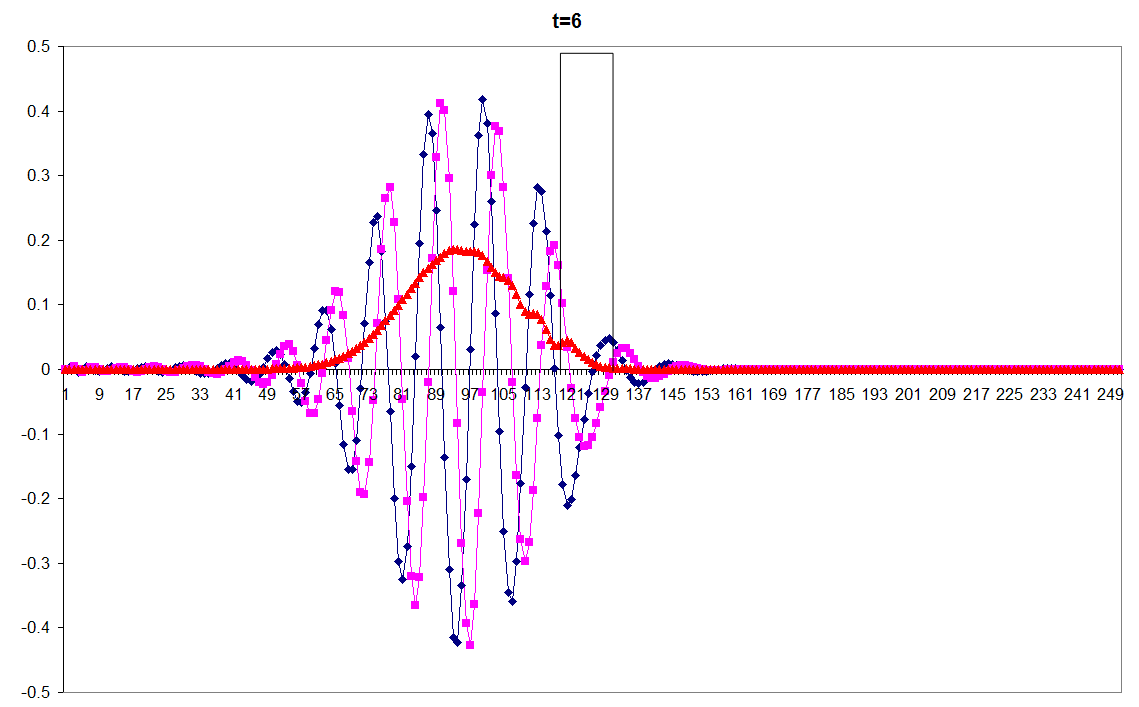}&
\includegraphics[width=0.5\textwidth]{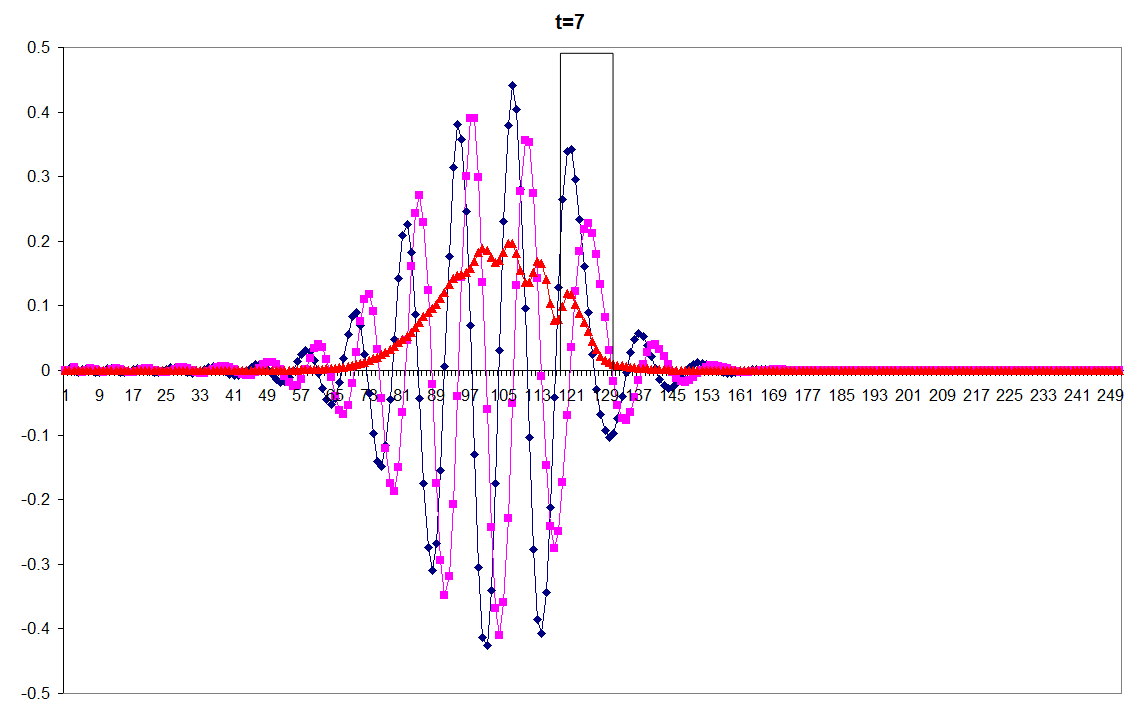}\\
\includegraphics[width=0.5\textwidth]{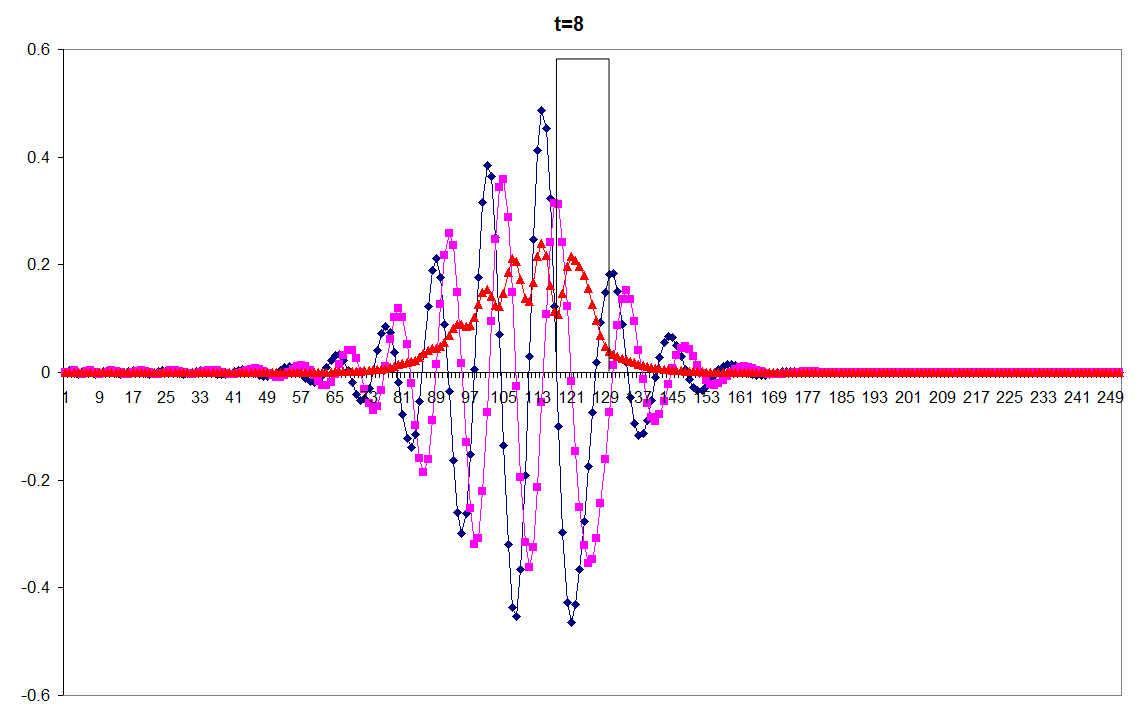}&
\includegraphics[width=0.5\textwidth]{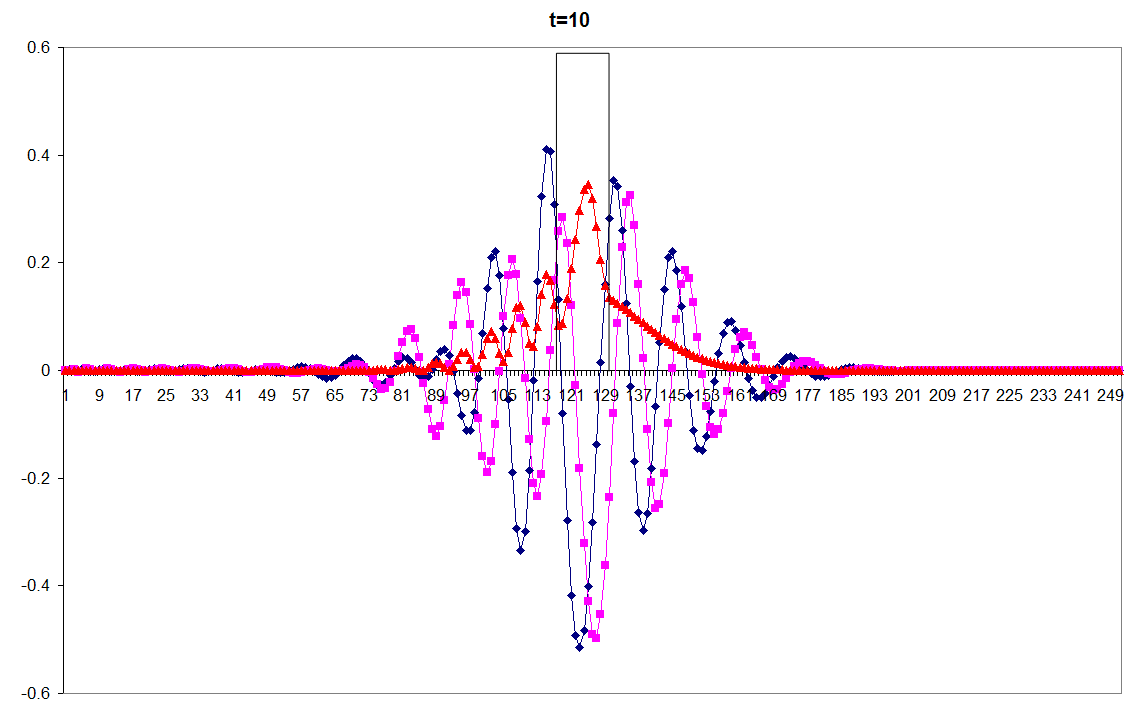}\\
\includegraphics[width=0.5\textwidth]{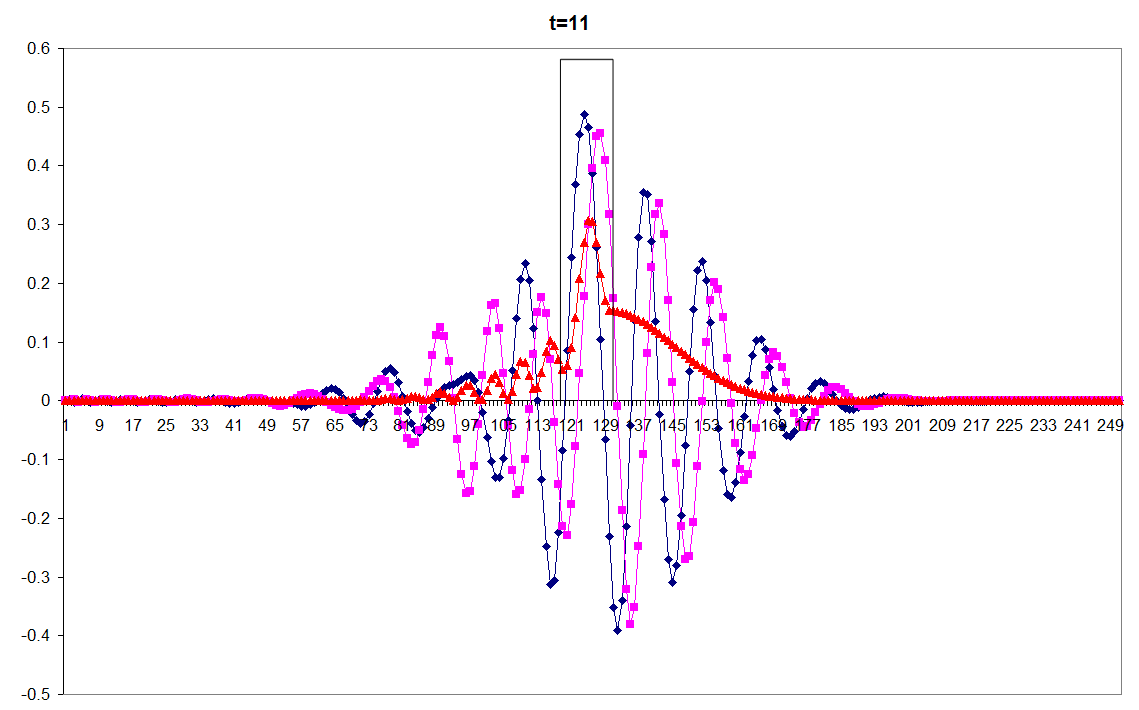}&
\includegraphics[width=0.5\textwidth]{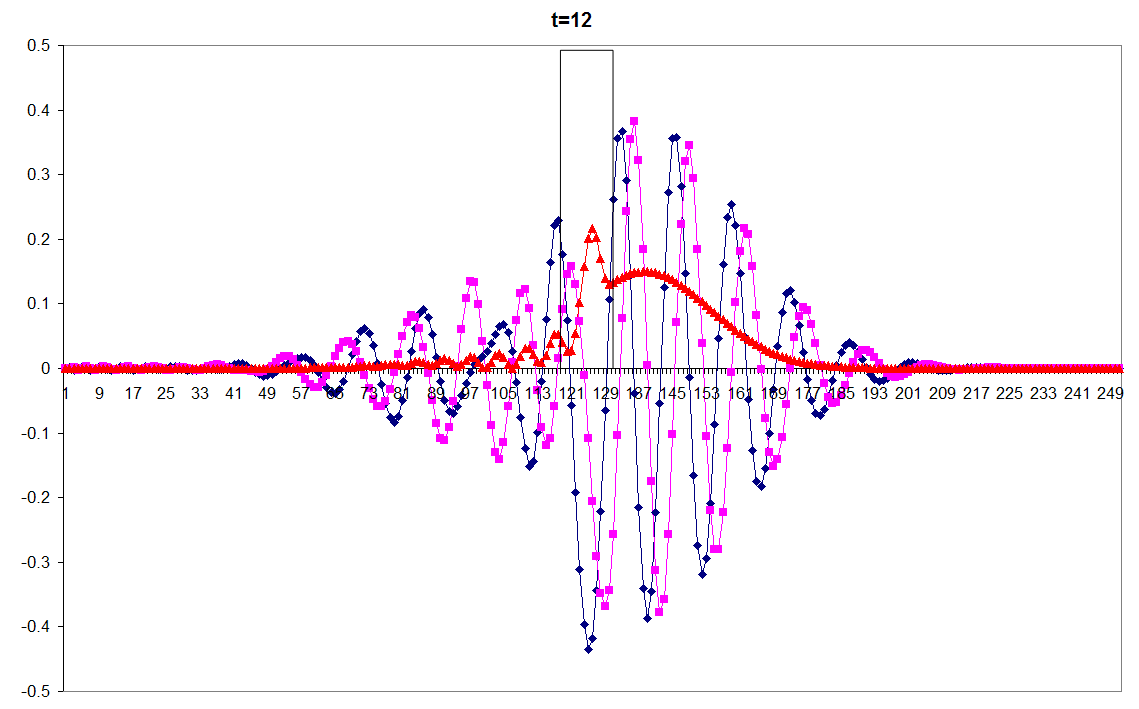}\\
\includegraphics[width=0.5\textwidth]{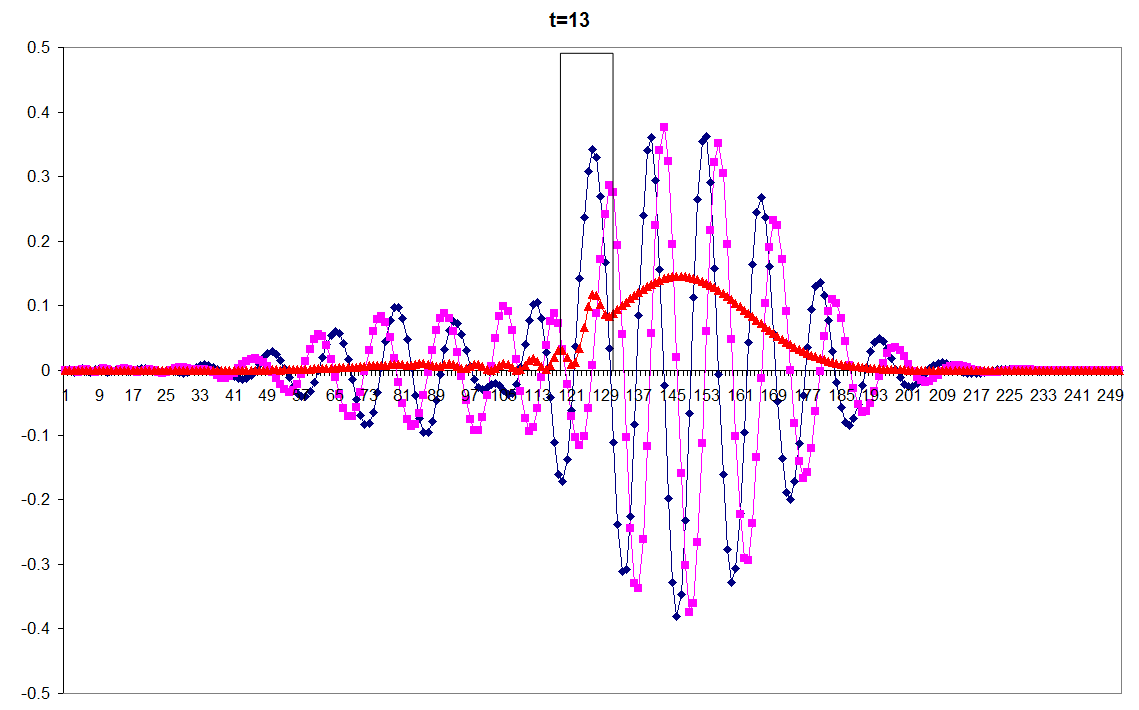}&
\includegraphics[width=0.5\textwidth]{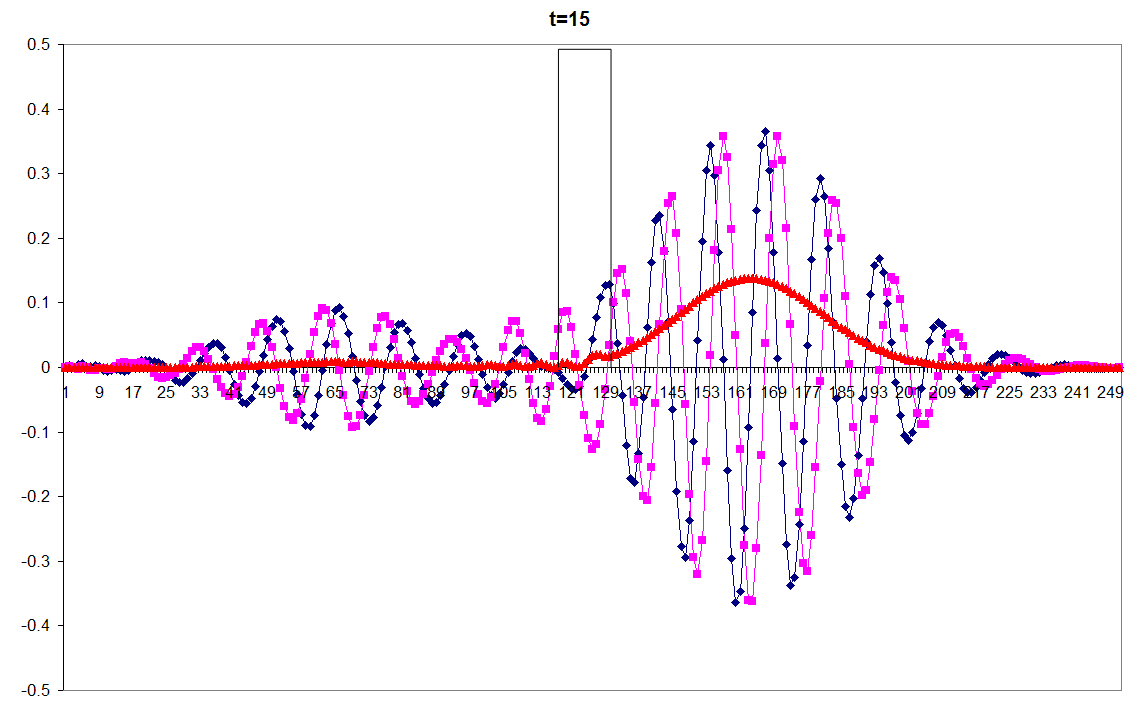}\\
\end{tabular}
\caption{Wave packet with initial wave vector $k_0=3$ trapped in a infinite well with barrier at elements $120\leq x_e \leq 130$.}
\label{infinite-packet-timestep-k=high-barrier}
\end{center}
\end{figure}

\begin{figure}
\begin{center}
\begin{tabular}{cc}
\includegraphics[width=0.5\textwidth]{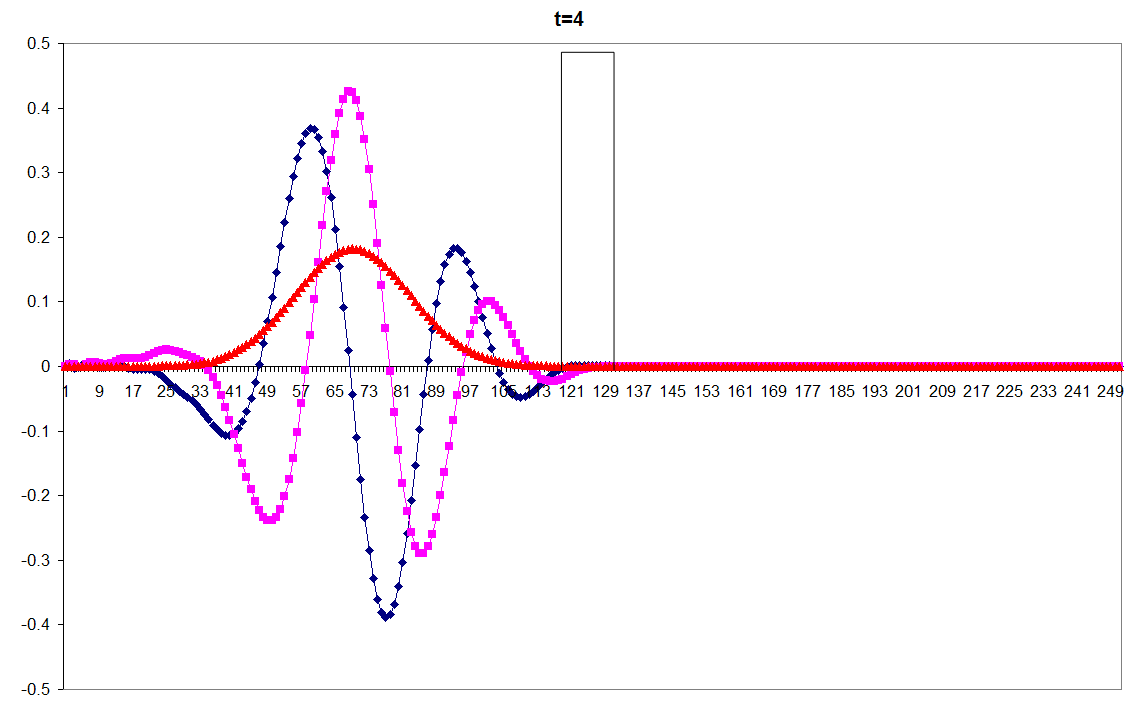}&
\includegraphics[width=0.5\textwidth]{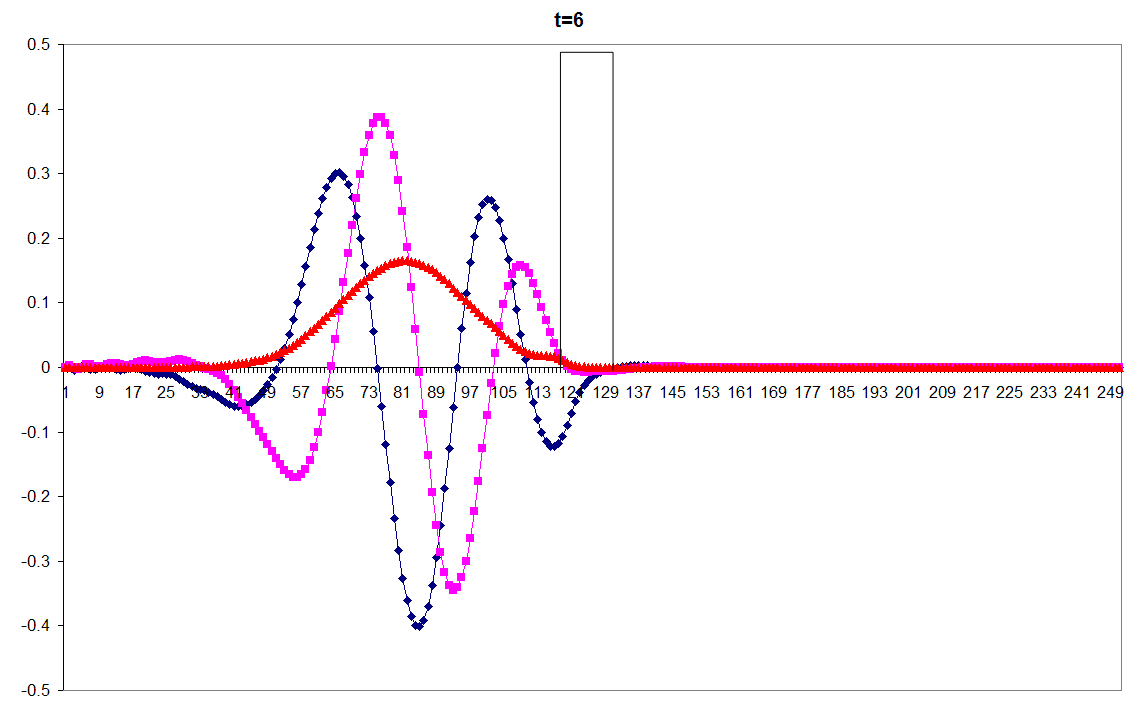}\\
\includegraphics[width=0.5\textwidth]{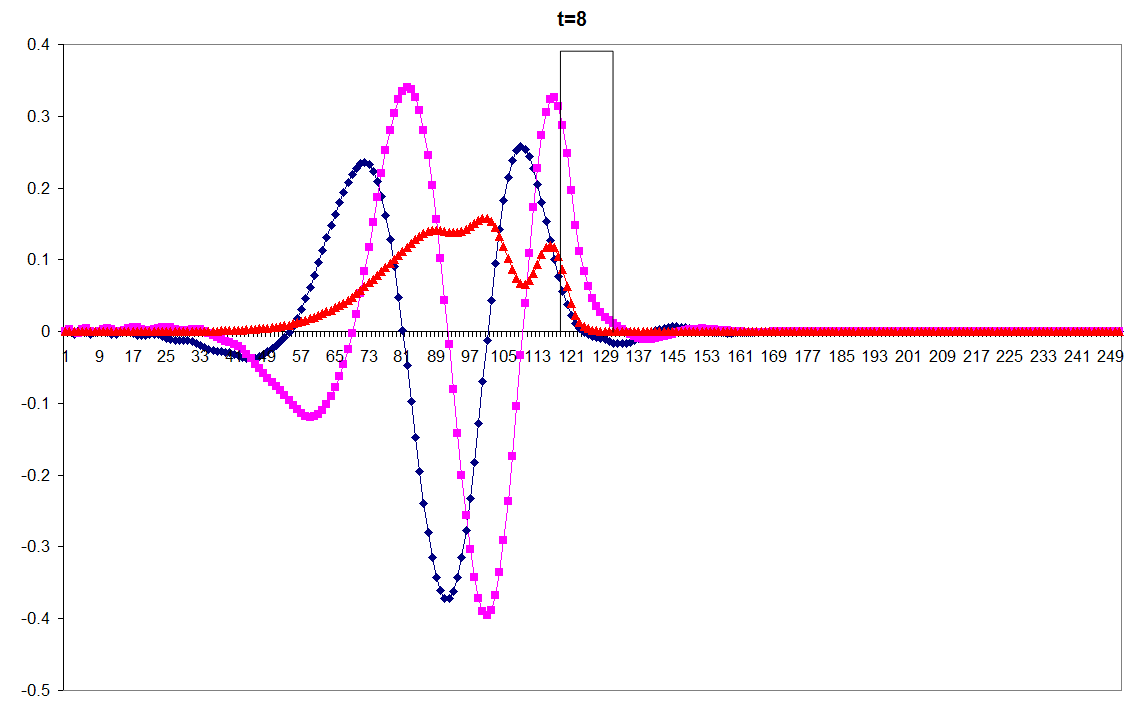}&
\includegraphics[width=0.5\textwidth]{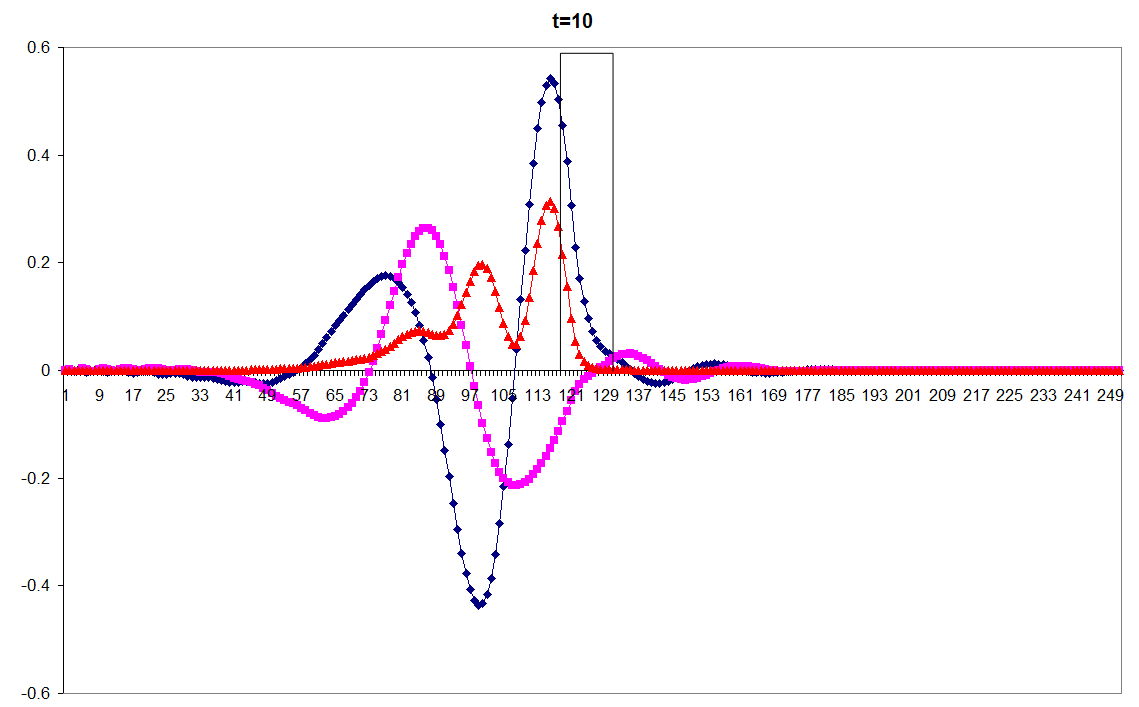}\\
\includegraphics[width=0.5\textwidth]{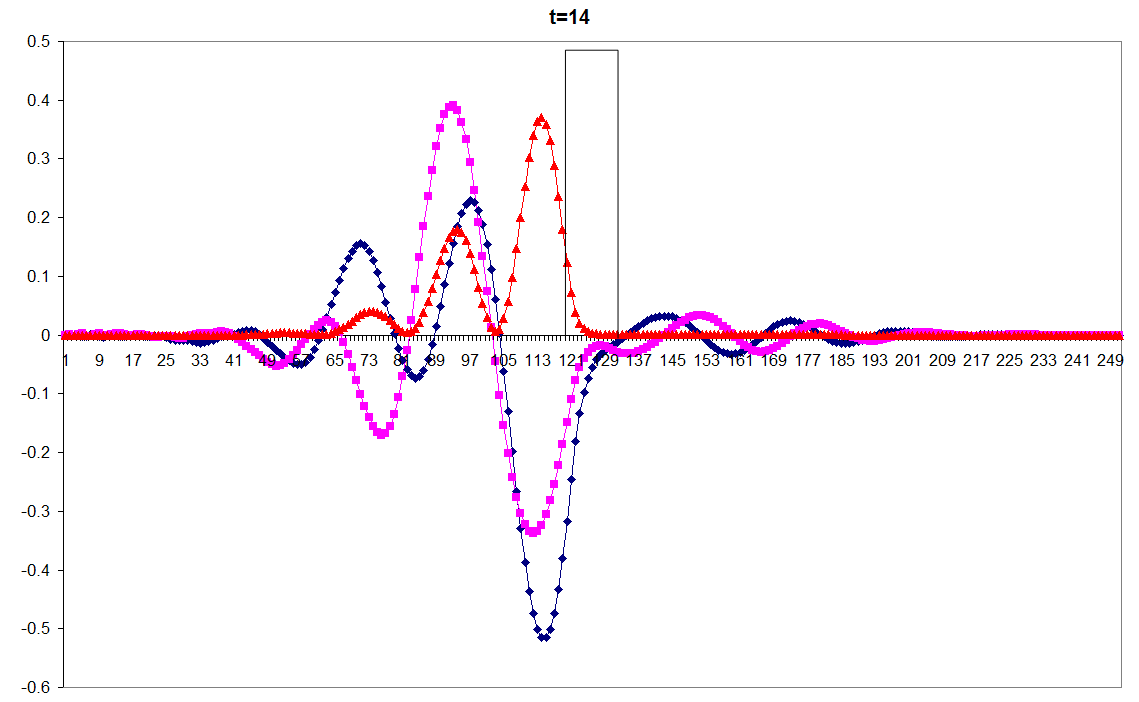}&
\includegraphics[width=0.5\textwidth]{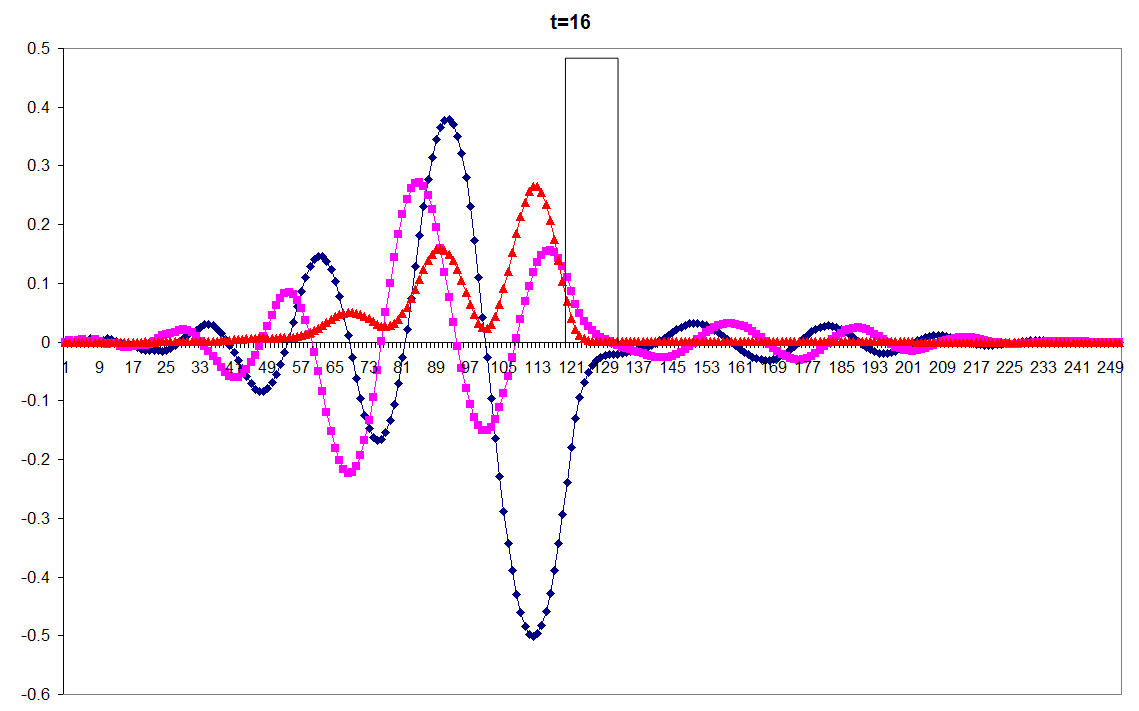}\\
\includegraphics[width=0.5\textwidth]{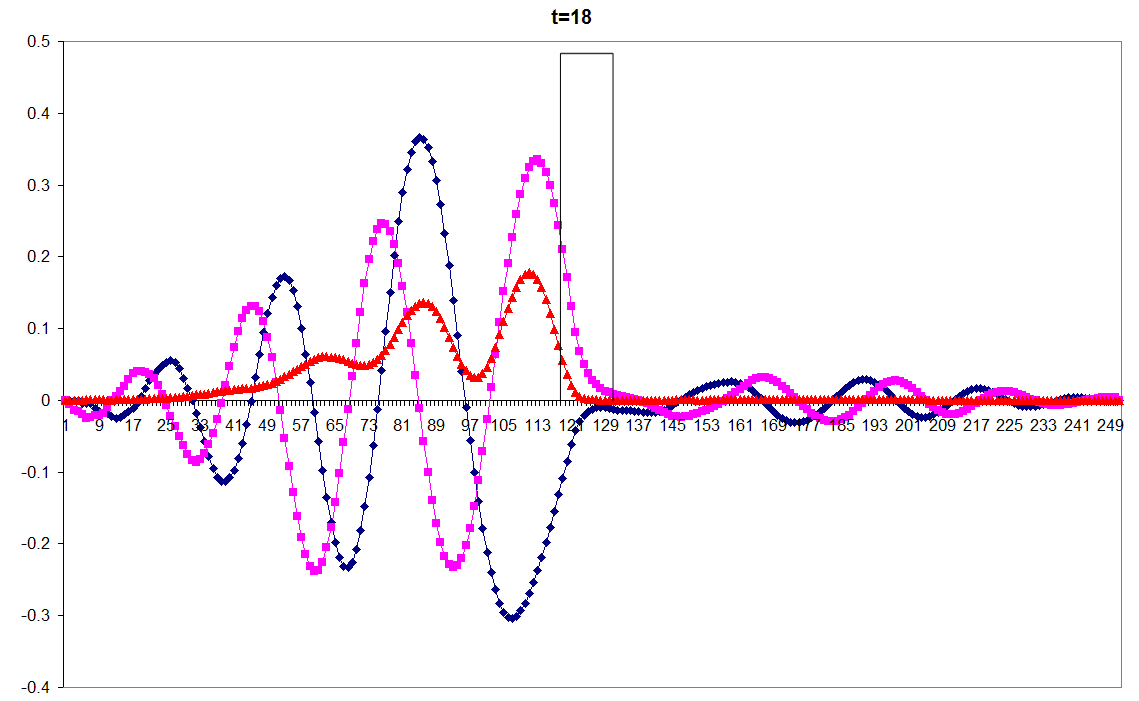}&
\includegraphics[width=0.5\textwidth]{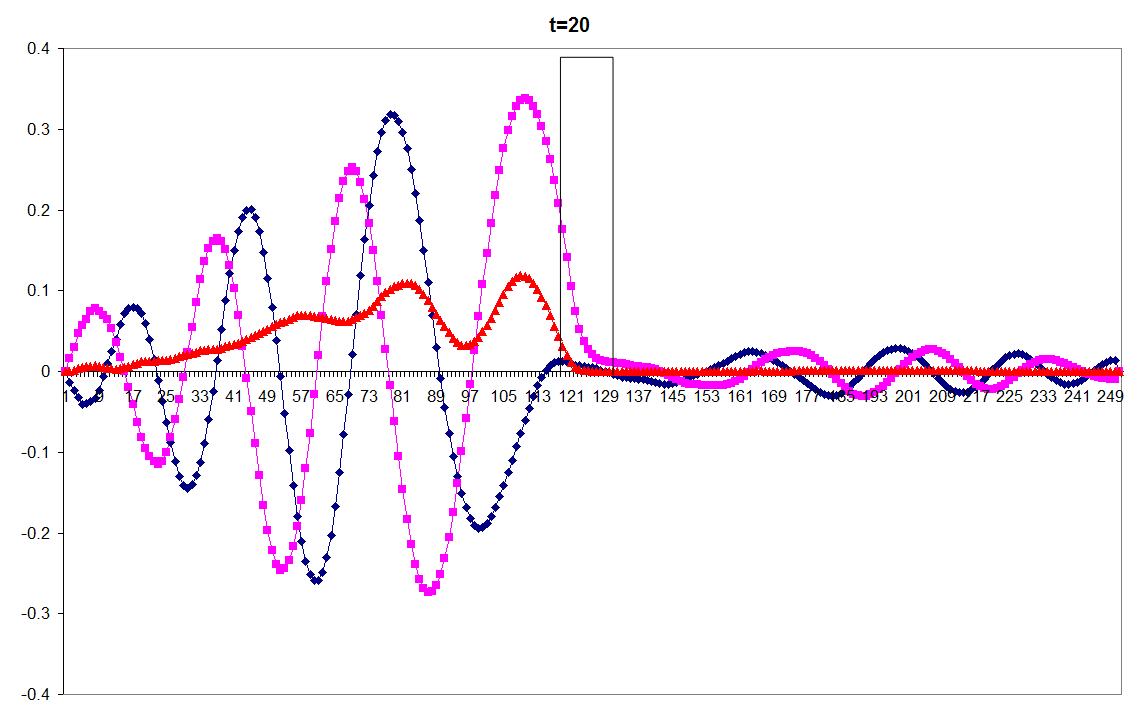}\\
\includegraphics[width=0.5\textwidth]{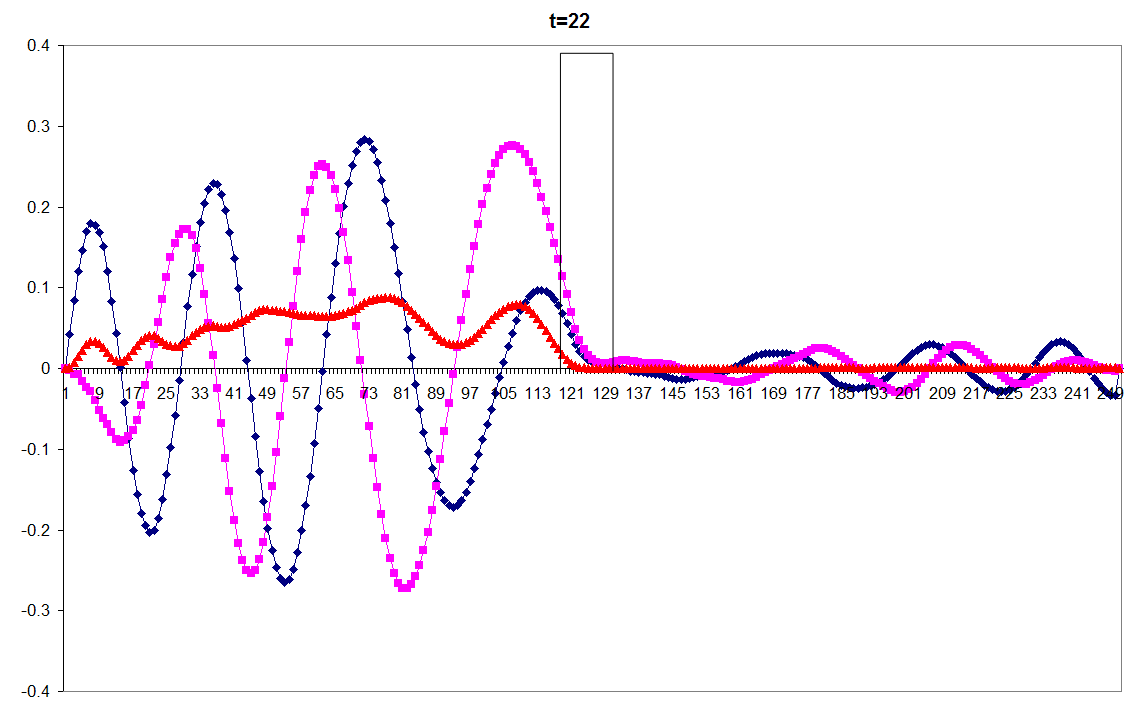}&
\includegraphics[width=0.5\textwidth]{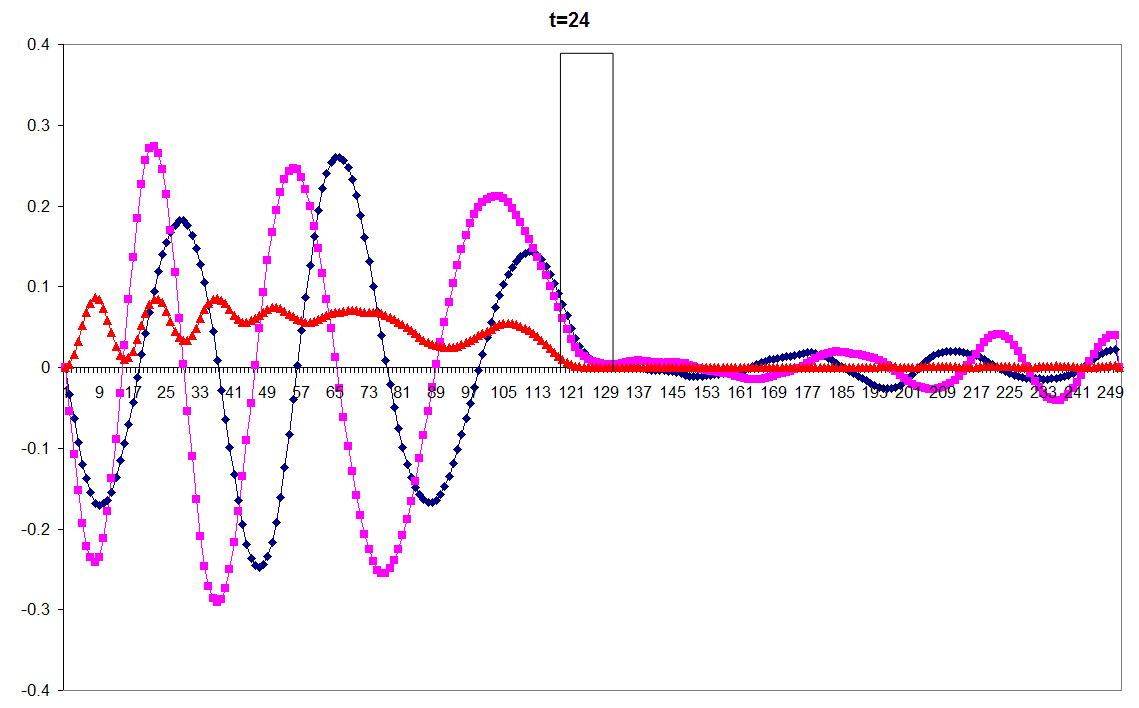}\\
\end{tabular}
\caption{Wave packet with initial wave vector $k_0=1$ trapped in a infinite well with barrier at elements $120\leq x_e \leq 130$.}
\label{infinite-packet-timestep-k=low-barrier}
\end{center}
\end{figure}
\begin{figure}
    \begin{center}
     \includegraphics[width=0.8\textwidth]{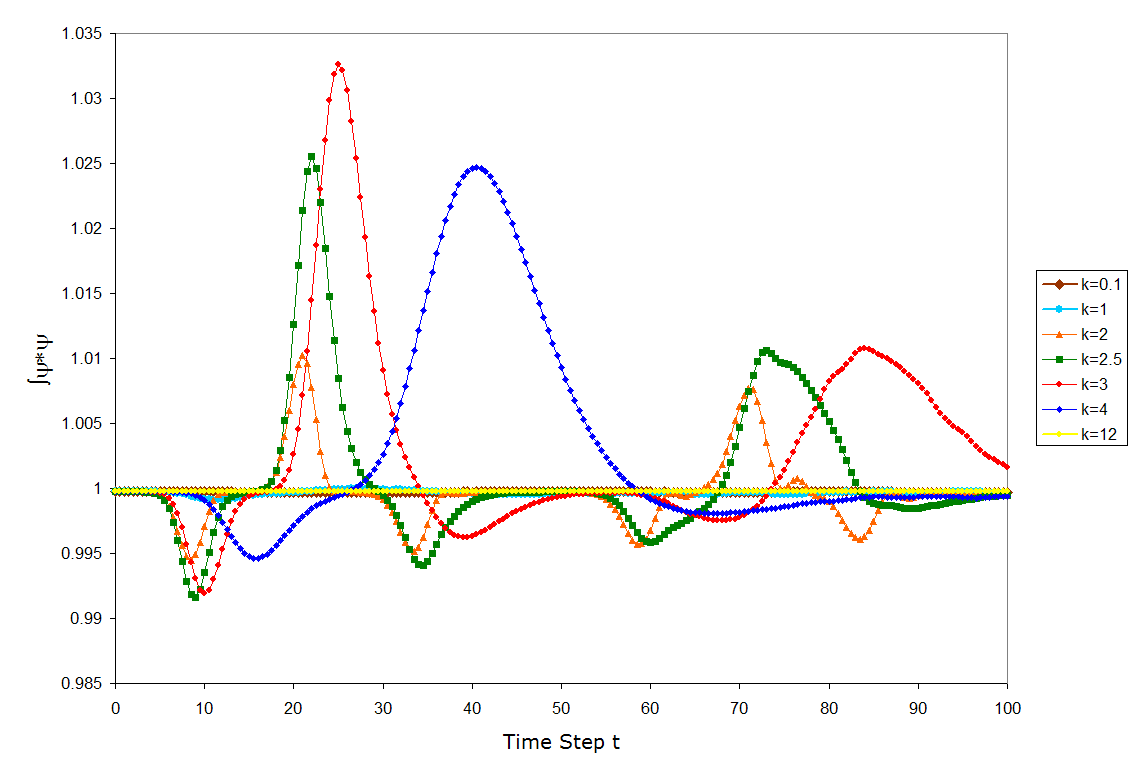}\\
        \caption{Conservation of area at each time step for varying initial wave vector $k_0$.}
        \label{area-conservation-barrier-k0}
    \end{center}
\end{figure}

\section{Space-Time Finite Element Analysis}
\subsection{Linear Continuous Space-Time Analysis}
\label{Linear Continuous Space-Time Analysis}
\subsubsection{Infinite Potential Well}
In order to apply space-time analysis we can follow the mathematical description for the discretisation of the diffusion equation in Sec.~\ref{linear-cont-discrete}.  As the \schro equation with $V=0$,
\begin{equation}
\frac{\partial \psi}{\partial t}-\gamma\frac{\partial^2\psi}{\partial x^2}=0 \qquad \gamma=i\frac{\hbar}{2m},
\end{equation}
is similar to Eqn.~(\ref{diffusion-eq}), we can simply write the discrete space time form of the \schro equation as
\begin{equation}
\textbf{0}=
\left\{
\left[
\begin{array}{cccc}
2&1&-1&-2\\
1&2&-2&-1\\
1&2&-2&-1\\
2&1&-1&-2\\
\end{array}
\right]- i\frac{\hbar\Delta t}{m\Delta x^2}
\left[
\begin{array}{cccc}
 2&-2&-1& 1\\
-2& 2& 1&-1\\
-1& 1& 2&-2\\
 1&-1&-2& 2\\
\end{array}
\right]
\right\}
\left[
\begin{array}{c}
\psi_1\\
\psi_2\\
\psi_3\\
\psi_4\\
\end{array}
\right],
\label{schro-space-time}
\end{equation}
where $\psi_1$ and $\psi_2$ are $j$ and $j+1$ nodal values respectively at $t_n$ and  $\psi_4$ and $\psi_3$ are $j$ and $j+1$ nodal values respectively at $t_{n+1}$ (Fig.~\ref{square-element}).  As $\psi_1$ and $\psi_2$ are known at each time step we can split Eqn.~(\ref{schro-space-time}) into two.  As we noted in Sec.~\ref{linear-cont-discrete}, using the first two rows we obtain an "explicit" time difference method, as the shape functions for $N_1$ and $N_2$ are weighted at $t_n$.  After some rearrangement we have
\begin{eqnarray}
\left\{
\left[
\begin{array}{cc}
2 & 1\\
1 & 2\\
\end{array}
\right]\right.
&+&\left.
i\frac{\hbar \Delta t}{m\Delta x^2}
\left[
\begin{array}{cc}
1 & -1\\
-1 & 1\\
\end{array}
\right]
\right\}
\left[
\begin{array}{c}
\psi_4\\
\psi_3\\
\end{array}
\right]
\nonumber\\
&=&
\left\{
\left[
\begin{array}{cc}
2 & 1\\
1 & 2\\
\end{array}
\right]-
i\frac{2\hbar\Delta t}{m\Delta x^2}
\left[
\begin{array}{cc}
1 & -1\\
-1 & 1\\
\end{array}
\right]
\right\}
\left[
\begin{array}{c}
\psi_1\\
\psi_2\\
\end{array}
\right].
\label{explicit-cont-spacetime}
\end{eqnarray}
We can do the same for the rows associated with the shape functions $N_3$ and $N_4$, giving an "implicit" time difference method, as this time the shape functions are weighted at $t_{n+1}$.  From these we obtain
\begin{eqnarray}
\left\{\left[
\begin{array}{cc}
2 & 1\\
1 & 2\\
\end{array}
\right]\right.&+&\left.
i\frac{2\hbar \Delta t}{m\Delta x^2}
\left[
\begin{array}{cc}
1 & -1\\
-1 & 1\\
\end{array}
\right]
\right\}
\left[
\begin{array}{c}
\psi_4\\
\psi_3\\
\end{array}
\right]\nonumber\\
&=&\left\{\left[
\begin{array}{cc}
2 & 1\\
1 & 2\\
\end{array}
\right]-
i\frac{\hbar\Delta t}{m\Delta x^2}
\left[
\begin{array}{cc}
1 & -1\\
-1 & 1\\
\end{array}
\right]
\right\}
\left[
\begin{array}{c}
\psi_1\\
\psi_2\\
\end{array}
\right]
\label{explicit-space-time-cont}
\end{eqnarray}
These equations can then be written as
\begin{eqnarray}
\left[\tilde{\textbf{A}}+i\tilde{\textbf{B}}\right]\bar{\psi}^{n+1}&=&\left[\tilde{\textbf{A}}-i2\tilde{\textbf{B}}\right]\bar{\psi}^{n},
\label{explicit}
\nonumber\\
\left[\tilde{\textbf{A}}+i2\tilde{\textbf{B}}\right]\bar{\psi}^{n+1}&=&\left[\tilde{\textbf{A}}-i\tilde{\textbf{B}}\right]\bar{\psi}^{n}
\label{implicit}
\end{eqnarray}
where
\begin{equation}
\tilde{\textbf{A}}=\left[
\begin{array}{cc}
2 & 1\\
1 & 2\\
\end{array}
\right] \qquad \tilde{\textbf{B}}=\frac{\hbar \Delta t}{m \Delta x^2}\left[
\begin{array}{cc}
1 & -1\\
-1 & 1\\
\end{array}
\right].
\end{equation}
We can note immediately that both these time step methods do not form unitary transformations, hence the conservation of probability property is not maintained\footnote{For a difference method $A\psi^{n+1}=B\psi^n$ to be unitary it must satisfy $\psi^{n+1\dagger}\psi^{n+1}=\psi^{n\dagger}(A^{-1}B)^{\dagger}(A^{-1}B)\psi^{n}=\psi^{n\dagger}\psi^{n}$}.  In Fig.~\ref{space-time-non-cayley} we have plotted the areas at each time step for these system of equations, and we can confirm that the area is not conserved.  For the explicit case the area explodes, and for the implicit case, even though it remains finite, it is heavily damped and so it drops to zero.
\begin{figure}
\begin{center}
\begin{tabular}{cc}
\includegraphics[width=0.5\textwidth]{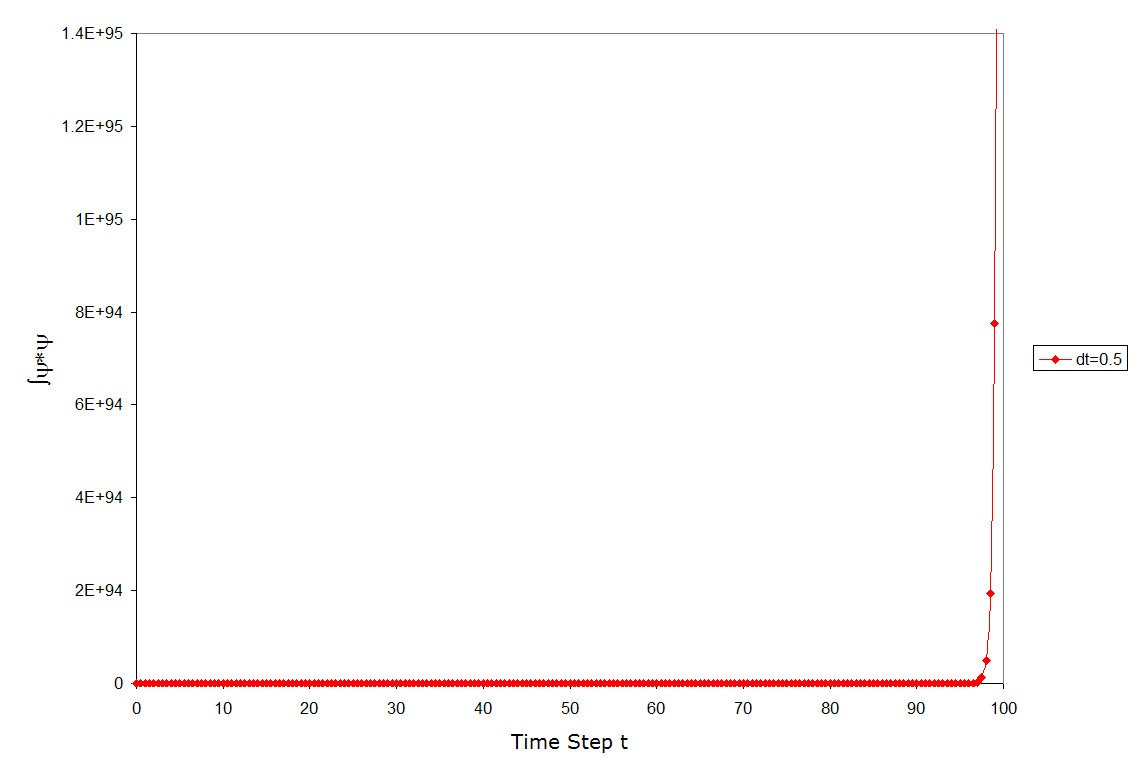}&
\includegraphics[width=0.5\textwidth]{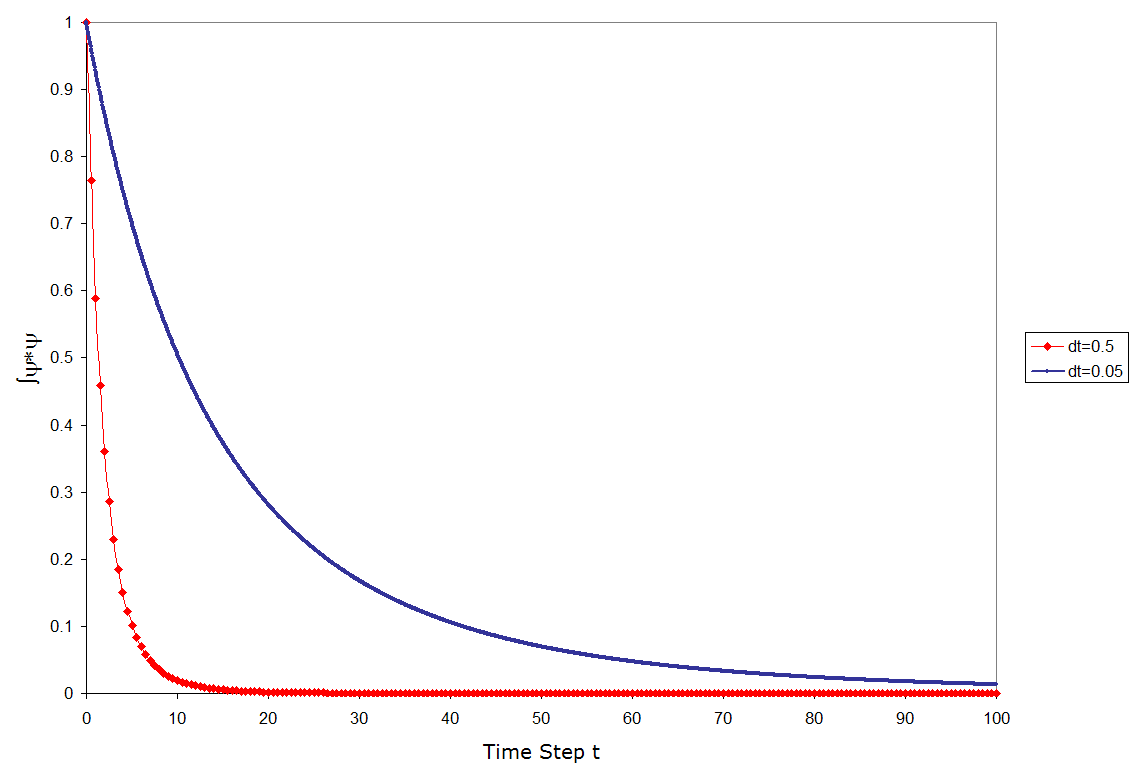}\\
\end{tabular}
\caption{Area after each time-step for explicit scheme (left) and implicit scheme (right).  Area explodes for the explicit case, and remains finite but not constant for the implicit case.}
\label{space-time-non-cayley}
\end{center}
\end{figure}
Also, if we combine Eqns.~(\ref{explicit}) and (\ref{implicit}), by summing them, we obtain
\begin{equation}
\left[\tilde{\textbf{A}}+i\frac{3}{2}\tilde{\textbf{B}}\right]\bar{\psi}^{n+1}=\left[\tilde{\textbf{A}}-i\frac{3}{2}\tilde{\textbf{B}}\right]\bar{\psi}^{n},
\label{a-32b}
\end{equation}
which is numerically identical to the equation derived by the FE and Crank-Nicolson methods in Eqn.~(\ref{cn-fd-schro}).  The reason for this is that when we add the two forms we are averaging the information from $t_n$ and $t_{n+1}$ and so we reproduce the Crank-Nicolson method (in particular for the linear elements.).

\subsubsection{Infinite Potential Well with Barrier}
Even though we have shown that working with the continuous space-time method is actually a disadvantage (due to a lack of conservation of probability) we will continue to construct the space-time potential term for the \schro equation as this will also be implemented into the discontinuous method.

In order to apply the space-time discretisation to the potential term we proceed as follows (assuming element sizes, $\Delta x$ and $\Delta t$, are constant over the space-time):
\begin{eqnarray}
-\frac{i}{\hbar}\int^{L}_0V(x)\phi\psi dx&=&
-i\frac{\Delta x\Delta t}{4\hbar}\sum_e V_e\int^{+1}_{-1}\bar{\phi}^{\dagger}\bar{N}^{\dagger}\bar{N}\bar{\psi}d\chi d\tau \nonumber\\
&=&
\bar{\phi}^{\dagger}\left\{-i\frac{\Delta x\Delta t}{4\hbar}\sum_e V_e\int^{+1}_{-1}\bar{N}^{\dagger}\bar{N}d\chi d\tau\right\} \bar{\psi},
\end{eqnarray}
where the space-time shape vector components, $\bar{N}=\left[N_1, N_2, N_3, N_4\right]$, are as in Eqn.~(\ref{space-time-shape-functions}).  After integrating over $\chi$ and $\tau$, we obtain the element equation
\begin{equation}
-i\frac{\Delta x\Delta t V_e}{36\hbar} 
\left[\begin{array}{cccc} 
4& 2& 1& 2\\
2& 4& 2& 1 \\
1& 2& 4& 2 \\
2& 1& 2& 4\\
\end{array}\right] 
\bar{\psi}
\end{equation}
To equate this to Eqn.~(\ref{schro-space-time}) we need to multiply this by the constant $-\frac{12}{\Delta x}$; this is due to the fact that in Eqn.~(\ref{full-equation-spacetime}) we divided the entire system by $\Delta x$ and multiplied by $-12$ in order to simplify the system.  After doing this the continuous space-time element equation is given by
\begin{eqnarray}
\textbf{0}=
\left\{
\left[
\begin{array}{cccc}
2&1&-1&-2\\
1&2&-2&-1\\
1&2&-2&-1\\
2&1&-1&-2\\
\end{array}
\right]\right.&-& \left.i\frac{\hbar\Delta t}{m\Delta x^2}
\left[
\begin{array}{cccc}
 2&-2&-1& 1\\
-2& 2& 1&-1\\
-1& 1& 2&-2\\
 1&-1&-2& 2\\
\end{array}
\right]\right.\nonumber\\
&-&
\left. 
i\frac{V_e \Delta t}{3\hbar} 
\left[\begin{array}{cccc} 
4& 2& 1& 2\\
2& 4& 2& 1 \\
1& 2& 4& 2 \\
2& 1& 2& 4\\
\end{array}\right]
\right\}
\left[
\begin{array}{c}
\psi_1\\
\psi_2\\
\psi_3\\
\psi_4\\
\end{array}
\right],
\label{schro-potential-space-time}
\end{eqnarray}
Now, reconstructing the explicit and implicit parts as before, we obtain
\begin{eqnarray}
\left[\tilde{\textbf{A}}+i\tilde{\textbf{B}}+i\tilde{\textbf{C}}\right]\bar{\psi}^{n+1}&=&\left[\tilde{\textbf{A}}-i2\tilde{\textbf{B}}-i2\tilde{\textbf{C}}\right]\bar{\psi}^{n},
\label{explicit-v}\\
\left[\tilde{\textbf{A}}+i2\tilde{\textbf{B}}+i2\tilde{\textbf{C}}\right]\bar{\psi}^{n+1}&=&\left[\tilde{\textbf{A}}-i\tilde{\textbf{B}}-i\tilde{\textbf{C}}\right]\bar{\psi}^{n}
\label{implicit-v}
\end{eqnarray}
where we have
\begin{equation}
\tilde{\textbf{A}}=\left[
\begin{array}{cc}
2 & 1\\
1 & 2\\
\end{array}
\right] \qquad \tilde{\textbf{B}}=\frac{\hbar \Delta t}{m \Delta x^2}\left[
\begin{array}{cc}
1 & -1\\
-1 & 1\\
\end{array}
\right]\qquad \tilde{\textbf{C}}=\frac{V_e\Delta t}{3\hbar}\left[
\begin{array}{cc}
2 & 1\\
1 & 2\\
\end{array}
\right].
\end{equation}
Just from observation we can conclude that these are not consistent with the conservation of probability.  However, as before, when we add Eqns.~(\ref{explicit-v}) and (\ref{implicit-v}) together we obtain
\begin{equation}
\left[\tilde{\textbf{A}}+i\frac{3}{2}\tilde{\textbf{B}}+i\frac{3}{2}\tilde{\textbf{C}}\right]\bar{\psi}^{n+1}=\left[\tilde{\textbf{A}}-i\frac{3}{2}\tilde{\textbf{B}}-i\frac{3}{2}\tilde{\textbf{C}}\right]\bar{\psi}^{n},
\label{explicit+implict-v}\\
\end{equation}
which is numerically identical to Eqn.~(\ref{cn-fd-schro-potential}).

\subsection{Linear Discontinuous Space-Time Analysis}
\label{Linear Discontinuous Space-Time Analysis}
\subsubsection{Infinite Potential Well}
By following the procedure described in Sec.~\ref{linear-discontinuous-discrete} we can write Eqn.~(\ref{schro-space-time}) in linear discontinuous form as
\begin{eqnarray}
\textbf{0}&=&
\left\{
\left[
\begin{array}{cccc}
2&-2&1&-1\\
2&-2&1&-1\\
1&-1&2&-2\\
1&-1&2&-2\\
\end{array}
\right]- i\frac{\hbar\Delta t}{m \Delta x^2}
\left[
\begin{array}{cccc}
 2&1&-2& -1\\
1& 2& -1&-2\\
-2& -1& 2&1\\
 -1&-2&1& 2\\
\end{array}
\right]\right.\nonumber\\
&-&\left.2\left[
\begin{array}{cccc}
 2&0&1&0\\
	0& 0& 0&0\\
1& 0& 2&0\\
 0&0&0& 0\\
\end{array}
\right]
\right\}
\left[
\begin{array}{c}
\psi^+_{j,n}\\
\psi^-_{j,n+1}\\
\psi^+_{j+1,n}\\
\psi^-_{j+1,n+1}\\
\end{array}
\right]
+
2\left[
\begin{array}{cccc}
0&2&0&1\\
0&0&0&0\\
0&1&0&2\\
0&0&0&0\\
\end{array}
\right]\left[
\begin{array}{c}
\psi^+_{j,n-1}\\
\psi^-_{j,n}\\
\psi^+_{j+1,n-1}\\
\psi^-_{j+1,n}\\
\end{array}
\right].\nonumber\\
\label{full-equation-spacetime}
\end{eqnarray}
Separating this into real and imaginary parts, as before, we obtain $8 \times 8$ matrix equation:
\begin{equation}
\left(\textbf{A}'-\textbf{B}'-\textbf{C}'\right)\bar{\psi}_{\alpha+1}=-\textbf{D}'\bar{\psi}_{\alpha},
\label{space-time-discontinuous-matrix-equation}
\end{equation}
where the vectors are given as
\begin{equation}
\bar{\psi}_{\alpha+1}=\left[\begin{array}{c}
Re[\psi^+_{j,n}]\\
Im[\psi^+_{j,n}]\\
Re[\psi^-_{j,n+1}]\\
Im[\psi^-_{j,n+1}]\\
Re[\psi^+_{j+1,n}]\\
Im[\psi^+_{j+1,n}]\\
Re[\psi^-_{j+1,n+1}]\\
Im[\psi^-_{j+1,n+1}]\\
\end{array}\right] \qquad 
\bar{\psi}_{\alpha}=\left[\begin{array}{c}
Re[\psi^+_{j,n-1}]\\
Im[\psi^+_{j,n-1}]\\
Re[\psi^-_{j,n}]\\
Im[\psi^-_{j,n}]\\
Re[\psi^+_{j+1,n-1}]\\
Im[\psi^+_{j+1,n-1}]\\
Re[\psi^-_{j+1,n}]\\
Im[\psi^-_{j+1,n}]\\
\end{array}\right],\\
\label{spacetime-vectors}
\end{equation}
and the matrices are given as
\begin{eqnarray}
\textbf{A}'&=&\tiny\left[
\begin{array}{ccccccccc}
2&0&-2&0&1&0&-1&0\\
0&2&0&-2&0&1&0&-1\\
2&0&-2&0&1&0&-1&0\\
0&2&0&-2&0&1&0&-1\\
1&0&-1&0&2&0&-2&0\\
0&1&0&-1&0&2&0&-2\\
1&0&-1&0&2&0&-2&0\\
0&1&0&-1&0&2&0&-2\\
\end{array}
\right]\normalsize\label{spacetime-1}\\
\textbf{B}'&=&\frac{\hbar\Delta t}{m \Delta x^2}\tiny
\left[
\begin{array}{cccccccc}
 0&-2&0&-1&0&2&0&1\\
 2&0&1&0&-2&0& -1&0\\
 0&-1&0&-2&0&1&0&2\\
1&0& 2&0& -1&0&-2&0\\
0&2&0&1&0&-2&0&-1\\
-2&0& -1&0& 2&0&1&0\\
0&1&0&2&0&-1&0&-2\\
-1&0&-2&0&1&0& 2&0\\
\end{array}
\right]\normalsize\\
\textbf{C}'&=&2\tiny\left[
\begin{array}{cccccccc}
 2&0&0&0&1&0&0&0\\
 0&2&0&0&0&1&0&0\\
	0&0&0&0&0& 0& 0&0\\
	0&0&0&0&0& 0& 0&0\\
1&0&0& 0& 2&0&0&0\\
0&1&0& 0&0& 2&0&0\\
	0&0&0&0&0& 0& 0&0\\
	0&0&0&0&0& 0& 0&0\\
\end{array}
\right]
\normalsize\\
\textbf{D}'&=&2\tiny
\left[
\begin{array}{cccccccc}
0&0&2&0&0&0&1&0\\
0&0&0&2&0&0&0&1\\
0&0&0&0&0&0&0&0\\
0&0&0&0&0&0&0&0\\
0&0&1&0&0&0&2&0\\
0&0&0&1&0&0&0&2\\
0&0&0&0&0&0&0&0\\
0&0&0&0&0&0&0&0\\
\end{array}
\right]\normalsize\label{spacetime-2}
\end{eqnarray}

These can then be assembled by summing the lower-right $4\times 4$ components of the first element to the top-left $4\times 4$ components of the second element, and so on for $n$ elements.  We can note from Eqn.~(\ref{space-time-discontinuous-matrix-equation}) that this system of equations is not unitary.

\subsubsection{Numerical Results}
Using the well specifications as in the Crank-Nicolson method ($-20\leq x \leq 20$) we obtained the wave-packet time evolution results.  In Figs.~\ref{spacetime-area-conservation-timestep} and \ref{spacetime-area-conservation-elements} it can be seen that even for the simple case, a wave-packet in an infinite potential well, there is a small amount of damping in the conservation of probability plots.  This damping can be controlled by varying the number of elements and the size of the time step.  Fig.~\ref{spacetime-area-conservation-timestep} shows that for $100$ elements there is significant damping for time steps $dt>0.1$; but for time steps $dt<0.05$ the damping becomes negligible, however such a small time step comes at the cost of greater computing time.  In Fig.~\ref{spacetime-area-conservation-elements} we have shown that the damping can also be controlled, to a lesser extent, by increasing the number of elements.  Using $50$ elements we have an almost $2\%$ loss of probability after $100$ time steps, however with $\geq100$ elements this loss reduces to $1\%$.  For $n$ elements we must again consider the computation time; for the discontinuous space-time method we have $8\times 8$ element matrices which gives global matrices of order $4n+4$.  Therefore, doubling the number of elements will increase the global matrices by a factor of $4$ which would in turn require more computation time.  
\begin{figure}
    \begin{center}
     \includegraphics[width=0.8\textwidth]{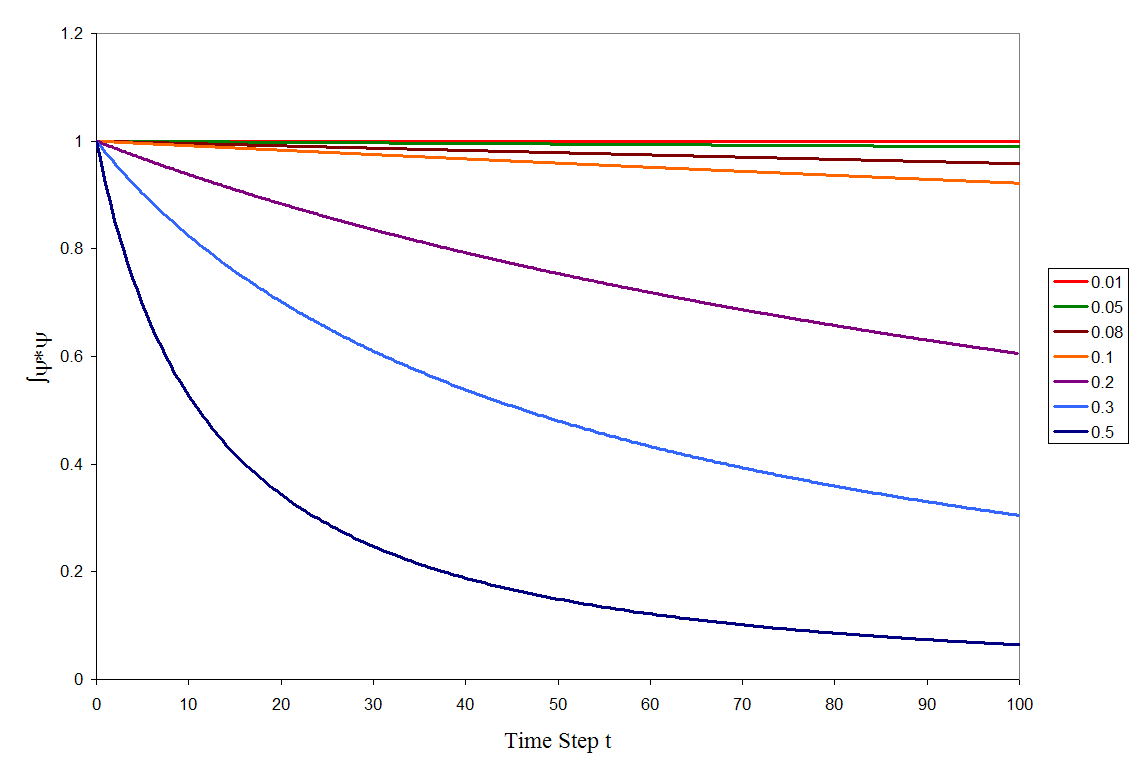}\\
        \caption{Conservation of probability at each time step for $100$ elements and varying size of timestep size: $dt$.}
        \label{spacetime-area-conservation-timestep}
    \end{center}
\end{figure}

\begin{figure}
    \begin{center}
     \includegraphics[width=0.8\textwidth]{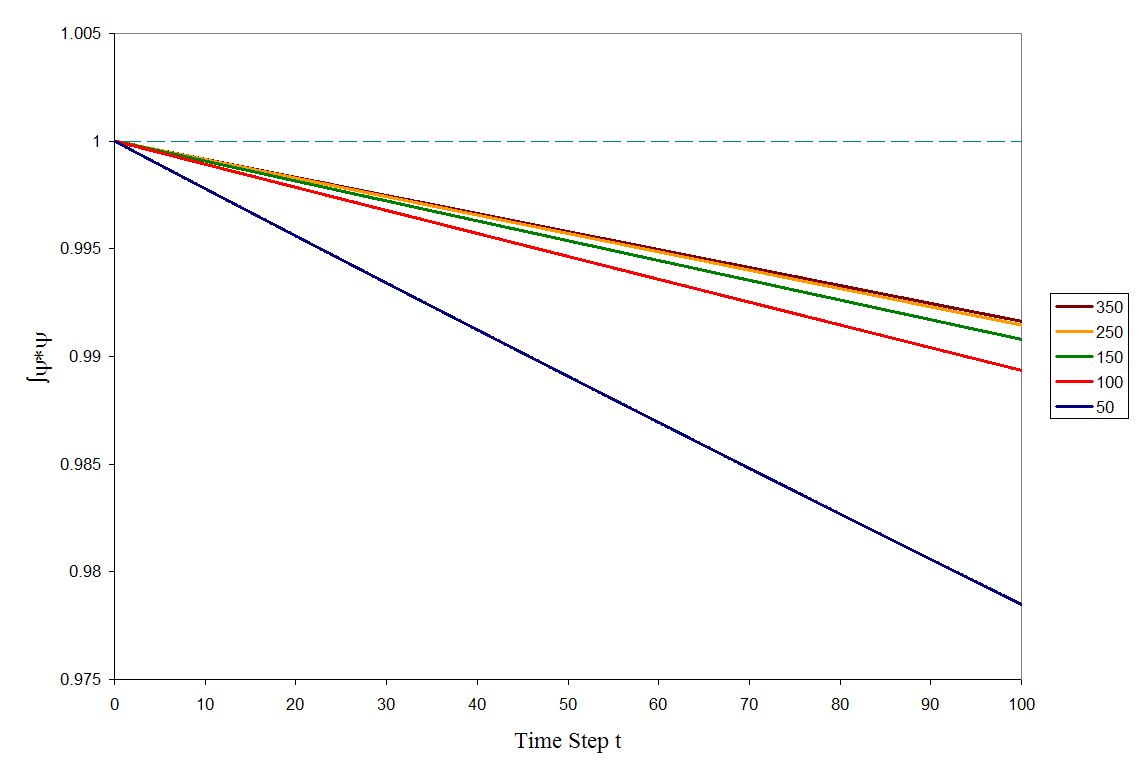}\\
        \caption{Conservation of probability at each time step for $dt=0.05$ and varying number of elements.}
        \label{spacetime-area-conservation-elements}
    \end{center}
\end{figure}
In Figs.~\ref{spacetime-prob-v-dt.png} and ~\ref{spacetime-prob-v-dt-close-up.png} we have plotted the total probability at timestep $t=100$ (for $100$ elements) against variations in time-step $dt$.  It can be seen that for smaller and smaller time steps the damping almost decreases to zero.  Also, in Fig.\ref{spacetime-prob-v-elements} we have plotted the total probability at timestep $t_n=100$ (for $dt=0.05$) against variations in number of elements.  This shows that changing the number of elements has very little affect after $100$ elements.  This implies, for the simple case of a packet in the infinite well without any barrier interactions, the time step size $dt$ has more of an effect on the damping than the number of elements used. 
\begin{figure}
    \begin{center}
     \includegraphics[width=0.8\textwidth]{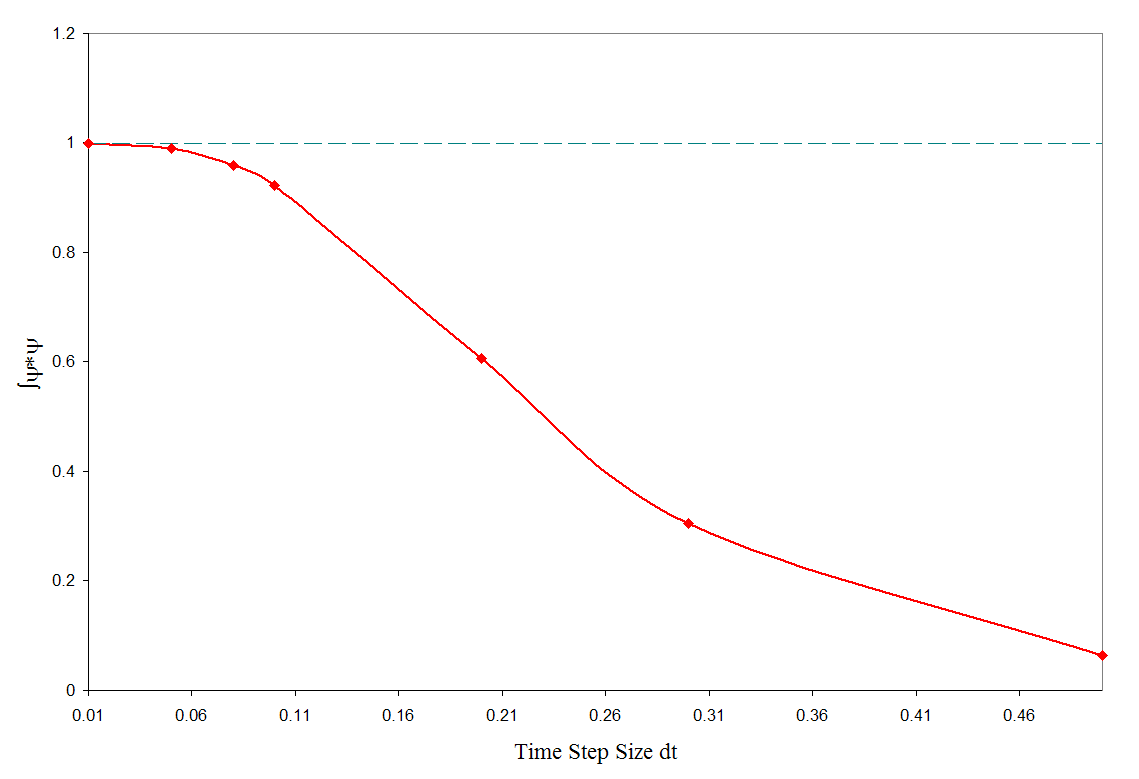}\\
        \caption{Conservation of probability at $t=100$ against $dt$ for $100$ elements.}
        \label{spacetime-prob-v-dt.png}
    \end{center}
\end{figure}

\begin{figure}
    \begin{center}
     \includegraphics[width=0.8\textwidth]{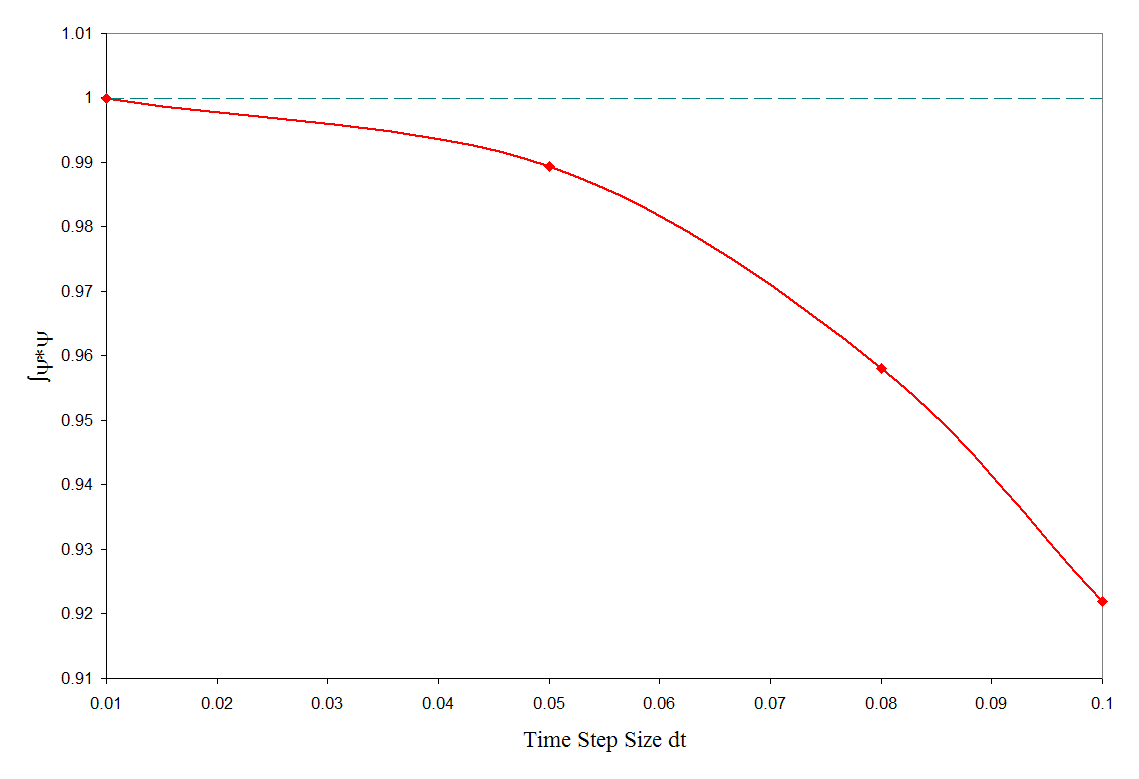}\\
        \caption{Close-up of conservation of probability at $t=100$ against $dt$ for $100$ elements.}
        \label{spacetime-prob-v-dt-close-up.png}
    \end{center}
\end{figure}

\begin{figure}
    \begin{center}
     \includegraphics[width=0.8\textwidth]{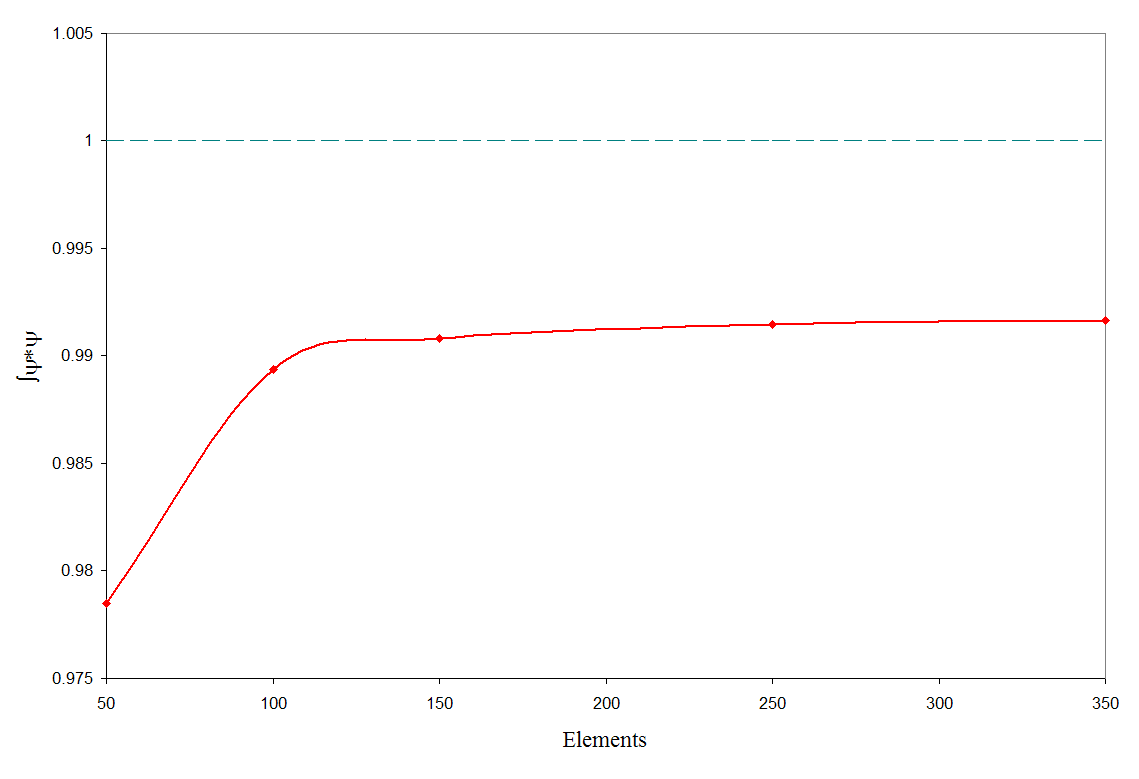}\\
        \caption{Conservation of probability at $t=100$ against number of elements for $dt=0.05$.}
        \label{spacetime-prob-v-elements}
    \end{center}
\end{figure}

In Fig.~\ref{spacetime-area-conservation-elements-timestep} we have made a comparison of the possible number of elements, which do not require large computation times, and varying timesteps $dt$.  We can see that using $100$ elements with $dt=0.05$ gives accepptable damping of $<1\%$, where if we use $dt=0.01$ we have almost no damping but the computation times are considerably larger. 

 \begin{figure}
    \begin{center}
     \includegraphics[width=0.8\textwidth]{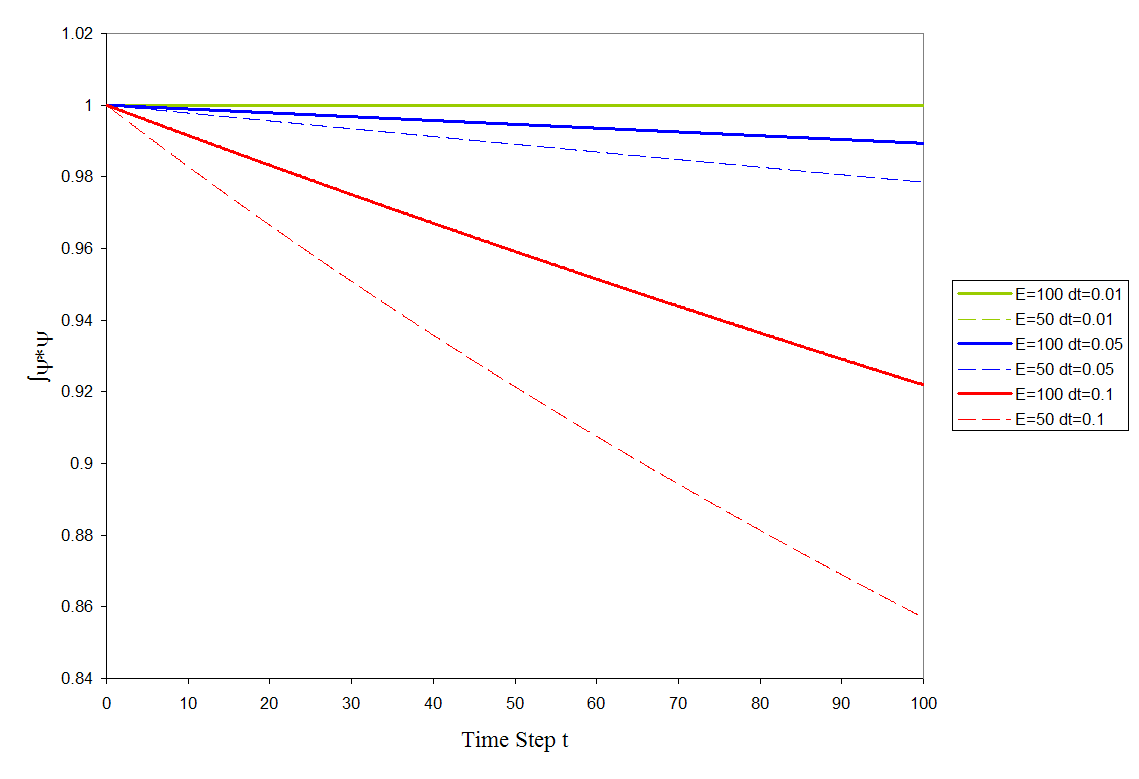}\\
        \caption{Conservation of probability at each time step for $50$ and $100$ elements and timesteps $dt=0.1$, $dt=0.05$, and $dt=0.01$.}
        \label{spacetime-area-conservation-elements-timestep}
    \end{center}
\end{figure}
When we compare the discontinuous space-time method to the Crank-Nicolson and Finite--Element method in Sec.~\ref{infinite-potential-well}, we can see that using the Crank-Nicolson method is not only computationally efficient but it also has zero damping i.e. it holds the unitary property of the \schro equation. 

\subsubsection{Infinite Potential Well with Barrier}
We can include a potential barrier into the discontinuous space-time method by using the continuous space-time potential term, given in Eqn.~(\ref{schro-potential-space-time}):
\begin{equation}
-i\frac{V_e \Delta t}{3\hbar} 
\left[\begin{array}{cccc} 
4& 2& 2& 1\\
2& 4& 1& 2 \\
2& 1& 4& 2 \\
1& 2& 2& 4\\
\end{array}\right],
\label{schro-discontinuous-potential-space-time}
\end{equation}
where the matrix has been rearranged to take into account the nodal numbering in the discontinuous method.  Writing this in real form we have the $8\times 8$ matrix
\begin{equation}
\textbf{V}'=\frac{V_e\Delta t}{3 \hbar}\tiny\left[
\begin{array}{ccccccccc}
0&-4&0&-2&0&-2&0&-1\\
4&0&2&0&2&0&1&0\\
0&-2&0&-4&0&-1&0&-2\\
2&0&4&0&1&0&2&0\\
0&-2&0&-1&0&-4&0&-2\\
2&0&1&0&4&0&2&0\\
0&-1&0&-2&0&-2&0&-4\\
1&0&2&0&2&0&4&0\\
\end{array}
\right].\normalsize\\
\end{equation}
We then have the full discontinuous space-time element equation:
\begin{equation}
\left(\textbf{A}'-\textbf{B}'-\textbf{C}'-\textbf{V}'\right)\bar{\psi}_{\alpha+1}=-\textbf{D}'\bar{\psi}_{\alpha},
\label{full-discontinuous-spacetime-potential}
\end{equation} 
where \textbf{A}', \textbf{B}', \textbf{C}' and \textbf{D}' are as before in Eqns.~(\ref{spacetime-1})--(\ref{spacetime-2}).

\subsubsection{Numerical Results}
As for the Crank-Nicolson method the barrier of height $V_e=2.5$ is implemented to the center of the well: $-0.8\leq x \leq 0.8$; then the assembly process is carried out as before: for $250$ elements we assemble Eqn.~(\ref{real-schro-barrier-crank-equation}) with $V_e=0$ from elements $1$ to $119$, then with $V_e=2.5$ from $120$ to $130$, then again with $V_e=0$ from $131$ to $250$.  Then using a timestep $dt$ and beginning with the initial wave-packet at $x_0=-13$ we obtain the time evolution of the initial wave-packet by the use of LU decomposition.

In Figs.~\ref{space-time-barrier-prob-element} and \ref{space-time-barrier-prob-dt} we have the conservation of probability plots for variations in number of elements and timestep size.  Fig.~\ref{space-time-barrier-prob-element} shows that by using larger number of elements we can resolve the wavefunction in more detail and thus reduce the fluctuations.  However, one point of interest is the fact that there are no peaks in the fluctuations, as was the case for the Crank-Nicolson method, Fig~\ref{area-conservation-barrier-space}.  

The main issue with the discontinuous space-time method is the damping and the computation time.  In  Fig.~\ref{space-time-barrier-prob-element} the damping can be seen to decrease by decreasing the timestep size ($dt$), however this in turn increases the computation time significantly.

\begin{figure}
    \begin{center}
     \includegraphics[width=0.8\textwidth]{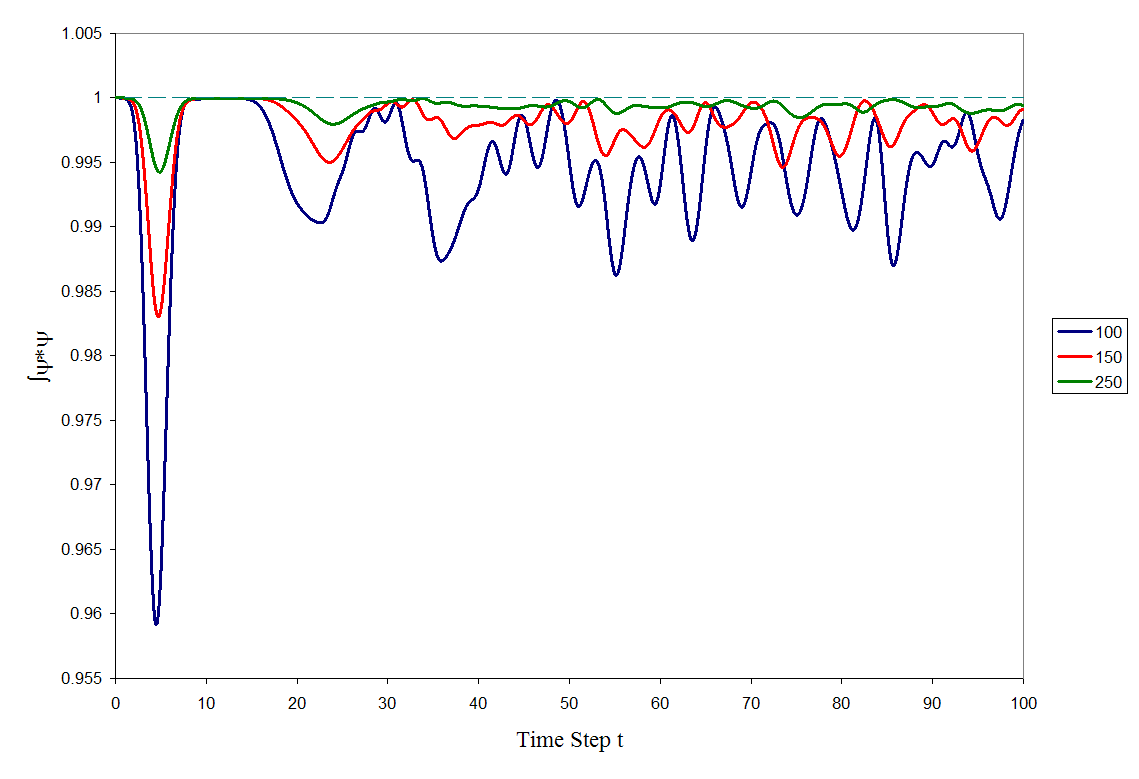}\\
        \caption{Conservation of probability at each timestep, for various number of elements (timestep size $dt =0.01$).}
        \label{space-time-barrier-prob-element}
    \end{center}
\end{figure}

 \begin{figure}
    \begin{center}
     \includegraphics[width=0.8\textwidth]{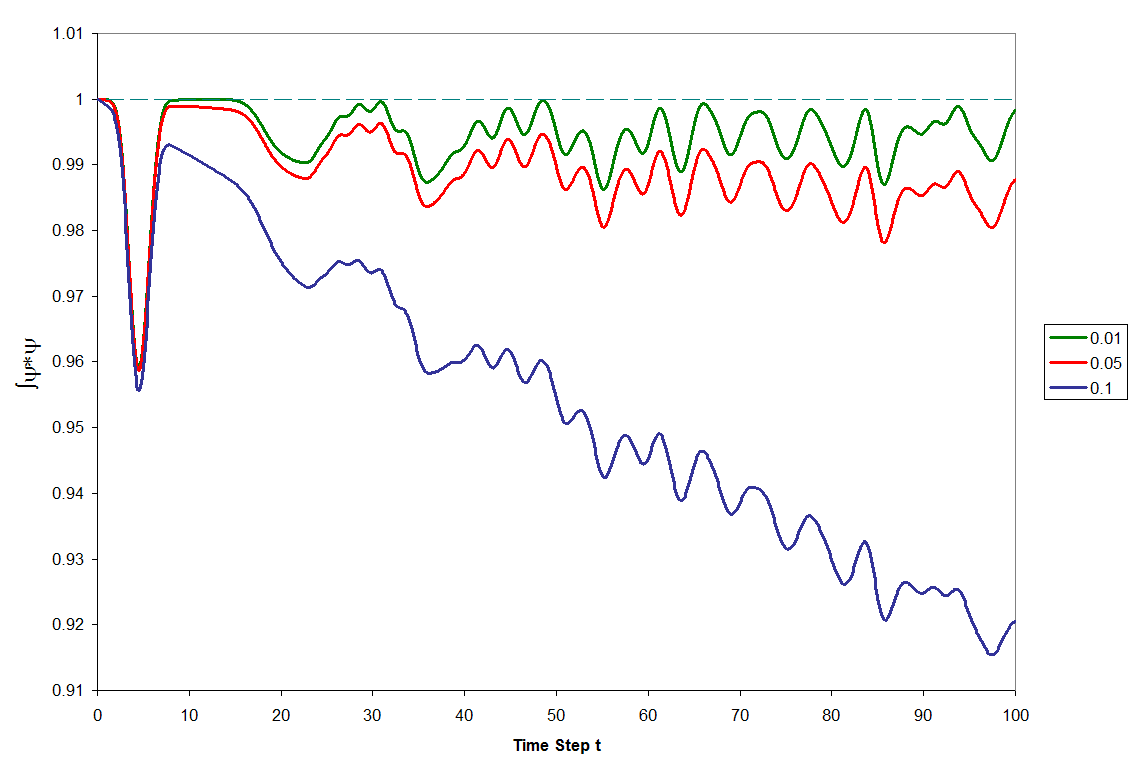}\\
        \caption{Conservation of probability at each timestep, for various timestep sizes $dt$ ($100$ elements).}
        \label{space-time-barrier-prob-dt}
    \end{center}
\end{figure}

In Fig.~\ref{dis-crank-space-time-barrier-conservation} it can be seen that for the same parameters ($250$ elements and $dt=0.05$) the Crank-Nicolson method conserves probability perfectly, whereas the discontinuous space-time method suffers from damping and thus falls by $1\%$ after $100$ timesteps.  However, the fluctuations assciated with the wavepackets interaction with the barrier and the well walls are slightly less in the space-time method compared to the Crank-Nicolson method.  In terms of computation time the Crank--Nicolson method is almost $5$ times faster than the space-time method for equivalent parameters.
 \begin{figure}
    \begin{center}
     \includegraphics[width=0.8\textwidth]{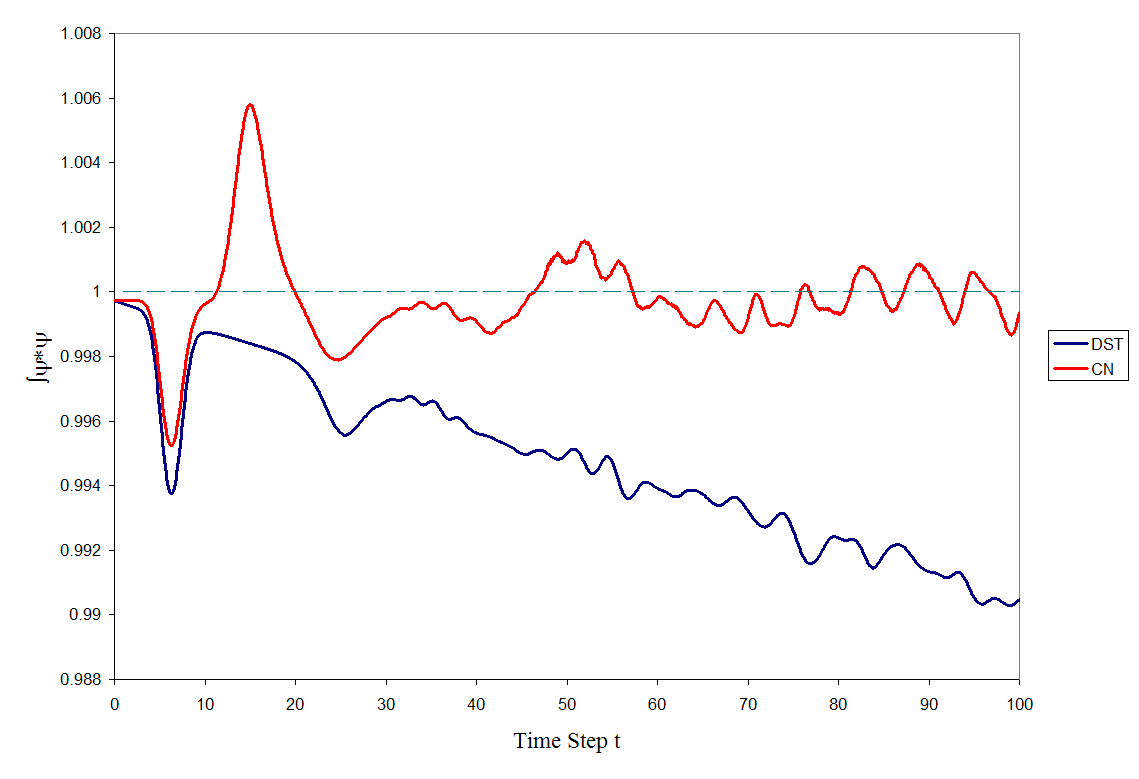}\\
        \caption{Conservation of probability at each timestep, for discontinuous space-time (DST) and Crank-Nicolson (CN) methods for $250$ elements and timestep $dt=0.05$.}
        \label{dis-crank-space-time-barrier-conservation}
    \end{center}
\end{figure}

}

{\typeout{Space Varying Potential $V(x)$}
\chapter{Wave-Packet in Sinusoidal Potential}
\label{space varying potential}
The Crank-Nicolson method turned out to be more efficient and accurate compared to the space-time methods.  An important limitation of the Crank-Nicolson time evolution method was when we model the wave function interacting with a change of potential.  This interaction results in fluctuations in the conservation of probability, however it was seen that these fluctuations can be controlled by increasing the number of spatial elements used.  In this chapter we will attempt to test this limitation by applying the Crank-Nicolson method to a particle in a periodic lattice potential,
\begin{equation}
V(x)=A \cos\left[k x\right],
\label{space-varying-potential}
\end{equation} 
where $A$ is the potential amplitude and $k$ is the wave number.  This periodic lattice potential is commonly used in solid state physics in order to simulate the atomic lattice structure within materials\footnote{For a detailed study of the solid state application and simplified solutions see \cite{solymar}}, Fig.~\ref{quantum-lattice-wire}.
\begin{figure}
    \begin{center}
     \includegraphics[width=1\textwidth]{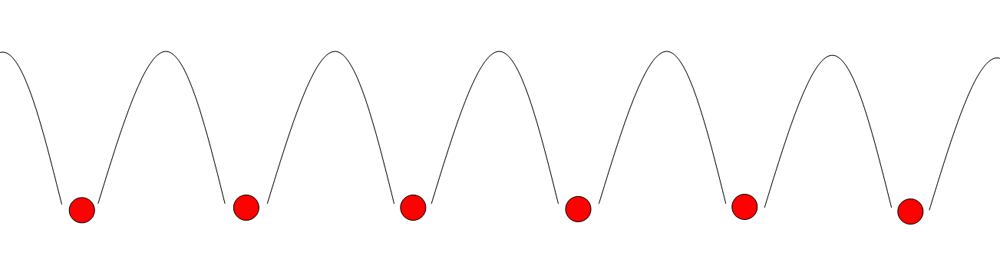}\\
        \caption{Model of a periodic lattice potential within a quantum wire.}
        \label{quantum-lattice-wire}
    \end{center}
\end{figure}
In terms of boundary conditions we could use periodic conditions, $\psi(x=0)=\psi(x=L)$, however in order to implement the changes into our previous work we will continue with the infinite well conditions, $\psi(x=0)=\psi(x=L)=0$.  This model of an infinite well with a sinusoidal potential could simulate a quantum wire where the potential represents the atomic lattice within the wire.  Thus, modelling the time evolution of a wave packet trapped within such a structure is a very simple "physical" test. 

\section{Construction of the model}
In order to incorporate this potential into our previous Crank-Nicolson/finite element method we could simply calculate the average value of Eqn.~(\ref{space-varying-potential}) within each element and then use these for $V_e$ in the element equation given in Eqn.~(\ref{standard-potential}).  In this way we obtain a constant FE approximation for the potential.  However, a more general method would be to recalculate the potential element term in Eqn.~(\ref{standard-potential}) using the potential given in Eqn.~(\ref{space-varying-potential}).  To obtain the best results we would need to use quadratic or higher order basis functions to take full advantage of this method of modelling the effects of the potential.  However, for simplicity we will continue using linear basis functions.  Therefore, using this general method and linear elements, as before, the potential element equation can be written as
\begin{eqnarray}
-\frac{i}{\hbar}\int^L_0 V(x)\phi \psi dx&=&-\frac{i}{2\hbar}\sum_e l_e \int^{1}_{-1}\bar{\phi}^{\dagger}V(\xi)\textbf{N}^{\dagger}\textbf{N}\bar{\psi}d\xi\nonumber\\
&=&-\frac{i}{8\hbar}\sum_e Al_e\int^1_{-1}\bar{\phi}^{\dagger}\cos\left[k\left(x_{e_1}\frac{1-\xi}{2}+x_{e_2}\frac{1+\xi}{2}\right)\right]\nonumber\\ 
& &\left[ \begin{array}{cc}
(1-\xi)(1-\xi)& (1-\xi)(1+\xi)\\
(1+\xi)(1-\xi)& (1+\xi)(1+\xi)\\
\end{array}
\right] \bar{\psi}d\xi,
\label{space-varying-potential-term}
\end{eqnarray}
where $x_{e_i}$ is the position of node $i$ of element $e$.  Carrying out the integral and simplifying we obtain the potential element matrix
\begin{equation}
\bar{\phi}^{\dagger}\left\{\frac{iAl_e}{8\hbar}\frac{8}{k^3(x_{e_1}-x_{e_2})^3}
\left[\begin{array}{cc}
C_{11}& C_{12}\\
C_{21}& C_{22}\\
\end{array}
\right]\bar{\psi}\right\},
\label{full-space-potential-equation}
\end{equation}
where 
\begin{eqnarray}
\tiny
C_{11}&=&(2-k^2(x_{e_1}-x_{e_2})^2)\sin\left[kx_{e_1}\right]-2(\sin\left[kx_{e_2}\right]+k(x_{e_1}-x_{e_2})\cos\left[kx_{e_1}\right])\nonumber\\ 
C_{12}=C_{21}&=&2\sin\left[kx_{e_2}\right]-2\sin\left[kx_{e_1}\right]+k(x_{e_1}-x_{e_2})(\cos\left[kx_{e_1}\right]+\cos\left[kx_{e_2}\right])\nonumber\\
C_{22}&=&(-2+k^2(x_{e_1}-x_{e_2})^2)\sin\left[kx_{e_2}\right]+2(\sin\left[kx_{e_1}\right]-k(x_{e_1}-x_{e_2})\cos\left[kx_{e_2}\right]).\nonumber\\
\normalsize
\end{eqnarray}
also, if $l_e=x_{e_1}-x_{e_2}$ we can write the full \schro element equation (\ref{full-schro-element-equation}) as
\begin{equation}
\left[
\begin{array}{cc}
	2 & 1 \\
	1 & 2 \\
\end{array}
\right]\dot{\bar{\psi}}+i\left\{\frac{6\hbar}{2ml_e^2}\left[
\begin{array}{cc}
	1 & -1 \\
	-1 & 1 \\
\end{array}
\right]+\frac{6A}{\hbar k^2l_e^3}
\left[\begin{array}{cc}
C_{11}& C_{12}\\
C_{21}& C_{22}\\
\end{array}
\right]\right\}\bar{\psi}.
\label{full-complicated-space-potential-equation}
\end{equation}
In operator form this expression is given as
\begin{equation}
\tilde{\textbf{A}}\dot{\bar{\psi}}+i\left\{\tilde{\textbf{B}}+\tilde{\textbf{C}}\right\}\bar{\psi}=\textbf{0},
\end{equation}
with $\tilde{\textbf{C}}$ representing the lattice-potential element matrix.  Now, following the steps as in Section~\ref{infinite-well-wave-packet-equation} we can write the Crank-Nicolson approximation as
\begin{equation}
\left(\tilde{\textbf{A}}+i\left\{\tilde{\textbf{B}} + \tilde{\textbf{C}}\right\}\right)   \bar{\psi}^{n+1}=
\left(\tilde{\textbf{A}}-i\left\{\tilde{\textbf{B}}+\tilde{\textbf{C}}\right\}\right)\bar{\psi}^{n}.
\end{equation}
Putting this in real form we have
\begin{equation}
\left(
\textbf{A}'+\alpha\textbf{B}'+\beta\textbf{C}'
\right)\bar{\psi}^{n+1}=
\left(
\textbf{A}'-\alpha\textbf{B}'-\beta\textbf{C}'
\right) \bar{\psi}^n,
\end{equation}
where \textbf{A}', \textbf{B}', $\alpha$ and $\beta$ are as in Eqn.~(\ref{real-schro-barrier-crank-equation}), but the potential term is 
\begin{equation}
\textbf{C}'=\left[\begin{array}{cccc}
0& -C_{11}& 0 &-C_{12}\\
C_{11}& 0 &C_{12}&0\\
0& -C_{21}& 0 &-C_{22}\\
C_{21}& 0 &C_{22}&0\\
\end{array}
\right],
\label{lattice-potential-matrix-term}
\end{equation}
and $\beta=\frac{6A}{\hbar k^2l_e^3}$.

\subsubsection{Wave-packet with $k\neq 0$}

When we use a wave-packet with $k_0=2$, $1000$ spatial elements and $dt=0.1$, combined with the potential parameters: $A=0.5$ and $k=\frac{2\pi}{10}$, we have a packet that begins to move to the right with a small amount of "diffusion" to the left, Fig~\ref{lattice-potential-k-2}. 
\begin{figure}
\begin{center}
\begin{tabular}{cc}
\includegraphics[width=0.5\textwidth]{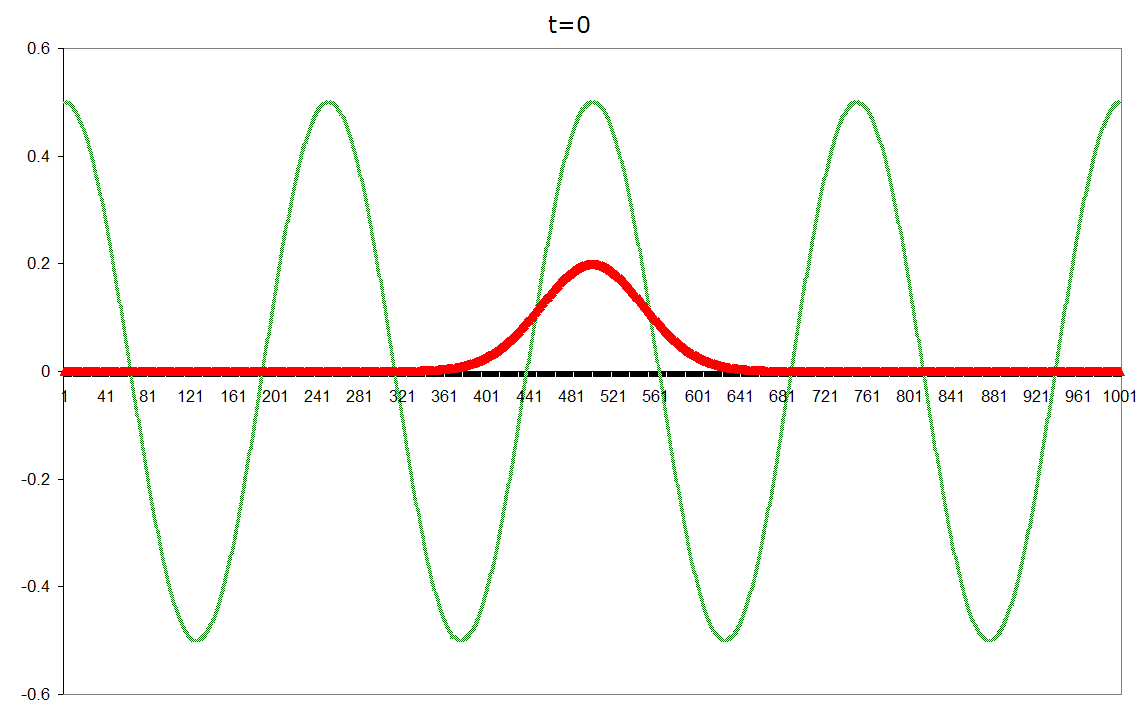}&
\includegraphics[width=0.5\textwidth]{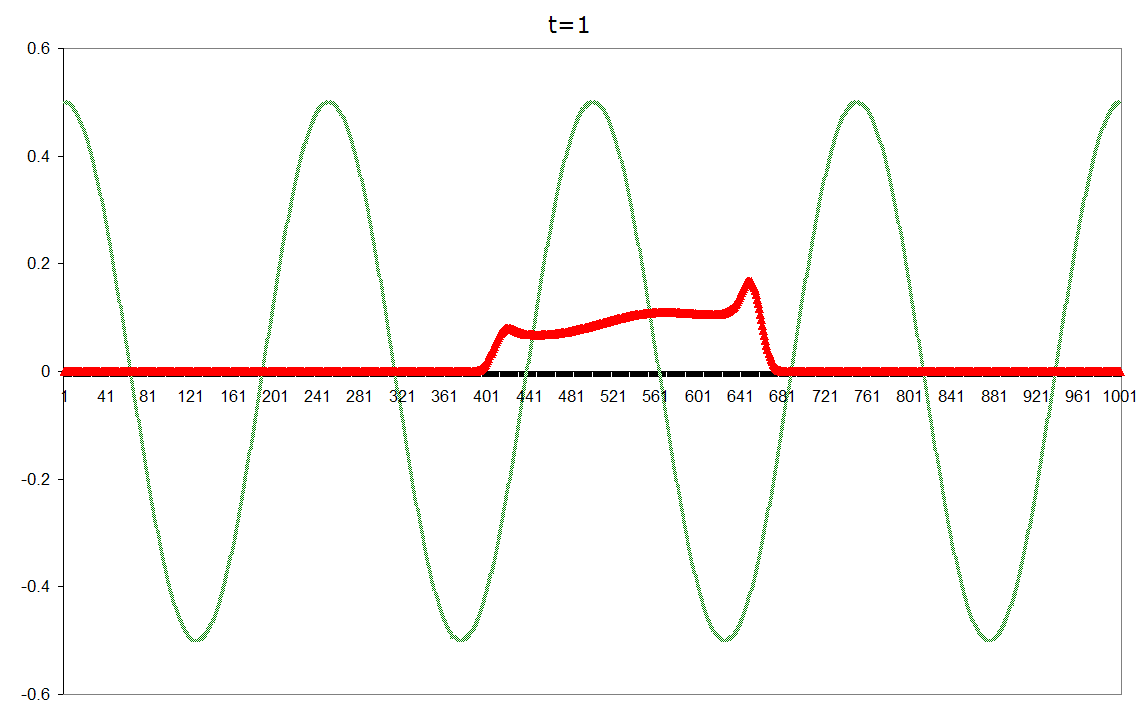}\\
\includegraphics[width=0.5\textwidth]{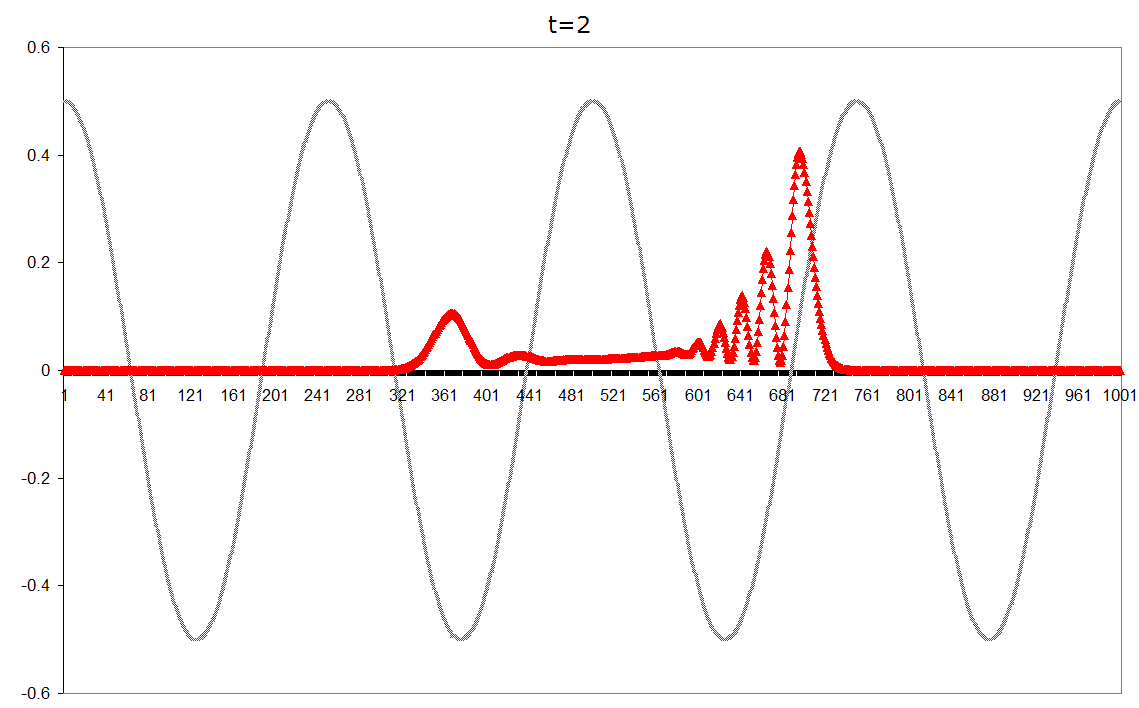}&
\includegraphics[width=0.5\textwidth]{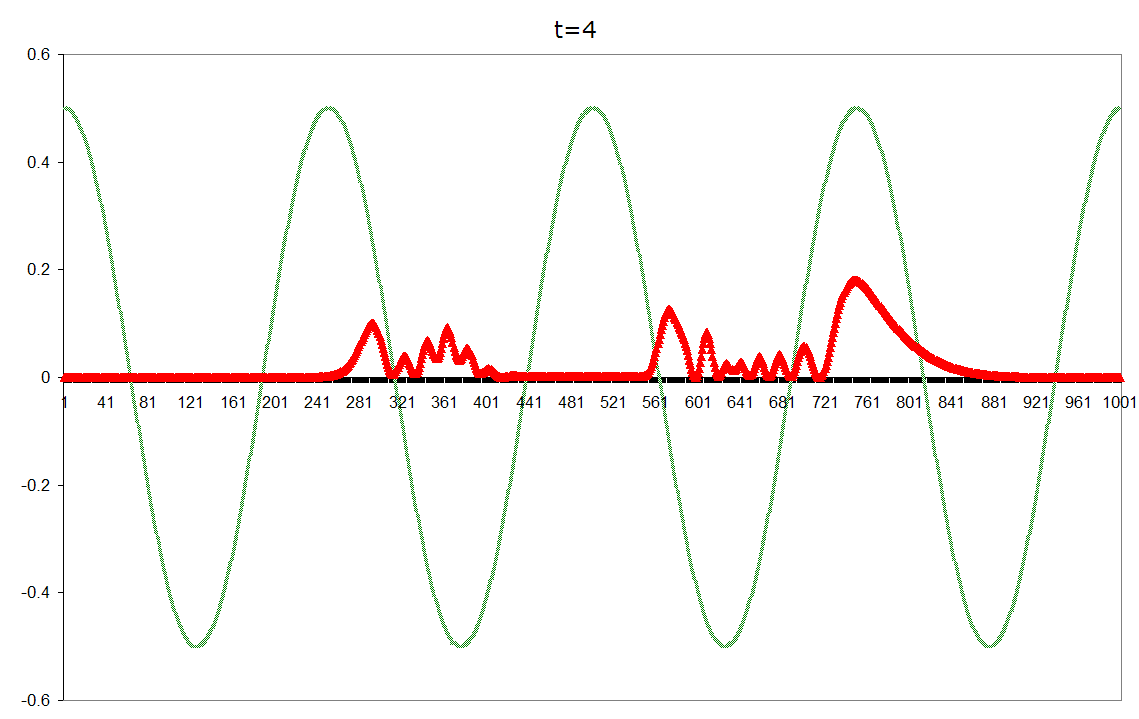}\\
\includegraphics[width=0.5\textwidth]{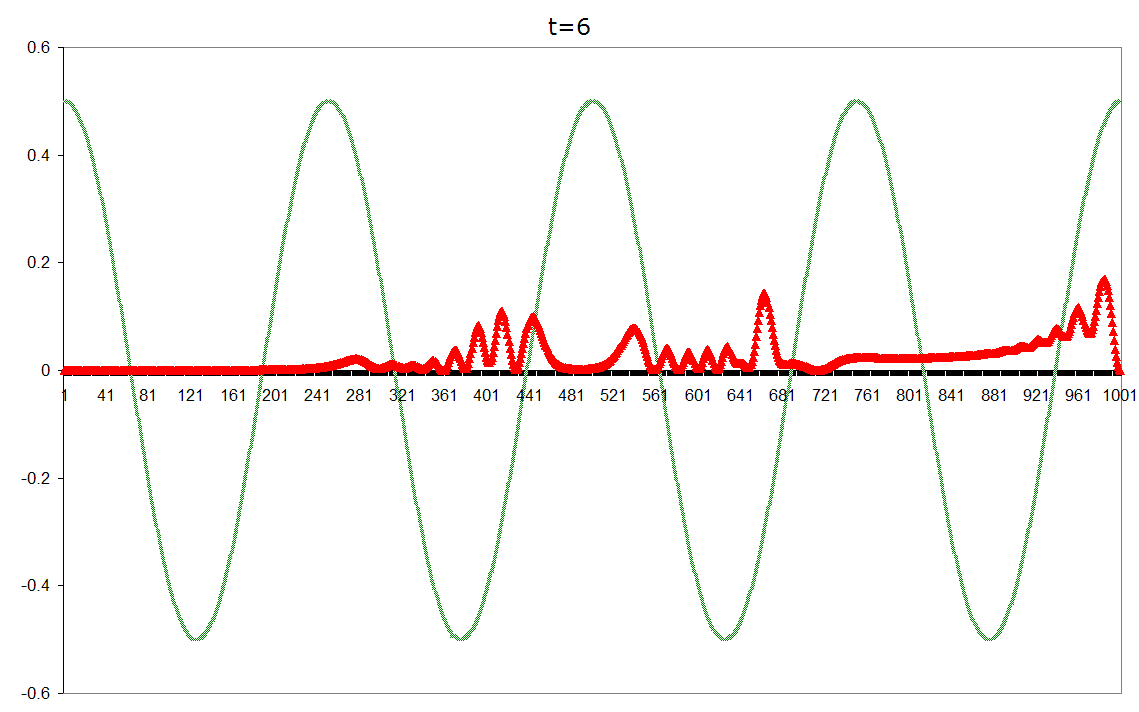}&
\includegraphics[width=0.5\textwidth]{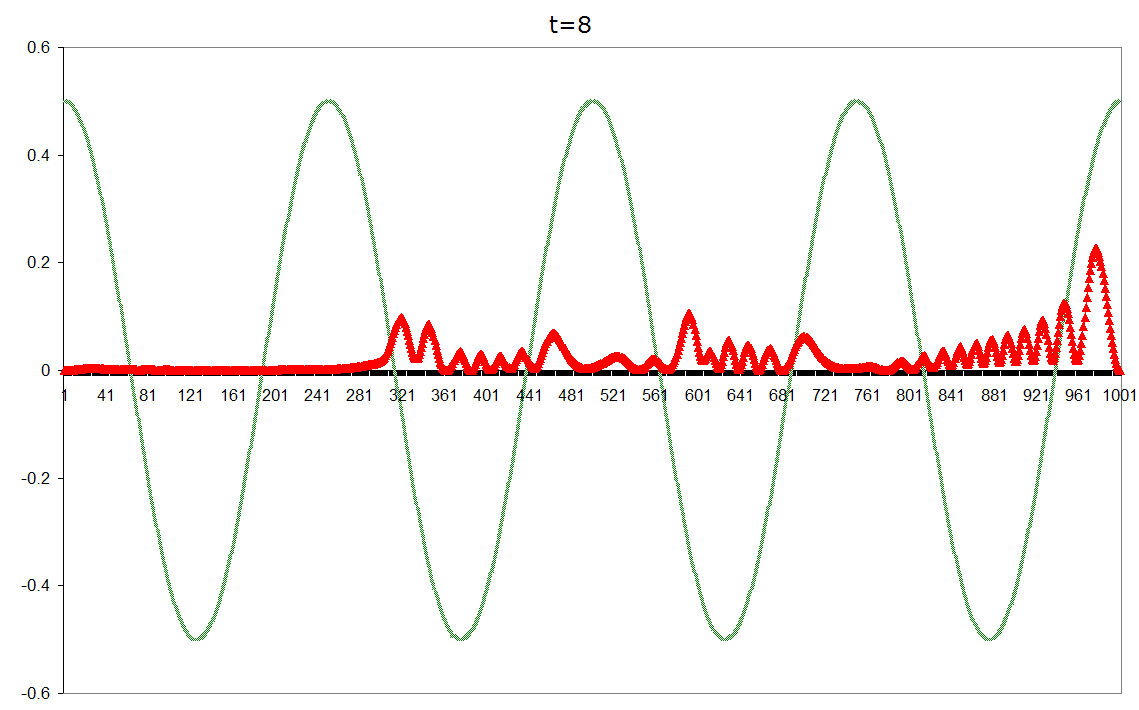}\\
\includegraphics[width=0.5\textwidth]{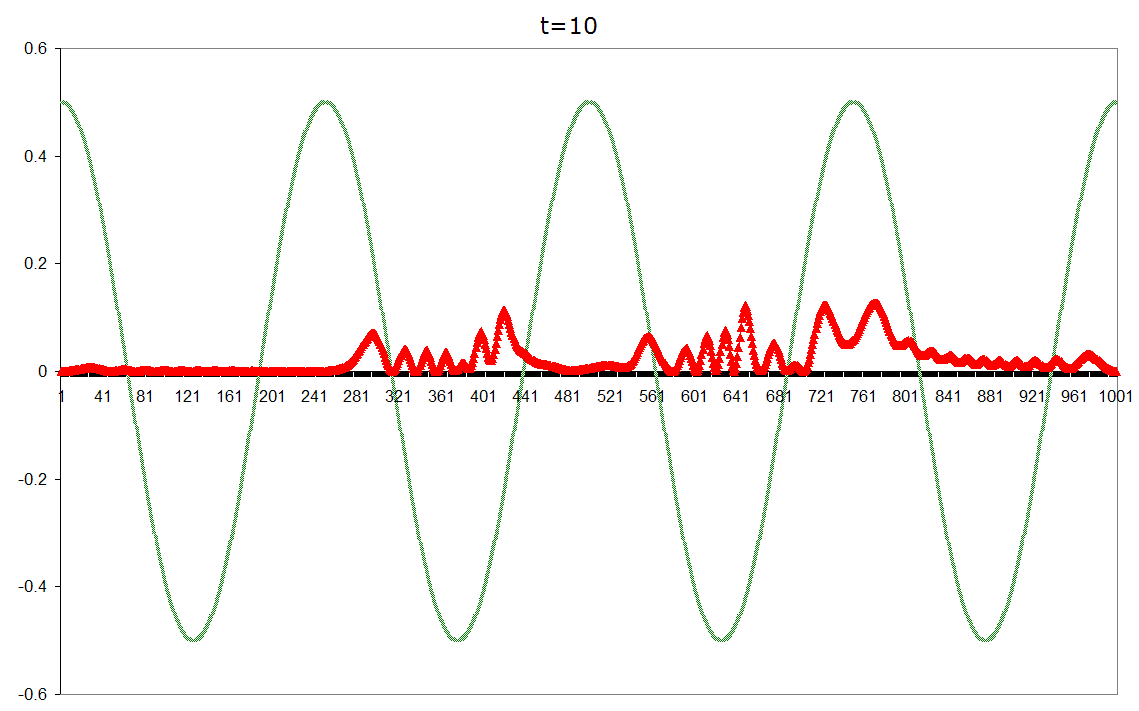}&
\includegraphics[width=0.5\textwidth]{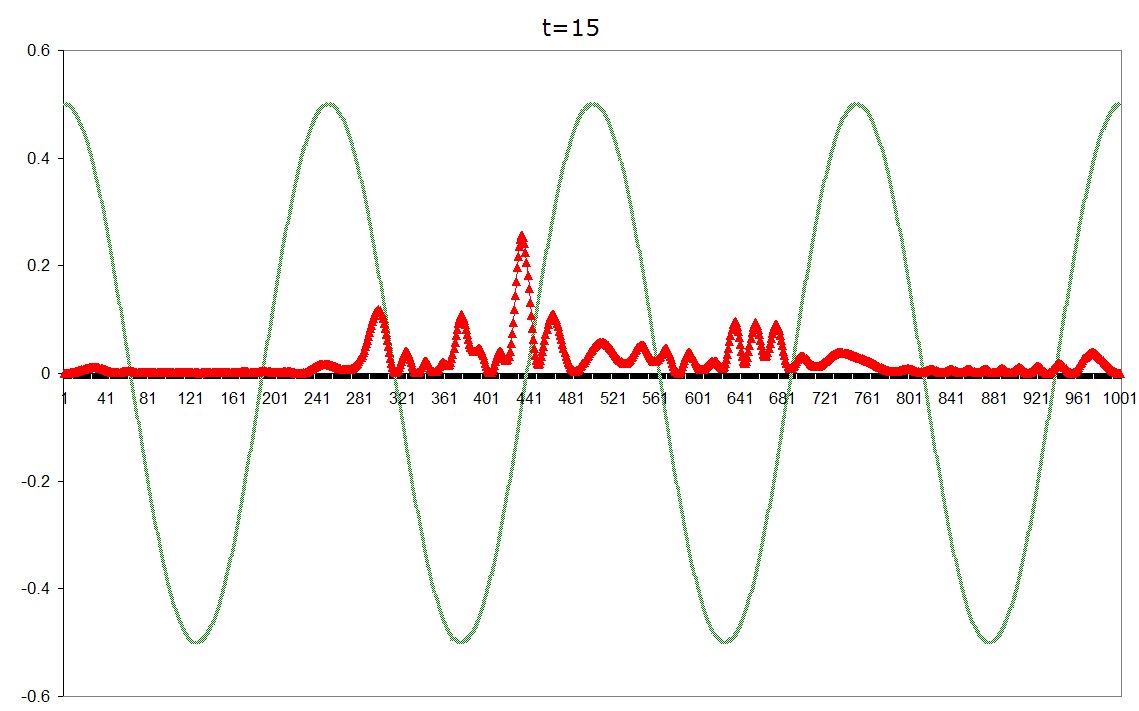}\\
\includegraphics[width=0.5\textwidth]{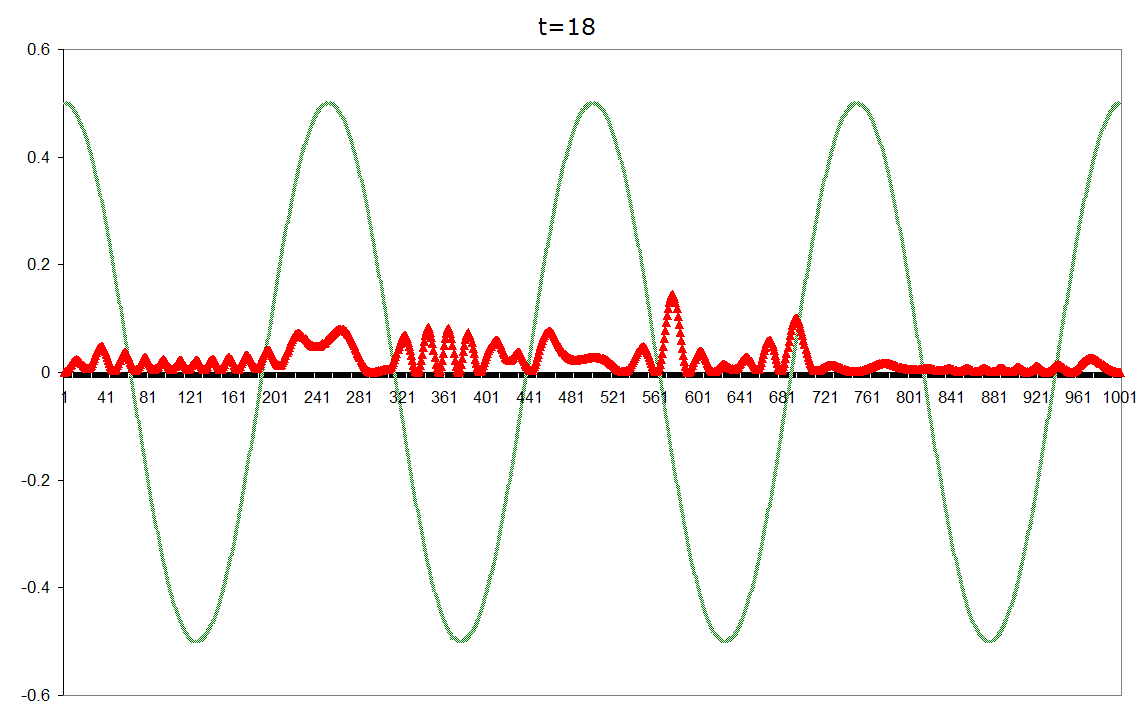}&
\includegraphics[width=0.5\textwidth]{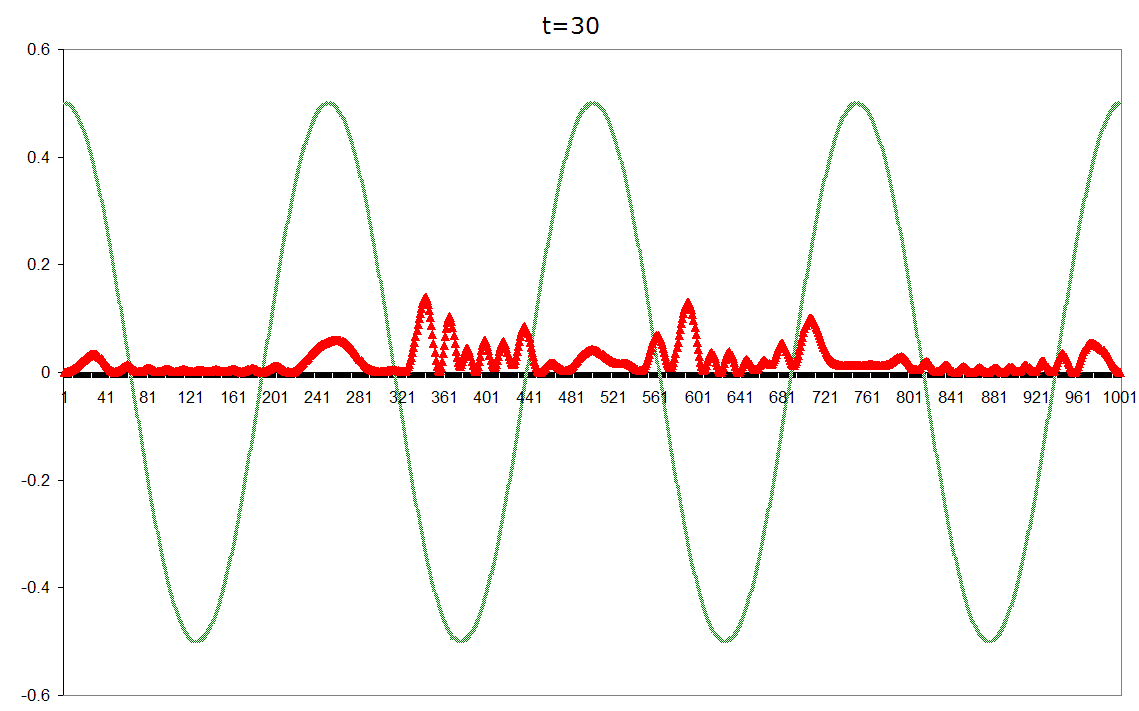}\\
\end{tabular}
\caption{Wave packet with $k_0=2$ spreads over the lattice-potential ($Re^2+Im^2$ are shown).}
\label{lattice-potential-k-2}
\end{center}
\end{figure}
As the packet interacts with many potential barriers in the lattice the probability conservation fluctuates and drops significantly after a very short time.  Using a timestep size $dt=0.1$ the probability, after $100$ timesteps, falls by $33\%$ for $500$ elements, $20\%$ for $1000$ elements and $12\%$ for $2000$ elements, Fig.~\ref{sinusoidal-vary-elements}.  From Fig.~\ref{sinusoidal-element-extrapolation} it can be seen that as we increase the number of elements the damping effect diminishes, and for larger number of elements the timestep size has little effect.  In Fig.~\ref{sinusoidal-vary-time-step} it can be seen that just reducing the timestep size has a worse effect, i.e. the errors due to a lack of resolving of the wave function at lower times are perpetuated along the time evolution, which results in highly unphysical results (probability $>1$).  Thus to obtain the best result we would need to use a time step of $dt=0.05$ (as reducing the timestep further would have little accuracy effect but would add greatly to the computation time) and use $\geq 3000$ spatial elements (more elements would add to the resolving power for complicated wave functions at later times). 
\begin{figure}
    \begin{center}
     \includegraphics[width=0.7\textwidth]{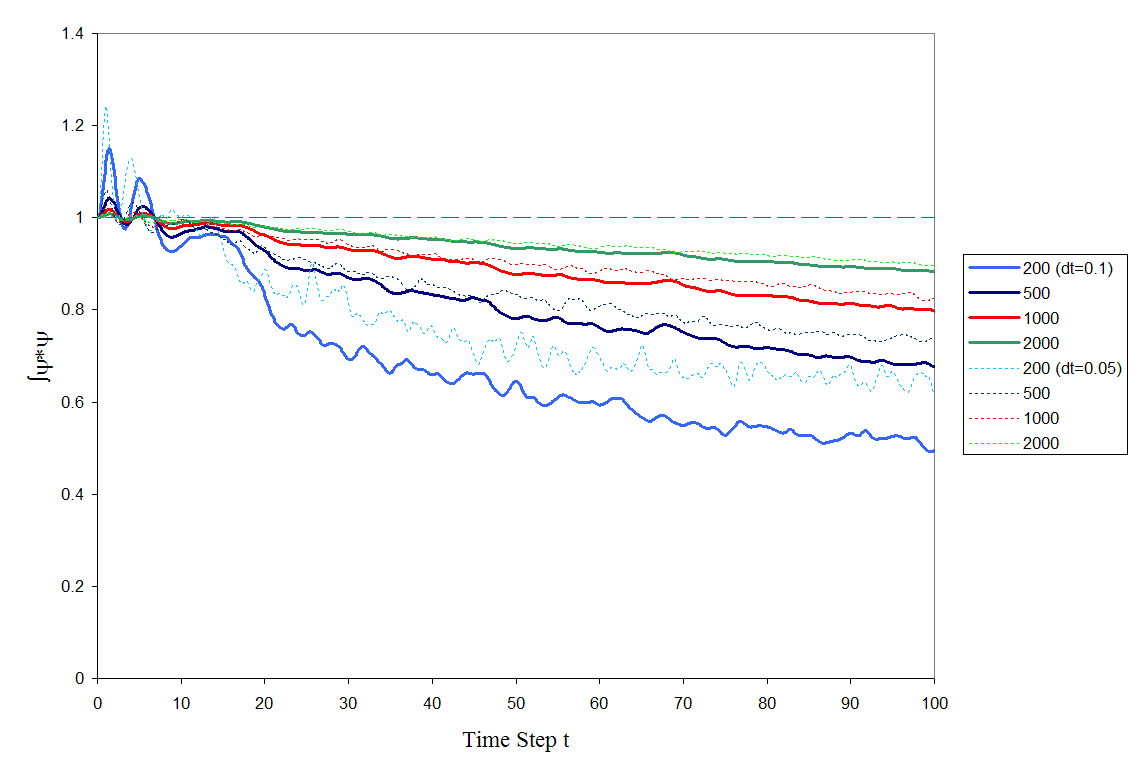}\\
        \caption{Conservation of probability for various number of elements with $dt=0.1$ (for packet with $k_0=2$.  Dashed line represent same results but for $dt=0.05$ }
               \label{sinusoidal-vary-elements}
    \end{center}
\end{figure}

\begin{figure}
    \begin{center}
	  \includegraphics[width=0.7\textwidth]{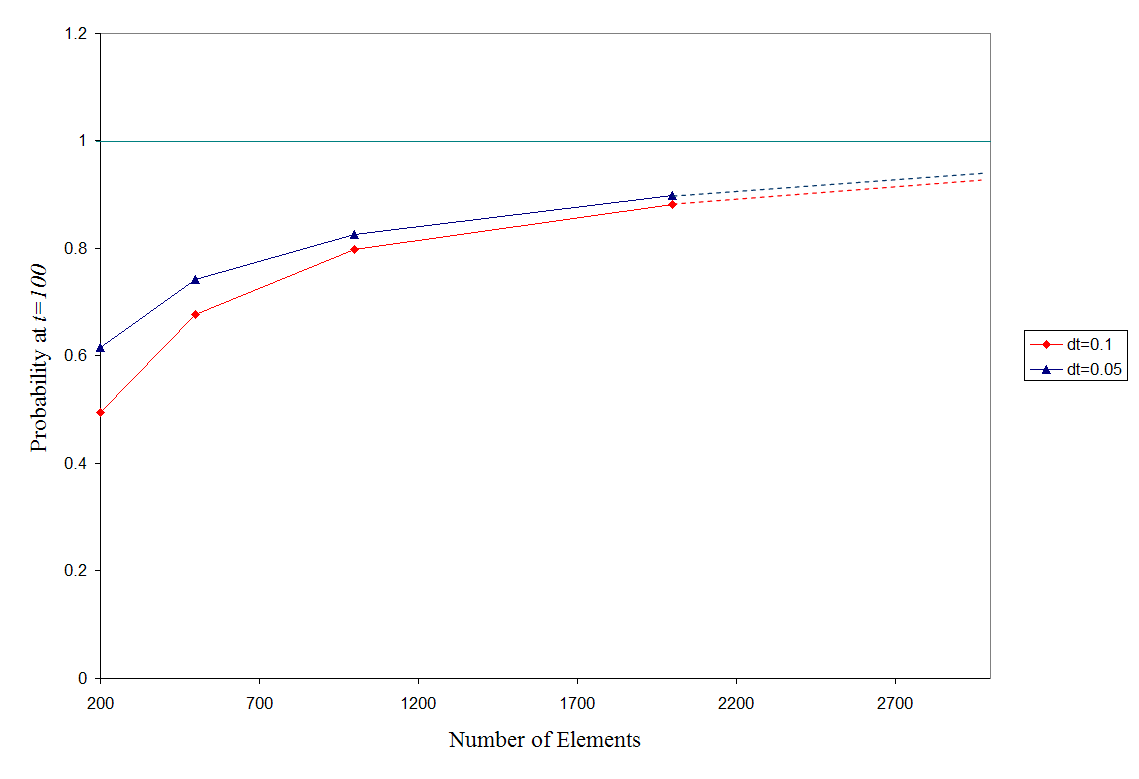}\\
        \caption{Probability at $t=100$ against number of elements.  Dashed line shows possible extrapolation. (for packet with $k_0=2$)}
        \label{sinusoidal-element-extrapolation}
    \end{center}
\end{figure}

\begin{figure}
    \begin{center}
	  \includegraphics[width=0.7\textwidth]{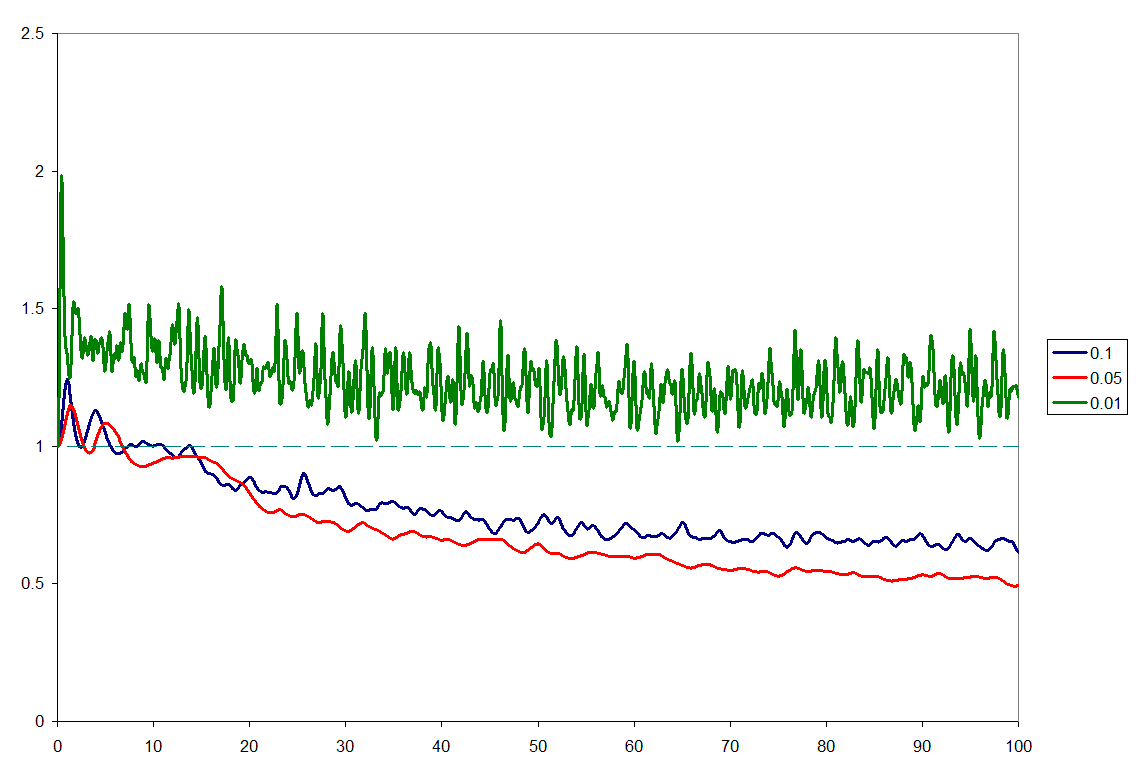}\\
        \caption{Probability conservation for various timesteps, $dt$ (for packet with $k_0=2$).}
        \label{sinusoidal-vary-time-step}
    \end{center}
\end{figure}
\newpage
\subsubsection{Wave-packet with $k=0$}
We also conducted the lattice potential simulation using a packet with $k_0=0$, The simulation results, for $1000$ elements and $dt=0.1$, are shown in Fig.~\ref{lattice-potential-k-0}.  It can be seen that the particle probability at the center of the well diffuses outwards from the central potential peak.  However, as the packet starts with $k=0$ its energy is too low for it to tunnel through the lattice potential, so the probability of finding it near the well edges is very small even after $t=100$ timesteps.  We can also note that the packet is more likely to be found within the central troughs -- either side of the central peak of the lattice-potential, which is due to the packet being attracted to the states of lowest potential energy it is easily able to access.  In this way we can confine the wave-packet for a long period within a certain location.  
\begin{figure}
\begin{center}
\begin{tabular}{cc}
\includegraphics[width=0.5\textwidth]{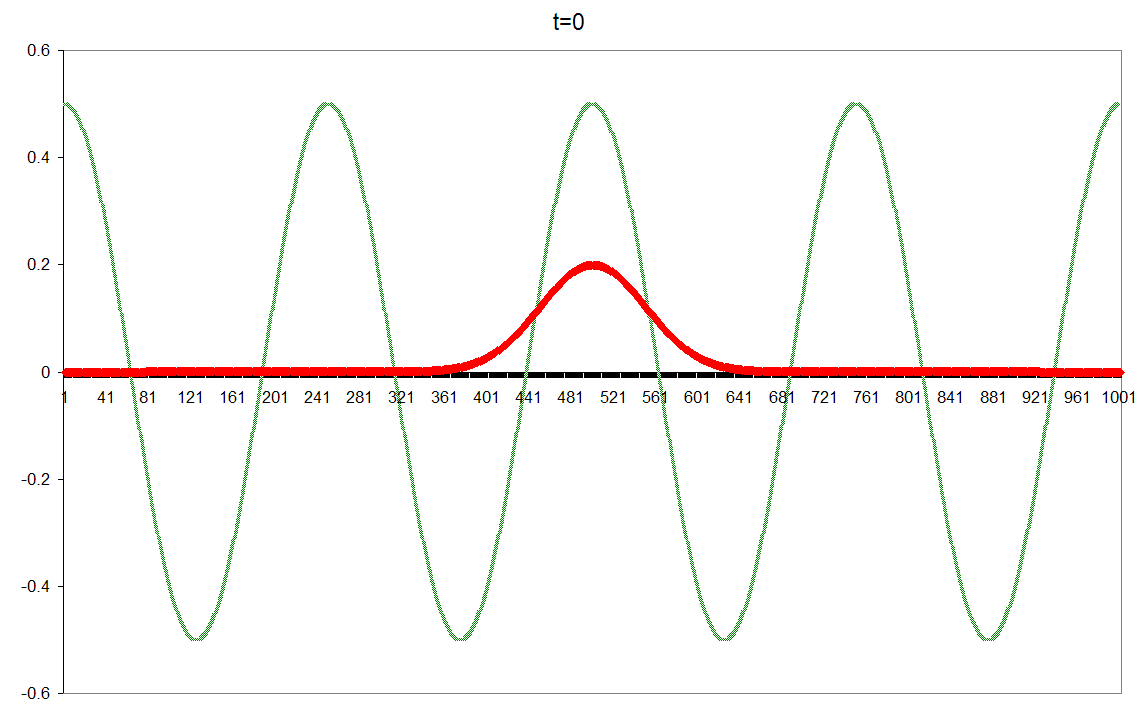}&
\includegraphics[width=0.5\textwidth]{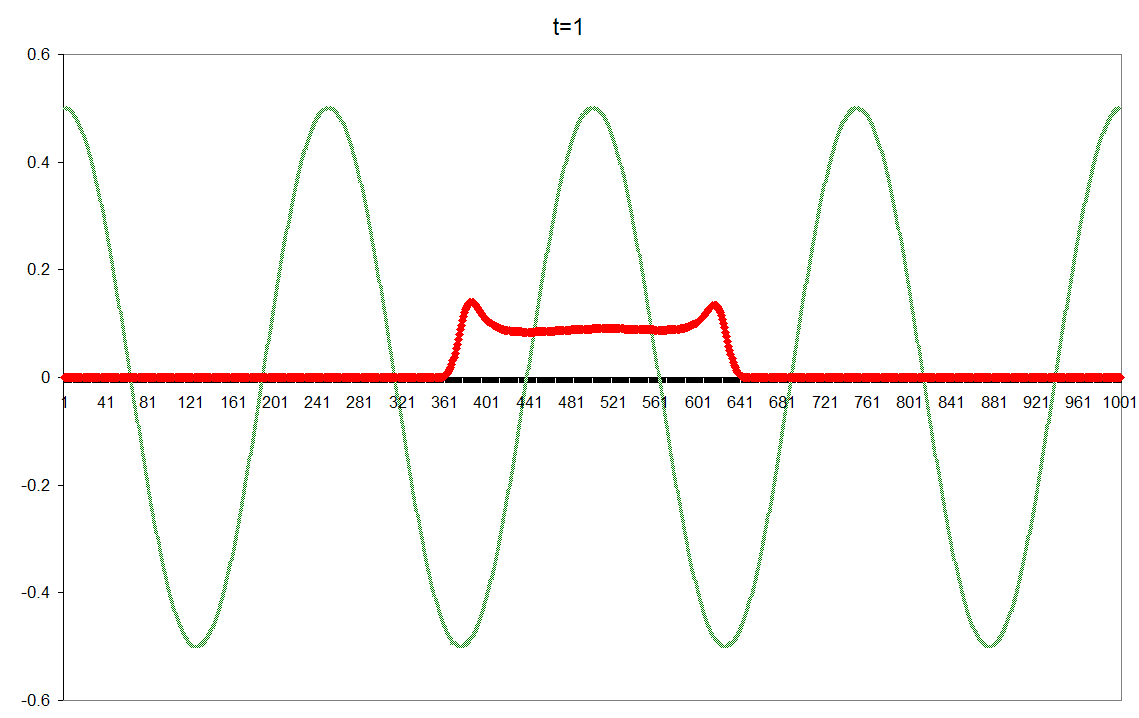}\\
\includegraphics[width=0.5\textwidth]{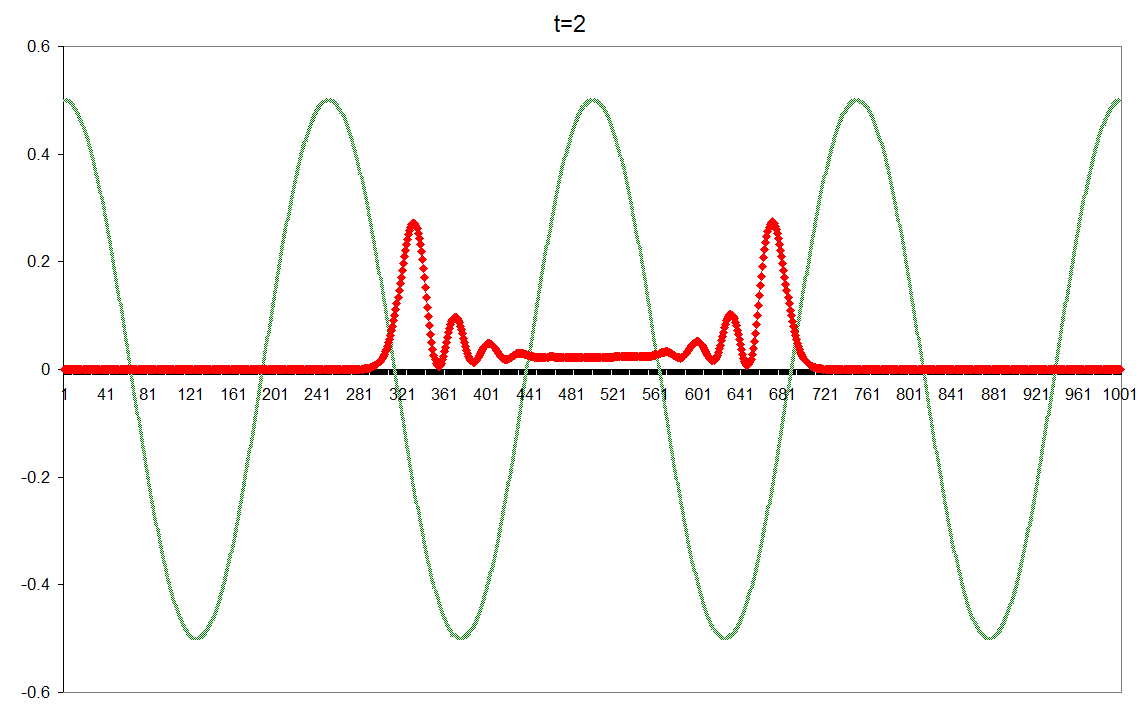}&
\includegraphics[width=0.5\textwidth]{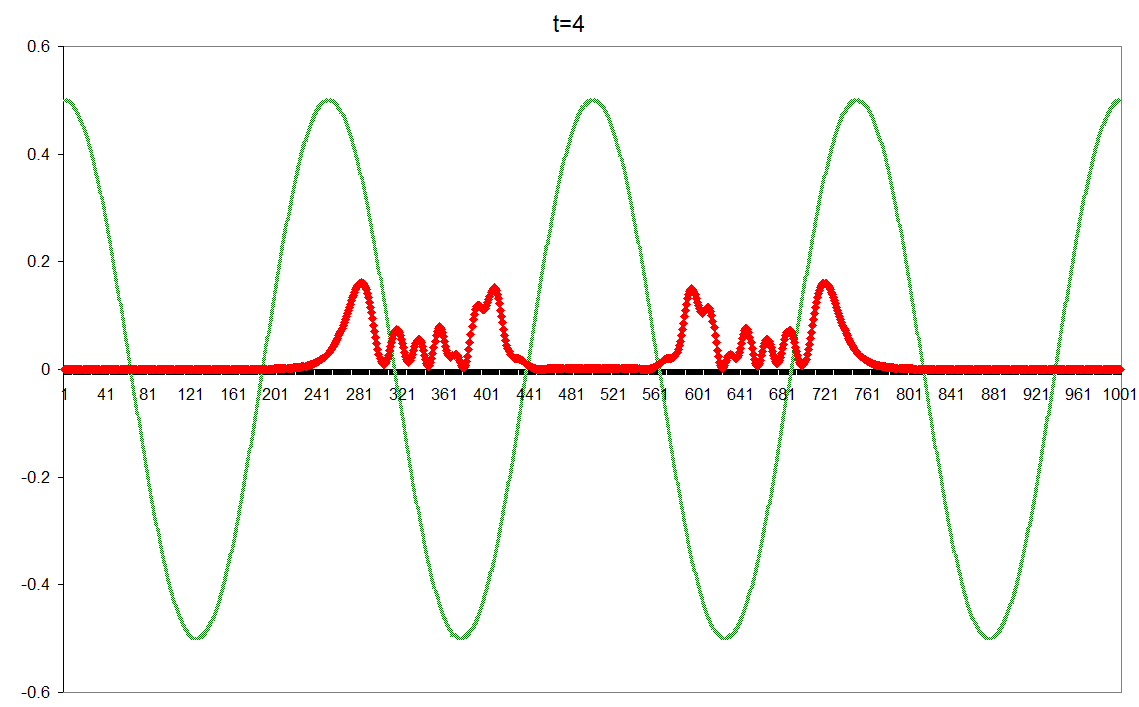}\\
\includegraphics[width=0.5\textwidth]{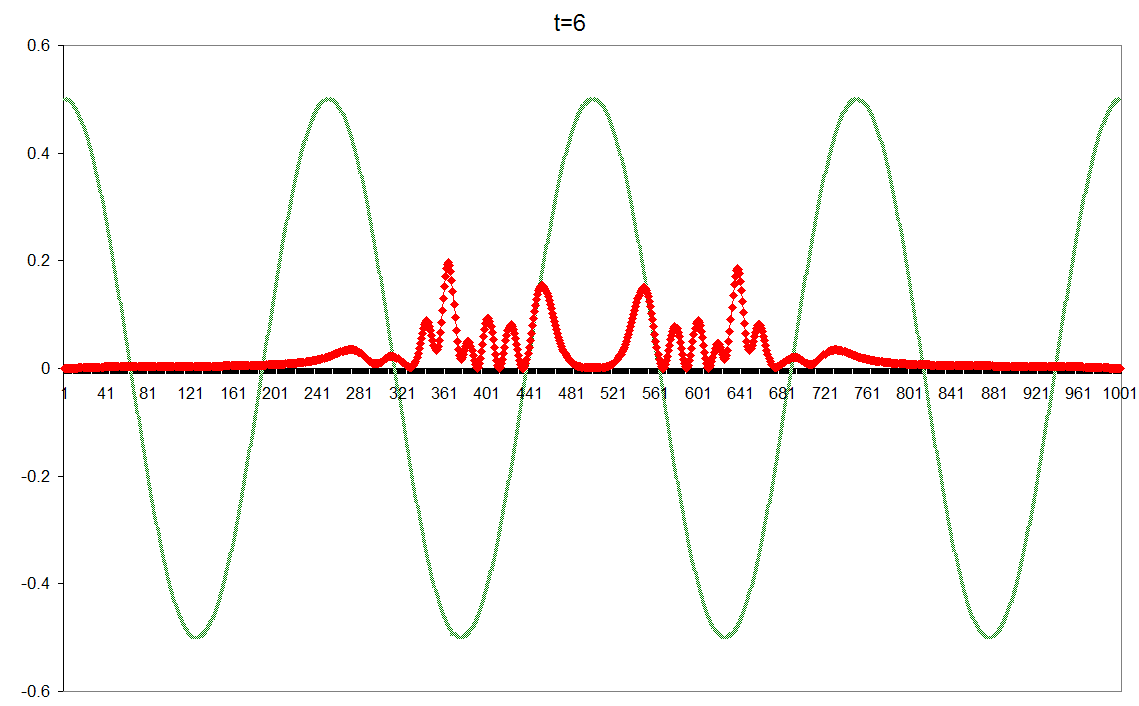}&
\includegraphics[width=0.5\textwidth]{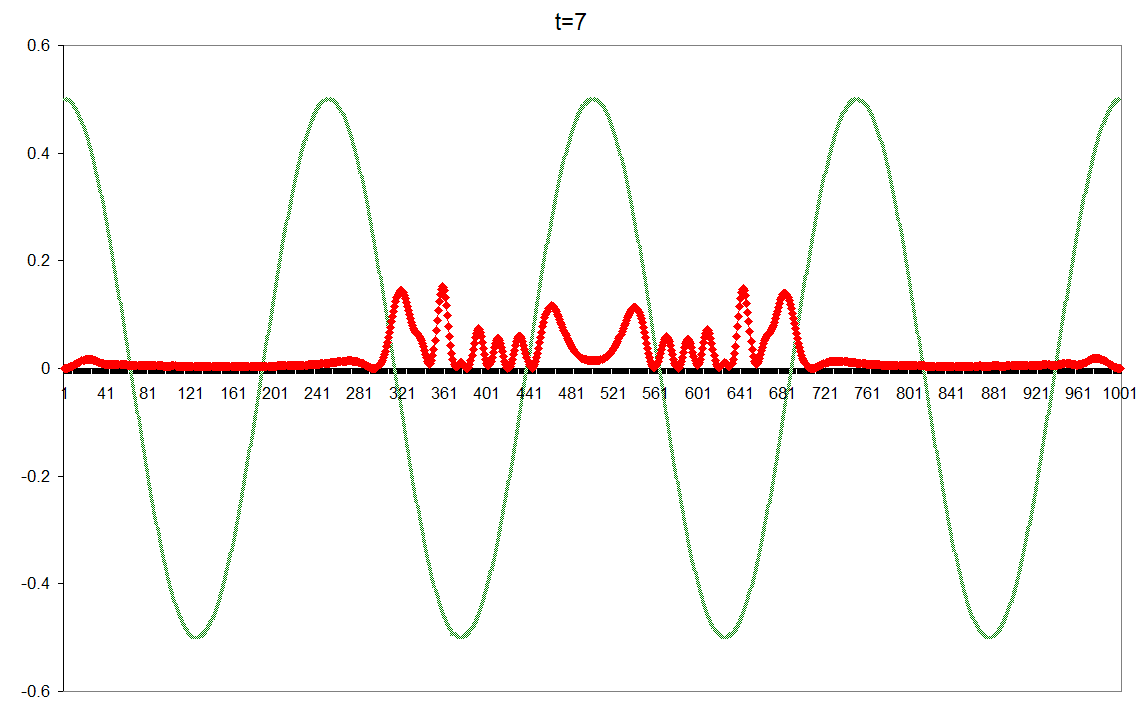}\\
\includegraphics[width=0.5\textwidth]{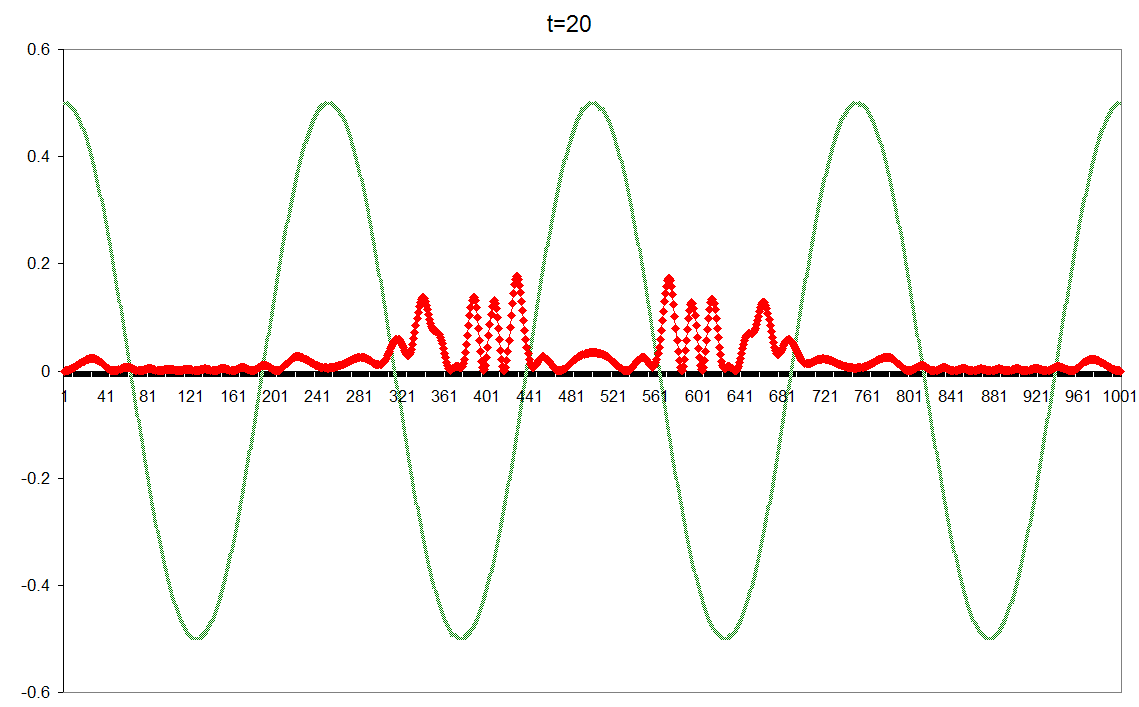}&
\includegraphics[width=0.5\textwidth]{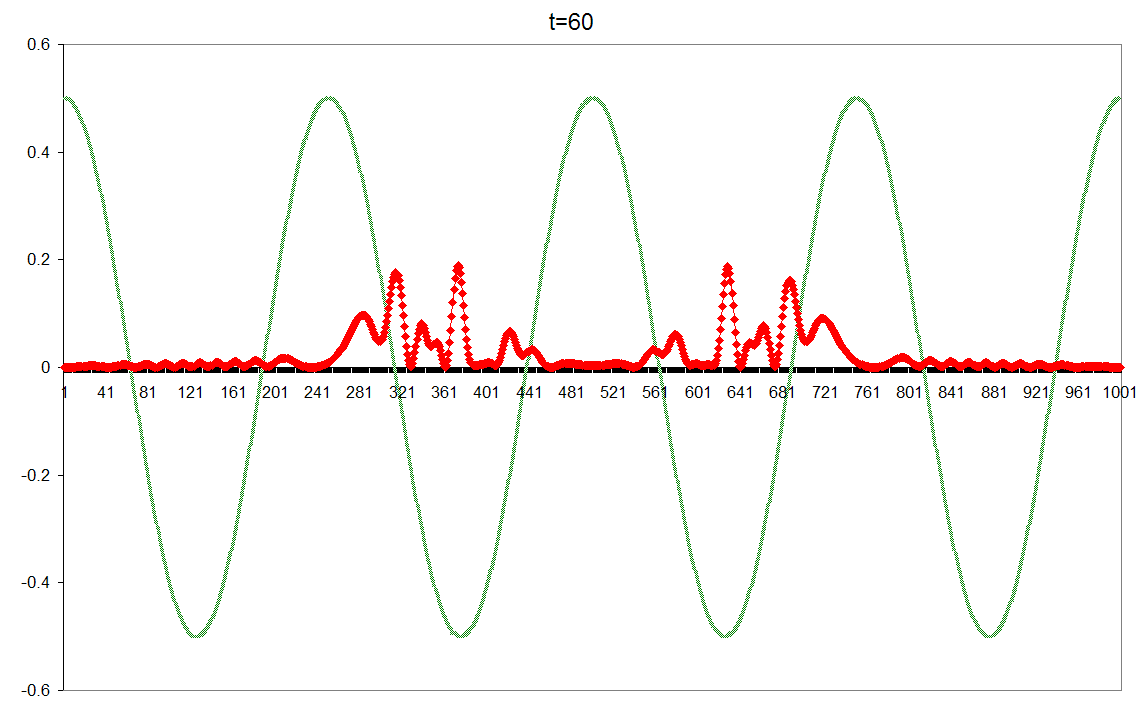}\\
\includegraphics[width=0.5\textwidth]{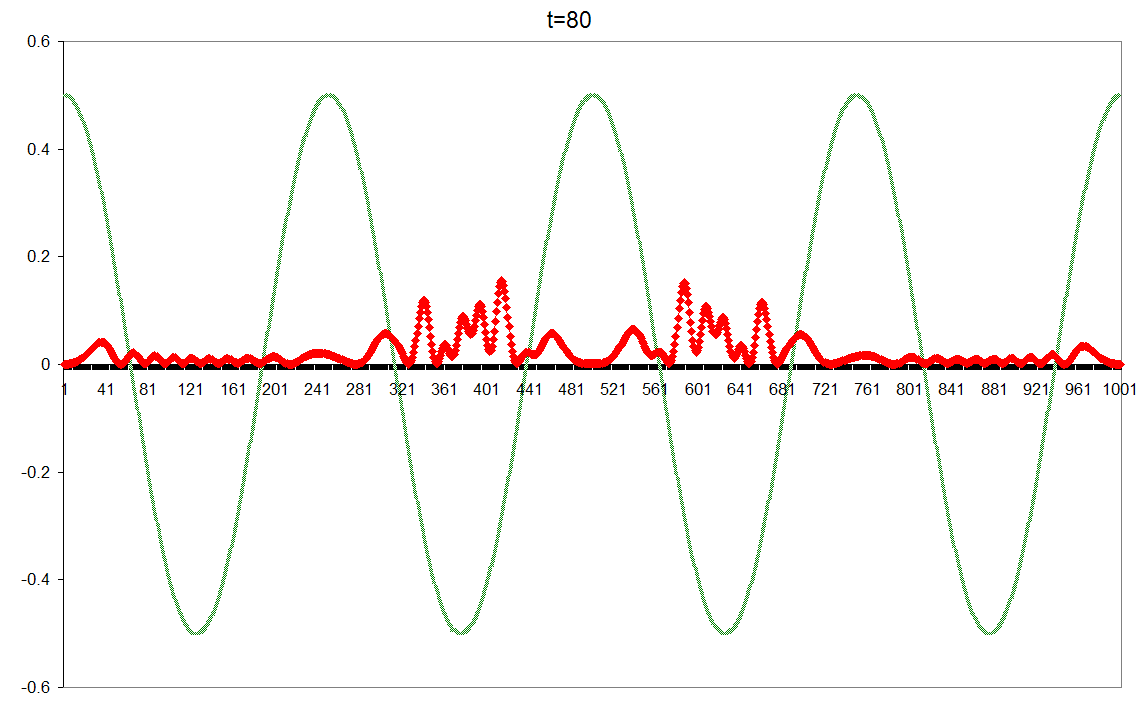}&
\includegraphics[width=0.5\textwidth]{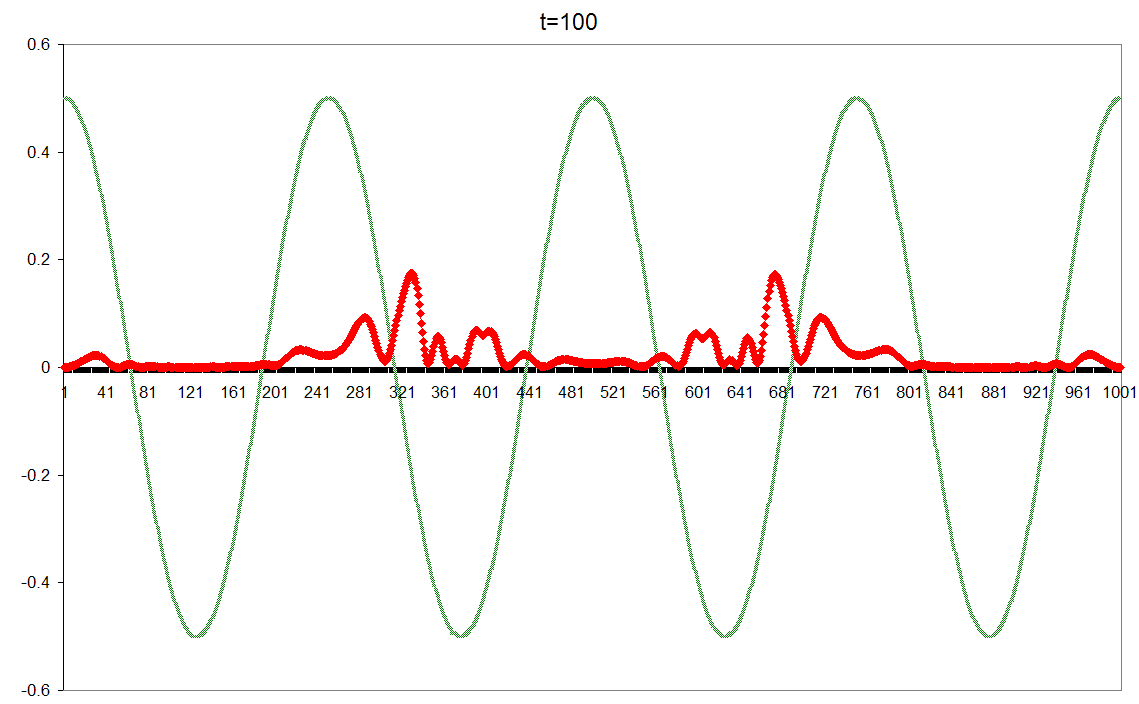}\\
\end{tabular}
\caption{Wave packet with $k_0=0$ is trapped within the central potential troughs ($Re^2+Im^2$ are shown).}
\label{lattice-potential-k-0}
\end{center}
\end{figure}
In Fig.~\ref{area-conservation-lattice-k=0} we have a plot of probability conservation.  This, again, shows that probability conservation falls slightly after $t=100$ timesteps, which as before can be controlled by increasing the number of spatial elements. 
\begin{figure}
    \begin{center}
     \includegraphics[width=0.7\textwidth]{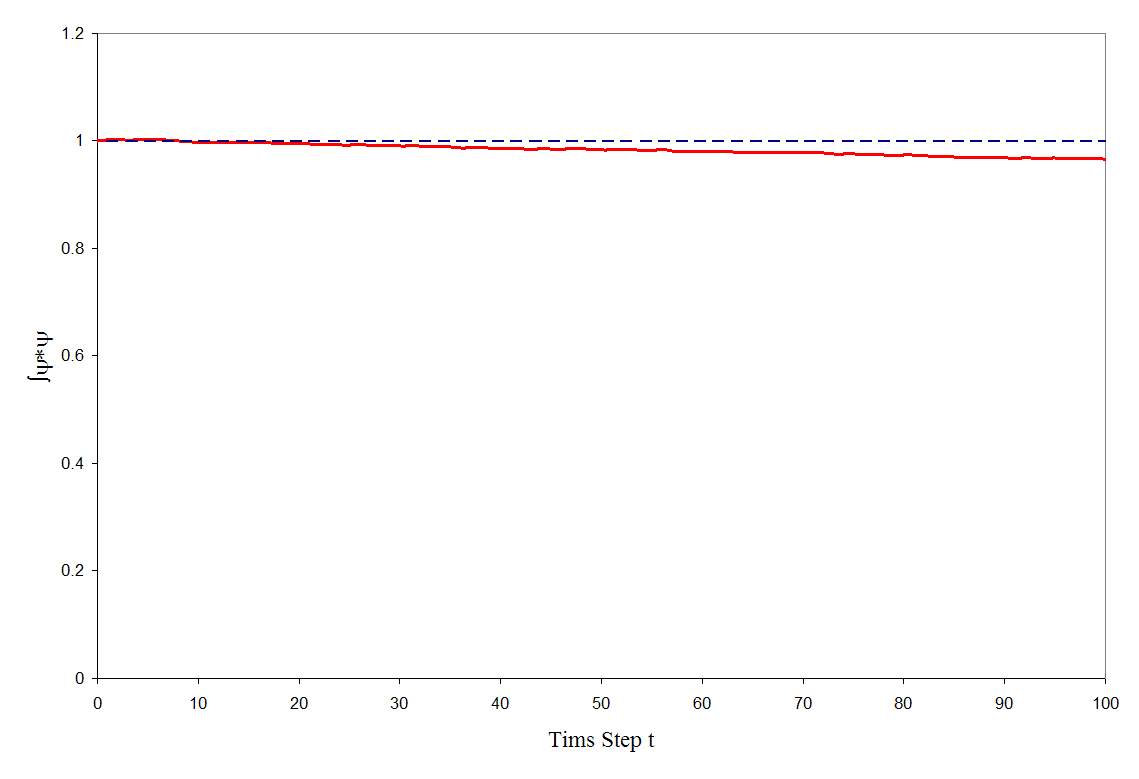}\\
        \caption{Conservation of area at each time step for packet with $k_0=0$ in lattice-potential (for 1000 elements).}
        \label{area-conservation-lattice-k=0}
    \end{center}
\end{figure}}

{\typeout{Summary}
\chapter{Summary}
\label{summary}

In Sec.~\ref{Crank-Nicolson and Finite Element Analysis} the \schro equation for a wave packet in an infinite well was numerically solved by the use of linear finite elements for spatial variation and Crank-Nicolson for time evolution.  Using a time step size $dt=0.5$ we found that the probability was perfectly conserved even using as few as $50$ elements, Fig~\ref{area-conservation-Infinite}.  However, it was seen that when a finite barrier is introduced within the well the conservation of probability is slightly "disturbed".  When the wave packet interacts with the barrier, and reflected and transmitted waves are introduced, the probability conservation shows fluctuations, Figs.~\ref{area-conservation-barrier-space} and \ref{area-conservation-barrier-time}.  These fluctuations are significantly reduced by increasing the number of elements, however reducing the timestep size below $dt=0.1$ has no effect.  Thus we can conclude that these fluctuations are a result of the greater amount of detail in the wave function as it interacts with the barrier, which then requires more elements to resolve its shape.  We also found that when a very high energy wave packet is used it can travel inside the well without noticing the barrier, hence no significant disturbances occur in its wave function, Fig.~\ref{area-conservation-barrier-k0}.  The same is true for a very low energy wave function, i.e. it is trapped between the infinite well wall and the finite barrier and so very little is transmitted, hence the low energy packet behaves as if it were trapped in a smaller infinite well.

In Sec.~\ref{Linear Continuous Space-Time Analysis} the \schro equation, for a wave packet in an infinite well, was solved by applying linear continuous finite elements in space and time.  In this method we obtain two non-unitary time stepping equations, Eqns.~(\ref{implicit}).  One is explicit, as it is weighted with the two basis function components at time step $t_n$, while the other is implicit, as it is weighted with the remaining basis function components at $t_{n+1}$.  As is expected the explicit case is unstable and it blows up, while the implicit case is heavily damped, Figs.\ref{space-time-non-cayley}.  However, we find that when we combine the explicit and implicit numerical equation we obtain the previous, Crank-Nicolson and finite element, time stepping equation.

In Sec.~\ref{Linear Discontinuous Space-Time Analysis} the \schro equation is solved by applying the linear discontinuous space-time finite element method.  Here we apply continuous finite elements in space and discontinuous finite elements in time.  We again obtain a numerical equation that has a non-unitary form, Eqn.~(\ref{full-equation-spacetime}).  We find that although the numerical solution obtained from this method suffers from damping it can be significantly controlled by decreasing the time step $dt$, and to a lesser extent by increasing the number of elements, Figs.~\ref{spacetime-prob-v-dt.png} -- \ref{spacetime-prob-v-elements}.  From Fig.~\ref{spacetime-area-conservation-elements-timestep} it can be seen that by using $100$ elements and a time step of $dt=0.01$ we have almost zero damping, however this comes at the cost of greater computation time.

In the case of introducing a finite potential barrier we find that the discontinuous space-time method slightly reduces the size of fluctuations, which like the Crank-Nicolson method can be controlled by the number of elements used, Fig.~\ref{space-time-barrier-prob-element}.  The main issue with the discontinuous space-time method is still the damping.  Although we can control this by reducing the timestep size, Fig.~\ref{space-time-barrier-prob-dt}, this however significantly increases the computational cost.  In Fig.~\ref{dis-crank-space-time-barrier-conservation} we have a direct comparison between the Crank-Nicolson and the discontinuous space-time methods for identical parameters: $dt=0.05$ and $250$ elements.  Even though the space-time method has less fluctuations it suffers from damping which causes a $1\%$ drop in probability after 100 timesteps. Another important factor is that the Crank-Nicolson method was five times faster in this test.

As the Crank-Nicolson method came out on top we went on to investigate the limits of this method: i.e. the extent of the fluctuations which occur when the wave function interacts with a barrier/change of potential in a "physical" example.  For this we modeled a particle in a quantum wire by using a wave packet in an infinite well containing a sinusoidal potential.  From this we find that as the wave packet constantly interacts with the sinusoidal potential a loss of probably occurs i.e. a type of "damping" effect, Fig.~\ref{sinusoidal-vary-elements}.  As for the simple barrier case, this is found to be an issue of resolving the wave function.  After many timesteps the packet goes through a lot of transmissions and reflections, which in turn produces a complicated wave function.   This complicated wave function then requires a greater number of elements to resolve the solution accurately.

After the comparison of the novel space-time finite element method and the Crank-Nicolson method for the numerical solution of the \schro equation we can conclude that in terms of accuracy, efficiency and convenience the Crank-Nicolson method is by far superior.
\\
\\

The logical extension to this work would be to go on and start to model a physical quantum device.  However, this will need to be done in discrete steps, as a foundation of numerical tools has to be developed.  The first step is to implement higher order spatial finite elements to model a wave function with greater accuracy, which could lead to a reduction of the fluctuations when the packet interacts with a change of potential.  The next step would be to model more than one particle, and even introduce the interaction with electromagnetic fields.  We will also need to implement different types of boundary conditions, for example periodic, open, absorbing, etc.  Finally this work would need to be extended to two dimensions.  Once these tools are developed we can apply them to model complicated quantum devices.}

{\typeout{Appendix A}
 \appendix{}
\chapter{Quantum Physics BackGround}
\label{background}

\section{Wave-Particle Duality}
Einstein showed that the momentum of a photon is:
\begin{equation}
p=\frac{h}{\lambda},
\label{einstein}
\end{equation}
where $p$ is the momentum, $h$ is the Planck's constant, and $\lambda$ is the wavelength.
This can be shown as follows.  The energy of a photon is given as $E=h\nu $ and its velocity as $c=\lambda \nu $, where $\nu$ is the photon frequency.  Combining these we obtain:
\begin{equation}
E=\frac{h c}{\lambda}
\end{equation}
Now using Einstein's mass energy relation from the theory of relativity, $ E = m c^2$, we have:
\begin{equation}
\lambda=\frac{h}{mc},
\label{lamb}
\end{equation} 
where $m$ is the relativistic mass of the photon\footnote{The rest mass of the photon is zero}.  Now noting that mass $m$ multiplied by velocity $c$ is momentum $p$, Eqn.~(\ref{lamb}) then becomes Eqn.~(\ref{einstein}).  Therefore, this way it can be shown that electromagnetic wave packets of evergy $h\nu$ have a particle like behaviour.  Since light can behave both as a wave, de Broglie reasoned in 1924 that matter also can exhibit this wave-particle duality. He further reasoned that matter would also obey Eqn.~(\ref{einstein}).  In 1927, Davisson and Germer observed diffraction patterns by bombarding metals with electrons, confirming de Broglie's proposition.

\section{Gaussian Wave Packets}
\label{gaussian-packet}
In terms of classical mechanics a particle's position and momentum can be measured exactly, however, in quantum mechanics this is not the case.  The Heisenberg uncertainty principle implies that the highest precision with which the position and momentum of a particle can be measured is given by the minimum uncertainty relation:
\begin{equation}
\Delta x \Delta P_x=\frac{\hbar}{2}
\end{equation}
The wave function that satisfies this uncertainty relation is a Gaussian wave packet,
\begin{equation}
\psi(x)=\left(\frac{1}{2\pi\sigma^2}\right)^{\frac{1}{4}}e^{ik_0 x}e^{-(x-x_0)^2/4\sigma^2}, 
\end{equation}
where $x_0$ denotes the center of the wave packet, $\hbar k_0$ is the mean momentum of the packet, and $\sigma$ is the uncertainty in the position of the particle ($\Delta x$).

In order to construct a numerical model of the wave packet we can separate the real and complex parts of the wave function:
\begin{equation}
\psi(x)=\left(\frac{1}{2\pi\sigma^2}\right)^{\frac{1}{4}}\left\{\cos(k_0 x) + i \sin(k_0 x) \right\} e^{-(x-x_0)^2/4\sigma^2}\nonumber
\end{equation}
This can now be written in the form of a vector equation,
\begin{equation}
\bar{\psi}(x)= 
\left[
\begin{array}{c}
	Re[\psi(x)] \\
	Im[\psi(x)] \\
\end{array}
\right]=
\left(\frac{1}{2\pi\sigma^2}\right)^{\frac{1}{4}}e^{-(x-x_0)^2/4\sigma^2}
\left[\begin{array}{c}
\cos(k_0 x)\\
\sin(k_0 x)\\
\end{array}
\right],
\end{equation}
This vector notation of the complex wavefunction is the most convenient form for numerical computation.  In Fig.~\ref{wavepacket} we have shown the real and imaginary plots of $\bar{\psi}$.
\begin{figure}
    \begin{center}
        \includegraphics[width=1\textwidth]{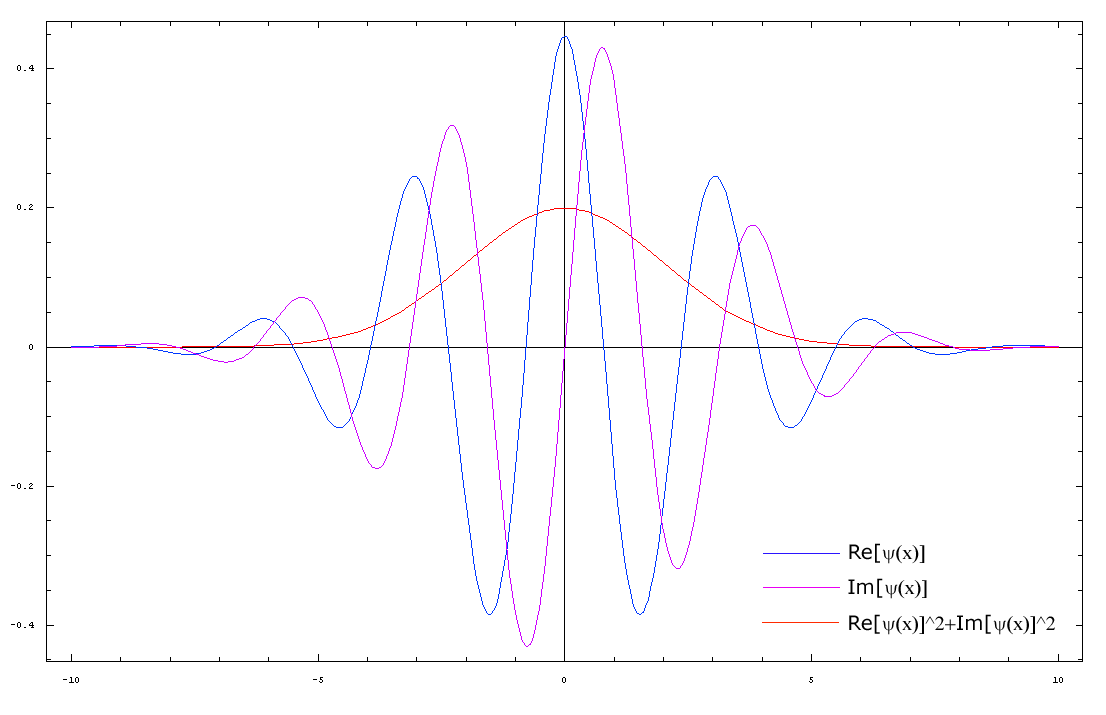}\\
        \caption{Real and Imaginary parts of $\bar{\psi}$, with $\sigma=2$, $k_0=2$, and $x_0=0$.}
        \label{wavepacket}
    \end{center}
\end{figure}
}

{\typeout{Appendix B}
 \chapter{C++ Code Samples}
\label{Code}
\section{Vector/Matrix Manipulation and the Vector Library}
In this work we created a sub-program using the C++ vector class library in order to handle the large matrices and nodal vectors of the finite-element equations.  One of the key advantages of vectors over arrays is their dynamic nature -- once a vector is defined we can adjust its size at any time in order to handle differing amounts of data.  Also writing a sub-program that constructs, handles and manipulates vectors would be a versatile tool which could be used for a variety of tasks.

The vector sub-program was created as a header file (vector.h), and the first thing we had to do was implement the vector library and define a vector and matrix type:

\tiny
\begin{verbatim}
#include <vector>
using namespace std;
typedef vector<double> vectors;
typedef vector<vectors> matrix;
typedef vectors::iterator vecit;
\end{verbatim}
\normalsize
\setlinespacing{1.66}
Once the vector and matrix types were defined we created a vector (vec) and Matrix (mat) classes, which incorporated member function (or methods) to take care of some important vector and matrix tasks: 

\tiny
\begin{verbatim}
	class vec:public vectors{					//Vector Class
	public:
		vec(){}     				            //General vector constructor
		vec(int i){	
			for(int k=0;k<i;k++){					//Constructor for vector with all zero components				
				push_back(0);
			}
		}
		void print();										//Function to output vector 
		void pop_front();								//Removes the first element of a vector (pop_back() exists in standard library)
		void pop_ends();								//Removes the first and last component (required for boundary conditions)
	};

	void vec::print(){
		int i=size();
		cout<<"\n\n";
		for(int k=0;k<i;k++){
			cout<<*(begin()+k)<<"     ";
		}
		cout<<"\n\n";
	}

	void vec::pop_front(){
		int i=size()-1;
		for(int k=0;k<i;k++){
			*(begin()+k)=*(begin()+k+1);
		}
		pop_back();
	}

	void vec::pop_ends(){
		pop_front();
		pop_back();
	}

	class mat:public matrix{					//Matrix Class
	public:
		mat(){}													//General matrix constructor
		mat(int i, int j){							//Constructor for matrix with all zero components	
			vec row;
			for(int k=0;k<j;k++){
				row.push_back(0);
			}
			for(int l=0;l<i;l++){
				push_back(row);
			}
		}
		void print();										//Function to output vector
		unsigned int rows();						//Outputs number of rows in matrix
		unsigned int cols();						//Outputs number of columns in matrix
		friend void pop_front();
		void pop_edges();								//Removes the outer components of matrix (required for boundary conditions)
	};

	unsigned int mat::rows(){
		return size();
	}

	unsigned int mat::cols(){
		return (*(begin())).size();
	}

	void mat::print(){
		int i=rows();
		int j=cols();
		vec row(j);
		cout<<"\n\n";
		for(int k=0;k<i;k++){
			for(int l=0;l<j;l++)cout<<(*(begin()+k))[l]<<"     ";
			cout<<"\n\n";
		}
	
	}

	void mat::pop_edges(){
		int i=size()-1;
		for(int k=0;k<i;k++){
			*(begin()+k)=*(begin()+k+1);
		}
		pop_back();
		pop_back();
		for(int j=0;j<i-1;j++){
			for(int k=0;k<i;k++){
			*((*(begin()+j)).begin()+k)=*((*(begin()+j)).begin()+k+1);
			}
			(*(begin()+j)).pop_back();
			(*(begin()+j)).pop_back();
		}
	}
\end{verbatim}
\normalsize
\setlinespacing{1.66}
In the vector.h library we also included some important functions and overloaded operators: 

\tiny
\begin{verbatim}
//Dot product of two vectors x and y of equal size and return a double value
        double operator*(vec &x, vec &y){
                if(x.size()!=y.size()){
                        cout<<x.size()<<"\n";
                        cout<<y.size()<<"\n";
                        cout<<"\n\n Vectors of unequal size"<<x.size()<<" and "<<y.size()<<"\n";
                        return 0;
                }
                else{
                        int max=x.size()-1;
                        double dot=0;
                        for(int i=0;i<=max;i++){
                                dot+=x[i]*y[i];
                        }
                        return dot;
                }
        }
        
//Matrix multiplication of two matrices a and b of sized ixj and jxk and return a matrix of size ixk
        mat operator*(mat &a, mat &b){
                if(a.cols()!=b.rows()){
                        cout<<"\n\n"<<"Matrix sizes not compatible: "<<a.rows()<<" x "<< a.cols()
                                <<" and "<<b.rows()<<" x "<<b.cols()<<"\n\n";
                        mat x(1,1);
                        return x;
                }
                else{
                        int maxi=a.rows();
                        int maxk=b.cols();
                        int maxj=a.cols();
                        vec row;
                        mat c;
                        c.clear();
                        double element;
                        for(int i=0;i<maxi;i++){
                                row.clear();
                                for(int k=0;k<maxk;k++){
                                        element=0;
                                        for(int j=0;j<maxj;j++)element+=(a[i][j]*b[j][k]);
                                        row.push_back(element);
                                }
                        c.push_back(row);
                        }
                return c;
                }
        }

//Vector addition of two vectors x and y of size i
        vec operator+(vec &x, vec &y){
                if(x.size()!=y.size()){
                        cout<<"\n\n"<<"Vectors sizes not compatible: "<<x.size()
                                <<" and "<<y.size();
                        vec z(1);
                        return z;
                }
                else{
                        vec z;
                        int maxi=x.size();
                        z.clear();
                        for(int i=0;i<maxi;i++){
                                z.push_back(x[i]+y[i]);
                                }
                        return z;
                        }
                }
        
//Matrix addition of two matrices a and b of size ixj
        mat operator+(mat &a, mat &b){
                if(a.cols()!=b.cols()&&a.rows()!=b.rows()){
                        cout<<"\n\n"<<"Matrix sizes not compatible: "<<a.rows()<<" x "<< a.cols()
                                <<" and "<<b.rows()<<" x "<<b.cols();
                        mat c(1,1);
                        return c;
                }
                else{
                        mat c;
                        vec row;
                        int maxi=a.rows();
                        int maxj=a.cols();
                        c.clear();
                        for(int i=0;i<maxi;i++){
                                row.clear();
                                for(int j=0;j<maxj;j++){
                                        row.push_back(a[i][j]+b[i][j]);
                                }
                                c.push_back(row);
                        }
                        return c;
                }
        }

//Vector subtraction of two vectors x and y of size i
        vec operator-(vec &x, vec &y){
                if(x.size()!=y.size()){
                        cout<<"\n\n"<<"Vectors sizes not compatible: "<<x.size()
                                <<" and "<<y.size();
                        vec z(1);
                        return z;
                }
                else{
                        vec z;
                        int maxi=x.size();
                        z.clear();
                        for(int i=0;i<maxi;i++){
                                z.push_back(x[i]-y[i]);
                                }
                        return z;
                        }
                }
                                
//Matrix subtraction of two matrices a and b of size ixj
        mat operator-(mat &a, mat &b){
                if(a.cols()!=b.cols()&&a.rows()!=b.rows()){
                        cout<<"\n\n"<<"Matrix sizes not compatible: "<<a.rows()<<" x "<< a.cols()
                                <<" and "<<b.rows()<<" x "<<b.cols();
                        mat c(1,1);
                        return c;
                }
                else{
                        mat c;
                        vec row;
                        int maxi=a.rows();
                        int maxj=a.cols();
                        c.clear();
                        for(int i=0;i<maxi;i++){
                                row.clear();
                                for(int j=0;j<maxj;j++){
                                        row.push_back(a[i][j]-b[i][j]);
                                }
                                c.push_back(row);
                        }
                        return c;
                }
        }

//Matrix transpose
        mat transpose(mat a){
                int i=a.rows();
                int j=a.cols();
                int k, l;
                mat t(j,i);
                for(k=0;k<i;k++){
                        for(l=0;l<j;l++){
                                t[l][k]=a[k][l];
                        }
                }
                return t;
        }

//Multiplication of a scalar with a matrix
        mat operator*(double &a, mat &b){
                        mat c;
                        vec row;
                        int maxi=b.rows();
                        int maxj=b.cols();
                        c.clear();
                        for(int i=0;i<maxi;i++){
                                row.clear();
                                for(int j=0;j<maxj;j++){
                                        row.push_back(a*b[i][j]);
                                }
                                c.push_back(row);
                        }
                        return c;
        }
        
//Matrix and vector multiplication
        vec operator*(mat &a,vec &v){
                int max=v.size();
                vec x(max);

                for(int i=0;i<max;i++){
                        for(int j=0;j<max;j++){
                                x[i]+=a[i][j]*v[j];
                        }
                }
                return x;
        }
\end{verbatim}

\section{LU Decomposition and Forward and Backward Substitution}
\normalsize
\setlinespacing{1.66}
In this work the global matrix equations,
\begin{equation}
\textbf{A}\cdot \textbf{X}=\left(\textbf{L}\cdot\textbf{U}\right)\cdot \textbf{X}=\textbf{B},\nonumber
\end{equation}
\normalsize
\setlinespacing{1.66}
are solved for the nodal values, $\textbf{X}$, by using simple LU decomposition as described in \cite{numerical}.  The biggest limitation with this simple method is that the entire $\textbf{A}$ matrix is used even though it is sparse diagonal and thus contains a large number of zeros.  A decision was made to use this method as it was easy to program, and the main aim of the work was to compare the results rather than actually build a fast program.  A library file, matrixop.h, was created to hold the decomposition and substitution routines so they could be easily installed into new simulations.  The LU decomposition routine takes a matrix $a[i][j]$ and replaces it by the LU decomposition of itself:

\tiny
\begin{verbatim}
//LU Decomposition
		void lu(mat &a, vec &indx, double &d){
			const double tiny=1.0e-20;
			int i,imax,j,k;
			double big,dum,sum,temp;
			int n=a.rows();
			vec vv(n);
			d=1.0;
			
			for(i=0;i<n;i++){
				big=0.0;
				for(j=0;j<n;j++){
					if((temp=fabs(a[i][j]))>big)big=temp;
				}
				if(big==0.0) cout<<"Singular matrix";
				vv[i]=1.0/big;
			}
			for(j=0;j<n;j++){
				for(i=0;i<j;i++){
					sum=a[i][j];
					for(k=0;k<i;k++) sum -= a[i][k]*a[k][j];
					a[i][j]=sum;
				}
				big=0.0;
				for(i=j;i<n;i++){
					sum=a[i][j];
					for(k=0;k<j;k++) sum -= a[i][k]*a[k][j];
					a[i][j]=sum;
					if((dum=vv[i]*fabs(sum))>=big){
						big=dum;
						imax=i;
					}
				}
				if(j!=imax){
					for(k=0;k<n;k++){
						dum=a[imax][k];
						a[imax][k]=a[j][k];
						a[j][k]=dum;
					}
					d=-d;
					vv[imax]=vv[j];
				}
				indx[j]=imax;
				if(a[j][j]==0.0) a[j][j]=tiny;
				if(j!= n-1){
					dum=1.0/(a[j][j]);
					for(i=j+1;i<n;i++) a[i][j] *= dum;
					}
				}
		}	
\end{verbatim}
\normalsize
\setlinespacing{1.66}
The routine for forward and backward substitution solves the equation $A\cdot X = B$. Here $a[i][j]$ is input as the LU decomposed version of the matrix $A$, and $b[j]$ is input as the right hand vector $B$.  Then on output the results for $X$ are returned in $B$:  

\tiny
\begin{verbatim}
//LU Substitution
		void luback(mat &a, vec &indx,vec &b){
				int i, ii=0,ip,j;
				double sum;
				int n=a.rows();
				
				for(i=0;i<n;i++){
					ip=indx[i];
					sum=b[ip];
					b[ip]=b[i];
					if(ii!=0){
						for(j=ii-1;j<i;j++) sum -= a[i][j]*b[j];
					}
					else if(sum != 0.0){
						ii = i + 1;
					}
					b[i]=sum;
				}
				for(i=n-1;i>=0;i--){
					sum=b[i];
					for(j=i+1;j<n;j++) sum -= a[i][j]*b[j];
					b[i]=sum/a[i][i];
				}
		}	
\end{verbatim}
\normalsize

\setlinespacing{1.66}}
\backmatter

 \setlinespacing{1.50}

\listoffigures
\end{document}